\documentclass[12pt]{article}
\pdfoutput=1
\usepackage{multirow}
\usepackage{calc}
\usepackage{amsmath,amssymb,amsthm,amscd}
\numberwithin{equation}{section}
\usepackage[bf]{caption}
\usepackage{longtable}
\usepackage{array}
\usepackage{enumerate}
\usepackage{float}
\usepackage{subfig}
\usepackage{multicol}
\usepackage{multirow}
\usepackage{epsfig}
\usepackage{graphicx}
\usepackage{pict2e}
\usepackage{color}
\usepackage{authblk}
\usepackage{cite}
\usepackage[all]{xy}
\usepackage{bbm}
\usepackage{hyperref}
\usepackage{multirow}
\usepackage[utf8]{inputenc}
\usepackage[T1]{fontenc}
\usepackage[english]{babel}
\usepackage{xspace}
\usepackage{pifont}
\usepackage{tikz}

\newcommand{\SU}[1]{\text{SU(#1)}\xspace}
\newcommand{\SO}[1]{\text{SO(#1)}\xspace}
\newcommand{\E}[1]{\ensuremath{\text{E}_{#1}}\xspace}

\def\cS{{\mathcal S}}

\newcommand{\one}[0]{\ensuremath{\mathbf{1} }\xspace}

\newcommand{\beq}{\begin{equation}}
\newcommand{\eeq}{\end{equation}}
\newcommand{\bea}{\begin{eqnarray}}
\newcommand{\eea}{\end{eqnarray}}
\newcommand{\Z}{\mathcal{Z}}
\newcommand{\Sa}{\mathcal{S}_7}
\newcommand{\Sb}{\mathcal{S}_9}
\newcommand{\CP}{\mathbb{P}^1}

\newcommand{\sou}{\text{SO(10)$\times$U(1)}}

\DeclareMathOperator{\tr}{\text{tr}}

\renewcommand{\tfrac}{\genfrac{}{}{}1}

\newcommand{\Sn}{\ensuremath{\mathcal{S}_9}}
\newcommand{\Ss}{\ensuremath{\mathcal{S}_7}}
\newcommand{\Kbi}{\ensuremath{K_B^{-1}}}

\usepackage[OT2,T1]{fontenc}
\DeclareSymbolFont{cyrletters}{OT2}{wncyr}{m}{n}		%Tate Shafarevich group 1
\DeclareMathSymbol{\Sha}{\mathalpha}{cyrletters}{"58}		%Tate Shafarevich group 2
\begin{document}
\baselineskip=15pt
\begin{titlepage}
\begin{flushright}
DESY 17-091\\
\end{flushright}

\begin{center}
\vspace*{ 2.0cm}
{\Large {\bf The Toric SO(10) F-Theory Landscape}}\\[12pt]
\vspace{-0.1cm}
\bigskip
\bigskip 
{
{{W.~Buchmuller}$^{\,\text{a}}$}, {{M.~Dierigl}$^{\,\text{a}}$}, {{P.-K.~Oehlmann}$^{\,\text{a}}$} and {{F.~Ruehle}$^{\,\text{b}}$}
\bigskip }\\[3pt]
\vspace{0.cm}
{\it 
 ${}^{\text{a}}$ Deutsches Elektronen-Synchrotron DESY,~Notkestr.~85,~22607~Hamburg,~Germany\\
${}^{\text{b}}$ Rudolf Peierls Centre for Theoretical Physics,~Oxford University,\\
1 Keble Road, Oxford, OX1 3NP, UK
}
\\[2.0cm]
\end{center}

\begin{abstract}
\noindent
Supergravity theories in more than four dimensions with grand unified gauge symmetries are an important intermediate step towards the ultraviolet completion of the Standard Model in string theory. Using toric geometry, we classify and analyze six-dimensional F-theory vacua with gauge group SO(10) taking into account Mordell-Weil U(1) and discrete gauge factors. We determine the full matter spectrum of these models, including charged and neutral SO(10) singlets.  Based solely on the geometry, we compute all matter multiplicities and confirm the
cancellation of gauge and gravitational anomalies independent of the base space. Particular emphasis is put on symmetry enhancements at the loci of matter fields and to the frequent appearance of superconformal points. They are linked to non-toric K\"ahler deformations which contribute to the counting of degrees of freedom. We compute the anomaly coefficients for these theories as well by using a base-independent blow-up procedure and superconformal matter transitions. Finally, we identify six-dimensional supergravity models which can yield the Standard Model with high-scale supersymmetry by further compactification to four dimensions in an Abelian flux background.
\end{abstract}

\end{titlepage}
\clearpage
\setcounter{footnote}{0}
\setcounter{tocdepth}{2}
\tableofcontents
\clearpage

% #################################
% #            Main part          #
% #################################
%%%%%%%%%%%%%%%%%%%%%%%%%%%%%%%%%%%%%%%%%%%%%%%%%%%%%%%%%%%%%%%%%%%%%%%%%%%%%%%%%%%%%%%%%%%%%%%%%%%%%%%%%%%%%%%%%%%%%%%%%%%%%%%%
\section{Introduction}
\label{sec:intro}

F-theory \cite{Vafa:1996xn,Morrison:1996na,Morrison:1996pp} provides a fascinating geometric picture of
fundamental forces and matter. Gauge interactions, matter fields and
their interactions are all encoded in the singularities of
elliptically fibered Calabi-Yau (CY)
manifolds: 
Codimension-one singularities determine non-Abelian
gauge groups, codimension-two singularities yield the representations
of matter fields \cite{Katz:1996xe,Bershadsky:1996nh} and codimension-three singularities their Yukawa
couplings \cite{Beasley:2008dc}.

Although the main ingredients of F-theory have been
known for two decades, significant progress towards realistic low
energy effective theories have only been made much later by searching
for F-theory vacua that incorporate higher-dimensional grand unified
theories (GUTs) \cite{Beasley:2008dc,Donagi:2008ca, Hayashi:2008ba,Beasley:2008kw,Donagi:2008kj}.
Making use of the geometry of del Pezzo surfaces and U(1)
fluxes of intersecting D7-branes, an interesting class of semi-realistic local
GUT models has been constructed (for reviews, see \cite{Heckman:2010bq,Weigand:2010wm,
Ibanez:2012zz}). These local models were then extended
to global GUT models which incorporate gravity, and
therefore the full geometry of the CY manifolds on which F-theory
is compactified \cite{Marsano:2009gv,Blumenhagen:2009yv,Grimm:2009yu}.

However, despite the remarkable progress in F-theory model building in recent
years, a number of important conceptual and phenomenological questions
still remain open. In fact, to the best of our knowledge, at present
there is no fully satisfactory F-theory GUT model, which would have to
account for symmetry breaking to the standard model gauge group, the 
matter content of the (supersymmetric) standard model, doublet-triplet
splitting, sufficiently suppressed proton decay,  supersymmetry
breaking and semi-realistic quark and lepton mass matrices. For
example, the
usually employed hypercharge flux breaking generically leads to massless
exotic states \cite{Dudas:2010zb,Palti:2012dd}, although this might be avoided in some models \cite{Krippendorf:2015kta} based on a classification of SU(5)$ \times $U(1) matter charges accomplished in \cite{Lawrie:2015hia}. Important progress has been made
towards implementing Wilson line breaking \cite{Marsano:2012yc} but
a realistic model still remains to be found. Note that interesting supersymmetric extensions of the Standard Model have also been obtained without the GUT paradigm \cite{Lin:2014qga, Cvetic:2015txa, Lin:2016vus, MayorgaPena:2017eda}. 

The present paper was motivated by a six-dimensional (6d) supergravity (SUGRA)
model with gauge group $\sou$ \cite{Buchmuller:2015jna}, based on previous work on
orbifold GUTs with Wilson lines\cite{Asaka:2001eh,Hall:2001xr,Asaka:2003iy,Braun:2006se}. 
U(1) gauge flux in the compact dimensions plays an important twofold
role. It generates a multiplicity
of quark-lepton generations, and it breaks supersymmetry \cite{Bachas:1995ik}.
Compactifying to four dimensions, this leads to
multiplets split with respect to either supersymmetry or the GUT symmetry,
a picture reminiscent of `split supersymmetry' \cite{ArkaniHamed:2004fb,Giudice:2004tc} or `spread
supersymmetry' \cite{Hall:2011jd}. From heterotic string compactifications it is
known that six-dimensional SUGRA theories can emerge as an
intermediate step in the compactification to four dimensions 
\cite{Kobayashi:2004ud,Forste:2004ie,Hebecker:2004ce,Buchmuller:2007qf}.
6d string vacua with GUT gauge symmetries have also been extensively studied in F-theory
(for reviews see, e.g. \cite{Taylor:2011wt,Johnson:2016qar}). 
It is then natural to ask whether models of this type  
can be embedded into F-theory or whether they belong to the
`swampland' \cite{Vafa:2005ui}. 
In this work we therefore classify a set of 6d global
F-theory models with gauge group SO(10) and  some additional
gauge factors of small rank.

Recently, F-theory was also used as an efficient tool to describe more exotic phenomena like tensionless strings in a consistent manner.
These sectors are realized in F-theory fibrations where the fiber develops a so-called (4,6,12) singularity in codimension-two in the base. In six dimensions these singularities have a physical interpretation 
in terms of superconformal field theories \cite{Seiberg:1996qx} related to
tensionless strings\cite{Seiberg:1996vs}. Following \cite{Anderson:2015cqy} we refer
to these singularities as superconformal points (SCP).
They can be viewed as
pairs of colliding singularities, which can be separated by blow-ups in the base. These blow-ups
yield new tensor multiplets and one obtains a CY manifold without SCPs
\cite{Bershadsky:1996nu}. The new tensors couple to the string
with a coupling strength given by the size of the blow-up cycle.
When the fiber is fully resolved in codimension one it
becomes non-flat over these codimension-two points \cite{Lawrie:2012gg,Braun:2013nqa,Borchmann:2013hta}.
This means that the dimension of the fiber jumps and contains higher
dimensional components. In  \cite{Lawrie:2012gg} it was then observed
that the presence of (4,6,12) singularities implies non-flatness of the resolved
fibration. These points are more likely to be present in theories with
large gauge groups, such as SO(10). 
Hence, as we are considering resolved SO(10) models, we indeed encounter many theories with superconformal points, present as non-flat fibers in codimension two.
In this analysis we also study these theories, i.e.\ matter
representations, anomaly cancellation and relations to other theories via tensionless string transitions \cite{Bershadsky:1996nu} in global F-theory models over an arbitrary base.

In the following we systematically study 6d F-theory with gauge
symmetry SO(10) and additional low-rank group factors. Our starting point is the
base-independent analysis of all toric hypersurface fibrations in \cite{Klevers:2014bqa},
together with the classification of all tops leading to non-Abelian
gauge groups in \cite{Bouchard:2003bu}. 
  Some global SO(10) models have
already been studied in \cite{Chen:2010ts,Tatar:2012tm,Baume:2015wia}. In our
work we extend this to all toric models with torus fibrations
described by a single hypersurface, which includes
fibrations with discrete groups, Mordell-Weil U(1)
factors \cite{Morrison:2012ei} and additional non-Abelian gauge groups
over arbitrary bases. \footnote{Independent of the construction a classification of $\mathbf{10}$-plet matter charges in SO(10)$\times$ U(1) theories was provided in \cite{Hayashi:2014kca}.}

In our analysis of the 6d F-theory vacua we determine the complete
massless matter spectra, including all SO(10) singlets and non-flat fibers points, i.e.\ SCPs, using geometric computations only. For all theories we confirm cancellation of gauge and gravitational anomalies and we provide the anomaly coefficients base-independently. 
We find a total number of 36 different models with additional U(1) symmetries up to rank three,
as well as $\mathbb{Z}_2$ and $\mathbb{Z}_3$ discrete symmetries that are of possible
phenomenological interest after further compactification to four
dimensions. In particular for the models with discrete gauge
symmetries, we compute singlet multiplicities and discrete charges of
SO(10) matter multiplets. Furthermore, we discuss
the connection of fibrations with different SO(10) tops via conifold transitions in the generic fiber.

In total around 80\% of the models contain SCPs, for which the fiber becomes non-flat.
For those theories we carry out a base-independent blow-up procedure and provide the
anomaly coefficients as well. In the computation of the full spectrum, we find an important new contribution
of non-toric K\"ahler deformations coming from the non-flat fiber points that correspond to 6d tensors and have to be taken into account for the correct counting of neutral singlets. Furthermore, we show that theories with non-flat fibers are 
connected to tops that have points  in the interior of a face which can often be reached via tensionless string transition from another top with no superconformal points. In these transitions we again show the appearance of non-toric K\"ahler deformations in the fiber in the computation of base-independent Euler numbers. In particular we discuss these transitions for the first time in global theories with additional (discrete) Abelian gauge factors over an arbitrary base.

Our analysis is strongly based on previous studies of SU(5) vacua
\cite{Krause:2011xj,Borchmann:2013jwa,Braun:2013yti,Braun:2013nqa,Cvetic:2013uta,Krippendorf:2014xba}.  
In Section~\ref{sec:6dvac} we describe the various steps of the calculation in detail for
one example, a torus given by the polygon $F_3$
\cite{Grassi:2012qw}, which allows for a U(1) factor.
Fibering the ambient space $X_{F_3}$ over 
 $\CP$, we construct a $K3$ manifold which can be tuned to have an SO(10)
singularity according to the Kodaira classification \cite{kodaira}. Resolving
this singularity with an SO(10) top produces five $\CP$s which, together
with the torus, show the intersection pattern of the extended SO(10)
Dynkin diagram. Particular emphasis is given to the symmetry
enhancements at codimension-two and codimension-three singularities,
which yield the loci of matter fields and Yukawa couplings.\footnote{Note that Yukawa couplings only occur in codimension 3 and hence do not appear in our 6d models; however, since our analysis is base-independent, we can classify these points with our methods as well.} We find
the standard pattern of extended Dynkin diagrams but also some 
non-Kodaira fibers generically present where matter curves
self-intersect \cite{Esole:2011sm, Marsano:2011hv}. 
To complete the analysis, the multiplicities of matter fields
are computed for the Hirzebruch base $\mathbb{F}_0 = \CP\times\CP$.

In Section~\ref{sec:bimultis} a base-independent analysis is performed and the matter
multiplicities are evaluated as intersection numbers on the base. A
challenging problem is the computation of the SO(10) singlet
spectra. We obtain the multiplicities of all charged and neutral singlets.
This is achieved by
unhiggsing the gauge group SO(10)$\times$U(1) to SO(10)$\times$U(1)$^2$ as an
intermediate step where the computations are feasible.

Section~\ref{sec:clasvac} contains the main result of the paper, the classification and analysis of
all 6d toric SO(10) vacua. We briefly review the structure of the fibers
describing a torus in different ambient spaces \cite{Grassi:2012qw}
and the SO(10) tops that can be added to the various polygons
\cite{Bouchard:2003bu}. 
In total there are 36 different models.
Using the techniques that were exemplified in Section~\ref{sec:6dvac} and \ref{sec:bimultis}, we then calculate all matter
representations, compute their multiplicities and confirm cancellation of all anomalies in a base-independent manner for each model.
 A complete list of these data is given in Appendix~\ref{App:classification}. An interesting outcome of our
classification is the frequent appearance of SCPs. We identify these points as an additional source of (1,1)-forms in the fiber, which is important for counting all neutral degrees of freedom. After the separation of the codimension-two colliding singularities via a blow-up in the base we confirm cancellation of all anomalies in these theories as well. Moreover, we discuss the connection of these theories 
via higgsings and tensionless string transitions.

Section~\ref{sec:phenomodel} is devoted to 6d supergravity models with gauge group $\sou$
which are phenomenologically promising. These models contain one
charged ${\bf 16}$-plet that yields the quark-lepton generations as
zero modes in an Abelian flux compactification, and additional uncharged ${\bf 16}$-plets
needed for $B-L$ breaking. In addition, these models have several neutral ${\bf 10}$-plets
which, via doublet-triplet splitting, yield two Higgs doublet
superfields in the 4d effective theory. We first consider the model in
\cite{Buchmuller:2015jna} and show that, after adding charged and
neutral SO(10) singlets, all anomalies can be canceled. This model,
however, is not contained in our classification and therefore belongs
to the `toric swampland'. On the other hand, variants of this model
with charged ${\bf 10}$-plets, which have additional vector-like matter,
can be obtained as 6d F-theory vacua.
 
A summary of our results and a brief discussion of unsolved
challenging problems are presented in Section~\ref{sec:summary}. Appendix~\ref{a:d6dvac} gives more
details required for a full understanding of the example discussed in Section~\ref{sec:6dvac} and \ref{sec:bimultis}. In Appendix~\ref{sec:WeierstrassForm} polynomials and divisor classes are given for the
fibers $F_2$ and $F_4$, as well as the expressions for the functions
$f$ and $g$ needed to obtain the elliptic curves in Weierstrass form. 
Appendix~\ref{App:classification} contains the data of the 36 models contained in our
analysis. Finally, in Appendix~\ref{ScanF3Top4} a list of phenomenologically viable
models is given.

%%%%%%%%%%%%%%%%%%%%%%%%%%%%%%%%%%%%%%%%%%%%%%%%%%%%%%%%%%%%%%%%%%%%%%%%%
\section{A 6d vacuum with gauge group \texorpdfstring{$\sou$}{SO(10)xU(1)}}
\label{sec:6dvac}

In this section we discuss the explicit geometric construction of a specific global F-theory model with gauge group SO(10)$\times$U(1). Moreover, we evaluate its matter spectrum, Yukawa couplings and anomaly coefficients in full detail.

%%%%%%%%%%%%%%%%%%
\subsection{Torus with non-trivial Mordell-Weil group}
\label{T6dvac}

Our starting point is an elliptic curve $\mathcal{E}$ with a Mordell-Weil
group of rank one, which yields a U(1) gauge group when fibered over an
appropriate base space. This is the case for the torus contained in the
two-dimensional toric ambient space $dP_1$ which can be parametrized by
four homogeneous coordinates $[u:v:w:e_1]$, with two independent 
$\mathbb{C}^*=(\mathbb{C}-\{0\})$ scale transformations modded
out.

\begin{figure}
\begin{center}
\begin{picture}(0,120)
\put(-80,0){\includegraphics[scale=0.8]{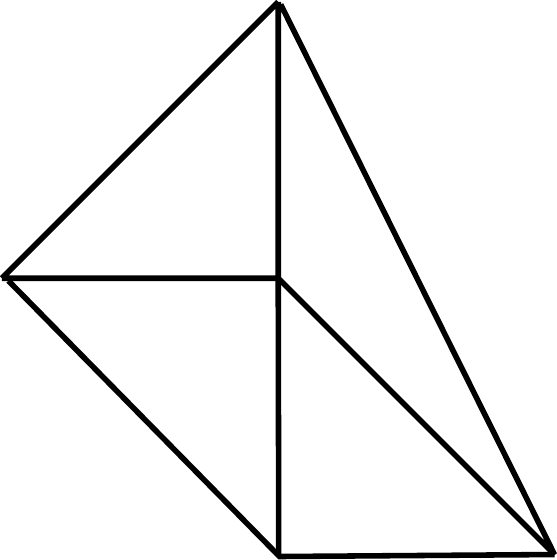}}
\put(-95,60){\Large $v$}
\put(-20,135){\Large $w$}
\put(-20,-12){\Large $e_1$}
\put(50,-12){\Large $u$}
\end{picture}
\end{center}
\caption{\label{F3Toric} The polygon $F_3$ describing a torus in the
  ambient space $dP_1$.}
\end{figure}
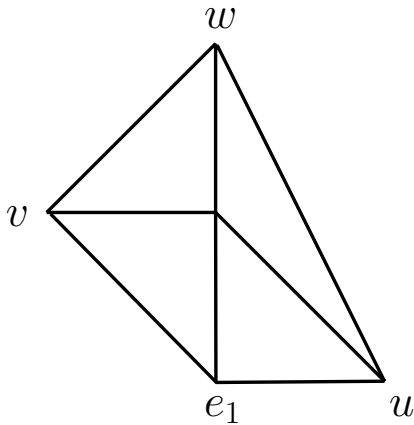

In order to obtain the elliptic curve $\mathcal{E}$ inside the ambient space
$dP_1$, one first chooses the corresponding toric ambient space  polygon \cite{Grassi:2012qw}, $F_{3}$, where each
homogeneous coordinate is associated with a two-dimensional vector
(see Figure~\ref{F3Toric}, Table~\ref{tab:1foldexample}).

One then constructs the dual
polygon $F_3^*$ which, together with $F_3$, defines the 
the polynomial $p_{F_3}$ (see Appendix~\ref{a:d6dvac}), 
\begin{align}\label{p1}
p_{F_3} = &\ s_1 u^3e_1^2 + s_2 u^2 ve_1^2  + s_3 u v^2 e_1^2  +  s_4 v^3e_1^2 +
  s_5 u^2 w e_1 \nonumber\\
&+  s_6 u v w e_1 +  s_7 v^2 w e_1 + s_8 u w^2 + s_9 v w^2 \,.
\end{align}
 This polynomial defines a torus in the toric ambient space,
\begin{align}\label{Y1}
\mathcal{E} = \{p_{F_3}=0\} \,,
\end{align}
with the coefficients $s_1, \ldots, s_9$ being generic complex
numbers. The vanishing of the homogeneous coordinates $u$, $v$, $w$ and $e_1$
defines four divisors
\begin{align}\label{torusdivisors}
D_u\,\,, D_v\,\,, D_w\,\,, D_{e_1}\,,
\end{align}
where $D_{x_i} = \{x_i = 0\}$.
 
\begin{table}
\begin{center}$
\begin{array}{|c|c|c|cc|}\hline
\text{coordinates} & \text{vertices} & \text{divisor classes} & [v] & [e_1] \\ 
\hline \hline
u & (1, -1) & [v] & 0 & 1 \\
v & (-1, 0) & [v] & 0 & 1  \\
w & (0, 1) & [ve_1]  & 1 & 0 \\
e_1 & (0, -1) & [e_1] & 1 & -1 \\ \hline
\end{array}$
\end{center}
\caption{\label{tab:1foldexample}Coordinates, vertices, divisor classes and
  intersection numbers (charges of $\mathbb{C}^*$-actions) for the two-dimensional  toric variety $dP_1$.}
 \end{table} 
Since the ambient space $dP_1$ is two-dimensional there are two linear
dependencies,
\begin{align}\label{lindepen1}
\sum_i Q_{ai}D_{x_i} \sim 0\,,\quad a=1,2\,,
\end{align}
where 
$Q_{ai}$ are the charges of the two $\mathbb{C}^*$-actions.
In Table~\ref{tab:1foldexample} we have also listed the intersection
numbers of the two divisor classes\footnote{We
  indicate divisor classes by brackets $[\cdot]$.} $[v]$ and $[e_1]$ with all
divisors. One easily verifies that these intersection numbers play the
role of the charges of the two $\mathbb{C}^*$-actions,
$x_i \rightarrow \lambda^{Q_{ai}} x_i$ with $\lambda \in
\mathbb{C}^*$, under which the polynomial $p_{F_3}$ transforms
homogeneously. 
Using the linear dependencies the two remaining divisor
classes can be expressed in terms of the two independent ones,
\begin{align}
[u] = [v]\,, \quad [w] = [v] + [e_1] = [ve_1]\,.
\end{align}

\begin{figure}
\begin{center}
\includegraphics[scale=0.7]{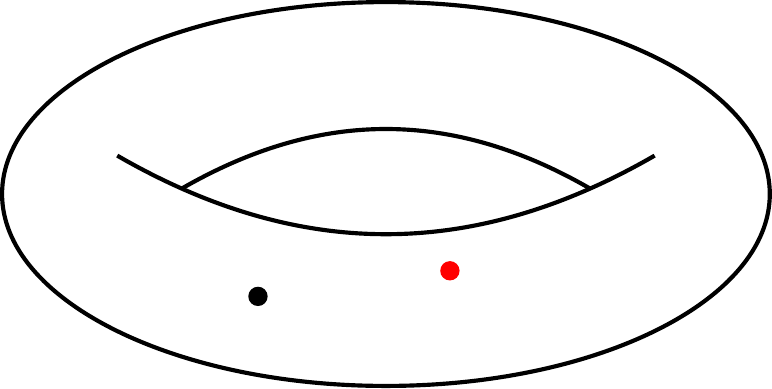}
\end{center}
\caption{\label{toruswithpoints} The elliptic curve $\mathcal{E}$ with
  two points, $\hat{s}_0$ and $\hat{s}_1$ (see text).}
\end{figure}

The torus $\mathcal{E}$ has a `zero-point' $P_0$  which is obtained as intersection\footnote{Such a torus is called elliptic
  curve.} with the divisor $D_{e_1}$. On this divisor $w$ can be set to one by a
$\mathbb{C}^*$-action (see Appendix~\ref{a:d6dvac}), which yields for the
coordinates of $P_0$
\begin{align}\label{s_0}
\hat{s}_0 = D_{e_1} \cap\mathcal{E}:\ [s_9, -s_8,1,0]\,.
\end{align}
There exists a second rational point $P_1$ on the torus (see Figure~\ref{toruswithpoints}), which can be
obtained from the a tangent $t_P$ at $P_0$ along the torus \cite{Klevers:2014bqa},
\begin{align}\label{s_1}
\hat{s}_1 = \{t_P = 0\}\cap \mathcal{E}\,, \quad t_P = s_8 u + s_9 v \,.
\end{align}
The Mordell-Weil group then defines an addition, $P_0 \rightarrow P_0 +
P_1 \rightarrow P_0 + 2P_1 \rightarrow \ldots$.
For a fibration of the torus over some base the points $\hat{s}_0 $
and $\hat{s}_1$ become 
functions of the base coordinates and define a divisor which corresponds
to the generator of a continuous U(1) symmetry.  The divisor
is obtained from $[\hat{s}_0]$ and $[\hat{s}_1]$ and can be
written as \cite{Morrison:2012ei}
\begin{align}\label{U1}
\sigma_1 \equiv [\hat{s}_1] - [\hat{s}_0] = [v] - [e_1]\,,
\end{align}
where we have represented the tangent $t_P$ by the divisor class $[v]$.

%%%%%%%%%%%%%%%%%%
\subsection[\texorpdfstring{$K3$}{K3} manifold]{\texorpdfstring{$\boldsymbol{K3}$}{K3} manifold}
\label{K3}

Now we describe the construction of a $K3$ manifold with an SO(10) singularity leading to the desired gauge group of the model as well as its resolution to a smooth CY twofold.

%%%%%%%%%
\subsubsection*{Singular limit}\label{sing6dvac}

We fiber the elliptic curve $\mathcal{E}$ over some
base space, in the simplest case a $\mathbb{P}^1$. This is achieved by
extending the two-dimensional ambient space to a three-dimensional
ambient space for which we choose a polyhedron with vertices
given in Table~\ref{tab:2foldsingular}. The homogeneous coordinates of
the base $\mathbb{P}^1$ are $z_0$ and $z_1$. 

Like for the ambient space $dP_1$, one can determine the dual
polytope and the polynomial $p_{\hat{Y}_2}$ that defines a CY twofold 
$\hat{Y}_2$ via $\hat{Y}_2 =
\{p_{\hat{Y}_2} = 0 \}$. The polynomial  $p_{\hat{Y}_2}$ turns out to consist of 28 monomials
corresponding to a smooth $K3$ manifold (see Appendix~\ref{a:d6dvac}). Tuning ten
of them to zero yields the following factorization (see \eqref{p1}):
\begin{equation}\label{p2a}
\begin{split}
s_1 &= d_1 z_0\,, \;  s_2 = d_2 z_0^2\,,  \;  s_3 = d_3 z_0^2\,, \;  s_4 = d_4
z_0^3\,, \;  s_5 = d_5\,, \\ 
 s_6 &=  d_6 z_0\,,  \;  s_7 = d_7 z_0\,, \;  s_8 =  d_8\phantom{z_0^2}\,, \;  s_9 = d_9\,.
\end{split}
\end{equation}
The coefficients $d_1,\ldots d_9$ are homogeneous functions of
the base coordinates, $d_i = d_i(z_0,z_1)$ (see Appendix
\ref{a:d6dvac})\footnote{More precisely, they are sections over the base $\mathbb{P}^1$.}. 

For the three-dimensional toric variety $dP_1$ fibered over $\CP$ one can define curves as
intersections of divisors with the CY twofold $\hat{Y}_2$. From the two
independent divisor classes $[v]$ and $[e_1]$, and the base
coordinate $[z_0]$, one obtains the curves\footnote{For simplicity, we denote divisors with the same name as the respective coordinates in equations for intersections.} $\mathcal{C}_v$,
$\mathcal{C}_{e_1}$ and $\mathcal{C}_b$:
\begin{align}\label{curveshatY2}
\begin{array}{l|l c}
\,\text{curves} &  \quad\quad\text{intersections} & \text{coordinate patch} \\ \hline
\quad\mathcal{C}_v & v \cap (d_1 e_1^2 z_0 + d_5 e_1 w +d_8 w^2) & u=1 \\
\quad\mathcal{C}_{e_1} & e_1 \cap ( d_8 u + d_9 v) & w=1\\
\quad\mathcal{C}_b & z_0 \cap( d_5 e_1 u^2 + d_8 u + d_9 v) & z_1=1 \\ 
\hline
\end{array}
\end{align}
In Table~\ref{tab:2foldsingular} we have listed the intersection
numbers of these curves with fiber and base divisors\footnote{For the calculations we
  used the coordinate patches listed in \eqref{curveshatY2}, which
  requires the choice of a Stanley-Reissner ideal (SRI), see Appendix~\ref{a:d6dvac}.},
\begin{align}
[v]\cdot \mathcal{C}_v\,,\quad [v]\cdot \mathcal{C}_{e_1}\,,\quad [v]\cdot \mathcal{C}_b\,, \ldots
\end{align}
Note that the self-intersections of the divisors on the smooth CY twofold are
$[v]\cdot \mathcal{C}_v = -2$ and $[e_1]\cdot \mathcal{C}_{e_1} = -2$. Hence,
these curves correspond to two $\mathbb{P}^{1}$s. 
\begin{table}
\begin{center}$
\begin{array}{|c|c|c|ccc|}\hline
\text{coordinates} & \text{vertices} & \text{divisor classes} & \mathcal{C}_v & \mathcal{C}_{e_1} & \mathcal{C}_b\\ 
\hline \hline
u & (1, -1,0) & [vz_0] & 0 & 2 & 2 \\
v & (-1, 0,0) & [v]  & -2 & 1  & 2 \\
w & (0, 1,0) & [ve_1z_0] & 1 & 0 & 3 \\
e_1 & (0, -1,0) & [e_1] & 1 & -2 & 1 \\ 
z_0 & (0,0,1) & [z_0] & 2 & 1 & 0 \\
z_1 & (-1,0,-1) & [z_0] & 2 & 1 & 0 \\
\hline
\end{array}$
\end{center}
\caption{\label{tab:2foldsingular}Coordinates, vertices, divisor classes and
intersection numbers (charges of $\mathbb{C}^*$-actions) for the three-dimensional
toric variety $dP_1$ fibered over $\CP$.}
 \end{table}

For the chosen values of the parameters of $p_{\hat{Y}_2}$,
and therefore the sections $s_i$,  the CY twofold $\hat{Y}_2$ develops
singularities that can be related to non-Abelian groups
according to the Kodaira classification. To study these singularities
one first uses a standard procedure \cite{An2001304} that brings the torus \eqref{p1} to Weierstrass form,
\begin{align}
F = -y^2 + x^3 + fx +g =0 \,.
\end{align}
Here $x$ and $y$ are certain functions of the homogeneous coordinates
$u$, $v$, $w$ and $e_1$, and the coefficients $f$ and $g$
are functions of the base coordinates $z_0$ and $z_1$, $d_i(z_0,z_1)$.
Expanding the sections $d_i$ (see \eqref{diY2}) around $z_0 = 0$,
$d_i(z_0,z_1) = d_i + \mathcal{O}(z_0)$, and using
Eqs.~\eqref{eq:fcubic} and \eqref{eq:gcubic},
we find
\begin{equation}\label{fgz0}
\begin{split}
f =& z_0^2 \left( -\frac{1}{3} d_5^2 d_7^2 + z_0 R_1 + \mathcal{O}(z_0^2)\right)\,, \\
g =& z_0^3 \left( -\frac{2}{27} d_5^3 d_7^3 +  z_0 R_2 + \mathcal{O}(z_0^2)\right)\,,
\end{split}
\end{equation}
where $R_1$ and $R_2$ are polynomials in the sections $d_i$.
The torus is singular at a point $(x,y)$ when
\begin{align}
F = \frac{\partial F}{\partial x} = \frac{\partial F}{\partial y} =
0\,.
\end{align} 
This is the case when the discriminant $\Delta$ vanishes,
\begin{align}\label{deltaz0}
\Delta= 4 f^3 + 27 g^2 = 0\,.
\end{align}
From Eqs.~\eqref{fgz0}, \eqref{deltaz0} and the expressions for
$R_{1,2}$ one obtains
\begin{align}
\Delta= z_0^7 (P + z_0 R+ \mathcal{O}(z_0^2) ) \,,\label{DeltaWS}
\end{align}
where
\begin{align}
P = -d_5^3 d_7^3 (d_3 d_5 - d_1 d_7)^2 d_9^2 \,, 
\end{align}
and $R$ is a generic polynomial of degree 16 in the $d_i$ .
Clearly, at the base coordinate $z_0 = 0$ the torus is singular at
$(x,y) = (0,0)$. The order (Ord) of this
codimension-one singularity is characterized by the power in $z_0$ of $f$, $g$ and
$\Delta$. From Eqs.~\eqref{fgz0} and \eqref{deltaz0} we
infer an $\text{Ord}(f,g,\Delta) = (2,3,7)$ singularity which
corresponds to the gauge group SO(10) \cite{kodaira}. 

For a higher-dimensional base the sections $d_i$ depend on additional
coordinates and can possibly vanish at certain points of the
base where in addition to $z_0$ one of the $d_i$'s vanishes.  According to
Eqs.~\eqref{fgz0} and \eqref{DeltaWS}
the vanishing of some $d_i$'s enhances the singularity of the torus at
$z_0 = 0$. Generically, this corresponds to larger symmetries according to the
Kodaira classification.\footnote{Note, however, that for these codimension-two singularities  Kodaira's classification does not necessarily apply.} The relevant cases are summarized in
Table~\ref{singularities}. These codimension-two singularities\footnote{Strictly speaking, a CY twofold has no codimension-two singularities. For simplicity we nevertheless use this notion since we shall later consider an application to a CY threefold.} will be analyzed in more detail in 
Section~\ref{sec:mattersplits}.
\begin{table}
\begin{center}$
\begin{array}{| c | c | c|}\hline
\text{locus} &  \text{Ord} (f, g , \Delta) & \text{fiber singularity} \\ \hline \hline
z_0=0 & (2,3,7) & SO(10) \\ \hline
z_0=d_9 = 0 & (2,3,8) & SO(12) \\ 
z_0=d_5 = 0 & (3,4,8) & E_6 \\  
z_0=d_7 = 0 & (3,4,8) & E_6 \\  
z_0=d_3 d_5 - d_1 d_7 = 0 & (2,3,8) & SO(12) \\ \hline
\end{array}$
\caption{\label{singularities} Codimension-one and codimension-two
  singularities of the CY twofold and the associated symmetry groups.}
\end{center}
\end{table}

%%%%%%%%%
\subsubsection*{Resolved $\boldsymbol{K3}$ manifold}
\label{smooth6dvac}

To analyze the gauge symmetries and the matter content encoded in the
singular $K3$ manifold, one has to resolve the singularities. This is
achieved by adding an SO(10) `top' \cite{Candelas:1996su} to the polytope following the
classification of \cite{Bouchard:2003bu}. Since the gauge group has rank $5$, five
new coordinates are introduced, which yield five additional
divisor classes,
\begin{align} 
\{D_1, \ldots, D_5\} = \{[f_2],[g_1],[g_2],[f_3],[f_4]\} \,.
\end{align}
The vertices of the polytope are listed in
Table~\ref{tab:2foldsmooth}. The dual polytope has 18 vertices (see
Appendix~\ref{a:d6dvac}), and the sections $s_i$ in the polynomial \eqref{p1} take
the form
\begin{align}\label{p2b}
\begin{array}{lll}
s_1 = d_1 z_0 f_2^2 f_4 g_1\,, & s_2 = d_2 z_0^2 f_2^2 f_3 f_4 g_1^2
g_2\,, & s_3 = d_3 z_0^2 f_2 f_3 g_1\,, \\ 
s_4 = d_4 z_0^3 f_2 f_3^2 g_1^2 g_2\,, & s_5 = d_5 f_2 f_4\,, & s_6 = 
d_6 z_0 f_2 f_3 f_4 g_1 g_2\,, \\ 
s_7 = d_7 z_0 f_3\,, & s_8 =  d_8 f_2 f_3 f_4^2 g_1 g_2^2\,, & 
s_9 = d_9 f_3 f_4 g_2 \,,
\end{array}
\end{align}
where now the sections $d_i$ depend on the coordinates $z_1$ and $z$,
\begin{align}\label{GUTdivisor}
d_i = d_i(z,z_1)\,,\quad z = z_0f_2g_1^2g_2^2f_3f_4 \,.
\end{align}  
The polynomial \eqref{p1} together with \eqref{p2b} defines a CY
twofold $Y_2$ where the singularities of the tuned CY twofold
$\hat{Y}_2$ have been resolved. The divisor $\{z_0 = 0\}$ is now
replaced by the divisor $\{z = 0\}$ which is the sum
of the base divisor $\{z_0 = 0\}$ and the fiber divisors $\{f_2 = 0\},
\ldots , \{g_2 = 0\}$, i.e. $\{z = 0\}$ differs from $\{z_0 = 0\}$ by
the sum of the SO(10) Cartan divisors. 
Correspondingly, for $Y_2$ the divisors $D_z$ 
and $D_{z_1}$ belong to the same divisor class (see Table~\ref{tab:2foldsmooth})
whereas for $\hat{Y}_2$ the divisors $D_{z_0}$ and $D_{z_1}$ belong to the
same class (see Table~\ref{tab:2foldsingular}).

The polynomial obtained from Eqs.~\eqref{p1} and \eqref{p2b} reads explicitly
\begin{equation}\label{pY2}
\begin{split}
p_{Y_2} =& d_1 u^3e_1^2 z_0f_2^2f_4g_1 + d_2 u^2ve_1^2
z_0^2f_2^2f_3f_4g_1^2g_2 + d_3 uv^2e_1^2 z_0^2f_2f_3g_1 \\
& + d_4 v^3e_1^2z_0^3f_2 f_3^2g_1^2g_2 + d_5 u^2we_1 f_2f_4  +
d_6 uvwe_1 z_0f_2f_3f_4 g_1g_2 \\
&+ d_7 v^2we_1 z_0f_3 +  d_8 uw^2f_2f_3f^2_4g_1g_2^2 + d_9 vw^2f_3f_4 g_2 \,.
\end{split}
\end{equation}   
It defines a $K3$ manifold with resolved SO(10) singularity and
will be the basis of the following calculations.

\begin{table}
\begin{center}$
\begin{array}{|c|c|c|cccccccc|}\hline
 \text{coordinates} & \text{vertices} & \text{divisor classes} & \mathcal{C}_v & \mathcal{C}_{e_1} & \mathbb{P}^1_{0}
 &\mathbb{P}^1_{1} &\mathbb{P}^1_{2} &\mathbb{P}^1_{3}
 &\mathbb{P}^1_{4} &\mathbb{P}^1_{5} \\ 
\hline\hline
u  & (1,-1,0) & [vz_0g_1g_2f_3]  & 0 & 2 & 0 & 1 & 0 & 0 & 0 & 1 \\
 v  & (-1,0,0) & [v] & -2 & 1 & 1 & 0 & 0 & 0 & 1 & 0 \\
 w  & (0,1,0) & [ve_1z_0g_2^{-1}f_4^{-1}]  & 0 & 0 & 0 & 0 & 0 & 1 & 0 & 1\\
 e_1  & (0,-1,0) & [e_1] & 1 & -2 & 1 & 0 & 0 & 0 & 0 & 0 \\
 f_2  & (1,0,1) & D_1 & 0 & 0 & 0 & -2 & 1 & 0 & 0 & 0 \\
g_1  & (1, 1,2) & D_2 & 0 & 0 & 1 & 1 & -2 & 1 & 0 & 0 \\
g_2  & (1,2,2) & D_3 & 0 & 0 & 0 & 0 & 1 & -2 & 1 & 1 \\
f_3  & (0, 1,1) & D_4 & 1 & 0 & 0 & 0 & 0 & 1 & -2 & 0 \\
f_4  & (1,1,1) & D_5 & 0 & 0 & 0 & 0 & 0 & 1 & 0 & -2 \\
z_0  & (0,0,1) & D_0 & 1 & 1 & -2 & 0 & 1 & 0 & 0 & 0 \\
z_1  & (-1,0,-1) & [z] & 2 & 1 & 0 & 0 & 0 & 0 & 0
 & 0\\ \hline
z & & [z] & 2 & 1 & 0 & 0 & 0 & 0 & 0 & 0 \\
\hline
\end{array}$
\end{center}
\caption{\label{tab:2foldsmooth}Coordinates, vertices, divisor classes and
intersection numbers (charges of $\mathbb{C}^*$-actions) for the three-dimensional
  toric variety $dP_1$ with SO(10) top, fibered over $\CP$. The divisor $[z]$ is given by $[z_0f_2g_1^2g_2^2f_3f_4]$.}
\end{table}

The presence of additional exceptional divisors changes the linear
dependencies. The divisor classes $[u]$ and $[w]$ can now be written
as (see Appendix~\ref{a:d6dvac})
\begin{align}
[u] &= [v] + D_0 + D_2 + D_3 + D_4 \,,\\
[w] &= [v] + [e_1] + D_0 - D_3 - D_4 \,.
\end{align}

The intersection of the divisors $D_v$, $D_{e_1}$, $D_0$, $\ldots$, $D_5$ with the CY twofold $Y_2$ defines a set of curves that 
are given in Table~\ref{curvesY2}. For each divisor certain
coordinates can be set to one (see Appendix~\ref{a:d6dvac}), which simplifies the
calculations\footnote{For a divisor $\{x_i = 0\}$ this is the case for
coordinates that appear together with $x_i$ in the Stanley-Reissner ideal.}. 
The intersection numbers of divisor classes $[x_i]$ and
curves $C_j = x_j \cap p_{Y_2}$ correspond to the intersections of two
divisors on the CY twofold,
\begin{align}
[x_i] \cdot C_j = C_i \cdot [x_j] = [x_i] \cdot [p_{Y_2}] \cdot [x_j]\,.
\end{align}
All intersection numbers are given in
Table~\ref{tab:2foldsmooth}. Note that self-intersection numbers of the
divisors on the CY twofold, i.e.\ the intersection numbers of the divisors
$[v]$,  $\ldots$,$D_5$ with the associated curves $\mathcal{C}_v$, $\ldots$,$\mathbb{P}^1_5$, are
all equal to $-2$. The curves $\mathcal{C}_v$ and $\mathcal{C}_{e_1}$ are
base $\mathbb{P}^1$s whereas $\mathbb{P}^1_0$ to $\mathbb{P}^1_5$ lie
in the fiber.

Particularly interesting are the intersections numbers of the divisors that were introduced to
resolve the SO(10) singularity. From Table~\ref{tab:2foldsmooth}
one reads off
\begin{align}
\label{eq:SO10Cartan}
D_i \cdot \mathbb{P}^1_j = \mathbb{P}^1_i \cdot D_j =
-(C_{SO(10)})_{ij} \,,\quad i,j = 1,\ldots 5\,,
\end{align}
\begin{table}
\begin{center}$
\begin{array}{|c|l l|}\hline
\text{curves} &  \text{intersections} & \text{coordinate patches} \\ \hline\hline
\mathcal{C}_v & v_{} \cap (d_1e_1^2z_0 + d_5we_1 + d_8w^2f_3) &(u,f_2,g_1,g_2,f_4)=1\\
\mathcal{C}_{e_1} & e_1 \cap (d_8u + d_9v) & (w,f_2,g_1,g_2,f_3,f_4)=1\\
\mathbb{P}^1_{0} & z_0 \cap( d_8 f_2 f_3 g_1 u + d_5 e_1 f_2 u^2 + d_9 f_3 v) & (w,f_4, g_2)=1 \\ 
\mathbb{P}^1_{1} & f_2 \cap (d_7 z_0 + d_9 f_4) & (v, e_1, w, f_3, g_2)=1 \\
\mathbb{P}^1_{2} & g_1 \cap (d_7 z_0 f_3 + d_5 f_2 f_4 + d_9 f_3 f_4 g_2) & (u,v,w,e_1)=1 \\
\mathbb{P}^1_{3} & g_2 \cap (d_3 f_3 g_1 + d_1 f_4 g_1 + d_7 f_3 w + d_5 f_4 w) & (u,v,e_1,z_0,f_2)=1 \\ 
\mathbb{P}^1_{4} & f_3 \cap (d_1 z_0 g_1 + d_5 w) & (e_1, u, f_2, f_4)=1 \\
\mathbb{P}^1_{5} & f_4 \cap (d_4 f_2 g_1^2 g_2 + d_3 f_2 g_1 u + d_7 w) & (v,e_1, z_0, f_3)=1 \\ \hline
\end{array}$
\caption{\label{curvesY2}Curves on the CY twofold $Y_2$ defined as intersections of divisors with $Y_2$. On each divisor  the coordinates enclosed in $(\ldots)$ can be set to one by $\mathbb{C}^*$-actions.} 
\end{center}
\end{table}

\begin{figure}
\begin{center}
\begin{picture}(20,80)
 \put(-50,2){\includegraphics[scale=0.65]{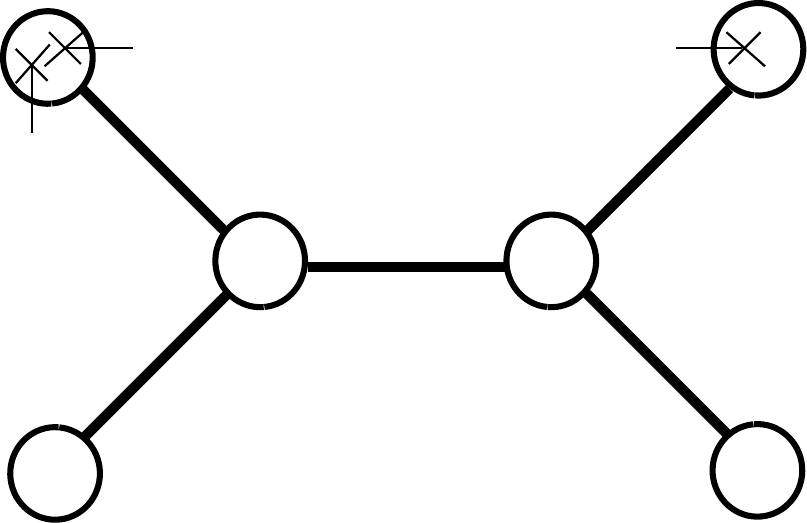}}
 \put(-70,90){ $\mathbb{P}^1_0 $}
  \put(-50,70){ $e_1 $}
   \put(-10,30){ $\mathbb{P}^1_2$}
    \put(44,30){ $\mathbb{P}^1_3$}
 \put(65,90){ $v $}
  \put(-25,90){ $v $}
    \put(100,90){ $\mathbb{P}^1_4$}
    \put(100,0){ $\mathbb{P}^1_5$}
    \put(-68,0){ $\mathbb{P}^1_1$}
 \end{picture}
\end{center}
\caption{\label{fig:DynkinSO10} The extended Dynkin diagram of SO(10)
obtained from intersections of divisors $D_I$ and curves
$\mathbb{P}^1_J$,  with $I,J \in \{0,\ldots,5\}$. Intersections of the torus divisors $[e_1]$ and $[v]$ are
also indicated.}
\end{figure}
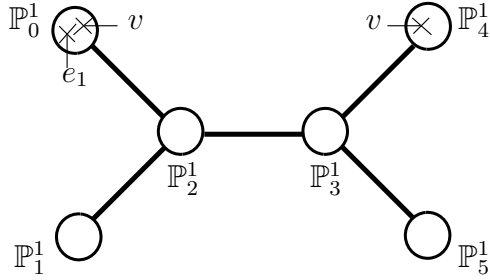
where 
\begin{align}
C_{SO(10)}= \left( 
\begin{array}{ccccc}   2& -1& 0& 0& 0\\ -1& 2& -1& 0& 0\\ 0& -1& 2&  -1& -1\\ 
0& 0& -1& 2& 0\\ 0& 0& -1& 0& 2                
\end{array}   
\right)
\end{align}
is the SO(10) Cartan matrix. Hence, the divisors $D_i$ are referred
to as Cartan divisors. Together with the divisor $D_0$ of the base
coordinate $z_0$, the Cartan divisors have the intersection numbers
corresponding to the extended Dynkin diagram of SO(10)
(see Table~\ref{tab:2foldsmooth}, Figure~\ref{fig:DynkinSO10}).
One easily verifies that the intersection numbers of the 8 curves
$\mathcal{C}_v$,...,$\mathbb{P}^1_5$ with the 11 divisors $[u]$,...,$[z_1]$
represent the charges of the 8 $\mathbb{C}^*$-scalings that leave
the CY twofold $Y_2$ invariant.

The exceptional torus divisor $[e_1]$ only intersects the affine node,
\begin{align}
[e_1] \cdot \mathbb{P}^1_I = (1,0,0,0,0,0)_I\,, \quad I= 0,\ldots,5\,,
\end{align}
whereas the divisor $[v]$ intersects the affine node and one Cartan
divisor,
\begin{align}
[v] \cdot \mathbb{P}^1_I = (1,0,0,0,1,0)_I\,, \quad I= 0,\ldots,5\,.
\end{align}

After the resolution of the singularity the divisor \eqref{U1} of the U(1)
symmetry has to be modified such that it is orthogonal to the SO(10) Cartan
divisors. It is then given by the Shioda map \cite{Morrison:2012ei},
\begin{align}
\sigma(\hat{s}_1) = [v]-[e_1] + ([v]-[e_1]) \cdot \mathbb{P}^1_i (C_{SO(10)}^{-1})_{ij} D_j\,,
\end{align}
where $C_{SO(10)}^{-1}$ is the inverse Cartan matrix,
\begin{align}
C_{SO(10)}^{-1}= \left(  
\begin{array}{ccccc}   
1& 1& 1& 1/2& 1/2 \\ 1& 2& 2& 1& 1 \\1& 2& 3& 3/2& 3/2 \\ 
1/2& 1&  3/2& 5/4& 3/4 \\ 1/2& 1& 3/2& 3/4& 5/4       
\end{array}    
\right)\,.
\end{align}

A straightforward calculation yields
\begin{align}
\sigma (\hat{s}_1) = [v] - [e_1]  + \frac{1}{4}\left(4 D_0 + 6 D_1 + 12 D_2 +
  14 D_3 + 9 D_4 + 7 D_5\right)  \,,
\end{align}
and one easily verifies 
\begin{align}
\sigma(\hat{s}_1) \cdot \mathbb{P}^1_i = 0 \,, \quad i = 1,\ldots,5\,,
\end{align}
i.e., the SO(10) roots do not carry U(1) charge, and the total
symmetry group is indeed \sou. On the other hand,
the affine node $\mathbb{P}^1_0$ has a nonvanishing intersection with
the U(1) divisor,
\begin{align}
\sigma(\hat{s}_1) \cdot \mathbb{P}^1_0 = 1 \,. 
\end{align}

\begin{figure}[t]
\begin{center}
\includegraphics[scale=0.8]{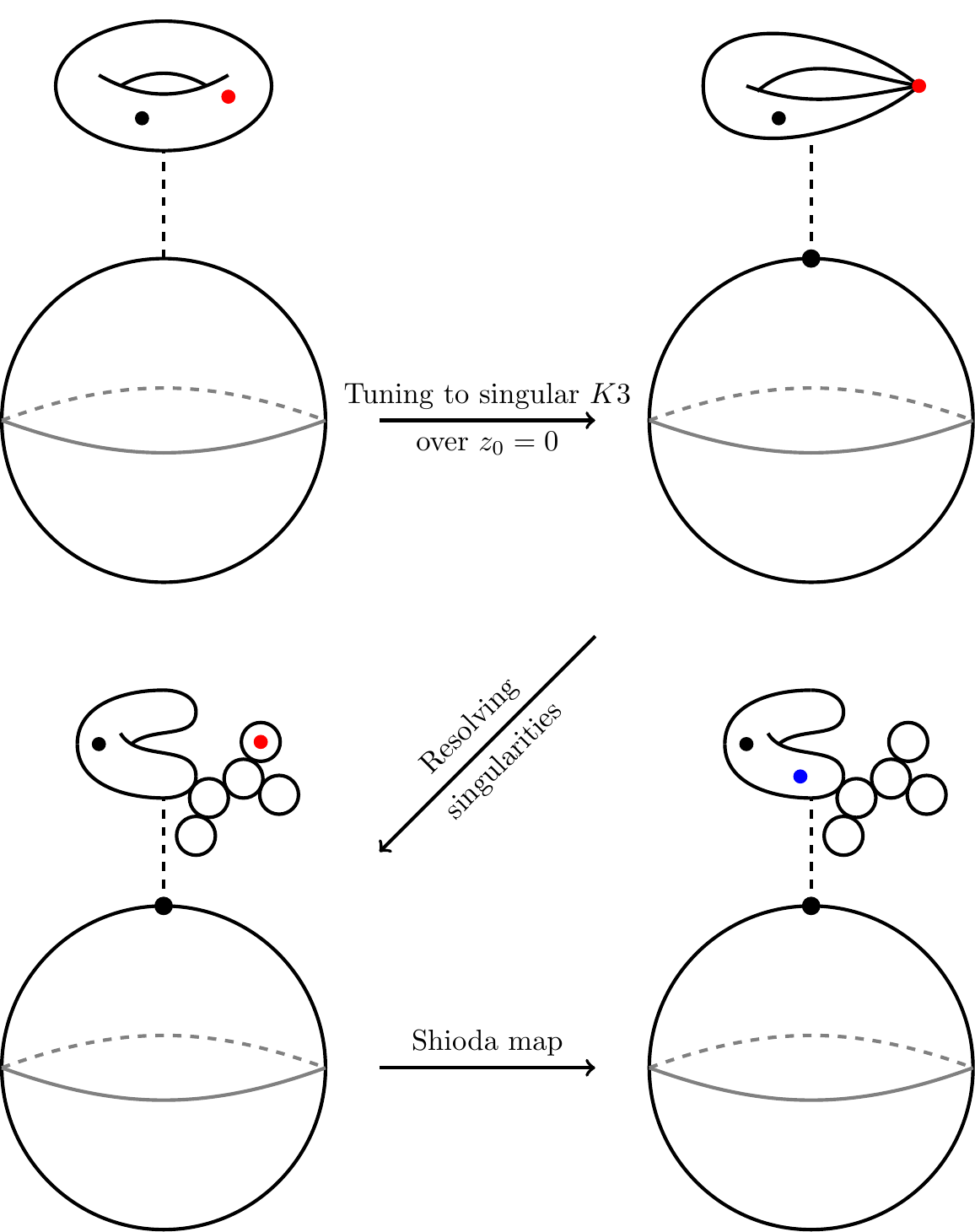}
\end{center}
\caption{\label{fig:tuning} The fibration of the torus $\mathcal{E}$ over
  $\CP$ (top left) turns via tuning into a singular $K3$
  manifold (top right). Resolution of the singularity generates five
  $\mathbb{P}^1$s with SO(10) intersections (bottom left); after the
  Shioda map has been carried out, the U(1) divisor intersects the affine node $\mathbb{P}^1_0$
 (bottom right).}
\end{figure}

The construction of a smooth CY twofold described in this
section is summarized in Figure~\ref{fig:tuning}: the fibration of a
torus with two points over a $\CP$ yields a $K3$ manifold that can be
tuned to have a codimension-one singularity corresponding to the group SO(10).
Adding an SO(10) top to the fibered ambient space this
singularity is resolved, leading to a smooth $K3$ manifold. The new
fiber consists of the torus and five additional $\CP$s which represent
the resolution of the singularity. Finally, the Shioda map
orthogonalizes the U(1) factor with respect to the SO(10) Cartan divisors.

%%%%%%%%%%%%%%%%%%
\subsection{Matter splits}
\label{sec:mattersplits}

In the following we investigate the codimension-two singularities in
the fiber, which occur at the four loci listed in Table~\ref{singularities}. 
Here the SO(10) symmetry enhances to SO(12) or E$_6$. One or more
$\mathbb{P}^1$s of the fiber split into several $\mathbb{P}^1$s whose
intersections correspond to the extended Dynkin diagram of the enhanced symmetry.

%%%%%%%%%
\subsubsection*{SO(12) matter locus $\mathbf{d_9 = 0} $} 

\begin{figure}
\begin{center}
 \begin{picture}(0,100)
 \put(-130,2){\includegraphics[scale=0.65]{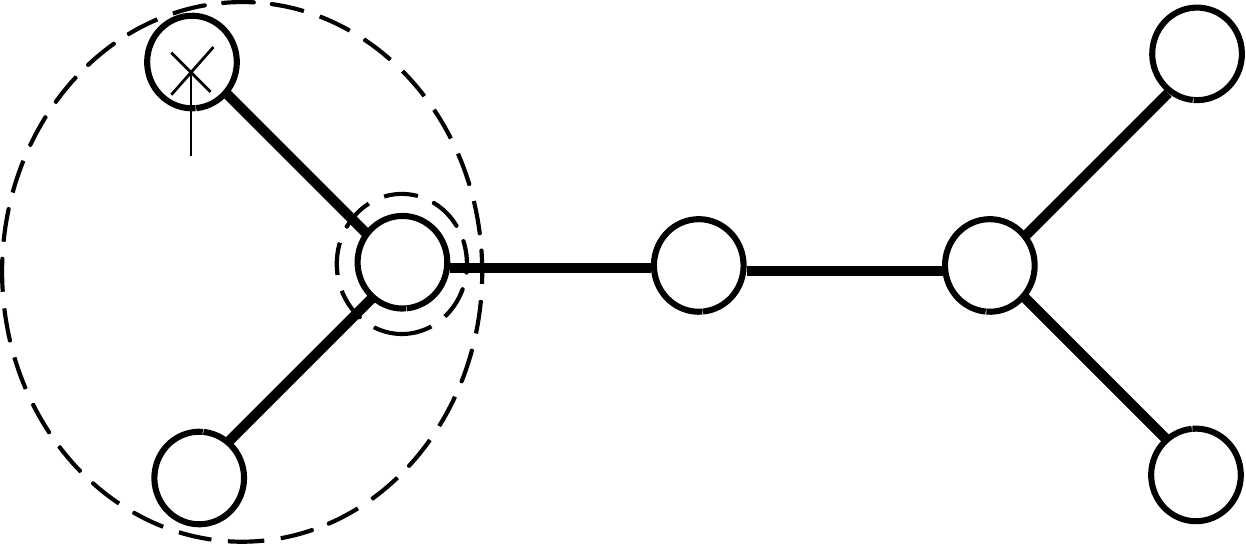}}
 \put(-100,65){$e_1$}
  \put(-140,100){$\mathbb{P}^1_{0,u}$}
    \put(-140,0){$\mathbb{P}^1_{0a}$}
      \put(-100,50){$\mathbb{P}^1_{0,2}$}
       \put(-10,30){$\mathbb{P}^1_2$}
       \put(50,30){$\mathbb{P}^1_3$}
        \put(110,100){$\mathbb{P}^1_4$}
         \put(110,0){$\mathbb{P}^1_5$}
 \end{picture}
 \caption{\label{fig:DynkinSO12d9}Extended Dynkin diagram of SO(12) at
   the $d_9=0$ locus. The affine node has an intersection with
   $[e_1]$. The dashed circles indicate SO(10) nodes before the
   matter split.}
\end{center}
\end{figure}
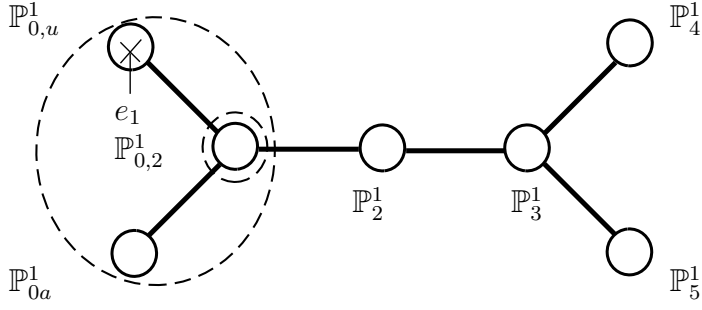

At this locus the equation for the curve $\mathbb{P}^1_0$ in
Table~\ref{curvesY2} changes. The polynomial representing the CY
twofold factorizes into three terms which means that $\mathbb{P}^1_0$
splits into three curves:
\begin{align}\label{splitd9}
\begin{array}{c|ll }
\text{nodes} & &  \text{nodes after split}    \\ \hline 
\mathbb{P}^1_0 & \rightarrow & z_0 \cap (d_5 u e_1  + d_8 g_1 f_3) + z_0 \cap f_2  + z_0 \cap  u \\
&& \equiv \mathbb{P}^1_{0a} + \mathbb{P}^1_{0,2} + \mathbb{P}^1_{0,u} \\ 
\mathbb{P}^1_1 && f_2 \cap (d_7 z_0)    \\ 
\mathbb{P}^1_2 && g_1 \cap    (d_7 z_0 f_3 + d_5 f_2 f_4)   \\ 
\mathbb{P}^1_3 && g_2 \cap   ( d_1 g_1f_4 + d_3 v^2 g_1 f_3 + d_5 w f_4 + d_7 v^2 w f_3)   \\ 
\mathbb{P}^1_4 && f_3 \cap (d_1 z_0g_1 + d_5 w)   \\ 
\mathbb{P}^1_5 && f_4 \cap  ( d_3 u f_2 g_1 + d_4 f_2 g_1^2 g_2 + d_7 w)  \\ \hline
\end{array}
\end{align}
The other five $\mathbb{P}^1$s are not affected. One easily verifies
that the new curves are again $\mathbb{P}^1$s by calculating the
corresponding self-intersection numbers:
\begin{equation}
\begin{split}
\mathbb{P}^1_{0a}: &\quad [z_0] \cdot [ue_1] \cdot ([z_0] - [f_2] - [u]) = -2\,, \\
\mathbb{P}^1_{0,2}: &\quad [z_0] \cdot [f_2] \cdot ([z_0] - [ue_1] - [u]) =
-2\,, \\
\mathbb{P}^1_{0,u}: &\quad [z_0]  \cdot [u] \cdot ([z_0] - [ue_1] - [f_2])
= -2\,,
\end{split}
\end{equation}
where the split of the original $\mathbb{P}^1_0$ has been taken into
account in the last divisor. One of the split $\mathbb{P}^1_0$s is
identical to one of the SO(10) $\mathbb{P}^1_i$s,
\begin{align}
\mathbb{P}^1_{0,2} = \mathbb{P}^1_2\,,
\end{align}
$\mathbb{P}^1_{0,u}$ intersects with the exceptional torus divisor,
\begin{align}
[e_1] \cdot \mathbb{P}^1_{0,u} =1\,,
\end{align}
whereas 
\begin{align}
[e_1] \cdot \mathbb{P}^1_{0a} =[e_1] \cdot \mathbb{P}^1_{0,2} = 0\,.
\end{align}
Hence, $\mathbb{P}^1_{0,u}$ is the new affine node. The intersections
of the new $\mathbb{P}_0^1$s are
\begin{equation}
\begin{split}
\mathbb{P}^1_{0,u} \cap \mathbb{P}^1_{0,2}: &\quad [z_0] \cdot [u]
\cdot [f_2] = 1\,,\\
 \mathbb{P}^1_{0,u} \cap \mathbb{P}^1_{0a}: &\quad [z_0] \cdot [ue_1]
\cdot [u] = 0\,,\\
\mathbb{P}^1_{0a} \cap \mathbb{P}^1_{0,2}: &\quad [z_0] \cdot [ue_1]
\cdot [f_2] = 1\,.
\end{split}
\end{equation}
Together with the SO(10) $\mathbb{P}^1$s one obtains the extended
Dynkin diagram of SO(12) (see Figure~\ref{fig:DynkinSO12d9}).
The curve $\mathbb{P}^1_{0a}$ extends the Dynkin diagram of SO(10)
to the Dynkin diagram of SO(12).   $\mathbb{P}^1_{0a}$ can be
identified as matter curve,
\begin{align}
\mathbb{P}^1_{0a} = \mathcal{C}^\omega_q\,,
\end{align}
with Dynkin label and U(1) charge,
\begin{align}
D_i\cdot \mathbb{P}^1_{0a} = \omega_i = (1,0,0,0,0)_i\,,  \quad 
q= \sigma (\hat{s}_i) \cdot \mathbb{P}^1_{0a} = 3/2\,,
\end{align}
corresponding to a ${\bf 10}$-plet of SO(10). Since $\omega$ is the highest
weight of the ${\bf 10}$ representation, all states can be obtained by
adding SO(10) $\mathbb{P}^1_i$s, $i=1\ldots 5$ to
$\mathbb{P}^1_{0a}$, which corresponds to the subtraction of roots
$\alpha_i$ from $\omega$ in the usual way.
 
%%%%%%%%%
\subsubsection*{E$_6$ matter locus $\mathbf{d_5 = 0} $}  

At this locus the curves $\mathbb{P}^1_0$, $\mathbb{P}^1_2$ and $\mathbb{P}^1_4$
split into two $\mathbb{P}^1$s each:
\begin{align}\label{splitE6a}
\begin{array}{c|ll }
\text{nodes} &&  \text{nodes after split}    \\ \hline
\mathbb{P}^1_0 & \rightarrow & z_0 \cap (d_8 uf_2 g_1 + d_9 v ) + z_0
\cap f_3 \\ 
&& \equiv \mathbb{P}^1_{0a} + \mathbb{P}^1_{0,4}\\
\mathbb{P}^1_1 && f_2 \cap (d_7 z_0 + d_9 f_4)   \\ 
\mathbb{P}^1_2 & \rightarrow & g_1 \cap (d_7 z_0 + d_9 g_2 f_4)  + g_1 \cap f_3  \\ 
&& \equiv \mathbb{P}^1_{2a}  + \mathbb{P}^1_{2,4} \\
\mathbb{P}^1_3 & &g_2 \cap  ( d_1 g_1 f_4  + d_3 g_1 f_3 + d_7 w f_3 )   \\ 
\mathbb{P}^1_4 & \rightarrow & f_3 \cap z_0  + f_3 \cap g_1  \\ 
&& \equiv \mathbb{P}^1_{0,4}  + \mathbb{P}^1_{2,4} \\
\mathbb{P}^1_5 && f_4 \cap  (d_3 u f_2 g_1 +  d_4 f_2 g_1^2 g_2 + d_7 w)   \\\hline
\end{array}
\end{align}
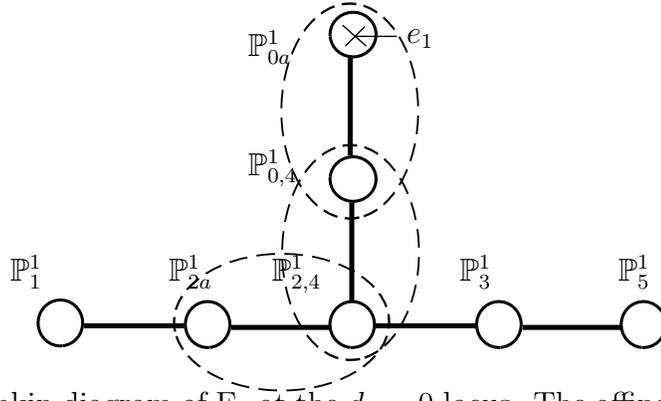
\begin{figure}
\begin{center}
 \begin{picture}(0,150)
 \put(-100,-12){\includegraphics[scale=0.65]{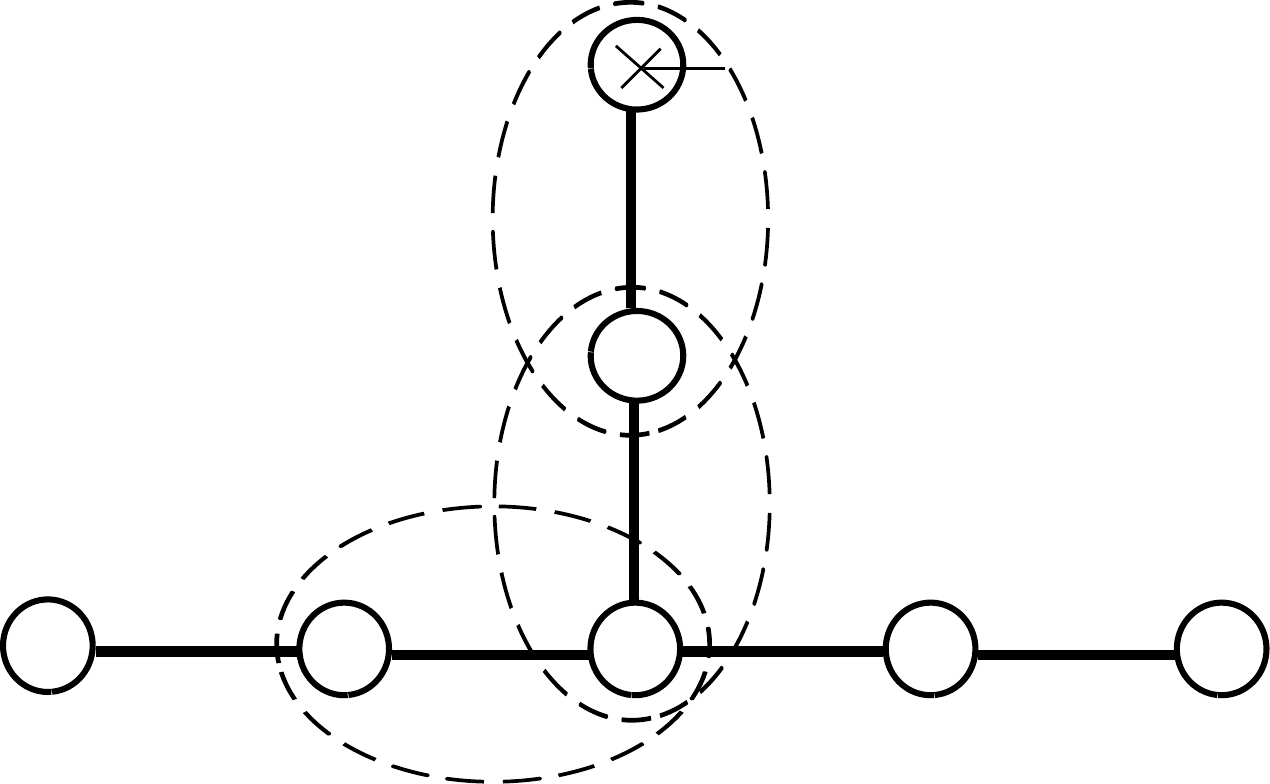}}
 \put(40,120){$e_1$}
   \put(-20,115){$\mathbb{P}^1_{0a}$}
    \put(-20,70){$\mathbb{P}^1_{0,4}$}
      \put(-110,30){$\mathbb{P}^1_{1}$}
     \put(-50,30){$\mathbb{P}^1_{2a}$}
     \put(-11,30){$\mathbb{P}^1_{2,4}$}
     \put(60,30){$\mathbb{P}^1_{3}$}
     \put(120,30){$\mathbb{P}^1_{5}$}
 \end{picture}
  \caption{\label{fig:DynkinE6d5}Extended Dynkin diagram of E$_6$ at
    the $d_5=0$ locus. The affine node intersects $[e_1]$. The dashed ellipses indicate SO(10) nodes before the
    matter split. Two of the E$_6$ nodes occur in the split of two
    SO(10) nodes.}
\end{center}
\end{figure}
The curve $\mathbb{P}^1_{0a}$ is the only node which has nonvanishing
intersection with $[e_1]$ and it therefore represents the new
affine node. The four new nodes all have nonvanishing U(1) charge
and the identification of the matter curve is unique up to complex
conjugation and the addition of SO(10) roots. We choose
\begin{align}\label{matterE6a}
\mathbb{P}^1_{0,4} = \mathcal{C}^\omega_q\,,
\end{align}
with Dynkin label and U(1) charge
\begin{align}\label{chargeE6a}
\quad D_i\cdot
\mathbb{P}^1_{0,4} = \omega_i = (0,1,0,-1,0)_i \,,\quad 
q = \sigma (\hat{s}_1) \cdot \mathbb{P}^1_{0,4} = 3/4\,,
\end{align}
corresponding to a ${\bf 16}$-plet of SO(10).
The remaining new $\mathbb{P}^1$s can then be written as linear
combinations of the matter curve and SO(10) roots,
\begin{equation}
\begin{split}
\mathbb{P}^1_{0a} & = -\mathcal{C}^\omega_{3/4} + \mathbb{P}^1_0\,, \\
\mathbb{P}^1_{2a} & = \mathcal{C}^\omega_{3/4} - \mathbb{P}^1_4 + \mathbb{P}^1_2\,, \\
\mathbb{P}^1_{2,4} & = -\mathcal{C}^\omega_{3/4} + \mathbb{P}^1_4\,. 
\end{split}
\end{equation}

It is straightforward to calculate the intersections of the curves
given in \eqref{splitE6a}. As a result one obtains the extended Dynkin
diagram of E$_6$ (see Figure~\ref{fig:DynkinE6d5}).

%%%%%%%%%
\subsubsection*{E$_6$ matter locus  $\mathbf{d_7 = 0}$}

Here again an enhancement from SO(10) to E$_6$ takes place. This
time the curves $\mathbb{P}^1_2$ and $\mathbb{P}^1_5$ split into two
and three $\mathbb{P}^1$s, respectively:
\begin{align}\label{splitE6b}
\begin{array}{c|ll }
\text{nodes} & &  \text{nodes after split}    \\ \hline
\mathbb{P}^1_0 && z_0 \cap (d_5 u^2 e_1 f_2 + d_8 u f_2 g_1 f_3 + d_9 v f_3)  \\ 
\mathbb{P}^1_1 && f_2 \cap   f_4 \equiv \mathbb{P}^1_{1,5}   \\ 
\mathbb{P}^1_2 & \rightarrow &   g_1 \cap (d_5 f_2 + d_9 g_2 f_3) +  g_1 \cap   f_4 \\ 
&& \equiv \mathbb{P}^1_{2a} + \mathbb{P}^1_{2,5} \\
\mathbb{P}^1_3 & & g_2 \cap   (d_1 g_1 f_4 + d_3 g_1 f_3 + d_5 w f_4)   \\  
\mathbb{P}^1_4 && f_3 \cap (d_1 z_0 g_1 + d_5 w)   \\ 
\mathbb{P}^1_5 & \rightarrow  & f_4 \cap  (d_3 u + d_4 g_1 g_2) + f_4
\cap f_2 + f_4 \cap g_1\\
&& \equiv \mathbb{P}^1_{5a} + \mathbb{P}^1_{1,5} + \mathbb{P}^1_{2,5}
\\ \hline 
\end{array}
\end{align}
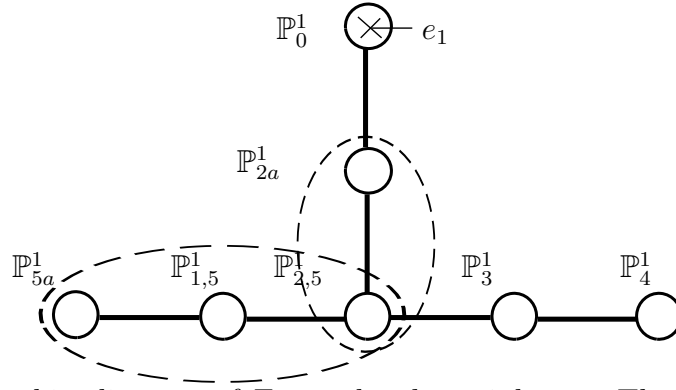
\begin{figure}
\begin{center}
 \begin{picture}(0,150)
 \put(-100,-10){\includegraphics[scale=0.65]{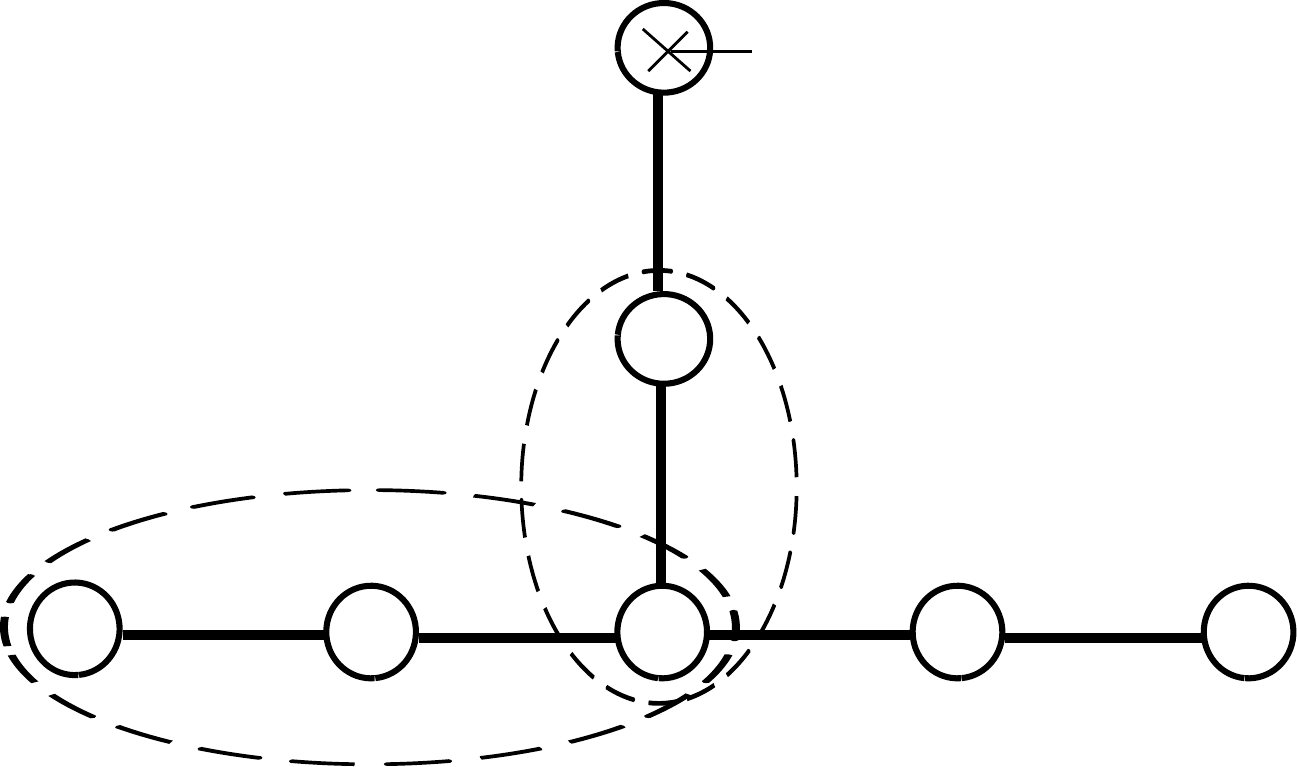}}
 \put(45,120){$e_1$}
   \put(-10,120){$\mathbb{P}^1_{0}$}
    \put(-25,70){$\mathbb{P}^1_{2a	}$}
      \put(-110,30){$\mathbb{P}^1_{5a }$}
     \put(-50,30){$\mathbb{P}^1_{1,5}$}
     \put(-11,30){$\mathbb{P}^1_{2,5}$}
     \put(60,30){$\mathbb{P}^1_{3}$}
     \put(120,30){$\mathbb{P}^1_{4}$}
 \end{picture}
  \caption{\label{fig:DynkinE6d7}Extended Dynkin diagram of E$_6$ at
    the $d_7=0$ locus. The affine node has an intersection with
    $[e_1]$. The dashed ellipses indicate SO(10) nodes before the
    matter split.}
\end{center}
\end{figure}
The affine node $\mathbb{P}^1_0$ is unaffected by the splits. The
identification of the matter curve is again unique up to complex
conjugation and addition of roots. We choose
\begin{align}
\mathbb{P}^1_{2a} = \mathcal{C}^\omega_q\,,
\end{align}
with Dynkin label and U(1) charge
\begin{align}
 D_i\cdot \mathbb{P}^1_{2a} = \omega_i = (0,-1,0,0,1)_i \,,\quad 
q = \sigma (\hat{s_1}) \cdot \mathbb{P}^1_{2a} = -1/4\,,
\end{align}
corresponding again to a ${\bf 16}$-plet of SO(10). The other two new
roots are linear combinations of the matter curve and SO(10) $\mathbb{P}^1$s,
\begin{align}
\mathbb{P}^1_{2,5} & = -\mathcal{C}^\omega_{-1/4} + \mathbb{P}^1_2\,, \\
\mathbb{P}^1_{5a} & = \mathcal{C}^\omega_{-1/4} - \mathbb{P}^1_1 - \mathbb{P}^1_2 + \mathbb{P}^1_5 \,.
\end{align}
Calculating the intersection numbers of all $\mathbb{P}^1$s one
finds again the extended Dynkin diagram of E$_6$. This is displayed in Figure~\ref{fig:DynkinE6d7}
where also the split of the SO(10) $\mathbb{P}^1$s is indicated by
dashed ellipses.

%%%%%%%%%
\subsubsection*{SO(12) matter locus $\mathbf{d_3 d_5 - d_1 d_7 = 0} $} 

Finally, at this locus a second enhancement of SO(10) to SO(12)
occurs. In this case only $\mathbb{P}^1_3$ splits into two $\mathbb{P}^1$s:
\begin{align}\label{splitSO10b}
\begin{array}{c|ll }
\text{nodes}  &&  \text{nodes after split}    \\ \hline 
\mathbb{P}^1_0 && z_0 \cap (d_5 u^2 e_1 f_2 + d_8 u f_2 g_1 f_3 + d_9 v f_3)  \\ 
\mathbb{P}^1_1 && f_2 \cap (d_7 z_0 + d_9 f_4)   \\ 
\mathbb{P}^1_2 && g_1 \cap  (d_5 f_2 f_4 + d_7 z_0 f_3 + d_9 g_2 f_3 f_4)  \\ 
\mathbb{P}^1_3 & \rightarrow & g_2 \cap (d_1 g_1 + d_5 w) + g_2 \cap( d_1 f_4 + d_3 f_3)  \\ 
&& \equiv \mathbb{P}^1_{3a} + \mathbb{P}^1_{3b} \\
\mathbb{P}^1_4 && f_3 \cap (d_1 z_0 g_1 + d_5 w) \\ 
\mathbb{P}^1_5 && f_4 \cap  (d_3 u f_2 g_1 + d_4 f_2 g_1^2 g_2 + d_7
w)   \\ \hline 
\end{array}
\end{align}
\begin{figure}
\begin{center}
 \begin{picture}(0,100)
 \put(-100,2){\includegraphics[scale=0.65]{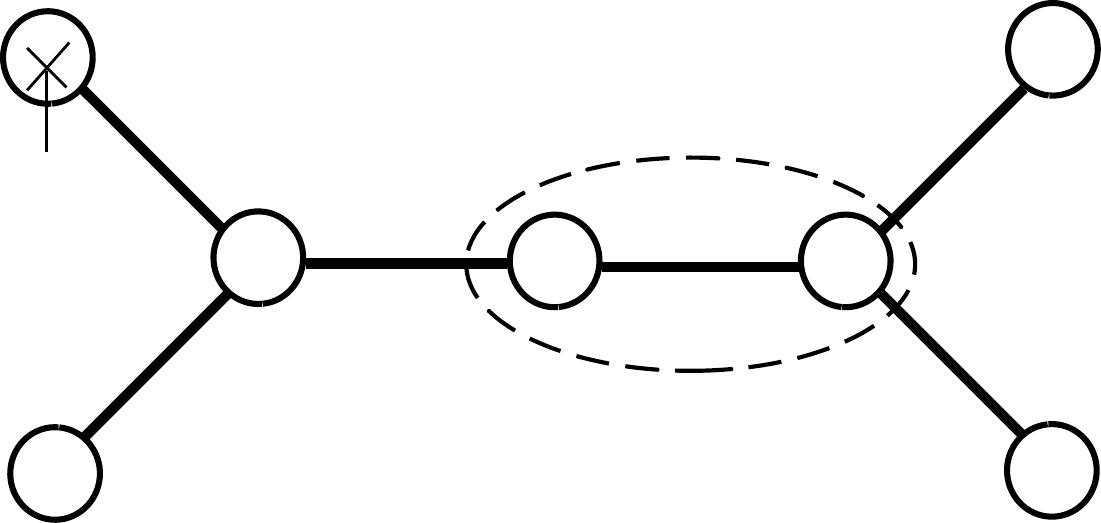}}
 \put(-100,65){$e_1$}
  \put(-115,100){$\mathbb{P}^1_{0}$}
    \put(-110,0){$\mathbb{P}^1_{1}$}
      \put(-50,70){$\mathbb{P}^1_{2}$}
       \put(-10,23){$\mathbb{P}^1_{3a}$}
       \put(50,23){$\mathbb{P}^1_{3b}$}
         \put(110,100){$\mathbb{P}^1_{4}$}
           \put(30,80){$\mathbb{P}^1_{3}$}
                  \put(110,0){$\mathbb{P}^1_{5}$}
 \end{picture}
 \caption{\label{fig:DynkinSO12nt} Extended Dynkin diagram of SO(12)
   at the $d_3 d_5 - d_1 d_7 =0$ locus. The affine node  intersects
   with $[e_1]$. The dashed ellipse indicates the SO(10) node $\mathbb{P}^1_3$.}
\end{center}
\end{figure}
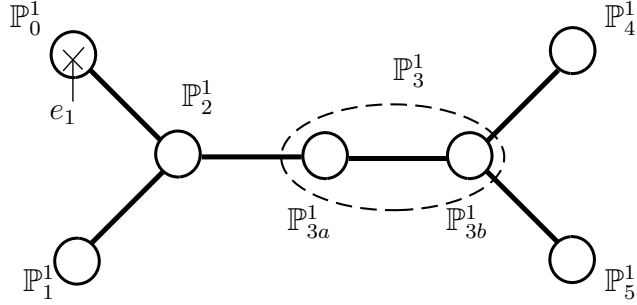
One immediately identifies, up to complex conjugation, the matter curve
\begin{align}
\mathbb{P}^1_{3a} = \mathcal{C}^\omega_q\,,
\end{align}
with Dynkin label and U(1) charge
\begin{align}
\quad D_i\cdot \mathbb{P}^1_{3a} = \omega_i = (0,1,-1,0,0)_i \,,
\quad q = \sigma (\hat{s}_1) \cdot \mathbb{P}^1_{3a} = -1/2\,,
\end{align}
corresponding to a ${\bf 10}$-plet of SO(10). The second new $\mathbb{P}^1$
is given by
\begin{align}
\mathbb{P}^1_{3b}  = -\mathcal{C}^\omega_{-1/2} + \mathbb{P}^1_3  \,.
\end{align}
The intersection number of all $\mathbb{P}^1$s yield the Dynkin
diagram displayed in Figure~\ref{fig:DynkinSO12nt}.

The symmetry enhancements and matter representations at all
four matter loci are summarized in Table~\ref{MatterSpectrum}.
\begin{table}[th]
\begin{center}$
\begin{array}{| c | c | c | c |}\hline
\text{locus} &  \text{Ord} (f, g , \Delta) & \text{fiber singularity} & \text{representation}  \\ \hline \hline
z_0 = d_9 = 0 & (2,3,8) & $SO$(12) & \mathbf{10}_{3/2} \\ \hline 
z_0 = d_5 = 0 & (3,4,8) & $E$_6  & \mathbf{16}_{3/4} \\ \hline 
z_0 = d_7 = 0 & (3,4,8) & $E$_6 & \mathbf{16}_{-1/4}  \\ \hline 
z_0 = d_3 d_5 - d_1 d_7 = 0 & (2,3,8) & $SO$(12) & \mathbf{10}_{-1/2}  \\  \hline
\end{array} $
\caption{\label{MatterSpectrum} Codimension-two loci, enhanced
  symmetry groups and matter representations.}
\end{center}
\end{table}

%%%%%%%%%%%%%%%%%%
\subsection{Yukawa couplings of SO(10) matter}

For completeness, we now consider Yukawa couplings of SO(10) matter.
Tuning two coefficients of the polynomial \eqref{p2b} to zero, one
finds further symmetry enhancements corresponding to codimension-three
singularities of a CY fourfold. At these loci three matter curves
intersect and Yukawa couplings are generated. 
 
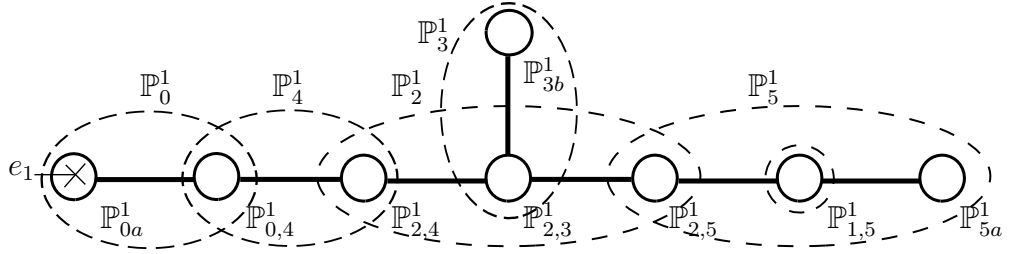
\begin{figure}[!ht]
\begin{center}
 \begin{picture}(100,100)
 \put(-100,2){\includegraphics[scale=0.65]{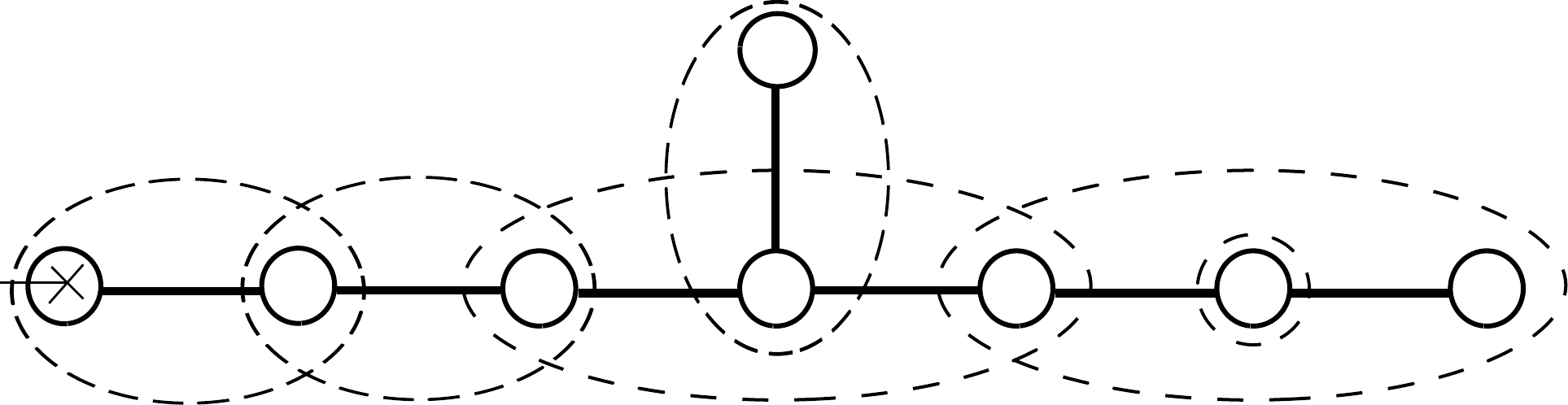}}
 \put(-110,30){$e_1$}
  \put(-75,10){$\mathbb{P}^1_{0a}$}
    \put(-60,60){$\mathbb{P}^1_{0}$}
      \put(-10,60){$\mathbb{P}^1_{4}$}
    \put(-20,10){$\mathbb{P}^1_{0,4}$}
              \put(35,10){$\mathbb{P}^1_{2,4}$}
               \put(85,10){$\mathbb{P}^1_{2,3}$}
  \put(140,10){$\mathbb{P}^1_{2,5}$}
    \put(200,10){$\mathbb{P}^1_{1,5}$}
        \put(250,10){$\mathbb{P}^1_{5a}$}
                \put(85,65){$\mathbb{P}^1_{3b}$}
                 \put(35,60){$\mathbb{P}^1_{2}$}
                 \put(45,80){$\mathbb{P}^1_{3}$}
                  \put(170,60){$\mathbb{P}^1_{5}$}
 \end{picture}
\caption{\label{fig:E7A} Extended Dynkin diagram of E$_7$ at
  the locus $z_0 = d_5 = d_7 = 0 $ with Yukawa coupling $\mathbf{16}_{3/4}
  \mathbf{16}_{-1/4} \mathbf{10}_{-1/2}$; the affine node intersects
  with $[e_1]$, the dashed ellipses indicate the SO(10)
  $\mathbb{P}^1$s before the matter splits.}
\end{center}
\end{figure}

Let us first consider the locus $z_0 = d_5 = d_7 = 0$, where also 
$d_1d_7-d_3d_5 = 0$. According to Table~\ref{MatterSpectrum}  
at this locus three matter curves intersect, which leads to
the Yukawa coupling $\mathbf{16}_{3/4} \mathbf{16}_{-1/4} \mathbf{10}_{-1/2}$.
It is instructive to study the matter splits, starting from
\eqref{splitE6a} or \eqref{splitE6b}. The result reads:
\begin{align}\label{splitE7}
\begin{array}{c|ll }
\text{nodes} &&  \text{nodes after split}    \\ \hline
\mathbb{P}^1_0 & \rightarrow & z_0 \cap (d_8 uf_2 g_1 + d_9 v ) + z_0
\cap f_3 \equiv \mathbb{P}^1_{0a} + \mathbb{P}^1_{0,4}\\
\mathbb{P}^1_1 & \rightarrow & f_2 \cap f_4 \equiv \mathbb{P}^1_{1,5}   \\ 
\mathbb{P}^1_2 & \rightarrow & g_1 \cap g_2 + g_1 \cap f_3 + g_1 \cap f_4   
\equiv \mathbb{P}^1_{2,3}  + \mathbb{P}^1_{2,4}  + \mathbb{P}^1_{2,5} \\
\mathbb{P}^1_3 & \rightarrow & g_2 \cap g_1 + g_2 \cap  ( d_1 f_4  + d_3 f_3) 
\equiv \mathbb{P}^1_{2,3} + \mathbb{P}^1_{3b} \\   
\mathbb{P}^1_4 & \rightarrow & f_3 \cap z_0  + f_3 \cap g_1   
\equiv \mathbb{P}^1_{0,4}  + \mathbb{P}^1_{2,4} \\
\mathbb{P}^1_5 & \rightarrow & f_4 \cap  (d_3 u +  d_4 g_1 g_2)  + f_4
\cap f_2 + f_4 \cap g_1 
\equiv \mathbb{P}^1_{5a} + \mathbb{P}^1_{1,5} + \mathbb{P}^1_{2,5} \\\hline
\end{array}
\end{align}
Now all six SO(10) $\mathbb{P}^1$s split, yielding eight $\mathbb{P}^1$s
some of which occur twice. These $\mathbb{P}^1$s provide links
between the chains of $\mathbb{P}^1$s into which the original SO(10)
$\mathbb{P}^1$s split. These links determine the intersection pattern
of all $\mathbb{P}^1$s and as a result one easily obtains the extended
Dynkin diagram of \E7 shown in Figure~\ref{fig:E7A}.
The affine node is again indicated by the intersection with the
divisor $[e_1]$, and the dashed ellipses indicate the SO(10)
$\mathbb{P}^1$s before the matter splits.

\begin{figure}[h!t]
\begin{center}
 \begin{picture}(100,130)
 \put(-100,2){\includegraphics[scale=0.65]{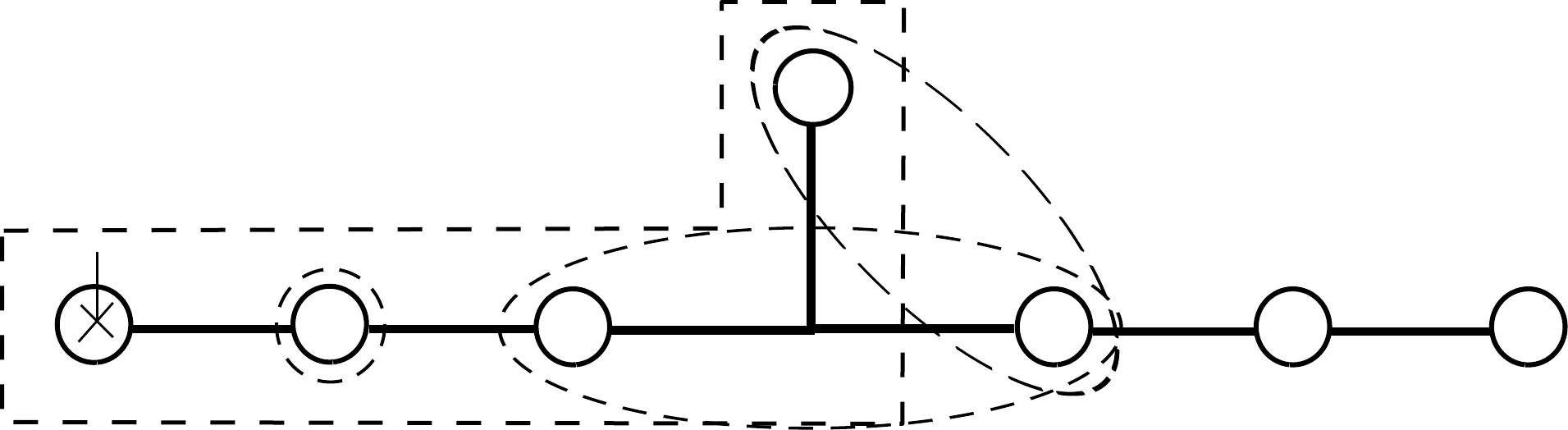}}
 \put(-90,40){$e_1$}
 \put(-95,10){$\mathbb{P}^1_{0,u}$}
  \put(-50,10){$\mathbb{P}^1_{0,1}$}
    \put(40,10){$\mathbb{P}^1_{0,2}$}
      \put(157,10){$\mathbb{P}^1_{2,4}$}
            \put(205,10){$\mathbb{P}^1_{3}$}
             \put(260,10){$\mathbb{P}^1_{5}$}
       \put(88,60){$\mathbb{P}^1_{0,4}$}
                    \put(-120,50){$\mathbb{P}^1_{0}$}
                             \put(-10,10){$\mathbb{P}^1_{1}$}
                     \put(140,60){$\mathbb{P}^1_{4}$}
                      \put(110,0){$\mathbb{P}^1_{2}$}
 \end{picture}
 \caption{\label{fig:E7B} $\mathbb{P}^1$ intersection pattern at $z_0 = d_7 = d_9 = 0$;
non-Kodaira singularity with Yukawa coupling $\mathbf{16}_{3/4}
\mathbf{16}_{3/4} \mathbf{10}_{-3/2}$.}
\end{center}
\end{figure}
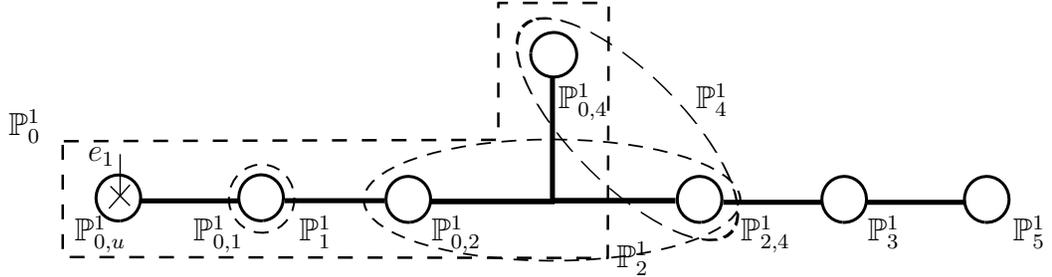

At the locus $z_0 = d_9 = 0$ we have chosen the representation
$\mathbf{10}_{3/2}$  as matter curve. Alternatively, we could have
chosen the complex conjugate representation $\mathbf{10}_{-3/2}$
as matter curve. In this case a Yukawa coupling $\mathbf{16}_{3/4} \mathbf{16}_{3/4} \mathbf{10}_{-3/2}$
can be generated at the locus $z_0 = d_5 = d_9=0$, where we
find a non-Kodaira fiber. The intersections of the
$\mathbb{P}^1$s are displayed in 
Figure~\ref{fig:E7B}, which is reminiscent of the extended E$_7$
Dynkin diagram, with a missing node in the middle. Such an
intersection pattern has previously been observed in \cite{Esole:2011sm}.

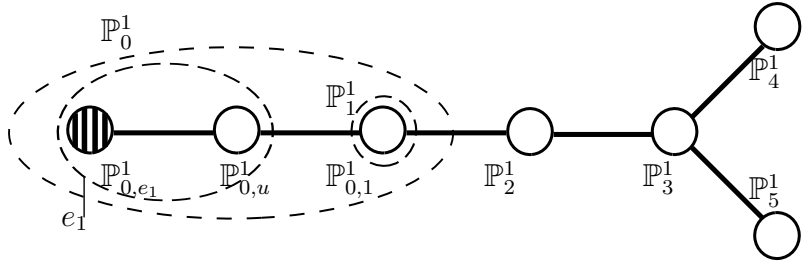
\begin{figure}[h!t]
\begin{center}
 \begin{picture}(100,100)
 \put(-100,2){\includegraphics[scale=0.65]{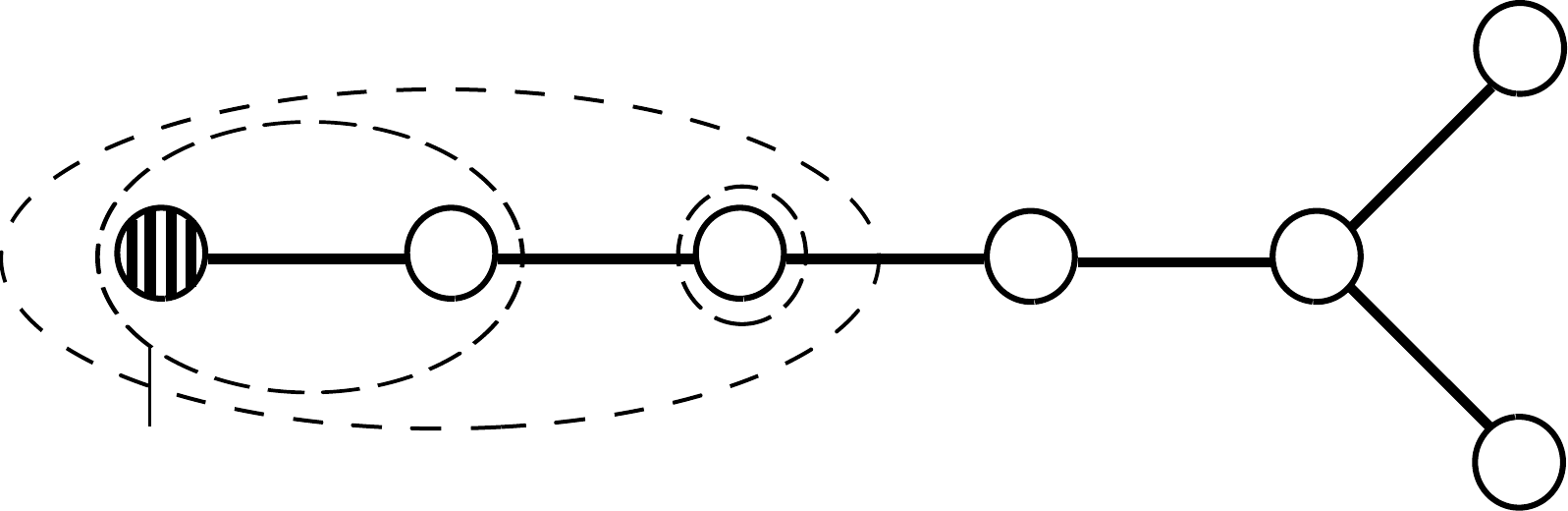}}
 \put(-80,15){$e_1$}
 \put(-65,30){$\mathbb{P}^1_{0,e_1}$}
  \put(-65,85){$\mathbb{P}^1_{0}$}
    \put(20,60){$\mathbb{P}^1_{1}$}
  \put(-20,30){$\mathbb{P}^1_{0,u}$}
    \put(20,30){$\mathbb{P}^1_{0,1}$}
        \put(80,30){$\mathbb{P}^1_{2}$}
                \put(140,30){$\mathbb{P}^1_{3}$}
                \put(180,70){$\mathbb{P}^1_{4}$}
                 \put(180,25){$\mathbb{P}^1_{5}$}
 \end{picture}
 \caption{\label{fig:DynkinSO14}  $\mathbb{P}^1$ intersection pattern at $z_0 = d_8 = d_9$;
non-Kodaira singularity with Yukawa coupling $\mathbf{10}_{-3/2}
\mathbf{10}_{-3/2} \mathbf{1}_{3}$}
\end{center}
\end{figure}

\begin{figure}[h!t]
\begin{center}
 \begin{picture}(100,140)
 \put(-100,2){\includegraphics[scale=0.65]{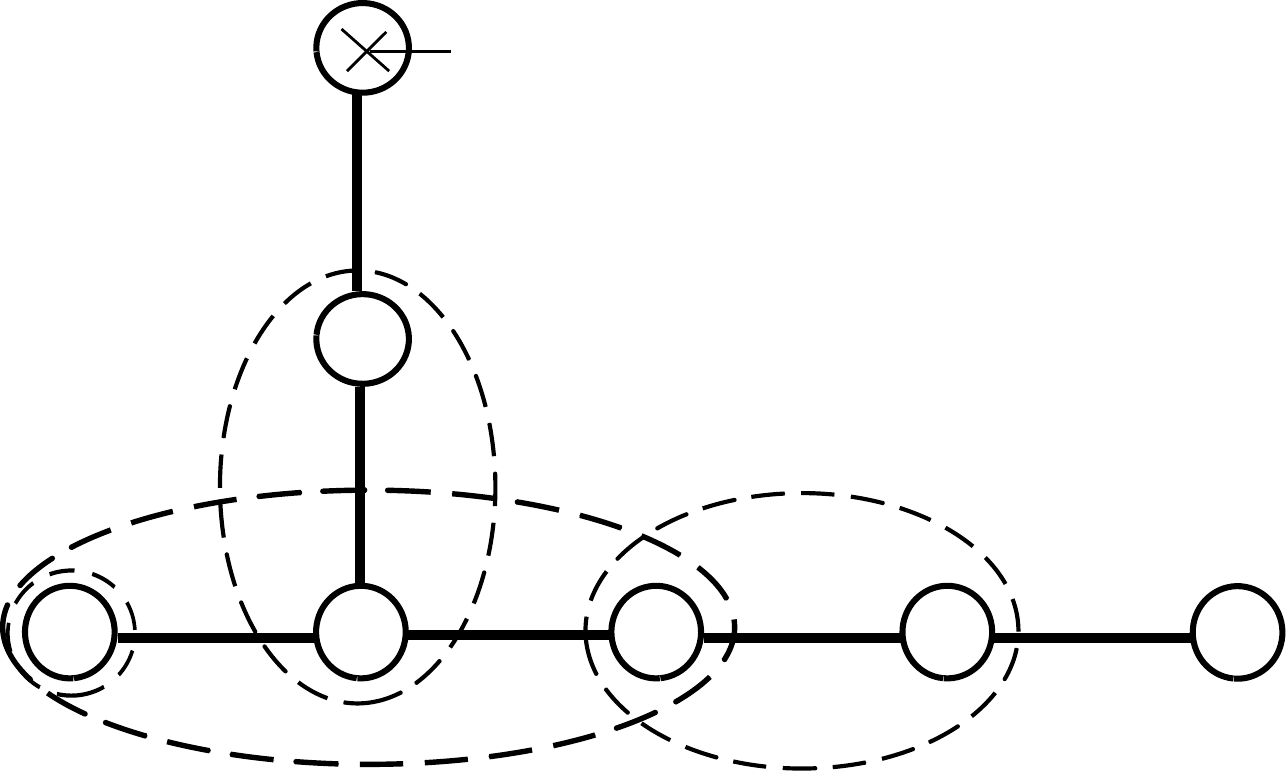}}
 \put(-15,135){$e_1$}
  \put(-15,80){$\mathbb{P}^1_{2a}$}
      \put(-70,15){$\mathbb{P}^1_{1,5}$}
          \put(-20,15){$  \mathbb{P}^1_{2,5}$}
\put(40,15){$ \mathbb{P}^1_{3,5}$}
\put(85,15){$ \mathbb{P}^1_{3b}$}
\put(140,15){$ \mathbb{P}^1_{4}$}
  \put(-75,80){$\mathbb{P}^1_{2}$}
  \put(-75,80){$\mathbb{P}^1_{2}$}
 \end{picture}
 \caption{\label{fig:DynkinE7C}  $\mathbb{P}^1$ intersection pattern
   at $z_0 = d_3d_5 -d_1d_7 = d_7 = 0$; non-Kodaira singularity with Yukawa coupling $\mathbf{16}_{-1/4}
\mathbf{16}_{-1/4} \mathbf{10}_{1/2}$}
\end{center}
\end{figure}
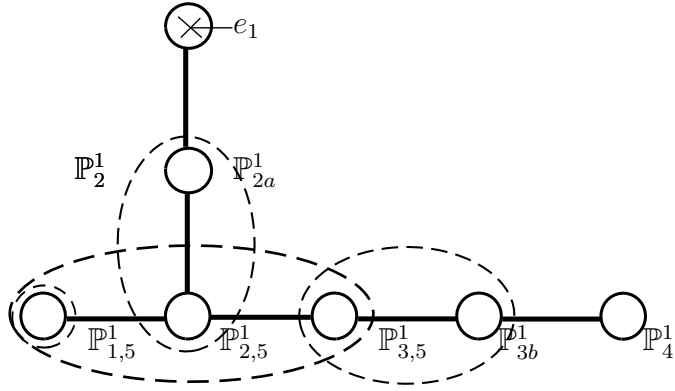

Anticipating $d_8 = d_9 = 0$ as locus of singlets with
charge three (see \eqref{ideals1}), we note that at $z_0 = d_8 = d_9 =
0$ a Yukawa coupling $\mathbf{10}_{-3/2} \mathbf{10}_{-3/2}\mathbf{1}_{3}$ can be generated. 
We again find a codimension-three singularity that is not of Kodaira
type. The intersection pattern of the $\mathbb{P}^1$s is shown in
Figure~\ref{fig:DynkinSO14}. It represents the (non-affine) Dynkin
diagram of SO(14). The torus divisor $[e_1]$ wraps the affine node
entirely, corresponding to an intersection number $-1$, which is
indicated by a striped circle.  A similar
pattern has previously been observed in \cite{Esole:2011sm}.
Finally, at the locus $z_0 = d_3d_5-d_1d_7 = d_7 = 0$ we find the intersection pattern
of $\mathbb{P}^1$s shown in Figure~\ref{fig:DynkinE7C}, which is the non-affine Dynkin diagram of \E7. The corresponding Yukawa coupling is $\mathbf{16}_{-1/4}
\mathbf{16}_{-1/4} \mathbf{10}_{1/2}$.

%%%%%%%%%%%%%%%%%%
\subsection{Calabi-Yau threefold and matter multiplicities}

\begin{figure}
\begin{center}
\hspace{0cm}{\includegraphics[width=0.43\textwidth]{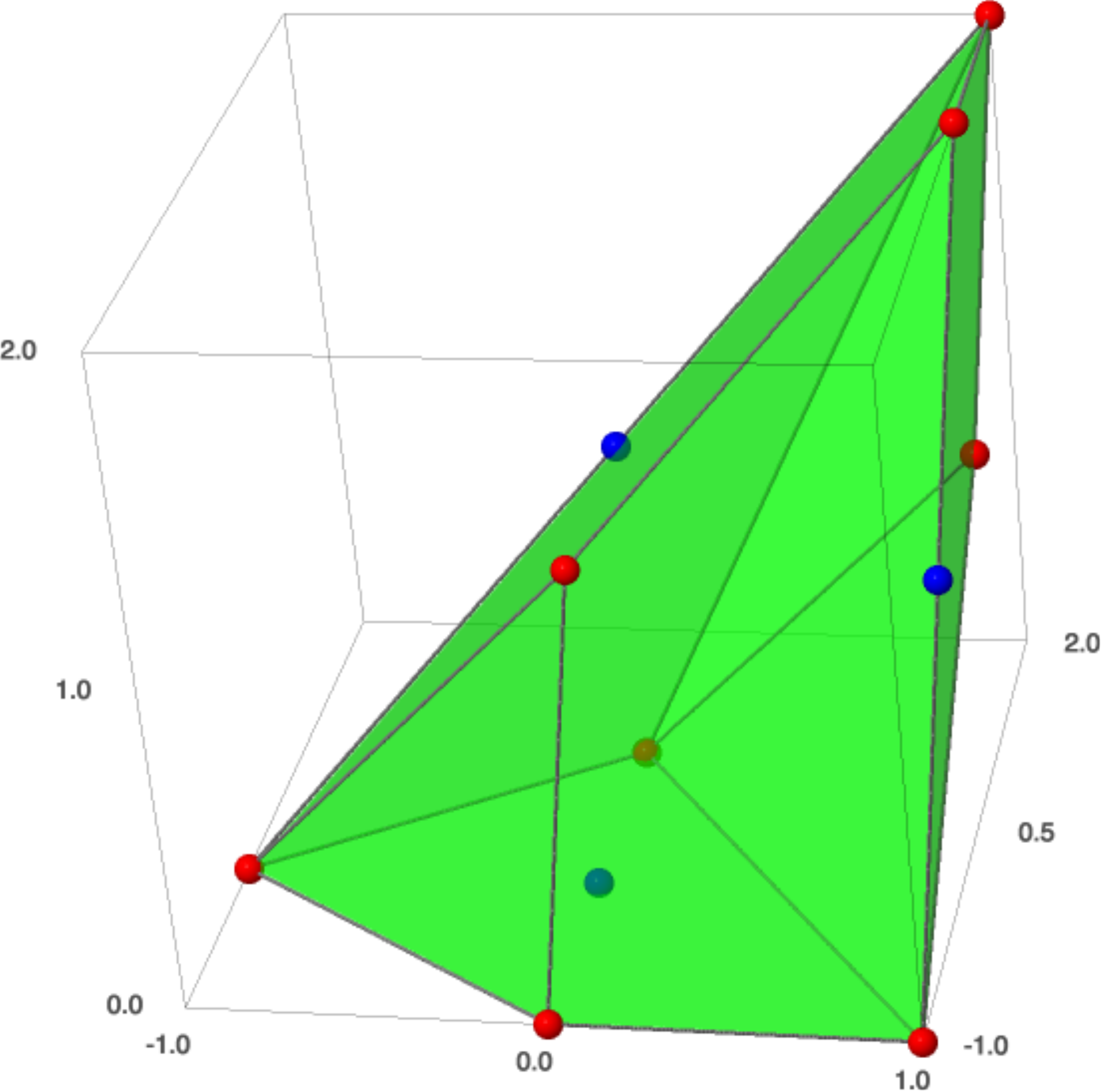}}\hspace{7cm}
\begin{picture}(0,180)
 \put(-140,100){  \begin{tabular}{c|c}
coordinates & vertices \\ \hline \hline
u& (1, -1, 0,0) \\
v & (-1, 0, 0 ,0) \\
w& (0, 1, 0,0) \\
$e_1$ & (0, -1, 0,0) \\ \hline
  $f_2$ & (1, 0, 1, 0	) \\
  $g_1$ & (1, 1, 2,0) \\
  $g_2$ & (1, 2, 2,0)\\
  $f_3$ &  (0, 1, 1,0) \\
  $f_4$ & (1, 1, 1,0) \\ 
 $z_0$ &  (0, 0, 1 , 0) \\
$z_1$ & (-1, 0, -1,0) \\ \hline
$z_2$ & (-2,0,0,1) \\
$z_3$ & (0,0,0,-1) \\ \hline
\end{tabular}} 
\end{picture}
\end{center}
\caption{\label{fig:SO10Top} The polytope formed by $F_3$ and an SO(10) top
  with vertices listed in the table. The polygon $F_3$ is visible at
  height $0$; four SO(10) vertices appear at height $1$ and the other
  two at height $2$. The vertices corresponding to $z_1$ and the
  coordinates $z_2$, $z_3$ of the second $\mathbb{P}^1$ are not shown in the figure.}
\end{figure}
So far we have analyzed the fibration of a torus over a $\mathbb{P}^1$
base space, which gave us the gauge group $\sou$ and which allowed us,
after tuning, to anticipate loci of matter and Yukawa couplings.
These all lie in the
hyperplane of the  GUT divisor $\mathcal{Z}$, which is the projection of $[z_0]$ to the base, and are furthermore characterized
by the vanishing of certain coefficients $d_i$ of the polynomial
$p_{Y_2}$. The $d_i$ are polynomials in the base coordinates
$z_0$ and $z_1$ (see~\eqref{diY2}). The multiplicities of the matter
fields are fixed once the twofold $Y_2$ is extended to a threefold
$Y_3$. In the following we shall consider the simplest case which
corresponds to adding a second $\mathbb{P}^1$ with coordinates $z_2$,
$z_3$. The coefficients of the polynomials $d_i(z_0,z_1)$ then also
depend on the additional coordinates $z_2$ and $z_3$.

For this specific case the full threefold geometry is given in
Figure~\ref{fig:SO10Top}. The polytope is defined in a four-dimensional
lattice $\mathbb{Z}^4$ with vertices $v_i = (v_i^1,\dots, v_i^4)$. A projection onto the two-dimensional base space
can be obtained by projecting onto the last two coordinates,
\begin{align}
(v_i^1,v_i^2,v_i^3,v_i^4) \xrightarrow[ ]{ \pi} (v_i^3,v_i^4) \, .
\end{align}
The base coordinates now correspond to vertices in a $\mathbb{Z}^2$ lattice,
$z_0: (1,0)$, $z_1: (-1,0)$, $z_2: (0,1)$, $z_3:
(0,-1)$, which form the toric diagram of  $\mathbb{P}^1 \times
\mathbb{P}^1 = \mathbb{F}_0$. This base space has two divisor classes,
\begin{align}
\mathcal{Z} \sim \pi_B ([z_0]) \equiv  H_1\,, \quad [z_2] \sim [z_3]
\equiv H_2\,, \label{P1divisors}
\end{align}
with the intersection numbers
\begin{align}\label{interbase}
H_1^2 = H_2^2 = 0 \, , \quad H_1 \cdot H_2 = 1\  .
\end{align}
From the dependence of the coefficients $d_i$ on the base coordinates
(see Eq.~\eqref{diY3}) one can read off the relations\footnote{A
  polynomial $d \sim z^p z_1^q z_2^r z_3^s$ implies the relation between the
  divisors $[d] \sim (p+q) H_1 + (r+s) H_2$. If $d$ is a sum of
  several monomials, the degree in $z_0z_1$ is always $p+q$ and in
  $z_2z_3$ always $r+s$ (see Eq.~\eqref{diY3}).}  between the
divisors $[d_i]$ and the base divisors $H_1$ and $H_2$. The divisors
$[d_i]$ are effective, i.e. they are linear
combinations of $H_1$ and $H_2$ with positive coefficients,
\begin{align}\label{diH12}
\begin{array}{llll}
 &[d_1] \sim 0 \,, & [d_2] \sim 2 H_2 \,, & [d_3] \sim H_1 + 4 H_2 \, , \\
 &[d_4] \sim H_1 + 6 H_2 \,, & [d_5] \sim H_1 \,, & [d_6] \sim H_1 + 2
 H_2 \, , \\
 &[d_7] \sim 2 H_1 + 4 H_2 \,, & [d_8] \sim H_1 & [d_9] \sim 2 H_1 + 2
 H_2 \, .    
\end{array}
\end{align}
Given the intersection numbers \eqref{interbase} one can also easily
calculate the genus of the GUT divisor $\mathcal{Z}$,
\begin{align}
g &= 1 - \tfrac{1}{2} ([K_B^{-1}] - \mathcal{Z}) \cdot \mathcal{Z}
\nonumber\\
& = 1 - \tfrac{1}{2} (2H_1 +2 H_2 - H_1) H_1 = 0 \,,
\end{align}
where we have used that the anticanonical divisor is given by
$[K_B^{-1}] = 2H_1 + 2H_2$. It is no surprise that the GUT divisor
$\mathcal{Z}$ is a genus-zero curve since it was just a point on the
one-dimensional base $\mathbb{P}^1$ of $K3$. It is an immediate
consequence that the considered model has no matter multiplets in 
the adjoint representation of SO(10).

Intersections of the GUT divisor with the four matter loci yield the
multiplicities of matter representations. For instance, for the
$\mathbf{10}$-plet at $z = d_9 = 0$, one has (cf.~\eqref{diH12})
\begin{align}
n[\mathbf{10}_{3/2}] = [d_9]\cdot\mathcal{Z} = (2H_1 +
2H_2)\cdot H_1 = 2 \,.
\end{align}
For the other SO(10) representations, and the charged and neutral singlets
one finds:
\begin{align}\label{H12multis}
\begin{array}{c|c|c}
\text{representation} & \text{locus} & \text{multiplicity}  \\ \hline
\mathbf{10}_{3/2}  & z = d_9 = 0 &  2 \\
\mathbf{16}_{3/4}  & z = d_5 = 0 & 0 \\
\mathbf{16}_{-1/4}  & z =d_7 = 0 & 4 \\  
\mathbf{10}_{-1/2}  & z = 0 = d_3d_5 - d_1d_7 & 4 \\ \hline
\mathbf{45} & z = 0 &  0 \\ \hline 
\mathbf{1}_1 & V(I_3) &  2 \\
\mathbf{1}_2 & V(I_2) &  36 \\
\mathbf{1}_3 & V(I_1) &  76\\
  \mathbf{1}_0  & & 51+1 \\
  T  & & 1 \\  \hline
 \end{array} 
\end{align}
A detailed, base-independent discussion of the singlet multiplicities
will be given in Section~\ref{sec:bimultis}.

\begin{figure}[t]
\begin{center}
\includegraphics[scale=0.8]{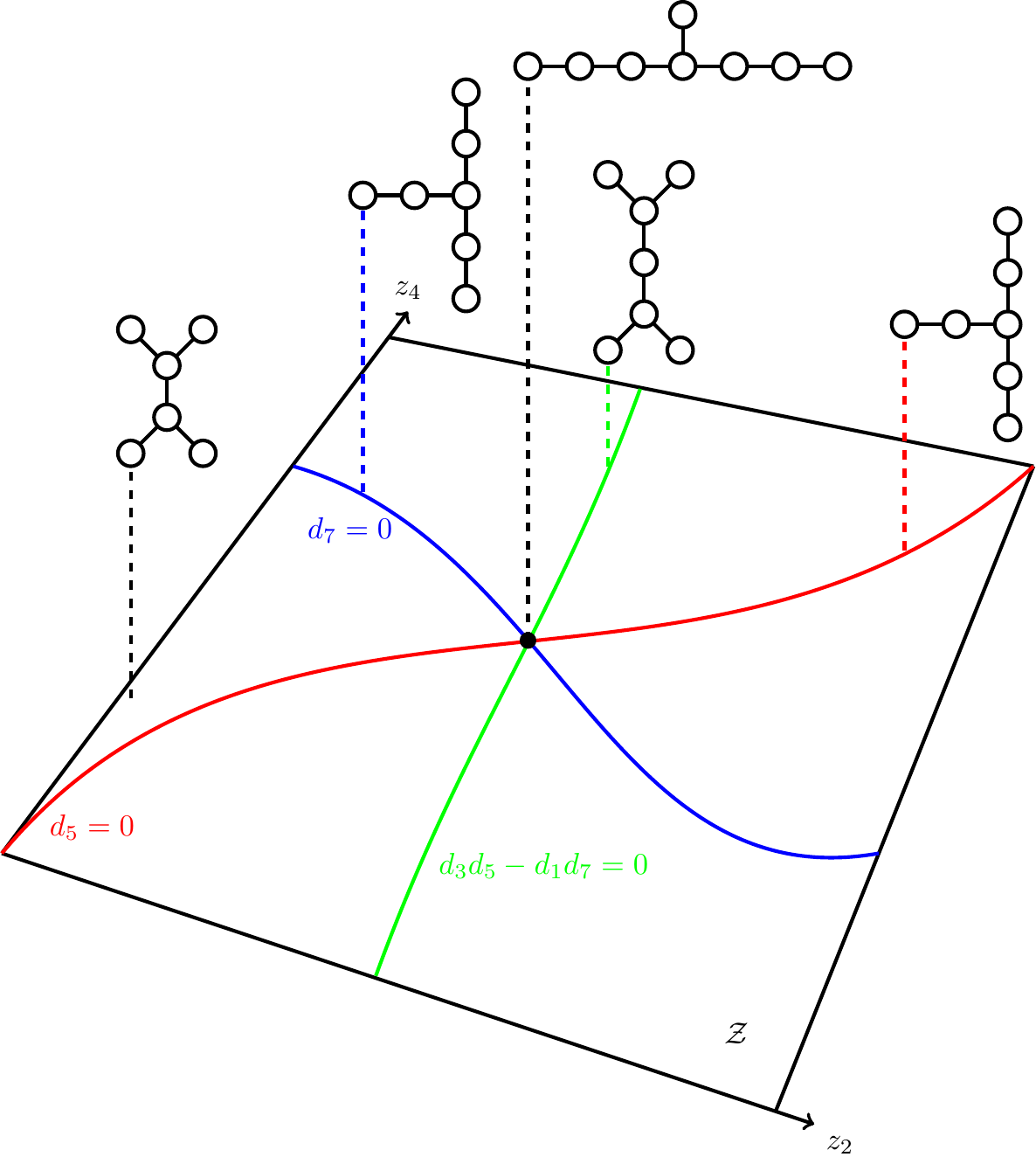}
\end{center}
\caption{\label{fig:Yukawa} In a complex three-dimensional space, SO(10)
  matter is represented by curves in the plane $\Z$, the
  locus of the SO(10) gauge fields. The curves of the matter fields
${\bf 16}_{3/4}$ ($z_0 = d_5 = 0$), ${\bf 16}_{-1/4}$ ($z_0 = d_7 =
0$) and ${\bf 10}_{-1/2}$ ($z_0 = d_1d_7 - d_3d_5 = 0$) intersect at
the codimension-three E$_7$ singularity  $z_0 = d_5 = d_7 = d_1d_7 - d_3d_5 = 0$.}
\end{figure}

SO(10) gauge fields live in a space of real dimension eight, defined by
a GUT divisor of codimension one. Similarly, 
SO(10) matter is located in a six-dimensional subspace defined by
the intersection of two divisors. 
Once we extend the CY threefold
considered so far to a CY fourfold, the matter points become matter
curves which can intersect in the compact dimensions,  leading to
the generation of Yukawa couplings. For the considered model this
generic pattern is illustrated in Figure~\ref{fig:Yukawa}.

%%%%%%%%%%%%%%%%%%
\subsection{Anomaly cancellation}
\label{sec:anomexample}

With the matter spectrum at hand we can check the vanishing of the irreducible anomalies as well as the factorization of the remaining anomaly polynomial\footnote{For the normalization of the anomaly polynomial as well as its factorization we use the conventions of \cite{Park:2011wv,Morrison:2012ei}.} $\mathcal{I}_8$. We compute the neutral singlets in \eqref{H12multis} from the Euler and Hodge numbers of the threefold,
\begin{align}
(h^{1,1},h^{2,1})_\chi = (9,51)_{-84} \, ,
\end{align}
which we evaluated using SAGE and which coincides with the general formula given in Appendix~\ref{App:classification}.
As expected, the nine K\"ahler deformations come from the six Cartan
generators of the gauge group, one from the torus and two from the base.

The theory contains $V = 46$ vector multiplets accounting for the
gauge fields of the group $\sou$, $H = 290$ hypermultiplets from the
charged and uncharged fields in \eqref{H12multis}, and a single tensor
multiplet, $T = 1$. The irreducible gravitational part of the anomaly
polynomial then reads
\begin{equation}
\begin{split}
\mathcal{I}_8 \supset& -\frac{1}{5760} (H - V + 29 T -273) \left( \text{tr} R^4 + \tfrac{5}{4} (\text{tr} R^2)^2 \right)  \\
=& - \frac{1}{5760} (290 - 46 + 29 - 273) \left( \text{tr} R^4 + \tfrac{5}{4} (\text{tr} R^2)^2 \right) = 0 \,,
\end{split}
\end{equation}
i.e., it vanishes for the given field content.

Similarly, we can evaluate the irreducible non-Abelian part of the
anomaly polynomial. Rewriting the traces of adjoint representation, $\text{Tr}$,
and spinor representation, $\text{tr}_{\mathbf{16}}$, in terms of
$\text{tr}_{\mathbf{10}} \equiv \text{tr}$,
\begin{equation}
\begin{split}
\tr_{\mathbf{16}} \tilde{F}^2 & = 2 \, \tr \tilde{F}^2 \,, \quad 
\tr_{\mathbf{16}} \tilde{F}^4  = - \tr \tilde{F}^4 + \tfrac{3}{4}
(\tr \tilde{F}^2)^2\,, \\
\text{Tr} \tilde{F}^2 &= 8 \, \tr \tilde{F}^2 \,, \quad
\text{Tr} \tilde{F}^4 = 2 \tr \tilde{F}^4 + 3 (\tr \tilde{F}^2)^2 \,,
\end{split}
\label{SOgrouptheory}
\end{equation}
we find
\begin{equation}
\begin{split}
\mathcal{I}_8 &\supset \frac{1}{24} \left( \text{Tr} \tilde{F}^4 - 4 \, \text{tr}_{16} \tilde{F}^4 - 6 \, \text{tr} \tilde{F}^4 \right) \\
&= \frac{1}{24}  (2  + 4  - 6) \, \text{tr} \tilde{F}^4 + \frac{1}{24} ( 3 - 3 ) ( \text{tr}\tilde{F}^2)^2= 0\,,
\end{split}
\end{equation}
where $\tilde{F}$ denotes the SO(10) field strength. We see that for the given matter spectrum even the reducible part vanishes. The remaining contributions to the anomaly polynomial evaluated with the spectrum \eqref{H12multis} are
\begin{align}
\mathcal{I}_8 = - \frac{1}{16} (\text{tr} R^2)^2 + \frac{1}{16} \text{tr} R^2 \, \text{tr} \, \tilde{F}^2 + \frac{99}{32} \text{tr} R^2 \, F^2 - \frac{3}{2} \text{tr} \tilde{F}^2 \, F^2 - \frac{153}{4} F^4 \,,
\label{AnomPolEx}
\end{align}
with U(1) field strength $F$. This anomaly polynomial factorizes,
\begin{align}
\mathcal{I}_8 = - \frac{1}{16} \left( \text{tr} R^2 - \text{tr} \tilde{F}^2 - \tfrac{51}{2} F^2 \right) \left( \text{tr} R^2 -24 F^2 \right) \,.
\end{align}
In the conventions of \cite{Park:2011wv} and Section~\ref{sec:bianomalycanc} this corresponds to the anomaly coefficients
\begin{align}
a = \begin{pmatrix} 2 \\ 2 \end{pmatrix} \,, \quad b = \begin{pmatrix} -1 \\ 0 \end{pmatrix} \,, \quad b_{11} = \begin{pmatrix} - \tfrac{51}{4} \\ -12 \end{pmatrix} \,,
\label{anomcoeff}
\end{align}
which match the expressions derived from the general approach in Section~\ref{sec:bianomalycanc}.

%%%%%%%%%%%%%%%%%%%%%%%%%%%%%%%%%%%%%%%%%%%%%%%%%%%%%%%%%%%%%%%%%%%%%%%%%
\section{Base-independent matter multiplicities}
\label{sec:bimultis}

In this section we extend the discussion for the specific base $B = \mathbb{F}_0$ given above to a general base space. This allows us to derive base-independent expressions for the matter multiplicities.

%%%%%%%%%%%%%%%%%%
\subsection{Counting singlets}
\label{sec:singletcount}

In Section~\ref{K3} we have discussed codimension-two singularities 
where the GUT symmetry SO(10) is enhanced to
SO(12) or E$_6$ and where matter multiplets are localized with
quantum numbers in the coset of SO(12)/SO(10) and E$_6$/SO(10),
respectively. However, already the torus fibration $\mathcal{E}$ admits
SU(2) singularities where 
SO(10) singlets occur as matter fields with
U(1) charge. Once the torus is fibered over a two-dimensional
base these singularities correspond to the loci of matter multiplets. 
For the tori corresponding to the $16$ ambient spaces,
codimension-two singularities have been
comprehensively analyzed in \cite{Klevers:2014bqa} and the U(1)
charges of the matter fields have been determined.
Adding an SO(10) top does not change the U(1) charges of the SO(10)
singlet fields but it does affect their multiplicities, which we shall
study in this section. 

Let us recall how one finds the loci of charged matter fields in
the case without an SO(10) top, following the discussion in \cite{Klevers:2014bqa}.
In a first step one has to
rewrite the torus in Weierstrass form. The elliptic curve obtained from
$F_3$ has one rational point with coordinates $(x_1,y_1,z_1)$. 
In this case the Weierstrass form can be written as \cite{Morrison:2012ei}
\begin{align}
F &= -y^2 + x^3 + fx +g =0 \,, \nonumber\\
f &= c - \hat{x}_1^2\,, \quad g = - c \hat{x}_1 - \hat{y}_1^2\,. \label{fgsing}
\end{align}
Here $c$ is a constant and we have used  a $\mathbb{C}^*$-action, 
$(x,y,z) \rightarrow (\hat{x}, \hat{y}, 1) = (x/z^2,y/z^3,1)$, such that the coordinates of the rational point are
$(\hat{x}_1, \hat{y}_1, 1)$.
A singular point $(x,y)$, with
\begin{align}
F = \frac{\partial F}{\partial x} = \frac{\partial F}{\partial y} =
0\,, \nonumber
\end{align}
occurs if the discriminant
\begin{align} \label{Deltasing}
\Delta &= 4 f^3 + 27 g^2 \nonumber\\
&= (4 c - \hat{x}_1^2) (c + 2 \hat{x}_1^2)^2  + 54
\hat{y}_1^2 c \hat{x}_1  + 27 \hat{y}_1^4     
\end{align}
vanishes. This happens for 
\begin{align}\label{singsing}
\hat{y}_1 = c + 2 \hat{x}_1^2 =  0 
\end{align}
at $(\hat{x},\hat{y}) = (\hat{x}_1,0)$. From Eqs.~\eqref{fgsing} and \eqref{Deltasing} we
infer that at this point the torus has an $\text{Ord}(f,g,\Delta) = (0,0,2)$
singularity associated with the group SU(2). As
expected we have a codimension-two singularity as in the case of the
matter loci discussed in Section~\ref{K3}.

For a fibration of the torus over a two-dimensional base the two
conditions \eqref{singsing} define matter curves. Replacing $c$ by the
function $f$ (see Eq.~\eqref{fgsing}), and going back to homogeneous coordinates,
the two conditions can be written as
\begin{align}
\label{eq:Codim2Sing}
y_1 = z_1^4 f + 3 x_1^2   =0\, . 
\end{align}
The coordinates $x_1$, $y_1$, $z_1$ and $f$ are known
functions of the coefficients $s_1,\ldots,s_9$ \cite{Klevers:2014bqa}. The conditions
\eqref{eq:Codim2Sing} imply that two polynomials, $P_1(s_i)$ and
$P_2(s_i)$, vanish. To find the corresponding roots it is helpful to consider $P_1$
and $P_2$ as generators of a codimension-two ideal and to decompose
this into irreducible prime ideals. One finds three prime ideals,
$I_1$, $I_2$ and $I_3$ whose zeros correspond to the loci $V(I_1)$,
$V(I_2)$ and $V(I_3)$ of singlets with charge $3$, $2$ and $1$, respectively.
For the loci corresponding to the prime ideals $I_1$ and $I_2$ one
obtains \cite{Klevers:2014bqa}: 
\begin{align}
\begin{array}{c|l}\label{ideals1}
\text{singlet} & \qquad\qquad\text{constraint}  \\ \hline
\mathbf{1}_3 & V(I_1): s_8 = s_9 = 0  \\ 
\mathbf{1}_2 & V(I_2): 
s_4 s_8^3 -s_3 s_8^2 s_9 + s_2 s_8 s_9^2 -s_1 s_9^3 \\ 
&\qquad\qquad =s_7 s_8^2 + s_5 s_9^2 - s_6 s_8 s_9=0 \\
&\qquad\qquad (s_8,s_9) \neq (0,0) 
\end{array}
\end{align}
Since the generators of the prime ideal $I_3$ are polynomials of high
order, the determination of the corresponding zeros is technically
nontrivial. This problem will be solved in the next section by
unhiggsing 
the $\sou$ fiber to an $\sou^2$ fiber.

In order to count the number of charge-two singlets one has to
determine how often the ideal $I_1$ is contained in $I_2$. This can be
done by means of
the resultant technique \cite{Cvetic:2013nia}. For the two polynomials
$Q_1(s_8,s_9)$ and $Q_2(s_8,s_9)$ of the ideal $I_2$ (see \eqref{ideals1})
one defines the resultant with respect to $s_8$,
 \begin{align}
 R = \textit{Res}_{s_8}( Q_1,Q_2) \, ,
 \end{align}
which is a polynomial in $s_9$. The resultant has the property that for every root $s_9 = \beta$ of $R$
there exists a value $s_8 = \alpha$ with $Q_1(\alpha,\beta) = Q_2(\alpha,\beta) = 0$. 
The explicit expression for the resultant reads
\begin{equation}
\begin{split}
 R = & s_9^6 (s_4^2 s_5^3 - s_3 s_4 s_5^2 s_6 + s_2 s_4 s_5 s_6^2 - s_1 s_4 s_6^3 + 
  s_3^2 s_5^2 s_7 - 2 s_2 s_4 s_5^2 s_7  - s_2 s_3 s_5 s_6 s_7 \\
& + 3 s_1 s_4 s_5 s_6 s_7 +
   s_1 s_3 s_6^2 s_7 + s_2^2 s_5 s_7^2 - 2 s_1 s_3 s_5 s_7^2 - s_1 s_2 s_6 s_7^2 + 
  s_1^2 s_7^3)\,.
\end{split}
 \end{equation}
Hence, $R$ has a root of order $6$ at $s_9 = 0$, with $Q_1(0,0) = Q_2(0,0) =
0$. Correspondingly, the ideal $I_1$ is contained six times in the
ideal $I_2$. The singlet multiplicities are determined by the
intersection numbers of the base divisors. For the base $\mathbb{F}_0 =
\CP\times \CP$ of the previous section one finds
$n[\mathbf{1}_3] = [s_8]\cdot [s_9]$, $n[\mathbf{1}_2] = [Q_1]\cdot [Q_2]
- 6 [s_8]\cdot [s_9]$.

Adding the SO(10) top to the fiber changes the singlet multiplicities. 
As discussed in Section~\ref{K3}, the coefficients $s_i$ now depend on the
base coordinates, $s_i = z_0^{n_i} d_i(z_0,z_1)$. This implies that
the ideals \eqref{ideals1} for the singlet localization are modified:
 \begin{align}\label{ideals2}
 \begin{array}{c|l|c}
 \text{singlet} & \qquad\qquad\text{constraint} & \text{multiplicity} \\ \hline
 \mathbf{1}_3  & V(I_1): d_8 = d_9 = 0 & 2 \\ 
 \mathbf{1}_2 & V(I_2): 
  d_4 d_8^3 z_0^2 -d_3 d_8^2 d_9 z_0 + d_2 d_8 d_9^2 z_0 -d_1 d_9^3  \\ 
 &\qquad\qquad  =d_7 d_8^2 z_0 + d_5 d_9^2 - d_6 d_8 d_9 z_0=0 \\
 &\qquad\qquad   (d_8,d_9) \neq 0, (z_0, d_5) \neq 0, (z_0, d_9) \neq 0  & 36
\end{array}
\end{align}
Now the two polynomials $Q_1$ and $Q_2$ of the ideal $I_2$ also depend
on $z_0$. Note that the two polynomials, after imposing the factorization, factor out powers of $z_0$ by which we have to divide, as this leads to unwanted solutions over $z_0 = 0$ where SO(10) charged multiplets are localized. In order to find the number of singlets with charge
$q=2$, we have to subtract the number of solutions of $d_8 = d_9 =0$ as
well as those of $z_0 = d_9 = 0$, $z_0 = d_5 = 0$, $z_0 =
d_7 = 0$ and $z_0 = d_3d_5 - d_1d_7 = 0$, which correspond to SO(10)
matter (see Table~\ref{singularities}). The evaluation of the resultant 
with respect to $z_0$ yields
\begin{align}\label{resz0}
\textit{Res}_{z_0} (Q_1,Q_2) =   d_8^2 d_9^3 \hat{R}\,,
\end{align}
where $\hat{R}$ is a non-factorizable polynomial of
degree five in the $d_i$. We conclude that the locus 
$z_0 = d_9 = 0$ is contained three times in the ideal $I_2$.\footnote{Note that also $z_0 = d_8 = 0$
appears to order two, but it does not correspond to an SO(10) matter
locus and therefore no subtraction is needed.}.
Analogously, one can calculate the resultant with respect to $d_9$, which is given by
\begin{align}\label{eq:resd9}
\textit{Res}_{d_9}(Q_1,Q_2) = & d_8^6 z_0^3(d_3^2 d_5^2 d_7 - 2 d_1 d_3 d_5 d_7^2 + d_1^2 d_7^3 + d_4^2 d_5^3 z_0  -  d_3 d_4 d_5^2 d_6 z_0\nonumber\\
  & - 2 d_2 d_4 d_5^2 d_7 z_0 - d_2 d_3 d_5 d_6 d_7 z_0 + 
   3 d_1 d_4 d_5 d_6 d_7 z_0 + d_1 d_3 d_6^2 d_7 z_0  \nonumber\\ 
   &+ d_2^2 d_5 d_7^2 z_0 -    d_1 d_2 d_6 d_7^2 z_0 + d_2 d_4 d_5 d_6^2 z_0^2 - d_1 d_4 d_6^3 z_0^2)
\end{align}
The factor $d_8^6z_0^3$ implies one solution $d_8 = d_9 = 0$ to order
six, as in the case without SO(10) top, and a second
solution $z_0 = d_9 = 0$ to order three, which is consistent with the
resultant \eqref{resz0}. For the base $\mathbb{F}_0 = \CP\times \CP$
the number of charge-two singlets is then given by
\begin{align}
n[\mathbf{1}_2] = [Q_1]\cdot [Q_2] - 6 [d_8]\cdot [d_9]  - 3 [z_0]\cdot[d_9] = 36\,,
\end{align}
where we have used Eqs.~\eqref{P1divisors} - \eqref{diH12}.

%%%%%%%%%%%%%%%%%% 
\subsection{Parametrizing the base dependence}
\label{sec:basedep}

So far we have expressed singlet multiplicities in terms of
intersection numbers of the base divisors $[d_i]$ and $[z_0]$. In order to
determine the multiplicities one has to specify a base and calculate the
intersection numbers. A convenient parametrization of the base
dependence has been given in \cite{Cvetic:2013nia,Cvetic:2013uta}.
It has been used in the classification of all toric hypersurface
fibrations \cite{Klevers:2014bqa}, and we shall also use it in our
analysis of all toric 6d F-theory vacua with SO(10) gauge symmetry.

The toric ambient space $X_{F_3}$ has four coordinates, $u$, $v$, $w$ and $e_1$,
and the polynomial $p_{F_3}$ depends on nine coefficients $s_1,\ldots
s_9$. For a fibration over a two-dimensional base all these quantities
become functions of the base coordinates. Equivalently, one can use
the two sections $s_7$ and $s_9$ to parametrize the base dependence. Furthermore, by means of two
$\mathbb{C}^*$-actions one can achieve that only two coordinates, $u$
and $v$, depend on $s_7$ and $s_9$, and that this dependence is linear.
In terms of divisors, one can demand
\cite{Cvetic:2013nia,Cvetic:2013uta}:
\begin{align}
[u] \rightarrow H + \mathcal{S}_9 - K_B^{-1}\,, \quad
[v] \rightarrow H + \mathcal{S}_9 - \mathcal{S}_7\,. \label{divisorshift}
\end{align}
Here $H$ is the hyperplane class of the ambient space $\mathbb{P}^2$ of the fiber $\mathcal{E}$,
$\mathcal{S}_{7,9} = [s_{7,9}]$ and $K_B^{-1}$ is the
anticanonical divisor class of the base, i.e.\ the sum of all base divisors.
The base dependence of the divisors $[s_i]$ is determined by the
Calabi-Yau condition, i.e.\ the vanishing of the first Chern class,
which reads in terms of divisors 
\begin{align}\label{CYc}
[u] + [v] + [w] + [e_1] + K_B^{-1} - [p_{Y_3}] = 3H + 2[e_1] - \Sa +
2 \Sb - [p_{Y_3}] \sim 0\,,
\end{align}
where the divisor $[p_{Y_3}]$ cuts out the CY threefold from the
ambient space. The polynomial $p_{Y_3}$ is a sum of nine terms all of
which have to belong to the same divisor class. Using \eqref{p1} for
the polynomial $p_{Y_3}$ and assuming a factorization with respect to $z_0$, i.e.\ $s_i = z_0^{n_i} d_i$,
one obtains from the Calabi-Yau condition \eqref{CYc} the relations
 \begin{align}\label{bidivisors}
 \begin{array}{cll}
& [d_1] \sim 3 K_B^{-1} - \mathcal{S}_7 - \mathcal{S}_9 - n_1\mathcal{Z}\,, &
 [d_2] \sim 2 K_B^{-1} - \mathcal{S}_9 - n_2 \mathcal{Z}\,, \\
& [d_3] \sim K_B^{-1} + \mathcal{S}_7 - \mathcal{S}_9 - n_3 \mathcal{Z}\,, &
[d_4] \sim 2 \mathcal{S}_7 - \mathcal{S}_9 - n_4 \mathcal{Z}\,, \\
& [d_5] \sim 2 K_B^{-1} - \mathcal{S}_7   - n_5 \Z\,, & 
[d_6] \sim K_B^{-1}   - n_6 \mathcal{Z}\,, \\ 
& [d_7] \sim  \mathcal{S}_7 -   n_7 \mathcal{Z}\,,  & 
[d_8] \sim K_B^{-1} - \mathcal{S}_7 + \mathcal{S}_9 - n_8 \mathcal{Z}\,, \\ 
& [d_9] \sim  \mathcal{S}_9 - n_9 \mathcal{Z} \,.
 \end{array}
 \end{align}
For $n_i = 0$ these relations reduce to the expressions for the sections
$s_i$ obtained in \cite{Cvetic:2013nia,Cvetic:2013uta}.
Given the loci of the matter field representations, we can now list
their multiplicities in terms of the base divisor classes\footnote{For simplicity we omit the dot indicating the intersection product of two divisor classes.} $K_B^{-1}$,
$\Sa$, $\Sb$ and $\Z$:
\begin{align}\label{bimultis}
\begin{array}{l|c|c}
\text{representation} & \text{locus} & \text{multiplicity}  \\ \hline
\qquad\mathbf{10}_{-1/2}  & z_0 = d_3 d_5 - d_1 d_7 = 0 & (3 K_B^{-1} - \mathcal{S}_9 - 2\Z)\Z \\
\qquad\mathbf{10}_{3/2}  & z_0 = d_9 = 0 &  \mathcal{S}_9\Z \\
\qquad\mathbf{16}_{3/4}  & z_0 = d_5 = 0 & (2 K_B^{-1}-\mathcal{S}_7)\Z  \\
\qquad\mathbf{16}_{-1/4}  & z_0 = d_7 = 0 & (\mathcal{S}_7 - \Z)\Z  \\ \hline 
\qquad\mathbf{45} & z = 0 &  1- (K_B^{-1}-\Z)\Z/2 \\ \hline 
\qquad\mathbf{1}_3 & V(I_1) & (K_B^{-1}-\mathcal{S}_7+\mathcal{S}_9)\mathcal{S}_9  \\ \hline
\qquad\mathbf{1}_2 & V(I_2) & \begin{array}{l}6 (K_B^{-1})^2 +
    K_B^{-1} (-5 \mathcal{S}_7 + 4 \mathcal{S}_9 - 2 \Z)\\ +\Sa^2 + \mathcal{S}_7 (2 \mathcal{S}_9 + \Z) - 
 \mathcal{S}_9 (2 \mathcal{S}_9 + 5 \Z) \end{array} \\ \hline
\qquad\mathbf{1}_1 & V(I_3) & \begin{array}{l} 12 (K_B^{-1})^2 + K_B^{-1} (8 \mathcal{S}_7 - \mathcal{S}_9 - 25
  \Z)\\- \Sb^2 - 4 \mathcal{S}_7^2 + 6 \Z^2 + 
 \mathcal{S}_7 (\mathcal{S}_9 + 4 \Z) \end{array} \\ \hline
\qquad T & &  9-(K_B^{-1})^2  \\  \hline
 \end{array}
 \end{align}
The loci $V(I_1)$ and $V(I_2)$ are given in \eqref{ideals2}, the locus
$V(I_3)$ will be determined in the following section. 

Given the base-independent multiplicities it is straightforward to
compute the matter multiplicities for the base $\mathbb{F}_0 = \CP\times \CP$
that we considered in the previous section. Comparing the expression
for the divisors given in Eqs.~\eqref{bidivisors} with
Eq.~\eqref{diH12} one obtains
\begin{align}
K_B^{-1} = 2H_1 + 2H_2 = \Sb \,, \quad \mathcal{S}_7 =3 H_1 + 4 H_2
\,, \quad  \Z = H_1 \,. 
\label{basechoiceexample}
\end{align}
With the intersection numbers \eqref{interbase} and the
base-independent expressions \eqref{bimultis}
one then finds the matter multiplicities listed in \eqref{H12multis}.

%%%%%%%%%%%%%%%%%%
\subsection[Unhiggsing the fiber to \texorpdfstring{SO(10)$\times$U(1)$^2$}{SO(10) x U(1) x U(1)}]{Unhiggsing the fiber to \texorpdfstring{SO(10)$\boldsymbol{\times}$U(1)$^2$}{SO(10) x U(1) x U(1)}}
\label{sec:ExampleUnhiggs}
 
Our computing power is not sufficient to directly evaluate the resultant of the
ideal $I_3$. Hence, we use an elegant alternative way to obtain the multiplicities of the charged singlets
via unhiggsing SO(10)$\times$U(1) to SO(10)$\times$U(1)$^2$. To
this end one enlarges the ambient space $dP_1$ to $dP_2$ by adding another blow-up point in the polygon of the fiber.
The additional vertex in Figure~\ref{fig:SO10Top} can be chosen as\footnote{In order to match the conventions in \cite{Klevers:2014bqa} we name $e_1: (1,0,0,0)$ and $e_2: (0,-1,0,0)$.} 
$(1,0,0,0)$. The polygon $F_3$ is then
changed to $F_5$. It is straightforward to determine the dual polytope
and the polynomial defining the torus, 
\begin{equation}\label{pf5}
\begin{split}
p_{F_5} = &s_1 u^3 e_1^2 e_2^2 + s_2 u^2 v e_1 e_2^2 + s_3 u v^2 e_2^2
+ s_5 u^2 w e_1^2 e_2 \\
&+ s_6 u v w e_1 e_2 + s_7 v^2 w e_2 + s_8 u w^2 e_1^2 + s_9 v w^2 e_1\, .
\end{split}
\end{equation}
Compared to \eqref{p1} the polynomial $p_{F_5}$ depends on the
additional coordinate and the term proportional to $s_4$ is
missing. This is due to the fact that the polygon dual to $F_5$ has
less vertices than the one dual to $F_3$, which leads to one term
less in the associated polynomial. The elliptic curve $\mathcal{E} =
\{p_{F_5} = 0\}$ has three toric rational points, i.e.\ intersections
with the hypersurface, which read in terms of  the coordinates $[u:v:w:e_1:e_2]$:
\begin{align}
\begin{split}\label{eq:sectionsF5}
\hat{s}_0 &= D_{e_2}\cap \mathcal{E}: \quad [s_9:-s_8:1:1:0]\,,\\
\hat{s}_1 &= D_{e_1} \cap \mathcal{E}: \quad [s_7:1:-s_3:0:1]\,,\\
\hat{s}_2 &= D_u \cap \mathcal{E}: \, \quad [0:1:1:s_7:-s_9]\,.
\end{split}
\end{align}

Again, the $s_i$ are specialized coefficients that depend on SO(10) fiber coordinates as well as on the base that we will specify in a moment. The Hodge numbers of the above elliptically fibered threefold with the SO(10) top are given by
\begin{align}
(h^{1,1},h^{2,1})_\chi = (10,48)_{-76} \, .
\end{align}
Hence, we indeed get one additional $(1,1)$-form corresponding to the additional U(1) that we traded for 8 complex structure moduli. 

Adding the SO(10) top and the base $\mathbb{F}_0 = \CP\times\CP$, the
coefficients $s_i$ become functions of the additional
coordinates. These are identical to the ones given in Eq.~\eqref{p2b},
except for $s_4$ which is now missing. In particular one again finds
the  $\text{Ord}(f,g,\Delta) = (2,3,7)$ singularity in the base
coordinate $z_0$ for the Weierstrass form of the tuned $K3$ manifold (see \eqref{fgz0}, \eqref{deltaz0}),
\begin{align}
 f &= z_0^2 \left( -\frac{1}{3} d_5^2 d_7^2 - \frac{1}{2} z_0 R_1 + \mathcal{O}( z_0^2)\right)\,, \nonumber\\
 g &= z_0^3 \left(-\frac{2}{27} d_5^3 d_7^3 + z_0 R_2
   +\mathcal{O}(z_0^2)\right)\,, \nonumber \\
\Delta &= z_0^7 \left( - d_5^3 d_7^3 (d_3 d_5 - d_1 d_7)^2 d_9^2 + z_0
  R + \mathcal{O}(z_0^2) \right)\,. \nonumber
\end{align}
As expected, the gauge group SO(10) is unchanged and also the SO(10)
matter multiplets occur at the same codimension-two loci. Due
to the three rational points we can now construct two Shioda maps,
corresponding to two U(1) factors. It is straightforward to compute
their intersections with the matter curves yielding their U(1)
charges. We obtain:
\begin{align}\label{reps2U1s}
\begin{array}{l|c}
\text{representation} & \text{locus}   \\ \hline
\qquad\mathbf{10}_{-1/2,0}  & z_0 = d_3 d_5 - d_1 d_7 = 0  \\
\qquad\mathbf{10}_{1/2,1}  & z_0 = d_9 = 0  \\
\qquad\mathbf{16}_{1/4,1/2}  & z_0 = d_5 = 0  \\
\qquad\mathbf{16}_{1/4,-1/2}  & z_0 = d_7 = 0  \\ \hline 
\end{array}
\end{align}

The loci of the charged singlets correspond to $\text{Ord}(f,g,\Delta) = (0,0,2)$
singularities associated with the group SU(2). Without SO(10)
top they have been determined in \cite{Klevers:2014bqa}. The effect of
the SO(10) top can be treated in the same way as for $F_3$. One finds six
charge combinations associated with six ideals $I_1, \ldots, I_6$ which 
determine the singlet loci. The ideals $I_4$, $I_5$, $I_6$ contain
several loci which have to be subtracted. The corresponding order can be determined by means of the
resultant method. A lengthy calculation for the loci of the six
charged SO(10) singlets yields:
\begin{align}
\begin{array}{c | c | c | c}
\text{rep} & \text{locus} & \text{contained loci} & \text{order} \\ \hline
\one_{1,-1} &  V(I_1): \{d_3 = d_7 = 0 \} & & \\ \hline
\one_{1,2} &  V(I_2): \{d_8 = d_9 = 0 \}  & &  \\ \hline
\one_{0,2} &   V(I_3): \{ d_9 = d_7 = 0 \}  & &  \\ \hline
\one_{1,1} &  \begin{array}{cc} V(I_4): &\{ d_1 d_9^2 + d_3 d_8^2
  z_0 - d_2 d_8 d_9 z_0 	 = \\ & d_7 d_8^2 z_0 - d_6 d_8 d_9 z_0 + d_5
  d_9^2= 0 \}  \end{array} & \begin{array}{c} 
 z_0 = d_9 = 0 \\ 
 d_9 = d_8 = 0      
\end{array} 
& \begin{array}{c} 2 \\ 4 \end{array} \\ \hline
\one_{1,0} & \begin{array}{cc} V(I_5): &\{  -d_3 d_6 d_7 +
  d_2 d_7^2 + d_3^2 d_9  = \\ &-d_3 d_5 d_7 + d_1 d_7^2 + d_3^2 d_8
  z_0 = 0 \}  \end{array} & 
\begin{array}{c} 
d_3 = d_7 = 0 
\end{array} &
\begin{array}{c} 4    \end{array} \\ \hline
\one_{0,1} & \begin{array}{cc} V(I_6): &\{d_5 d_7^3 d_8 d_9^2
  - d_5 d_6 d_7^2 d_9^3 + d_3 d_5 d_7 d_9^4 \\
 & + (d_1 d_7^2 d_9^4 d_8^3 - 2 d_6 d_7^3 d_8^2 d_9 + d_6^2 d_7^2 d_8 d_9^2 \\ 
 & + 2 d_3 d_7^2 d_8^2 d_9^2 - 2 d_3 d_6 d_7 d_8 d_9^3 + d_3^2d_8 d_9^4) z_0 \\ 
 & = d_5 d_7^3 d_9^2 + (d_7^4 d_8^2 - d_6 d_7^3 d_8 d_9\\ 
 & + d_3 d_6 d_7 d_9^3 - d_2 d_7^2 d_9^3 - d_3^2 d_9^4  z_0= 0 \}   \end{array}  & 
\begin{array}{c}   
z_0 = d_7 = 0 \\ z_0 = d_9 = 0 \\ d_3 = d_7 = 0 
\\ d_7 = d_9 = 0 
\\  d_8 = d_9 = 0   
\end{array} &
\begin{array}{c} 1 \\ 4 \\ 4 \\ 20 \\ 8 \end{array} \\ \hline
\end{array}
\label{chargedsingletmultiplicities}
\end{align}
The loci listed in the third column have to be subtracted with the
associated orders from $V(I_4)$, $V(I_5)$ and $V(I_6)$, respectively, to
obtain the loci of the charged singlets. This involves a set of
loci $\{d_i = d_j = 0\}$. Together with Eq.~\eqref{bidivisors} one
obtains the base-independent singlet orders given in
Table \eqref{eq:F5top2} of Appendix~\ref{App:classification}. 

Let us exemplify this procedure for the $\one_{1,0}$ singlets
explicitly. Summing up the intersection numbers of the divisor classes
at the various loci and subtracting the intersection numbers of the
loci they contain with the appropriate multiplicities as given by the orders obtained from the resultants,
one finds
\begin{align}
\one_{1,0}: \quad ( [d_3]+[d_6]+[d_7])([d_3]+[d_5]+[d_7]) -
4[z_0][d_7] \,.
\end{align}
Using the relations \eqref{bidivisors} one obtains the
base-independent multiplicity
\begin{align}
\one_{1,0}: \quad 6 (\Kbi)^2 - 2 \Ss^2 + \Sn^2 + 4 \Z^2 + 3 \Sn \Z + \Kbi (4 \Ss - 5 \Sn - 14 \Z) + \Ss (\Sn + 2 \Z) \, .
\end{align}
This is the multiplicity listed in Table~\ref{eq:F5top2} of Appendix~\ref{App:classification}.

Vacuum expectation values of singlets $\one_{1,-1}$ can higgs the CY
threefold with fiber $F_5$ to the one with fiber $F_3$
\cite{Klevers:2014bqa}. The symmetry U(1)$\times$U(1) is then broken to a
single U(1) with unbroken charge
\begin{align}
q = q_1 + q_2\,.
\end{align}
The SO(10) matter representations listed in \eqref{reps2U1s} then
become the ones given in \eqref{bimultis}. For the charged singlets one
has $\one_{1,2} \rightarrow \one_3$, \{$\one_{0,2}, \one_{1,1}\}
\rightarrow \one_2$ and \{$\one_{1,0}, \one_{0,1}\} \rightarrow
\one_1$. Using the relations \eqref{interbase}, \eqref{diH12} and the
intersection numbers given in Table~\ref{eq:F5top2}, we obtain the singlet
multiplicities listed in \eqref{H12multis}. Finally, from the
$\one_{1,-1}$ matter states with multiplicity
\begin{align}
n[\one_{1,-1}]=[d_3] [d_7]  = 4 \, , 
\end{align}
we obtain the three additional complex structure moduli in the higgsed geometry, after subtracting the Goldstone mode.
 
%%%%%%%%%%%%%%%%%%%%%%%%%%%%%%%%%%%%%%%%%%%%%%%%%%%%%%%%%%%%%%%%%%%%%%%%%
\section{Analysis of 6d toric SO(10) vacua}
\label{sec:clasvac}

In this section we discuss the general algorithm of our analysis
for all SO(10) tops listed in \cite{Bouchard:2003bu}. This procedure is exemplified  in
Section~\ref{sec:6dvac} and \ref{sec:bimultis} for the fiber ($F_3$, top 1).

%%%%%%%%%%%%%%%%%%
\subsection{Polytopes and tops of SO(10)}
\label{sec:polytop}
 
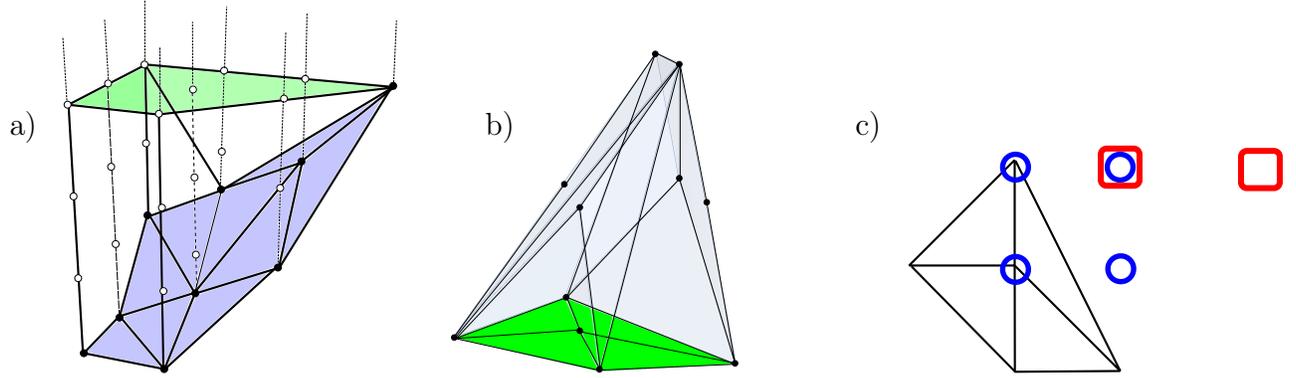
\begin{figure}
\begin{center}
\begin{picture}(0,170)
\put(-80,0){\includegraphics[scale=0.4]{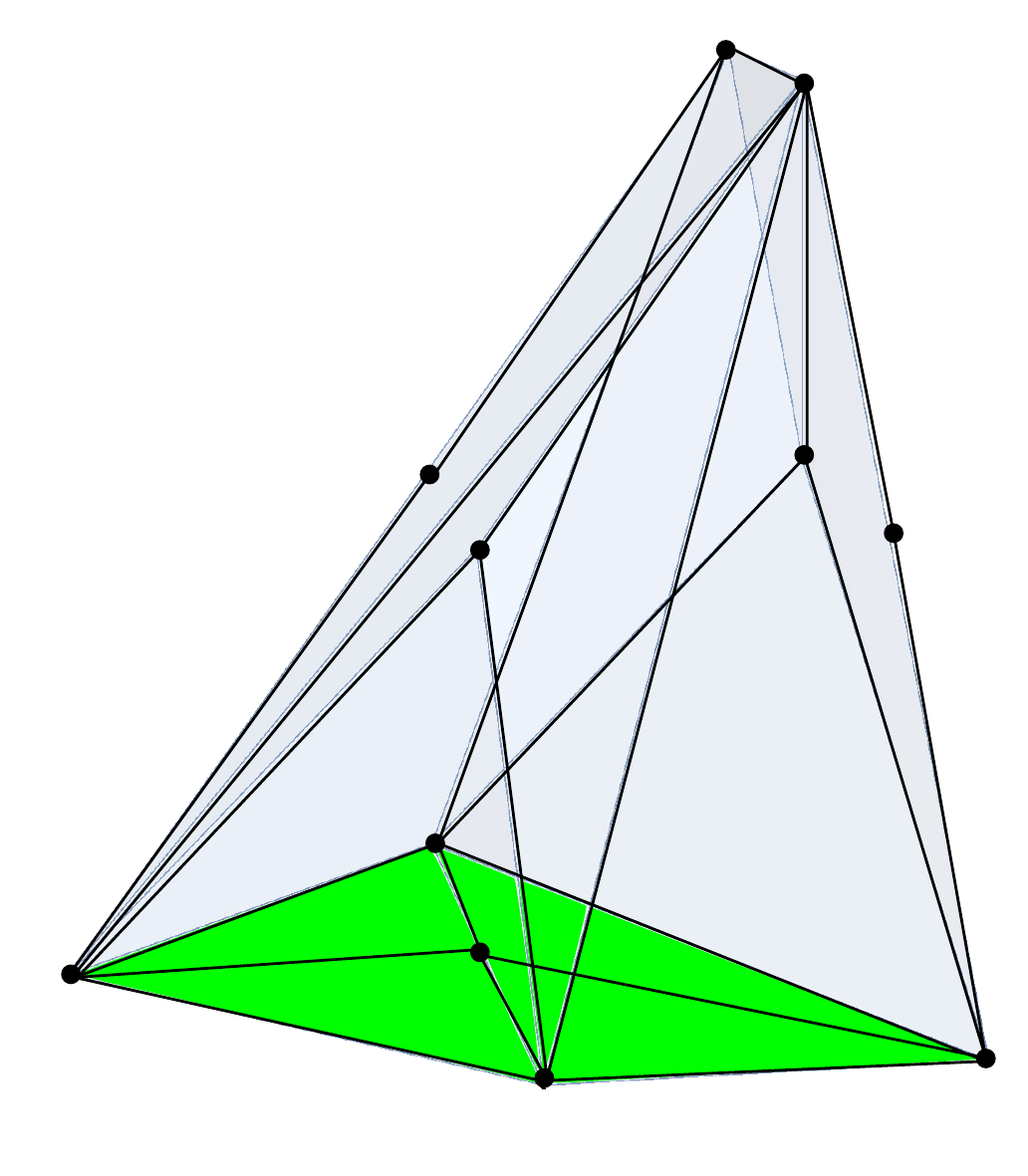}}
\put(100,10){\includegraphics[scale=0.5]{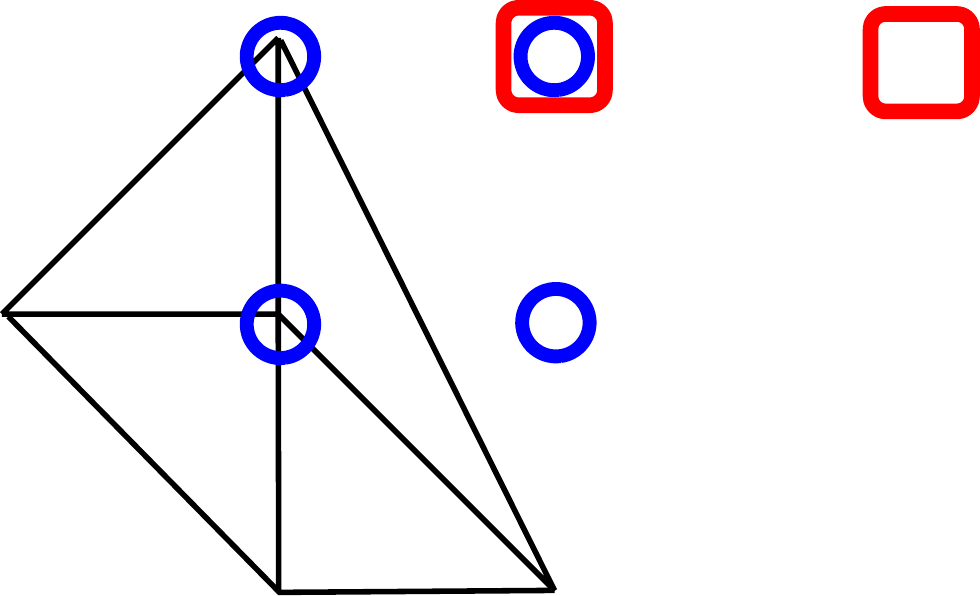}}
\put(-220,10){\includegraphics[scale=0.5]{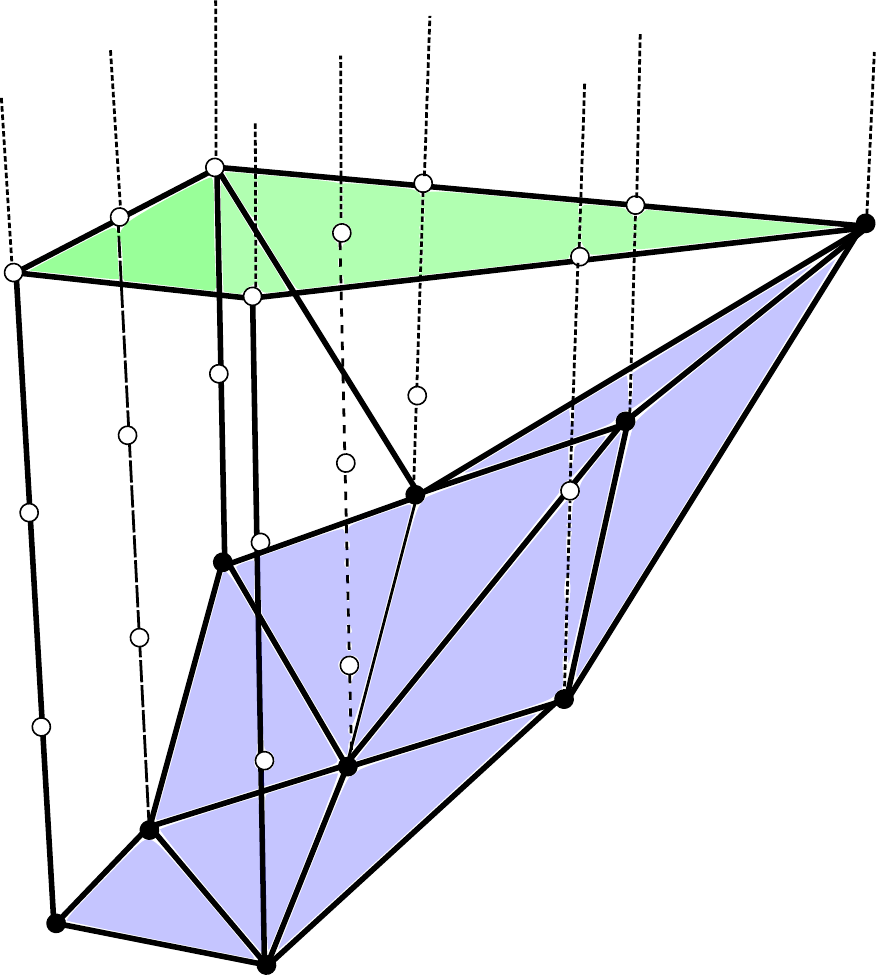}}
\put(-240,100){a)}
\put(-60,100){b)}
\put(80,100){c)}
\end{picture}
\end{center}
\caption{\label{fig:F3top1prism}Example ($F_3$, top 1). The dual prism is given in a) with $F_3^*$ at generic height (shaded in green); b) depicts the top polytope with $F_3$ at height 0 (shaded in green); c) is a 2d projection of b) where blue circles denote vertices at height 1 and red squares vertices at height 2.}
\end{figure}

Let us consider an elliptically fibered $K3$ hypersurface in a 3d
ambient space given by a reflexive lattice polytope $\Delta$ with
vertices $v_i = (v_i^1,v_i^2,v_i^3)$. From this point of view, a top  is a half-lattice polytope $\Diamond$ that
is obtained by slicing the $K3$ polytope $\Delta$ into two halves, a top
and a bottom. The top and bottom thus have a common face $F_0$ 
that contains the
origin as an inner point. This face $F_0$ itself is the ambient space
of a CY one-fold, i.e.\ a torus, which is the fiber over a
generic point in the base $\mathbb{P}^1$ of the full $K3$. Hence,
independent of the base the subpolytope  $F_0$ of $\Delta$ at height $v^3 = 0$ always encodes the generic torus fiber. The points at height $v^3>0$, however, represent resolution divisors $D_i$ that project onto the same point $z_0 = 0$ of the base $\mathbb{P}^1$,
\begin{align}
D_i  \xrightarrow{~\pi~}  \{ z_0 = 0 \} \, .
\end{align}
These $D_i$ are the resolution divisors of an ADE singularity in the fiber over the base locus $\pi_B ([z_0]) \sim \Z$. 
For a reflexive polytope we can define a top $\Diamond$ as a
lattice polytope whose vertices satisfy certain inequalities,
 \begin{align}\label{top}
\Diamond = \{ v \in \mathbb{Z}^3:
 \langle m_0 , v \rangle \geq 0 \, , \; \langle  m_i , v\rangle \geq -1\}\, , 
 \end{align}
for some  $ m_i  \in \mathbb{Z}^3 $.
By means of a GL$(3, \mathbb{Z})$ transformation we can always set $m_0=(0,0,1)$.
The face $F_0$ is a two-dimensional polygon at height zero given by the restriction
 \begin{align}
 F_0 = \{   v \in \Diamond: \langle m_0, v \rangle = 0 \} \, .
 \end{align}
For each reflexive polytope $\Delta$ of the $K3$ ambient space, there
is a dual polytope $\Delta^*$ defined by
 \begin{align}
 \Delta^* = \{ m \in \mathbb{Z}^3: \langle m, v \rangle \geq -1\; \forall \, v \in \Delta    \} \, .
 \end{align}
Analogously, one defines the dual $\Diamond^*$ of the top
$\Diamond$,
\begin{align}
\Diamond^*: \{ m \in \mathbb{Z}^3 : 
\langle m,v \rangle \geq -1\; \forall \, v \in \Diamond \} \,. \label{topdual}
\end{align}
For vertices\footnote{The relation to the notation in \cite{Bouchard:2003bu} 
is: $v_i^1 = \bar{x}_i$, $v_i^2 = \bar{y}_i$, $v_i^3 = \bar{z}_i$,
$m_j^1 = x_j$, $m_j^2 = y_j$, $m_j^3 = z_j$.} $v_s = (v^1_s,v^2_s,0) \in F_0$ one has $m^1v^1_s +
m^2v^2_s \geq 0$, which yields the two-dimensional dual $F_0^*$ of
$F_0$. Other vertices $v_t \in \Diamond$, $v_t \not\in F_0$ yield the
inequalities
\begin{align}
m^3 v^3_t \geq -1 - m^1v^1_t + m^2v^2_t \,.
\end{align}
With $v^3_t > 0$, this implies a lower bound for the third component
of $m$: $m^3 \geq m^3_{\text{min}}(m^1,m^2)$. Since there is no upper bound
on $m^3$, the dual of the top $\Diamond^*$ has the form of a prism with a cross section
given by $F_0^*$.
To summarize, a top $\Diamond$ over some polygon $F_i$ is dual to a half-infinite extended prism $\Diamond^*$ with $F_i^*$ at generic height and unique minimal height vertices $m^3_{\text{min}}(m^1,m^2)$
(see~Figure~\ref{fig:F3top1prism}).
In this way all tops have been classified in terms of $F_0^*$ and the
$m^3_{\text{min}}$ values of the half-open prisms \cite{Bouchard:2003bu}.

As first step of our analysis we use Eq.~\eqref{top} 
to construct the tops corresponding to the Lie algebra $D_5$, as listed in
\cite{Bouchard:2003bu}\footnote{Note that we included two tops over $F_8$ and one over $F_{12}$ which were classified as $B_5$ tops in \cite{Bouchard:2003bu}. However, a careful analysis of the dual edges shows that they correspond to the gauge algebra $D_5$.}. Next, we compute the gauge group and matter spectrum, as explained in the example in Section~\ref{sec:6dvac}. In the following
we describe the general algorithm and mention possible complications that occur in some models.

%%%%%%%%%
\subsubsection*{Base completion of the top}

We construct CY threefolds as hypersurfaces in toric varieties $\mathbb{P}_{F_i}^B$ with the fibration structure
\begin{align}
\begin{array}{ll}
\mathbb{P}_{F_i,\text{top}} \longrightarrow &\mathbb{P}_{F_i}^B (\mathcal{S}_7, \mathcal{S}_9, \mathcal{Z}) \, . \\
& \left\downarrow\rule{0cm}{0.5cm}\right. \pi \\
& B 
\end{array}
\end{align}
Here, the three divisor classes $\mathcal{S}_7$, $\mathcal{S}_9$ and
$\mathcal{Z}$ parametrize the fibration of the fibers
$X_{F_i}$ over a base with $\mathcal{Z}$ being the GUT divisor.

The hypersurface equation for the CY can be obtained using Batyrev's
construction \cite{Batyrev:1994hm}. 
From the top and its dual one obtains the polynomial for a non-compact
CY twofold,
\begin{align}\label{pDiamond}
p_{\Diamond} =\sum_{m_j \in \Diamond^*} d_j \prod_{v_i \in \Diamond}
x_i^{\langle m_j, v_i \rangle + 1 } = \sum_{m_j \in \Diamond^*} d_j
\left(\prod_{v_s \in F_0} x_s^{ \langle m_j,v_s\rangle+1}\right)  
\left(\prod_{v_t \in \Diamond , v^3 >0} x_t^{ \langle
    m_j,v_t\rangle+1}\right) \,.
\end{align}
The partial factorization shows that the structure of the
 hypersurface equation, which defines a torus in $F_0$ given by the coordinates $x_s$,  is preserved
(see \cite{Borchmann:2013hta}).  
For the top  coordinates $x_t$
with vertices $v_t^3 > 0$   we introduce the notation
\begin{align}
v_t: \{z_0, f_i , g_j \} \text{  for  } \{ z_0: \, (0,0,1) \,, \enspace f_i: \, (v^1_i ,v^2_i,1) \,, \enspace g_j: \, (v^1_j ,v^2_j,2) \} \, .
\end{align}
Here $D_0 = \{z_0 = 0\}$ is the base divisor whose dual curve
$\mathbb{P}^1_0$ corresponds to the affine node, whereas the
coordinates $f_i$ are at height $v_i^3 = 1$ and the $g_j$ are the `inner'
\SO{10} roots of height $v_j^3 = 2$. Note that these heights correspond to
the Dynkin multiplicities of the roots.
In our calculations we add a trivial bottom i.e.\ a vertex at height $v^3 = -1$ that completes the top
$\Diamond$ to a reflexive polytope $\Delta$, with a dual reflexive polytope
$\Delta^*$. The infinite sum over the vertices of $\Diamond^*$ now
becomes a finite sum over the vertices of $\Delta^*$,
\begin{align}\label{pDelta2}
p_{\Delta} =\sum_{m_j \in \Delta^*} d_j \prod_{v_i \in \Delta}
x_i^{\langle m_j, v_i \rangle + 1 } = \sum_{m_j \in \Delta^*} d_j
\left(\prod_{v_s \in F_0} x_s^{ \langle m_j,v_s\rangle+1}\right)  
\left(\prod_{v_t \in \Delta , v_t \not\in F_0} x_t^{ \langle
    m_j,v_t\rangle+1}\right) \,.
\end{align}
The hypersurface equation $p_\Delta = 0$ defines a compact CY twofold.
The  partial  factorization of the coordinates related to the equation that defines a torus in  $F_0$ is again
preserved. In Appendix~\ref{a:d6dvac} it is shown that this structure  also
remains once the base is extended to higher dimensions.

The base-dependence of the sections $d_i$ can be fixed by considering
the case without top, see Section~\ref{sec:basedep}.
The inclusion of the top then adds the additional coordinates $f_i$,
$g_j$ and the corresponding divisors $D_i$ that shift the fiber
coordinates. Using linear equivalences one has
\begin{align}
[x] \rightarrow [x] - \sum a_i D_i \, .
\end{align}
We take the point $z_0: (0,0,1)$ with divisor $D_0$, whose dual $\CP_0$ we choose as the affine node of the SO(10) Dynkin diagram. $D_0$ satisfies the linear equivalence
(see \eqref{GUTdivisor})
\begin{align}
\label{eq:ExtendedDynkinDivisor}
D_0 = - \sum_i l_i D_i \, + [z] , 
\end{align}
where $l_i$ are the Dynkin multiplicities and $[z]$ the divisor corresponding to $z = z_0 f_2 g_1^2 g_2^2 f_3 f_4$, respectively.

Finally, we want to fix the dependence of the sections $s_i$ on the
SO(10) GUT base divisor $\mathcal{Z}$ after inclusion of the top. Setting  
all $f_i =g_j= 1$, the sections take the form $s_i = z_0^{n_i} d_i$
which yields (see Section~\ref{sec:basedep})
\begin{align}
\label{eq:topFact}
 [s_i] \sim [d_i]
  + n_i \mathcal{Z}  \,.
\end{align}
The factors $n_i$ give the vanishing orders of the $s_i$ in $\mathcal{Z}$ and characterize
the spectrum of the top uniquely. 

In this work we mainly study torus fibers that are
cubic curves and blow-ups thereof. The generic 
cubic polynomial, corresponding to a torus in the polytope $F_1$, is
given by
\begin{equation}\label{cubic}
\begin{split}
p_{F_1} = 
&s_1 u^3 + s_2 u^2 v + s_3 u v^2 + s_4 v^3 + s_5 u^2 w \\
&+ s_6 u v w + s_7 v^2 w + s_8 u w^2 + s_9 v w^2 + s_{10} w^3 \, ,
\end{split}
\end{equation}
which has one monomial more than $p_{F_3}$.
Using adjunction, the parametrization of the torus divisors (see Section~\ref{sec:basedep}) is given by
\begin{align}
[u] \sim H + \mathcal{S}_7 -  K_B^{-1}  \, , \quad [v] \sim H +
\mathcal{S}_9 - \mathcal{S}_7 \, , \quad [w] \sim H \nonumber
\end{align}
with $H$ being the hyperplane class of the ambient space $\mathbb{P}^2$ of the fiber, and 
the  $s_i$ are given by \eqref{eq:topFact}. The base divisor classes of the sections $d_i$ are
given by 
\begin{align}\label{bidivisors2}
 \begin{array}{cll}
& [d_1] \sim 3 \Kbi - \mathcal{S}_7 - \mathcal{S}_9 - n_1\mathcal{Z}\,, &
 [d_2] \sim 2 \Kbi - \mathcal{S}_9 - n_2 \mathcal{Z}\,, \\
& [d_3] \sim \Kbi + \mathcal{S}_7 - \mathcal{S}_9 - n_3 \mathcal{Z}\,, &
[d_4] \sim 2 \mathcal{S}_7 - \mathcal{S}_9 - n_4 \mathcal{Z}\,, \\
& [d_5] \sim 2 \Kbi - \mathcal{S}_7   - n_5 \Z\,, & 
[d_6] \sim \Kbi   - n_6 \mathcal{Z}\,, \\ 
& [d_7] \sim  \mathcal{S}_7 -   n_7 \mathcal{Z}\,,  & 
[d_8] \sim \Kbi - \mathcal{S}_7 + \mathcal{S}_9 - n_8 \mathcal{Z}\,, \\ 
& [d_9] \sim  \mathcal{S}_9 - n_9 \mathcal{Z} \,, & [d_{10}] \sim 2 \mathcal{S}_9 - \mathcal{S}_7 - n_{10} \mathcal{Z} \, . 
 \end{array}
 \end{align}
Other fibers that are related to the cubic curve by a conifold
transition can be obtained by setting the respective section $d_i$ to
zero and using the above relations for the remaining divisors, as
discussed in more detail in Section~\ref{sec:trans}. However, the polygons $F_2$
and $F_4$ and their tops lead to biquadric and quartic polynomials, respectively, which cannot be reached by a transition from $p_{F_1}$ directly.
Hence, those curves differ in their general structure and we summarize their factorization, base-dependence and Weierstrass forms in Appendix~\ref{sec:WeierstrassForm}.

%%%%%%%%%%%%%%%%%%
\subsection{Spectrum computation}
\label{sec:speccomp}

The spectrum computation can be split into several steps (see Section~\ref{sec:6dvac} and \ref{sec:bimultis} for a detailed example). Here, we summarize the process and add some comments on several features that appear in different models.

%%%%%%%%%
\subsubsection*{SO(10) matter loci and charges}

Since the loci and charges of singlet matter fields are at $z_0 \neq 0$ we can use the results of \cite{Klevers:2014bqa} for them. For the \SO{10} charged matter, located over $\Z$ in the base, it is most beneficial to map the curve into its singular Weierstrass form, i.e.\ to use the expressions given in Appendix~\ref{sec:WeierstrassForm} and impose the factorization of $(f,g,\Delta)$ in the base coordinate $z_0$, which yields expressions of the form
\begin{equation}
\begin{split}
f =& z_0^2 \left(\mathcal{B}^2 \mathcal{C}^2 +  \mathcal{C}  z_0 R_1 + \mathcal{O}(z_0^2)\right)\,, \\
g =& z_0^3 \left( \mathcal{B}^3 \mathcal{C}^3 +  \mathcal{C}^2 z_0 R_2 + \mathcal{C} z_0^2 R_3 + z_0^3 R_4 + \mathcal{O}(z_0^4)   \right)\,, \\
\Delta =& z_0^7 \left(\mathcal{A}^2 \mathcal{B}^3 \mathcal{C}^5 +  \mathcal{C}^4 z_0 R_5 + \dots + z_0^5 R_9 + \mathcal{O}(z_0^6)  \right)\,,
\end{split}
\end{equation}
with $\mathcal{A},\mathcal{B},\mathcal{C}$ being reducible matter ideals relevant for the vanishing order in codimension two, and the $R_i$ being  some irreducible polynomials.
This corresponds to an \SO{10} locus over $\Z$ with matter loci given by the irreducible components of the $\mathcal{A},\mathcal{B},\mathcal{C}$, which we denote with a subscript $\mathcal{A}_i$, $\mathcal{B}_i$ and $\mathcal{C}_i$ with associated loci $V$. If those components vanish together with $z_0 = 0$, we obtain enhanced singularities and matter representations of SO(10), given by (see Section~\ref{sec:6dvac}):
\begin{align}
\begin{array}{c | c | c | c | c}
 & (f,g,\Delta) \text{ at } V \text{ over } z_0 = 0 & \text{ fiber type} & \text{rep} & \text{multiplicity} \\ \hline
\mathcal{A}_i=0 & (2,3,8) & I_2^* & \mathbf{10}_i & [\mathcal{A}_i] \mathcal{Z} \\ 
\mathcal{B}_i=0 & (3,4,8) & III & \mathbf{16}_i & [\mathcal{B}_i] \mathcal{Z}  \\
\mathcal{C}_i=0 & (4,6,12) & \text{non-min} & \text{SCP} & [\mathcal{C}_i] \mathcal{Z}
\end{array}
\end{align}
The 6d multiplicities are given by the intersection of the divisor classes associated with $z_0$ and the irreducible components $\mathcal{A}_i$, $\mathcal{B}_i$, $\mathcal{C}_i$ in the base. The type $\mathcal{C}$ ideals yield points with non-minimal singularities that are associated to SCPs. Note that the above factorization only allows us to deduce the non-Abelian representations. However, different components of the same type of ideal can have different Abelian charges, which  cannot be read off directly from the singular Weierstrass form.

The Abelian charges are obtained by imposing vanishing of the irreducible components $\mathcal{A}_i, \mathcal{B}_j$ in the resolved fiber and studying the splitting of the $\mathbb{P}^1_i$ curves into several irreducible $\mathbb{P}^1$'s that we identify with the matter nodes,
\begin{alignat}{2}
\mathcal{E} &\xrightarrow{~f_i = 0~}&&~\mathbb{P}^1_{i}\,, \\
\mathbb{P}^1_{i} &\xrightarrow{~\mathcal{A}_j; \,\mathcal{B}_j= 0~}&&~\mathbb{P}^{1}_{m_1} + \mathbb{P}^{1}_{m_2} + \ldots  \,,
\end{alignat}
and a subsequent evaluation of the intersection with the Shioda maps.

%%%%%%%%%
\subsubsection*{Shioda map and matter charges}

First, we identify the SO(10) Cartan matrix from the intersection of the resolution divisors $(C_{\SO{10}})_{i j}$ as in \eqref{eq:SO10Cartan} in a given triangulation.
Then, we compute an orthogonal basis of U(1) divisors using the Shioda map. In order to use the matter charges of SO(10) singlets as computed in \cite{Klevers:2014bqa}, we choose the zero-sections $\hat{s}_0$ and Mordell-Weil generators $\hat{s}_i$ as in \cite{Klevers:2014bqa} and include the SO(10) divisors as
\begin{align}
\sigma (\hat{s}_i) = [\hat{s}_i]- [\hat{s}_0] +  (\hat{s}_i-\hat{s}_0)\cdot \mathbb{P}^1_j (C^{-1}_{\SO{10}})_{jk} D_k \, ,
\label{eq:Shioda}
\end{align}
which orthogonalizes the U(1) generators and all other non-Abelian group factors, such that
\begin{align*}
\sigma (\hat{s}_i) \cdot \mathbb{P}^1_j = 0 \, \quad \forall \, i,j \, .
\end{align*}
We note that the appearance of the inverse of the non-Abelian Cartan matrix $C^{-1}$ generically leads to fractionally charged non-Abelian representations. This is the manifestation of a non-trivial embedding of some $\mathbb{Z}_n$ center of the non-Abelian gauge group $G$ into the U(1) factor \cite{Cvetic:2017epq, Grimm:2015zea, Grimm:2015wda}.
Hence, such a factor can lead to non-trivial group quotients and a global gauge group $G_X$ of the form
\begin{align}
G_X=\frac{\text{U(1)} \times G}{ \mathbb{Z}_p} \, ,
\label{eq:globalgauge}
\end{align}
where $p$ is some divisor of $n$ which we determine momentarily.
This fact is most easily seen by recalling that the U(1) charges $q$ of massless matter representations can be written in the form (see \eqref{eq:Shioda})
\begin{align}
\label{eq:shiodasimple}
q = l +  m \lambda \, .
\end{align}
Here, $l$ and $m$ are integers and $\lambda$ denotes the $\mathbb{Z}_n$ center charges of the representations quantized in units of $1/n$. It is readily confirmed that the U(1) charge spacing within the same $G$ representation is integral. In addition, if there is some non-trivial greatest common divisor $p$ of  $m$ and $n$ it might happen that only a $\mathbb{Z}_p$ subgroup of the full $\mathbb{Z}_n$ center of $G$ is modded out. Due to the form of the U(1) charge generator \eqref{eq:shiodasimple}, we can identify a quotient operator  $g_{p}$ by solving for the integer $l$ and exponentiating  as
\begin{align}
g_{p} = e^{2 \pi i (m \lambda  - q)}  \, .
\end{align}
This is a $\mathbb{Z}_p$ operator, that is constructed to be single-valued for all representations of the total gauge group and therefore can be viewed as the generator of the $\mathbb{Z}_p$ quotient appearing in the denominator of \eqref{eq:globalgauge}.

In our SO(10) analysis, we have the following center charges $\lambda$ of representations
\begin{align}
\begin{array}{c|cccc}
 &  \mathbf{1} & \mathbf{10} & \mathbf{16} & \mathbf{45}\\ \hline
\lambda & 0 & 1/2 & 1/4 & 0 \\
\end{array} \, .
\end{align}
The presence of the spinor representation reminds us that we actually have a Spin(10) group instead of an SO(10) group, whose $\mathbb{Z}_2$ center acts on the $\mathbf{10}$ representation. Note that Spin(10) has a $\mathbb{Z}_4$ center under which the spinor representation carries the minimal charge, which reflects the fact that it is a double cover of SO(10).

Hence, in the classification of global gauge groups, we can have two non-trivial cases: One where the full $\mathbb{Z}_4$ center is modded out and one, where only a $\mathbb{Z}_2$ subgroup of the center is modded out.  
All three cases do appear frequently and are identified via the charge of the spinor representations as in the following examples:
\begin{align}
\label{eq:GGexamples}
\begin{array}{c | c l c}
\text{top} & \text{spinor rep} & \text{global gauge group} \\ \hline
(F_3, \text{top 4}) & \mathbf{16}_ {0} & \text{SO(10)} \times \text{U(1)} \\
(F_3, \text{top 6}) & \mathbf{16}_ {1/2} & \text{SO(10)} \times \text{U(1)} / \mathbb{Z}_2 \\
(F_3, \text{top 1}) & \mathbf{16}_ {-1/4} & \text{SO(10)} \times \text{U(1)} / \mathbb{Z}_4 \\
\end{array}
\end{align}
In addition to the elliptic fibration, we also have genus-one fibrations based on $F_1, F_2 $ and $F_4$ that do not have sections but only multi-sections \cite{Braun:2014oya,Morrison:2014era, Anderson:2014yva, Cvetic:2015moa} that intersect the torus $\mathcal{E}$ several times; we denote this multiplicity by $k$
\begin{align}
[s^{(k)}] \cdot \mathcal{E} = k \, .
\end{align}
Those theories can be connected to U(1) theories via an unhiggsing similar to Section~\ref{sec:ExampleUnhiggs}. The higgsing process reveals the presence of a discrete symmetry $\mathbb{Z}_k$ induced by a Higgs field with non-minimal U(1) charge $k$. The $\mathbb{Z}_k$ generators corresponding to the multi-section $s^{(k)}$ can also be orthogonalized with respect to other non-Abelian group factors using a modified Shioda map,
\begin{align}
\sigma (s^{(k)}) =  [s^{(k)}] +   [s^{(k)}]\cdot \mathbb{P}^1_j  (C_{\SO{10}}^{-1})_{jk} D_k \, .
\end{align}
As for the gauge group U(1) above, the discrete gauge factors can mix with the center of the SO(10), leading to a modification of the global gauge group similar to \eqref{eq:globalgauge}.

We want to remark that theories based on the polytope $F_2$ are genus-one fibrations that generically admit a U(1) and $\mathbb{Z}_2$ gauge factor. Here the additional rational sections appears only in its Jacobian\footnote{This has also been observed in self-mirror genus-one fibrations with torsional sections in the context of complete intersection fibers \cite{Oehlmann:2016wsb}.}.
In this case, the Shioda map is generated by the difference of two linear inequivalent multi-sections.

After having fixed and orthogonalized all generators, the weight $\omega$ as well as U(1) and discrete charges $q_j$, $q^{(k)}$ of the matter state located on $\mathbb{P}^1_m$ can be computed by the intersections
\begin{align}
( \omega_i )_{q_j,q^{(k)}} =   ( \mathbb{P}^1_{m} \cdot D_i)_{(\mathbb{P}^1_{m} \cdot \sigma(\hat{s}_j)) , (\mathbb{P}^1_{m} \cdot \sigma(s^{(k)}))} \,.
\end{align}
These are also the conventions used for the charges given in Appendix~\ref{App:classification}.
 
%%%%%%%%%
\subsubsection*{Matter multiplicities of uncharged singlets}

The CY manifold admits a number of moduli which manifest themselves in the matter spectrum. For F-theory on a torus-fibered CY threefold $Y_3$ over a two dimensional base $B$ we have \cite{Vafa:1996xn,Morrison:1996na,Morrison:1996pp}
\begin{align}
\label{eq:6dDOFCorrespondence}
T=h^{1,1}(B)-1\,, \enspace  \text{rank}(G_{Y_3})=h^{1,1}(Y_3)-n_{\text{SCP}}-h^{1,1}(B)-1\,, \enspace H_\text{neut}=h^{2,1}(Y_3)+1\,.
\end{align}
Note that the rank of the total gauge group $G_X$ has been corrected by the appearance of SCPs, which we will explain in Section~\ref{sec:SCP}.
The number of uncharged singlets can be inferred from the Euler number $\chi(Y_3)$ of the CY threefold 
\begin{align}
H_{\text{neut}} =  h^{1,1}(B) + \text{rank}(G_{Y_3}) + n_{\text{SCP}}+2- \tfrac{1}{2} \chi(Y_3) \, .
\label{eq:neutsinglets}
\end{align}
As in the rest of our analysis, we want to perform the computation in a base-independent way and express everything in terms of the Chern classes of the base, the classes $\mathcal{S}_7$ and $\mathcal{S}_9$ that parametrize the fibration and the GUT divisor $\mathcal{Z}$. To do this we adapt the methods of \cite{Cvetic:2013uta}.

It proves beneficial for the following discussion to introduce the three sets of divisors: 
\begin{itemize}
  \item $\mathfrak{D}^{\text{torus}}$ contains all divisors dual to the points of the toric diagram at height zero of the top (i.e.\ $u,v,w,e_1$ in the example of Section~\ref{T6dvac}).
  \item $\mathfrak{D}^{\text{top}}$ contains the divisors in the top at height one and above that are not interior to facets (i.e.\ $z_0,f_2,f_3,f_4,g_1,g_2$ in the example of Section~\ref{smooth6dvac}).
  \item $\mathfrak{D}^{\text{facets}}$ contains the divisors in the top at height one and above that are interior to facets\footnote{Facets are codimension-one faces.} (these give rise to SCPs and appear in the example of Section~\ref{sec:SCP}).
\end{itemize}
First, we use that the total Chern class $c(V)=1+c_1(V)+c_2(V)+\ldots$ of a toric variety $V$ is given in terms of the product of all toric divisors\footnote{In the following we often use the equivalence between divisor classes and their dual $(1,1)$-forms.}, $c(V)=\prod_{m} (1+D_{m})$. The individual Chern classes $c_a(V)$ correspond to the terms of appropriate degree (i.e.\ those with $a$ divisors in this expansion). In order to express this in a base-independent way, we separate the contributions from fiber and base as
\begin{align}
\label{eq:TotalChernClass}
c(V)=\sum_{a=0}^{\text{dim}(B)} c_a(B) \prod_\alpha (1+D_\alpha) \frac{1}{1+ [z]}\,,
\end{align}
The first factor parametrizes the result in terms of the Chern classes of the base. The second factor includes all toric divisors in the top, $\mathfrak{D}^{\text{top}}\,\cup\,\mathfrak{D}^{\text{facets}}$. The third factor takes into account that the divisor class of $D_0$ that corresponds to the extended node of the Dynkin diagram already contains the GUT divisor $\mathcal{Z}$ of the base, cf.\ \eqref{eq:ExtendedDynkinDivisor}. This factor is defined via its formal expansion around $[z]=0$.

Since the CY $Y_3$ is given as the anticanonical hypersurface in the toric variety, we can express the Chern classes $c_a(Y_3)$ of the CY in terms of the toric Chern classes $c_a(V)$ using adjunction,
\begin{align}
\label{eq:CxGeneric}
c(Y_3)=1+c_1(Y_3)+c_2(Y_3)+\ldots=\frac{c(V)}{1+c_1(V)}\,,
\end{align}
where the last term is defined as above by its formal expansion. From this we can extract the term for the third Chern class $c_3(Y_3)$ and compute the Euler number as the integral thereof, 
\begin{align}
\chi(Y_3)=\int_{Y_3} c_3(Y_3) \,  . 
\end{align}
The expression obtained from \eqref{eq:CxGeneric} can be further simplified and written in terms of an integration over the base only by making use of the intersection ring defined by the top and the fact that a $k$-section of the fibration intersects the fiber in $k$ points. We thus reduce the polynomial $c_3(Y_3)$ in the quotient ring obtained from a quotient of the polynomial ring by the linear equivalence and the Stanley-Reissner ideal~(SRI)\footnote{This computation can be done conveniently in SAGE by defining the quotient ring and using degree reverse lexicographic ordering for the Groebner basis computation in the division algorithm to obtain an expression that is linear in the sections in the quotient ring.}.

More precisely, we use the linear equivalences to express the divisors in $\mathfrak{D}^{\text{torus}}$ in terms of the base divisors $c_1(B) \sim \Kbi$, $\mathcal{S}_7$, $\mathcal{S}_9$ that parametrize the fibration, the GUT divisor $\mathcal{Z}$, and their shift by the blow-up divisors of the top $\mathfrak{D}^{\text{top}}\,\cup\,\mathfrak{D}^{\text{facets}}$. Then we identify those divisors in $\mathfrak{D}^{\text{torus}}$ that correspond to sections $[\hat{s}_i]$, as these can be used to rewrite the integral over $c_3(Y_3)$ in terms of an integral over the base only, since
\begin{align}
\int_{Y_3} D_i^{\text{b}} D_j^{\text{b}} [\hat{s}_i] =\int_B D_i^{\text{b}} D_j^{\text{b}} \,,
\end{align}
for any base divisors $D_i^{\text{b}}$, $D_j^{\text{b}}$.

Next we use the properties of the intersection ring. First, we choose a triangulation of the top to obtain the fiber part of the SRI. For the base part we use the following generic intersection properties \cite{Cvetic:2013uta}:
\begin{align}
\label{eq:IntersectionGeneric}
D_i^{\text{b}} D_j^{\text{b}} D_k^{\text{b}} = 0\,, &\qquad\qquad D_i^{\text{b}} D_j^{\text{b}} D_k^{\text{t}} =0\,, & D_i^{\text{f}} D_j^{\text{f}} D_k^{\text{f}} =0\,,\nonumber\\
\quad D_i^{\text{b}} D_j^{\text{b}} D_k^{\text{f}} =0\,,&\qquad\qquad D_i^{\text{b}} D_j^{\text{t}}  D_k^{\text{t}} = -  (C_{\text{SO(10)}})_{jk} \mathcal{Z} D_i^{\text{b}} \,,& ([\hat{s}_i]^2  + \Kbi [\hat{s}_i])D_i^{\text{b}}=0\,,
\end{align}
where $D_i^{\text{t}}\in\mathfrak{D}^{\text{torus}}$ and $D_i^{\text{f}}\in\mathfrak{D}^{\text{facets}}$. The first three properties are true simply because the codimension of their intersection in the base exceeds its dimension. The fourth property follows from the fact that the facet points miss the anticanonical hypersurface. The fifth property makes use of the fact that the intersection of the resolution divisors of the top is the negative of the Cartan matrix $C$ of the associated gauge group, which in our case is SO(10). The last property is a direct consequence of adjunction, $(K_{Y_3}+ [\hat{s}_i])|_{[\hat{s}_i]}=K_{[\hat{s}_i]}$,
where $K_{Y_3}=0$ for CYs and $K_{[\hat{s}_i]}= - \Kbi$ since $\hat{s}_i$ is a section. Note that in the case of multi-sections this is no longer true. Hence, for polytopes $F_1$, $F_2$ and $F_4$ which have only multi-sections, we perform the computation completely in the ambient space by using that the CY is the anticanonical hypersurface, 
\begin{align}
\chi(Y_3)=\int_{Y_3} c_3(Y_3)=\int_V c_3(Y_3) c_1(V)= \int_V \big(c_3(V)-c_1(V) c_2(V)\big) c_1(V) \,,
\end{align}
where the last step follows again from adjunction.

In order to illustrate the computation, we present the steps in more detail for the example of Section~\ref{T6dvac}, i.e.\ ($F_3$, top 1). First, we find the total Chern class of $V$ \eqref{eq:TotalChernClass},
\begin{align}
\begin{split}
c(V)=\frac{1}{1+ [z]}&(1+c_1(B)+c_2(B))(1+ [u])(1+ [v])(1+ [w])\times\\
&(1+ [e_1])(1+ [z_0])(1+ [f_2])(1+ [f_3])(1+ [f_4])(1+ [g_1])(1+ [g_2])\,.
\end{split}
\end{align}
From this we extract the first Chern class and compute the total Chern class of $Y_3$ using \eqref{eq:CxGeneric}. We refrain from giving this lengthy expression explicitly. Next, we take the part corresponding to $c_3(Y_3)$ and reduce it in the quotient ring of the polynomial ring generated by the divisor classes modulo the equivalences
\begin{align}
\begin{split}
[z_0] - [f_2] - [f_3] - [f_4] - 2 [g_1] - 2 [g_2] &\equiv [z] \,, \\
[u] - H - \Sn + \Kbi + [f_2] + [f_3] + [g_1] + [g_2] &\sim 0 \,, \\
[v] - H - \Sn + \Ss - [e_1] + [f_2] + 2 [f_3] + [f_4] + 2 [g_1] + 3 [g_2] &\sim 0 \,, \\
[e_1] \equiv [\hat{s}_0] \,, \quad H = [w] &\equiv [s^{(3)}] \,.
\end{split}
\end{align}
We parametrize the base dependence of the divisors in $\mathfrak{D}^{\text{torus}}$ following \cite{Klevers:2014bqa}. Due to the blow-ups encoded in the top, the original linear equivalences now also contain divisors from $\mathfrak{D}^{\text{top}}$. This model has a zero section $\hat{s}_0$ corresponding to the toric divisor $[e_1]$ and a 3-section $s^{(3)}$ corresponding to $[w]$. On top of the linear equivalence ideal, we also have the SRI which can be used to further simplify the expression, where
\begin{align}
\text{SRI}=\text{SRI}_{\text{top}}\cup \text{SRI}_{\text{base}}\,.
\end{align}
We can obtain $\text{SRI}_{\text{top}}$ from any fine star triangulation\footnote{Note that while the intersection ring of the top changes, the physics is invariant with respect to the choice of a triangulation.} of the toric top, e.g.\
\begin{align}
\begin{split}
\text{SRI}_{\text{top}}=\{&v e_1, v z_0, v f_2, v g_1, u w, u z_0, u f_4, u g_1, u g_2, e_1 f_3, w f_3,\\ 
& z_0 f_3,f_3 f_4, w f_2, w g_1, w g_2, e_1 f_4, f_2 f_4, e_1 g_2, z_0 g_2, f_2 g_2, e_1 g_1\}\,.
\end{split}
\end{align}
In order to keep the computation independent of the base, we choose for $\text{SRI}_{\text{base}}$ a generic SRI which solely originates from codimension counting, i.e.\ we use the first four properties of \eqref{eq:IntersectionGeneric}. With these simplifications we obtain the expression 
\begin{align}
\chi(Y_3)= p_1 + p_2 +p_3 +p_4 \,,
\end{align}
with
\begin{align}
\begin{split}
p_1&=-6 \Kbi [f_4] [g_1] - 5 \Kbi [g_1]^2 - 8 \Kbi [g_1] [g_2] - 2 \Kbi [g_2]^2 - [g_1]^2 \mathcal{S}_7 \\  
&\phantom{=\;}- 6 [g_1] [g_2] \mathcal{S}_7 - 3 [g_2]^2 \mathcal{S}_7 - 4 [f_4]^2 \mathcal{S}_9 - 7 [f_4] [g_1] \mathcal{S}_9 - 2 [g_1]^2\mathcal{S}_9 + 7 [g_1] [g_2] \mathcal{S}_9\\
&\phantom{=\;} + 7 [g_2]^2 \mathcal{S}_9 + 7 [f_4]^2 \mathcal{Z} + 18 [f_4] [g_1] \mathcal{Z} + 9 [g_1]^2 \mathcal{Z} -  2 [g_1] [g_2] \mathcal{Z} - 9 [g_2]^2 \mathcal{Z}\,,\\[3mm]
p_2 &= [s^{(3)}] (-8 (\Kbi)^2 + 8 \Kbi \mathcal{S}_7 - 2 \mathcal{S}_7^2 - \Kbi \mathcal{S}_9 + 2 \mathcal{S}_7 \mathcal{S}_9 - 2 \mathcal{S}_9^2\\
&\phantom{=\;S_p} + 11 \Kbi \mathcal{Z} - 11 \mathcal{S}_7 \mathcal{Z} + 5 \mathcal{S}_9 \mathcal{Z} -  5 \mathcal{Z}^2)\,,\\[3mm]
p_3 &= [\hat{s}_0] (8 (\Kbi)^2 - 16 \Kbi \mathcal{S}_7 + 2 \mathcal{S}_7^2 + 9 \Kbi \mathcal{S}_9 - 2 \mathcal{S}_7 \mathcal{S}_9\\
&\phantom{=\;S_r} - 19 \Kbi \mathcal{Z} + 21 \mathcal{S}_7 \mathcal{Z} - 10 \mathcal{S}_9 \mathcal{Z} + 11 \mathcal{Z}^2)\,,\\[3mm]
p_4 &= [\hat{s}_0]^2 (8 \Kbi  - 10 \mathcal{Z})\,.
\end{split}
\end{align}
In order to simplify $p_1$, we use the fifth property of \eqref{eq:IntersectionGeneric}, i.e.\ that the divisors of $\mathfrak{D}^{\text{top}}$ intersect as given by the (negative) Cartan matrix of \SO{10} over $\mathcal{Z}$. In expressions $p_2$ and $p_3$ we use that $s^{(3)}$ and $\hat{s}_0$ are 3- and 1-sections, such that the terms in bracket contribute three and one times, respectively. Finally, in order to simplify $p_4$, we use the last property of \eqref{eq:IntersectionGeneric} to get an expression linear in $[\hat{s}_0]$, which can then be treated as in $p_3$. After these steps, we obtain the final expression in terms of base intersections,
\begin{align}
\begin{split}
\chi(X)= &-24 (\Kbi)^2 + 8 \Kbi \mathcal{S}_7 - 4 \mathcal{S}_7^2 + 6 \Kbi \mathcal{S}_9 + 4 \mathcal{S}_7 \mathcal{S}_9 - 6 \mathcal{S}_9^2\\
& + 30 \Kbi \mathcal{Z} - 10 \mathcal{S}_7 \mathcal{Z} + 10 \mathcal{S}_9 \mathcal{Z} - 20 \mathcal{Z}^2\,.
\end{split}
\label{eq:EulerF3top1}
\end{align}
We collect the results for all tops of all polytopes in Appendix~\ref{App:classification}.

%%%%%%%%%
\subsubsection*{Matter multiplicities of charged singlets}

Lastly, we compute the multiplicity of the charged \SO{10} singlet states by reading off the induced factorization of the top for the singlet matter ideals $I_{k}$ given in \cite{Klevers:2014bqa}. Since these ideals are often rather unwieldy, 
we refer to \cite{Klevers:2014bqa} for their explicit expressions in most of the cases.

We start by considering a fibration without a top, where the vanishing of a codimension-two ideal $I_{k}$ defines the locus of some singlet matter field. After the inclusion of the top this ideal is changed to $\hat{I_{k}}$. It happens regularly that powers of the base coordinate $z_0$ factor out (see Section~\ref{sec:singletcount}) of the two polynomials
\begin{align}
I_{k} = \{ Q_1, Q_2 \}   \rightarrow \hat{I_{k}}= \{ z_0^m \hat{Q}_1, z_0^n \hat{Q}_2     \}                    \, .
 \end{align} 
In such a case, we have to subtract the factored codimension-one loci $z_0 = 0$ with orders $m$ and $n$  to obtain the reduced ideal
 \begin{align}
 \hat{I}_{k,\text{red}} = \{\hat{Q}_1 ,\hat{Q}_2 \} \, .
 \end{align}
Secondly, the vanishing of the ideal $V( \hat{I}_{k})$ often includes simpler ideals  $V( \hat{I}_{r,\text{red}})$ associated to other matter states  that we have to subtract in order not to overcount. These subtractions have been carried out in \cite{Klevers:2014bqa}  for all SO(10) singlets, but need to be corrected in the presence of SO(10) matter and SCP loci.

The subtraction can be carried out by using resultant techniques (see Section~\ref{sec:singletcount}). For this the polynomials $\hat{Q}_1(x,y)$ and $\hat{Q}_2(x,y)$ of $ \hat{I}_{k,\text{red}} $ are considered as functions on $ \hat{I}_{r,\text{red}} = (x,y)$. We compute the resultant of $\hat{Q}_1$ and $\hat{Q}_2$ with respect to $x$ as 
 \begin{align}
 R(y) = \textit{Res}_x( \hat{Q}_1,\hat{Q}_2) \, ,
 \end{align}
 as the determinant of the Silvester matrix in $x$. 
The resultant polynomial $R(y)$ has eliminated the variable $x$ and vanishes over the locus $y=x$ for which $\hat{Q}_1=\hat{Q}_2=0$ is satisfied. Hence if $R(y)$ factorizes as 
 \begin{align}
 R(y) = y^{n_y}  \hat{R}(y) \, ,
 \end{align}
the resultant vanishes at $y=x=0$ to order $n_y$. Similarly we can take the resultant of $\hat{Q}_1,\hat{Q}_2$ with respect to the $y$ variable:
  \begin{align}
 R(x) = \textit{Res}_y( \hat{Q}_1,\hat{Q}_2) =x^{n_x}  \hat{R}(x) \, .
 \end{align}
It is important to remark that $n_x \neq n_y$, and hence there is an ambiguity which variable to take. Throughout this work, we always subtracted min$(n_x, n_y)$ in case of this ambiguity, which turns out to be consistent with anomaly cancellation.

The base-independent multiplicity $m(\hat{I}_k)$ of some singlet field, given by the ideal $V( \hat{I}_{k,\text{red}})$, is finally given by the intersection of its divisor classes minus the multiplicity of other ideals $\hat{I}_i = (x_i, y_i)$ times their resultant orders
\begin{align}
m(\hat{I}_k) = [\hat{Q}_1] [\hat{Q}_2] - \sum_{V_{\hat{I}_{i,\text{red}}}}  \text{min}(n_{x_i}, n_{y_i}) [x_i] [y_i] \,.
\end{align}

This completes the computation of the full matter spectrum for an arbitrary top. All spectra can be found in Appendix~\ref{App:classification} for all SO(10) tops given in \cite{Bouchard:2003bu}. We complete the
discussion of the 6d SUGRA models by considering anomaly cancellation of theories without SCPs in the following.

%%%%%%%%%%%%%%%%%%
\subsection{Base-independent anomaly cancellation}
\label{sec:bianomalycanc}

In this section we analyze the base-independent anomaly cancellation for the models described above, exemplifying the general procedure for ($F_3$, top 1) with gauge group SO(10)$\times$U(1). Other models with possible additional non-Abelian factors can be treated analogously and all anomaly coefficients are given in Appendix \ref{App:classification}. Here, we discuss models without SCPs; a discussion of anomaly cancellation for models with SCPs after a blow-up in the base is given in Section \ref{sec:anomSCP}. For our investigation we use the relation between the anomaly coefficients and the second base cohomology $H_2(B, \mathbb{Z})$, see e.g.\ \cite{Park:2011ji}, in connection with the parametrization of the base-dependence in terms of $\Ss$, $\Sn$, $\Kbi$, and $\Z$.

Denoting the SO(10) field strength by $\tilde{F}$ and the Abelian field strengths by $F$, the 6d anomaly polynomial\footnote{We use the notations and conventions of \cite{Park:2011wv, Park:2011ji}. $H$, $V$, and $T$ denote the number of hyper-, vector, and tensor multiplets, respectively. $n[\mathbf{R}_I]$ is the multiplicity of hypermultiplets in the SO(10) representation $\mathbf{R}_I$ with U(1) charges $q_{I}$ and $\mathcal{M}_I$ is given by $n[\mathbf{R}_I] \text{dim}(\mathbf{R}_I)$.} for gauge group SO(10)$\times$U(1) is
\begin{equation}
\begin{split}
\mathcal{I}_8 =& - \frac{1}{5760} (H - V + 29 T - 273) \left( \tr{R^4} + \tfrac{5}{4} (\tr{R^2})^2 \right) - \frac{1}{128} (9 - T) (\tr{R^2})^2 \\
& - \frac{1}{96} \tr{R^2} \Big( \text{Tr} \tilde{F}^2 - \sum_I n[\mathbf{R}_I] \, \tr_{\mathbf{R}_I} \tilde{F}^2 \Big) + \frac{1}{24} \Big( \text{Tr} \tilde{F}^4 - \sum_I n[\mathbf{R}_I] \, \tr_{\mathbf{R}_I} \tilde{F}^4 \Big) \\
& + \frac{1}{96} \sum_{I} \mathcal{M}_I \, q_{I}^2 \, \tr R^2 F^2 - \frac{1}{4} \sum_{I} n[\mathbf{R}_I] \, q_{I}^2 \, (\tr_{\mathbf{R}_I}\tilde{F}^2) F^2 - \frac{1}{24} \sum_{I} \mathcal{M}_I \, q_{I}^4 \, F^4 \,,
\end{split}
\end{equation}
where $\text{Tr}$ and $\tr_{\mathbf{R}}$ is the trace in the adjoint representation and representation $\mathbf{R}$ of SO(10), respectively. The sum with respect to $I$ runs over the charged hypermultiplets in the matter spectrum. A term of the form $(\tr_{\mathbf{R}} \tilde{F}^3) F$ is absent since SO(10) does not have a third order Casimir operator. Rewriting all traces in terms of $\tr_{\mathbf{10}} \equiv \tr$, see \eqref{SOgrouptheory}, we can split the anomaly polynomial into an irreducible part
\begin{equation}
\begin{split}
\mathcal{I}_8^{\text{irred}} =& - \frac{1}{5760} (H + 29 T - 317) \left( \tr{R^4} + \tfrac{5}{4} (\tr{R^2})^2 \right) \\
& + \frac{1}{24} (2 - 2 n[\mathbf{45}] + n[\mathbf{16}] - n[\mathbf{10}]) \tr \tilde{F}^4 \,,
\end{split}
\end{equation}
and a reducible part
\begin{equation}
\begin{split}
\mathcal{I}_8^{\text{red}} =& - \frac{1}{128} (9 - T) (\tr{R^2})^2 - \frac{1}{96} (8 - 8 n[\mathbf{45}] - 2 n[\mathbf{16}] - n[\mathbf{10}]) \tr R^2 \, \tr \tilde{F}^2 \\
& + \frac{1}{24} \left(3 - 3 n[\mathbf{45}] - \tfrac{3}{4} n[\mathbf{16}] \right)(\tr \tilde{F}^2)^2 + \frac{1}{96} \sum_{I} \mathcal{M}_I \, q_{I}^2 \, \tr R^2 F^2 \\
& - \frac{1}{4} \sum_{I} n[\mathbf{R}_I] \, q_{I}^2 (\tr_{\mathbf{R}_I}\tilde{F}^2) F^2  - \frac{1}{24} \sum_{I} \mathcal{M}_I \, q_{I}^4 \, F^4 \,.
\end{split}
\end{equation}
Note that $n[\mathbf{16}]$ includes both $\mathbf{16}$ and $\overline{\mathbf{16}}$-plets.

The irreducible part has to vanish for the matter spectrum of a consistent theory, leading to a relation between the number of different multiplets and SO(10) representations. The reducible part can be canceled by the Green-Schwarz mechanism \cite{Green:1984sg, Green:1984bx} if it factorizes as
\begin{align}
\mathcal{I}_8^{\text{red}} = - \frac{1}{32} \Omega_{\alpha \beta} X_4^{\alpha} X_4^{\beta} \,,
\end{align}
where the individual factors have to be of the form \cite{Park:2011wv, Park:2011ji}
\begin{align}
X_4^{\alpha} =  \tfrac{1}{2} a^{\alpha} \tr R^2 + b^{\alpha} \tr \tilde{F}^2 + 2 b^{\alpha}_{11} F^2 \,.
\end{align}
The matrix $\Omega_{\alpha \beta}$ is an SO$(1,T)$ metric specifying the contributions and transformations of the various 2-form fields in the generalized 6d version of the Green-Schwarz mechanism \cite{Sagnotti:1992qw, Sadov:1996zm}. It can be identified with the intersection matrix of the base divisors, see e.g.~\cite{Park:2011ji}. Let $\{ H_{\alpha} \}$ be a basis for $H_2(B, \mathbb{Z})$ such that we can express an arbitrary base divisor $D$ as
\begin{align}
D = \sum_{\alpha} d^{\alpha} H_{\alpha} \in H_2 (B, \mathbb{Z}) \,.
\end{align}
The intersection of two base divisors $D$ and $\tilde{D}$ is thus given by
\begin{align}
D \cdot \tilde{D} = \Omega_{\alpha \beta} \, d^{\alpha}  \tilde{d}^{\beta} \,,
\end{align}
with
\begin{align}
\Omega_{\alpha \beta} = H_{\alpha} \cdot H_{\beta} \,.
\end{align}
Since the number of tensor multiplets $T$ in models without SCPs is given in terms of the anticanonical class of the base $\Kbi$, one can identify the gravitational anomaly coefficient $a^{\alpha}$ as the coefficient vector of the anticanonical class of the base \cite{Morrison:1996na, Morrison:1996pp, Kumar:2010ru, Kumar:2010am},
\begin{align}
\Kbi = \sum_{\alpha} a^{\alpha} H_{\alpha} \,.
\end{align}
We will denote this relation of anomaly coefficients and base divisor classes by e.g.\ $a \sim \Kbi$.

Similarly, the $\SO{10}$ anomaly coefficient, which we denote by $b^{\alpha}$, can be identified with the GUT divisor $\Z$ in the base
\begin{align}
\Z = \sum_{\alpha} b^{\alpha} H_{\alpha} \,.
\end{align}
An analogous description holds for additional non-Abelian gauge groups that might generically appear due to the use of a certain ambient space for the fiber, see e.g.\ \cite{Klevers:2014bqa}. They are included in Appendix \ref{App:classification}.

Finally, also the Abelian anomaly coefficients can be associated with a geometrical meaning using the N\'eron-Tate height pairing involving the Shioda map $\sigma$ defining the Abelian group factor \cite{Park:2011ji, Morrison:2012ei},
\begin{align}
b_{11} \sim - \pi_B (\sigma (\hat{s}_1) \cdot \sigma (\hat{s}_1)) \,.
\end{align}
For more than one Abelian gauge factor the anomaly coefficients $b_{ij}$ can be derived analogously by using the corresponding Shioda maps $\sigma(\hat{s}_i)$, i.e.\ $b_{ij} \sim - \pi_B (\sigma (\hat{s}_i) \cdot \sigma (\hat{s}_j))$. With this connection to the base geometry we can express the complete factorized anomaly polynomial in terms of an intersection product of the base divisor classes $\Kbi$, $\Z$, $\Ss$, and $\Sn$.

We next elucidate the geometric concepts by generalizing the anomaly cancellation for ($F_3$,~top~1) discussed in Section \ref{sec:anomexample} to a base-independent formulation and show their equivalence after setting $B = \mathbb{F}_0$ and a making a specific choice for $\Ss$, $\Sn$, and $\Z$.

First we verify that the irreducible gravitational anomaly is indeed canceled. With the relation between the Euler number and the number of neutral singlets \eqref{eq:neutsinglets} as well as the base-independent expression for $\chi$ derived in \eqref{eq:EulerF3top1}, we find\footnote{Note that hypermultiplets in the adjoint representation only contribute as $\text{dim}(\textbf{Adj}) - \text{rank}(G)$ degrees of freedom for the corresponding gauge group in order to avoid overcounting.}
\begin{align}
H - V + 29 T - 273 &= \sum_I \mathcal{M}_I - 30 (\Kbi)^2 - \tfrac{1}{2} \chi (X) - 40 = 0 \,,
\end{align}
where we used the base-independent charged matter spectrum given in \eqref{bimultis}.

With the number of multiplets consistent with the irreducible gravitational anomaly we can evaluate the reducible part
\begin{align}
\mathcal{I}_8 \supset - \frac{1}{128} (9 - T) \big( \tr R^2 \big)^2 = - \frac{1}{32} \left( \tfrac{1}{2} \Kbi \right)^2 \big( \tr R^2 \big)^2 \,.
\end{align} 
For the irreducible $\SO{10}$ anomaly we need the base independent number of hypermultiplets in the various $\SO{10}$ representations. For the chosen top these are given by (see Table \eqref{bimultis})
\begin{equation}
\begin{split}
n[\mathbf{10}] &= (3 \Kbi - 2 \Z) \Z \,,\\
n[\mathbf{16}] &= (2 \Kbi - \Z) \Z \,, \\
n[\mathbf{45}] &= 1- \tfrac{1}{2} (\Kbi - \Z) \Z \,.
\end{split}
\end{equation}
The irreducible part of the non-Abelian anomaly is given by
\begin{equation}
\begin{split}
\mathcal{I}_8 &\supset \frac{1}{24} \left( 2 - 2 n[\mathbf{45}] + n[\mathbf{16}] - n[\mathbf{10}] \right) \tr \tilde{F}^4 \\
&= \frac{1}{24} \left( (\Kbi - \Z) \Z + (2 \Kbi - \Z) \Z - (3 \Kbi - 2\Z) \Z \right) \tr \tilde{F}^4= 0 \,.
\end{split}
\end{equation}
It vanishes independently of the chosen base as has to be the case for a well-defined theory. The reducible non-Abelian anomaly is given by
\begin{equation}
\begin{split}
\mathcal{I}_8 &\supset \frac{1}{24} \left( 3 - 3 \, n[\mathbf{45}] - \tfrac{3}{4} \, n[\mathbf{16}] \right) \big( \tr \tilde{F}^2 \big)^2 \\
&= \frac{1}{24} \left( \tfrac{3}{2} (\Kbi - \Z) \Z - \tfrac{3}{4} (2 \Kbi - \Z) \Z \right) \big( \tr \tilde{F}^2 \big)^2 = - \frac{1}{32} \, \Z^2  \big( \tr \tilde{F}^2 \big)^2 \,,
\end{split}
\end{equation}
which is exactly of the form expected from the relation with $H_2 (B, \mathbb{Z})$, since
\begin{align}
\Z^2 = \Omega_{\alpha \beta} \, b^{\alpha} b^{\beta} \,.
\end{align}
For the mixed anomaly involving gravity and the non-Abelian degrees of freedom we find
\begin{align}
\mathcal{I}_8 \supset - \frac{1}{96} \left( 8 - 8 n[\mathbf{45}] - 2 n[\mathbf{16}] - n[\mathbf{10}] \right) \tr R^2 \, \tr \tilde{F}^2 = \frac{1}{32} \Kbi \Z \big( \tr R^2 \, \tr \tilde{F}^2 \big)
\end{align}
Hence, we see that the non-Abelian and gravitational part of the anomaly polynomial factorize in the appropriate way and can be written as
\begin{align}
\mathcal{I}_8 \supset - \frac{1}{32} \left( \tfrac{1}{2} \Kbi \, \tr R^2 - \Z \tr \tilde{F}^2 \right)^2 \,.
\label{nonAbeliangravanom}
\end{align}
Even though we performed the calculation for a specific top, the factorization of the $\SO{10}$ and gravitational anomalies works in the same way for all models without SCPs and the form of \eqref{nonAbeliangravanom} is universal for all models with gauge group SO(10). 

Similar treatments can be performed after the inclusion of the U(1) factor. The complete anomaly polynomial for ($F_3$, top 1) can be factorized in terms of the base divisor classes as
\begin{align}
\mathcal{I}_8 = - \frac{1}{32} \left( \tfrac{1}{2} \Kbi \, \tr R^2 - \Z \, \tr \tilde{F}^2 + \tfrac{1}{2} (-24 \Kbi + 8 \Ss - 16 \Sn + 5 \Z) F^2 \right)^2 \,,
\end{align}
and we find the U(1) anomaly coefficient
\begin{align}
-6 \Kbi + 2 \Ss - 4 \Sn + \tfrac{5}{4} \Z = \sum_{\alpha} b_{11}^{\alpha} H_{\alpha} \,.
\end{align}
Note that $b_{11}$ coincides with the base-independent anomaly coefficient derived in \cite{Klevers:2014bqa} up to a correction term depending on the GUT divisor $\Z$ that originates from the orthogonalization of the U(1)  with respect to the Cartan divisors of the $\SO{10}$. Hence, the base independent anomaly coefficients for ($F_3$, top 1) are given by
\begin{align}
a \sim \Kbi \,, \enspace b \sim - \Z \,, \enspace b_{11} \sim - (6 \Kbi - 2 \Ss + 4 \Sn - \tfrac{5}{4} \Z) \,.
\end{align}

In order to verify the above expressions we calculate the anomaly coefficients of the specific model discussed in Section \ref{sec:anomexample} using the general base-independent expressions. Choosing the base divisor classes that parametrize the base \eqref{basechoiceexample} whose second homology basis $H_2 (\mathbb{F}_0, \mathbb{Z})$ is given by $\{H_1, H_2\}$, we can calculate the anomaly coefficients explicitly using the intersection matrix for $B = \mathbb{F}_0$ given by
\begin{align}
\Omega_{\alpha \beta} = \begin{pmatrix} 0 & 1 \\ 1 & 0 \end{pmatrix}\,.
\end{align}
We find
\begin{align}
a = \begin{pmatrix} 2 \\ 2 \end{pmatrix} \,, \quad b = \begin{pmatrix} -1 \\ 0 \end{pmatrix} \,, \quad b_{11} = \begin{pmatrix} - \tfrac{51}{4} \\ -12 \end{pmatrix} \,,
\end{align}
reproducing the coefficients in \eqref{anomcoeff}.

Similarly, we can analyze all the models with other gauge groups including the SO(10) top. All irreducible anomalies vanish base-independently and the remaining reducible part is factorizable. The anomaly coefficients in terms of base divisors for the models without SCPs are given by the universal expressions
\begin{align}
a \sim \Kbi \,, \quad b \sim -\Z
\end{align}
for the gravitational and non-Abelian SO(10) part. The remaining anomaly coefficients depend on the specific model. However, up to an overall sign and a contribution due to the SO(10) gauge group the coefficients match the expressions derived in \cite{Klevers:2014bqa}. For models with additional non-Abelian factors $G$ one has to include the corresponding anomalies. Again, the anomaly coefficients $b_{G}$ are related to the base divisor $D_G$ where the gauge group is located, i.e.~$b_G \sim - D_G$. The complete set of Abelian and non-Abelian anomaly coefficients is included in Appendix \ref{App:classification}.

The description above works in a straightforward fashion for all SUGRA models that do not have SCPs. However, the latter appear rather frequently in our analysis. Therefore, we discuss them in the following section and analyze the anomaly cancellation after a blow-up in the base which resolves the corresponding codimension-two singularity in Section~\ref{sec:anomSCP}. 

%%%%%%%%%%%%%%%%%%
\subsection{Theories with superconformal matter points}
\label{sec:SCP}

In many of the models we are considering, we have codimension-two points where the 
SO(10) divisor $\mathcal{Z}$ intersects another curve $\{d_s=0\}$ in the base, possibly with multiplicity $n_{\text{SCP}}= \mathcal{Z}  [d_s]$, such that the Weierstrass coefficients $(f,g,\Delta)$ vanish to orders $(4,6,12)$.
These points have also been encountered in resolved Tate models  \cite{Lawrie:2012gg}, where it was observed that over these points the fiber becomes non-flat. Non-flatness refers to the phenomenon that the fiber dimension jumps and includes higher dimensional components/curves.

\begin{figure}
 \centering
 \begin{tikzpicture}[scale=1.3]
   \draw[very thick] (-1,2) to[out = -15, in = 15] (-1,-2);
   \draw[very thick] (1,2) to[out = 195, in = 165] (1,-2);
   \node at (-1.2, 2.2) {$[d_s]$};
   \node at (1.2, 2.2) {$\Z$};
   \draw[very thick, fill=black] (0,1) circle (0.08);
   \draw[very thick, fill=black] (0,-1) circle (0.08);
   \node at (0.6,1) {SCP$_{1}$};
   \node at (0.6,-1) {SCP$_{2}$};  
      \draw[very thick] (-1+4,2) to[out = -70, in = 70] (-1+4,-2);
   \draw[very thick] (1+4,2) to[out = -110, in = 110] (1+4,-2);
   \draw[very thick] (-1.2+4,1) -- (1.2+4,1);
   \draw[very thick] (-1.2+4,-1) -- (1.2+4,-1);
   \node at (-1.2+4 , 2.2) {$[\hat{d}_s]$};
   \node at (1.2+4, 2.2) {$\hat{\Z}$};
   \node at (1.5+4,1) {$E_1$};
   \node at (1.5+4,-1) {$E_2$};
 \end{tikzpicture}
 \caption{Left: depiction of SCPs at the two simple intersection points of the two divisors $[d_s]$ and $\Z$; right: intersection points separated by two blow-up divisors $E_{1,2}$}
 \label{fig:blowupofSCP}
\end{figure}
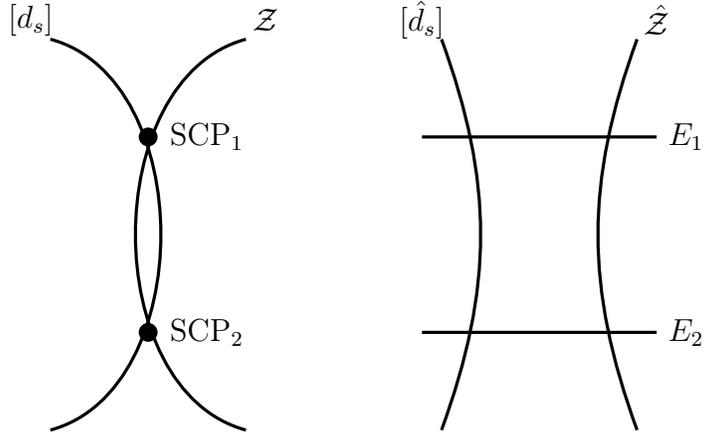

These points have a physical interpretation in terms of strings that become tensionless over those points \cite{Seiberg:1996vs} which contribute additional degrees of freedom to the theory. This can be seen by blowing up the intersection points of the divisors $\mathcal{Z}$ and $ [d_s]$ in the base as 
depicted in Figure~\ref{fig:blowupofSCP}. These blow-ups remove the non-flat fiber points and introduce additional 6d tensor multiplets. The vacuum expectation value (vev) of the scalar component $\langle s \rangle$ of the tensor multiplet encodes the size of the blow-up mode and parametrizes the coupling constant $\langle s \rangle = 1/g_s$ of the tensionless string that becomes strong in the blow-down limit \cite{Seiberg:1996qx} when the curves collide again.
 
These singularities are rather frequent in our theories, which can intuitively be understood from the fact that SO(10) needs a divisor with a $(2,3,7)$ singularity and therefore a large tuning already to begin with. Hence, a second divisor can easily bring the resulting codimension-two singularity to the critical value of $(4,6,12)$. Indeed, around 80\% of the analyzed models admit SCPs and in the following we study them and their interplay with the additional gauge symmetries.

If one is interested in theories without SCPs, there are two possibilities to get rid of those points:
\begin{itemize}
\item Choose a base where the relevant intersections vanish, i.e.\ $\mathcal{Z} [d_s]=0 \,$. 
\item Blow-up the intersections points as in Figure~\ref{fig:blowupofSCP}.
\end{itemize}
In the following we consider generic bases that include SCPs but use the second option to smoothly interpolate to a theory without SCPs and confirm anomaly cancellation in Section~\ref{sec:anomSCP}.

Similarly to the gauge group, the presence of SCPs is encoded in the structure of the top as well. For the SO(10) tops, we have seen that we need at least 6 vertices, two at height two and four at height one, that correspond to the divisors dual to the six roots of the affine SO(10) Dynkin diagram. However, we also have the option to consider a top with five vertices at height one, which are placed such that one of them lies in the interior of a face. As an example, ($F_5$, top 3) is depicted in Figure~\ref{fig:F5top3Poly}. In such a case, the divisor associated to the fifth vertex does not intersect the CY and therefore does not contribute an SO(10) root at codimension one, as also observed in \cite{Borchmann:2013jwa}.

%%%%%%%%%
\subsubsection*{Base independent blow-ups}

The SCPs are resolved by adding exceptional divisors $E_i$ in the base. In our case this implies that the fiber over the new exceptional divisors $E_i$ is smooth and one does not encounter additional gauge group factors after performing the blow-up, which resolves the base space $\hat{B}$. The exceptional divisors have the intersection form 
\begin{align}
E_i \cdot E_j = - \delta_{ij} \,.
\end{align}
Moreover, we can define the map \cite{Bershadsky:1996nu}
\begin{align}
\beta_* : H_2(\hat{B}, \mathbb{Z}) \rightarrow H_2(B, \mathbb{Z}) \,,
\end{align}
i.e., the push-forward of base divisor classes under the blow-down map $\beta$. This map preserves the intersection form and its kernel is generated by the exceptional divisors $E_i$, with $i \in \{1, \dots, r\}$. Moreover, we denote by $D^* = \beta_*^{-1} (D)$ the full preimage of $D \in H_2 (B, \mathbb{Z})$.

The anticanonical class of the resolved base $\hat{B}$ is modified as
\begin{align}
K_{\hat{B}}^{-1} = (\Kbi)^* - \sum_i E_i \,,
\end{align}
from which we can derive the number of tensor multiplets $\hat{T}$ in the blown-up base $\hat{B}$ with respect to the number of tensor multiplets $T$ of $B$,
\begin{align}
\hat{T} = 9 - (K_{\hat{B}}^{-1})^2 = 9 - (\Kbi)^2 - \sum_{i,j} E_i \cdot E_j = T + r \,.
\end{align}
As expected, $\hat{T}$ is increased by the number of exceptional divisors introduced during the blow-up procedure. Similarly, also the other base divisor classes get modified,
\begin{align}
\Z^* = \hat{\Z} + \sum_i n_{\Z,i} E_i \,, \quad \Ss^* = \hat{\Ss} + \sum_i n_{7,i} E_i \,, \quad \Sn^* = \hat{\Ss} + \sum_i n_{9,i} E_i \,,
\end{align}
where the integer parameters $n_{\Z,i}$, $n_{7,i}$, and $n_{9,i}$ depend on the specific model. The blow-up is performed in such a way that the base-independent intersections determining the matter multiplicities (see Appendix \ref{App:classification}) remain of the same form with hatted divisors. However, the intersections corresponding to the SCPs vanish.

%%%%%%%%%
\subsubsection*{Example: A theory with SCP and its resolution}

We consider ($F_5$, top 3) which is depicted in Figure~\ref{fig:F5top3Poly}. Its generic spectrum and multiplicities are summarized in \eqref{eq:F5top3} of Appendix~\ref{App:classification}.
\begin{figure}
\begin{center}
\begin{picture}(0,130)
\put(70,0){\includegraphics[scale=0.5]{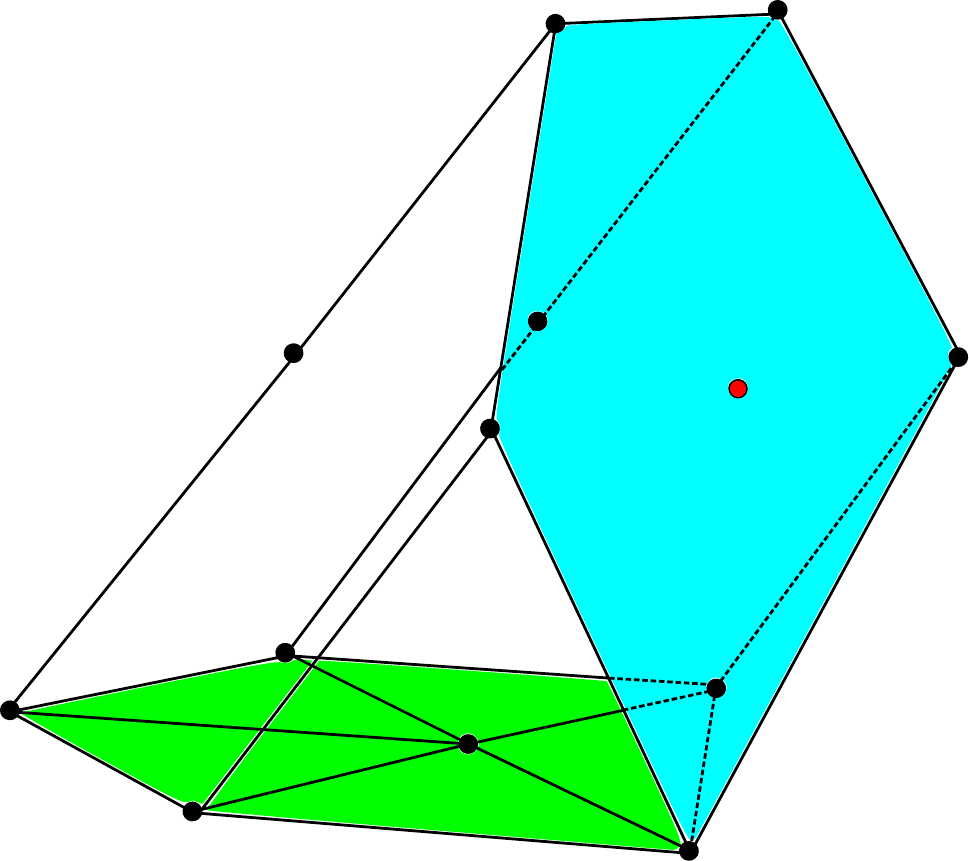}}
\put(-170,10){\includegraphics[scale=0.6]{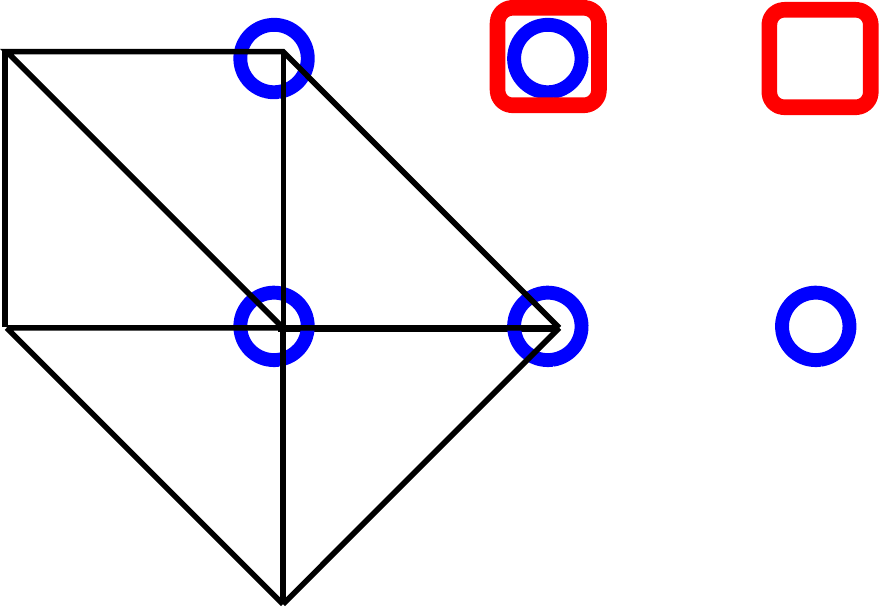}}
\put(70,100){b)}
\put(-190,100){a)}
\end{picture}
\end{center}
\caption{\label{fig:F5top3Poly}Two depictions of the polytope ($F_5$, top 3) that has SCPs. A projection of the top is depicted in a) with vertices at height one in blue and those at height two in red. The vertex at (1,0,1) lies in a face of the top. This face is shaded in blue in Figure b) where the interior vertex is marked in red.}
\end{figure}
This top yields the gauge group SO(10)$\times$U(1)$^2$ together with some SCPs. Here, the SCPs are generically localized over the intersection of $\mathcal{Z}$ and $[d_5]$ and are counted by
\begin{align}
n_{\text{SCP}} = (2 K_B^{-1} - \mathcal{S}_7) \Z \, .
\end{align}
To be explicit, we construct a concrete threefold $Y_3$ with base $\mathbb{F}_0=\mathbb{P}^1 \times \mathbb{P}^1$, where we denote the two divisor classes by $H_1$ and $H_2$.
We realize the CY as the anticanonical hypersurface in the polytope $\Delta$ with vertices:
 \begin{align}
 \begin{array}{c|c}
  \begin{tabular}{c|c}
coordinates & vertices \\ \hline  
$u$& (-1, 1, 0,0) \\
$v$ & (0, -1, 0 ,0) \\
$w$& (1, 0, 0,0) \\
$e_1$ & (0, 1, 0 ,0) \\
$e_2$& (-1, 0, 0,0) \\ \hline
$z_1$ & (0, 0, -1,0) \\ 
$z_2$ & (0,-1,0,1) \\
$z_3$ & (0,0,0,-1) \\  
\end{tabular}
    & 
   \begin{tabular}{c|c}
coordinates & vertices \\ \hline 
  $f_1$ & (0, 1, 1, 0	) \\
  $g_1$ & (1, 1, 2,0) \\
  $g_2$ & (2, 1, 2,0)\\
  $f_2$ &  (1, 1, 1,0) \\
  $f_3$ & (2, 0, 1,0) \\
     $\widehat{f_4}$ & (1, 0, 1,0) \\
       $z_0$ & (0, 0, 1,0) \\  
       & 
\end{tabular} \\
 \end{array}
 \end{align} 
With the general expressions in Appendix~\ref{App:classification} this amounts to choosing 
\begin{align}
  \mathcal{S}_9 = \Kbi = 2H_1 +2 H_2,  \quad \mathcal{S}_7 = 2H_1 + 3H_2 \, ,
\end{align}
The spectrum can then be easily computed, by using the intersections on $\mathbb{F}_0$ \eqref{interbase}  together with the Hodge and Euler numbers:
\begin{align}
\label{eq:EulerF5SCP}
(h^{1,1}(Y_3),h^{2,1}(Y_3))_\chi = (11,37)_ {-52}\, . \,  
\end{align}
We find exactly one SCP by computing the intersection
\begin{align}
[d_5]\mathcal{Z} =  (2H_1 + H_2) H_1 = 1 \, .
\end{align}
The rest of the spectrum can be computed similarly and is given by
\begin{align}
\begin{array}{c|c}
\begin{array}{c|c}
\text{representation}& $multiplicity$ \\ \hline
 \mathbf{10} _{-1/2,0} & 3 \\
  \mathbf{16} _{-1/4,1/2} & 3   \\
  \mathbf{10} _{1/2,1} & 2 \\
 n_{\text{SCP}} & 1 \\ \hline 
  \mathbf{45}_{0,0} & 0 \\
 \mathbf{1}_{0,0} & 38 \\
 T & 1 \\ 
\end{array}
&
\begin{array}{c|c}
$representation$&$ multiplicity $\\ \hline
 \mathbf{1}_{1,-1} & 3 \\
 \mathbf{1}_{1,2}& 6  \\
 \mathbf{1}_{0,2}& 8\\
 \mathbf{1}_{-1,-1}& 26 \\
 \mathbf{1}_{1,0}& 57 \\
 \mathbf{1}_{0,1}& 29  \\  
 & \\
\end{array}
\end{array}
\end{align}
When computing the coefficient of the irreducible gravitational anomaly by adding all perturbative degrees of freedom contained in hyper-, vector and tensor multiplets above, we find
\begin{align}
H - V + 29 T - 273 = -29 \,,
\end{align}
i.e.\ a mismatch of 29 degrees of freedom that enter the irreducible gravitational anomaly with the same chirality as the tensors before the inclusion of the SCP. This already hints at the fact that we get $n_{\text{SCP}}$ additional tensors on top of the usual $T=h^{1,1}(B)-1$, cf.~\eqref{eq:6dDOFCorrespondence}.

Next we want to consider the blow-up geometry. First, we deform the polynomial $d_5$ to factorize as
\begin{align}
d_5 \rightarrow z_3 \hat{d_5} \, ,
\end{align}
to enforce the SCP to lie on the toric locus $z_0 = z_3 = 0$. We resolve this locus by performing a blow-up of the ambient space, i.e.\ by adding the vertex
\begin{align}
\hat{e}_1: (1,-1,1,1) 
\end{align}
to the 4d polytope $\Delta$. This changes the base from $\mathbb{F}_0$ to $dP_2$ 
and shifts the classes as
\begin{align}
\mathcal{Z} \sim H_1 - E_1 \,, \enspace [z_1]\sim H_1 \,, \enspace [z_2] \sim H_2 - E_1 \,, \enspace [z_3] \sim H_2 \, . 
\end{align}
with $E_1 = [e_1]$.
Indeed, we have removed the SCP, as the vertex of $E_1$ subdivides the cone in $\mathbb{F}_0$ which is spanned by the vertices of $z_0$ and $z_3$. 
For the model at hand, the blow-up corresponds to a shift in the base divisor classes determining the matter multiplicities as
\begin{align}
K_B^{-1} \sim    2H_1 +2H_2 -E_1 \,, \enspace \mathcal{S}_7 \sim 2  H_1 + 3 H_2-E_1 \,, \enspace   \mathcal{S}_9 = 2 H_1 + 2 H_2 \, .
\end{align}
Inserting the intersections
\begin{align}
H_1^2 = 0 \,, \enspace H_2^2 = 0 \,, \enspace H_1 E_1 = 0 \,, \enspace H_2 E_1 = 0 \,, \enspace  E_1 E_1 = -1 \, ,
\end{align}
into the general expressions given in Appendix \ref{App:classification}, we confirm that the 
spectrum indeed stays invariant. However, now the SCPs have been removed and $\hat{T} = h^{1,1}(B)-1 =2$.

%%%%%%%%%
\subsubsection*{Euler and Hodge numbers in theories with SCPs}

Let us investigate the Euler and Hodge numbers of this theory in more detail as given in  \eqref{eq:EulerF5SCP}. Naively, the K\"ahler moduli of the CY threefold are
 \begin{align}
 \label{eq:H11Naive}
 h^{1,1}(Y_3) = \text{rank}(G_{Y_3}) + h^{1,1}(B) + 1  \,.
 \end{align}
From this counting, we would have anticipated only ten K\"ahler moduli but instead we find eleven. This additional  
K\"ahler modulus is non-toric and is accounted for by the point that lies in the face, via the Batyrev formula \cite{Batyrev:1994hm}
\begin{align}
h^{1,1}(Y_3) = \underbrace{ l (\Delta) - 4 - \sum_\Gamma l^\circ (\Gamma) }_{h^{1,1}_{\text{toric}}}+ \underbrace{ \sum_\Theta l^\circ(\Theta)   l^\circ(\Theta^*)  }_{h^{1,1}_{\text{nt}}} \, .
\end{align}
Here, $l (\Delta)$ is the number of points in the polytope $\Delta$, $\Gamma$ are its edges and $\Theta$ denotes codimension-two faces in $\Delta$ whereas $\Theta^*$ is its dual face in $\Delta^*$ of dimension one. $l^\circ$ counts the points in the relative interior of its argument. Hence, for a regular SO(10) top, all points are vertices that are not in a face and all divisors are toric, with toric K\"ahler deformations associated to them.
However, the presence of a point in $\Theta$ of the fiber leads generically to non-toric $(1,1)$-forms for a base where its dual face contains non-trivial points as well.

Indeed, in our example the additional non-toric K\"ahler deformation is associated to the point $\widehat{f_4}: (1,0,1,0)$ which is an interior point of the face
\begin{align}
\Theta_f = \{ (0,-1,0,0), (0,  0, 1, 0), (2,  0, 1, 0), (1,  1, 2, 0), (2,  1, 2, 0), (1,  0, 1, 0) \} \,,
\end{align}
and accounts for the SCP.
This picture is also consistent from the perspective of the resolved elliptic curve that reads
\begin{equation}
\begin{split}
p_{(F_5,\text{ top 3})} & = d_5 e_1^2 e_2 f_1 f_2 u^2 w + d_8 e_1^2 f_1 f_2^2 f_3^2 \widehat{f_4} g_1 g_2^2 u w^2 + d_9 e_1 f_2 f_3^2 \widehat{f_4} g_2 v w^2 \\
& + d_6 e_1 e_2 f_1 f_2 f_3 \widehat{f_4} g_1 g_2 u v w z_0 +  d_7 e_2 f_3 \widehat{f_4} v^2 w z_0 + d_1 e_1^2 e_2^2 f_1^3 f_2^2 \widehat{f_4} g_1^3 g_2^2 u^3 z_0^2  \\
& + d_2 e_1 e_2^2 f_1^2 f_2 \widehat{f_4} g_1^2 g_2 u^2 v z_0^2 +  d_3 e_2^2 f_1 \widehat{f_4} g_1 u v^2 z_0^2 \, .
 \end{split}
\end{equation}
Over the locus $d_5 = 0$ the fiber becomes reducible,
\begin{align}
p_{(F_5,\text{ top 3})}|_{d_5 = 0} =\widehat{f_4} \, p_3(u,v,w) \,,
\end{align}
where $p_3 (u,v,w)$  is a degree-three polynomial in the fiber coordinates $u,v,w$, which parametrize a smooth torus away from the SO(10) divisor $\mathcal{Z}$. In addition, we find a $\mathbb{P}^1$, given by $\widehat{f_4}$, which intersects the CY exactly over the SO(10) divisor $\mathcal{Z}$. Hence, again, from this perspective the interior point of a face of the SO(10) top , $\widehat{f_4}$, yields the non-flat fiber component over the collision points of $\mathcal{Z}$ and $[d_5]$. 

Performing the blow-up of the SCP in the base by adding the divisor $E_1$ removes the non-toric $(1,1)$-form and introduces a toric one in the base as discussed above.
 
From these observations, we conclude that the naive  counting \eqref{eq:H11Naive} of Hodge numbers for a CY threefold $Y_3$ has to be modified to include the contribution of non-toric K\"ahler deformations that
are represented by the non-flat fiber components, i.e.\ the SCPs, as
 \begin{align}
 \label{eq:SCMHodge}
\chi (Y_3) =& 2 (h^{1,1}_{\text{toric}}(Y_3) + h^{1,1}_{\text{nt}}(Y_3) - h^{2,1}(Y_3)) \, , \\
h^{1,1}(Y_3) =& \text{rank}(G_{Y_3}) + h^{1,1}(B) + 1 + n_{\text{SCP}}  \, .
\end{align}
Using this identification we can compute the neutral matter spectrum, corrected by the SCPs from the Euler numbers and generic rank of the gauge group, as
\begin{align}
H_{\text{neut}} = h^{2,1}(Y_3) + 1 = h^{1,1}(B)+n_{\text{SCP}}+2 + \text{rank}(G_{Y_3}) - \tfrac{1}{2} \chi(Y_3) \,,
\end{align}
justifying the expression in \eqref{eq:neutsinglets}.
This formula will prove very useful in the superconformal matter transitions that we consider in the next section.

%%%%%%%%%%%%%%%%%%
\subsection{Transitions between theories}
\label{sec:trans}

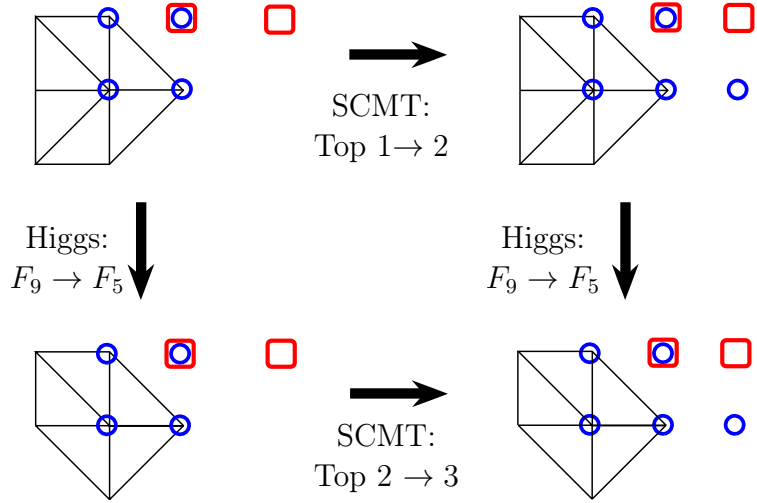
\begin{figure}
\begin{center}
 \begin{picture}(0,220)
\put(-020,140){\begin{tabular}{c} SCMT: \\ Top 1$\rightarrow 2$ \end{tabular}  }
\put(-020,15){\begin{tabular}{c} SCMT: \\ Top 2 $\rightarrow 3$ \end{tabular}  }
\put(-135,90){\begin{tabular}{c} Higgs: \\$F_9 \rightarrow F_5$ \end{tabular}  }
\put(45,90){\begin{tabular}{c} Higgs: \\$F_9 \rightarrow F_5$ \end{tabular}   }
 \put(-120,2){\includegraphics[scale=0.35]{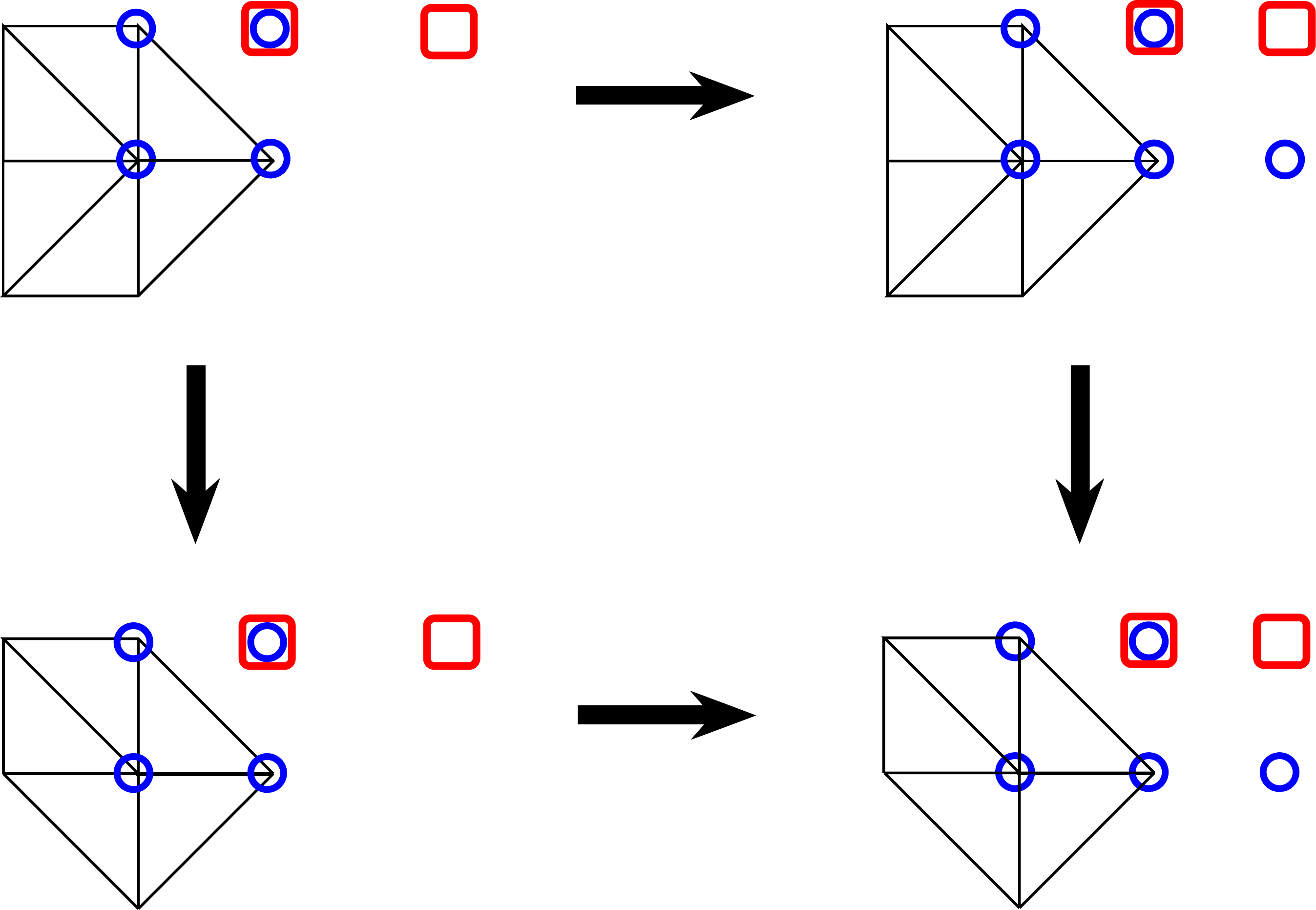}}
 \end{picture}
\end{center}
\caption{\label{fig:TopTransitions} Example for transitions between four different tops: Superconformal matter transitions (SCMTs) add a point in a face of a top and yield a model with SCPs. Higgs transitions remove a vertex of the base polytope $F_0$ and reduce the rank of the additional gauge groups besides SO(10).}  
\end{figure}

By considering the explicit structure of the tops it becomes apparent that many models are related by toric blow-ups/blow-downs, i.e.\ by the addition/removal of a vertex in the top $\Diamond$. These toric blow-ups also have a physical interpretation that depends on the height of the vertex. 
This leads to a distinction of two different transitions, which are depicted for a specific example in Fig.~\ref{fig:TopTransitions}.
 \begin{itemize}
 \item {\bf Higgs transition:} Such a transition occurs when the change in the vertex occurs at height $v^{3}=0$. This is the locus of the ambient space polygon $F_0$ of $\Diamond$, which encodes the generic fiber while the SO(10) vertices remain unchanged. Hence, while the SO(10) gauge group stays unaffected, the generic gauge group encoded in $F_i$ does change. Concretely, consider two tops ($F_i$, top A) and ($F_j$, top B) where ($F_j$, top B) has one vertex less.
It is well known that such a transition corresponds to a conifold transition since we first blow-down a divisor of the generic fiber and then resolve with a complex structure deformation by adding another monomial with base-dependent section $d_{(i,j)}$ in the residual generic fiber 
coordinates, consistent with the $\mathbb{C}^*$-actions.

Physically, we can describe this geometric process by a hypermultiplet $h$ in ($F_i$, top A) that gets a vev $\langle h \rangle \neq 0$, as long as there are enough hypermultiplets to satisfy the D-flatness conditions in six dimensions. This can also be seen from the difference in the Euler numbers of the two theories:
\begin{align}
\chi_{(F_{i},\text{ top A})}-\chi_{(F_{j},\text{ top B})}= 2 n[h] \, ,
\end{align}
where $n[h]$ is the multiplicity of the Higgs multiplets in ($F_{i}$, top A) that get a vev. As we blow down one vertex, we lose one K\"ahler modulus and therefore reduce the rank of the total gauge group by one. In addition, we find 
\begin{align}
\Delta h^{2,1} = n[h] -1
\end{align}
additional neutral singlets which give the additional D-flat directions, minus the Goldstone bosons of the Higgs multiplets, which is conform with the physical intuition.\\
Similarly, we can match the multiplicities of the charged spectrum after the higgsing from matter multiplicities that we had before: if we have two hypermultiplets $\mathbf{R}$ and $\mathbf{R}^\prime$ under the gauge group of theory ($F_{i}$, top A) that become the same representation $\widetilde{\mathbf{R}}$ under the gauge group in ($F_{j}$, top B) (up to charge conjugation), we expect the resulting multiplicity
\begin{align}
n[\widetilde{\mathbf{R}} ] = n[\mathbf{R} ] + n[\mathbf{R}^\prime ] \, .
\end{align}
Higgs transitions are very helpful for theories with tops over $F_1, \dots, F_4$, as in these models some singlet loci are complicated such that the resultant becomes unfeasible to compute on a conventional computer. Unhiggsing to a theory with simpler ideals can help to deduce the matter spectrum, as we exemplified in Section~\ref{sec:ExampleUnhiggs}.

 \item {\bf Superconformal matter transition:}  In such a situation, we introduce a blow-up in the SO(10) top at height $v^3=1$ in the top $\Diamond$. The blow-up is introduced such that one former vertex now lies in a face and therefore does not intersect the CY hypersurface in codimension one, i.e.\ the associated divisor does not introduce another Cartan generator. In addition, the transition reduces the total spectrum and induces additional non-flat fiber points at codimension two. Those transitions correspond to superconformal matter transitions \cite{Anderson:2015cqy} or tensionless string transitions \cite{Bershadsky:1996nu} and are discussed in more detail in Section~\ref{sec:SCPtrans}.
\end{itemize}

%%%%%%%%%%%%%%%%%%
\subsection{Transitions to theories with superconformal matter}
\label{sec:SCPtrans}

Similar to the Higgs transitions, we can perform a match of degrees of freedom between two tops ($F_i$, top A) and ($F_{i}$, top B) where the latter has a point in a face and thus SCPs. As the two fibrations differ only by a blow-up in the fiber, we expect these theories to be related by a smooth transition which corresponds to a physical process. 

These transitions are very useful when we consider anomaly cancellation of theories with SCPs by relating them to theories that have a well-defined SUGRA description.
This can be done by tracking the charged matter spectrum in the transition and relating missing charged multiplets to SCPs in the new theory.
To illustrate that point, we consider the transition between  ($F_5$, top 2)  and  ($F_5$, top 3). 

%%%%%%%%%
\subsubsection*{Example of SCP transition}

We start with the model ($F_5$, top 2) that has no SCPs. In terms of the sections $d_i$ of the fiber in ($F_5$, top 2) we can reach the model ($F_5$, top 3) simply by tuning $d_1$ such that it factors out one more SO(10) divisor 
\begin{align}
d_1^{(F_5, \text{top 2})} \rightarrow d_1^{(F_5, \text{top 3})} z_0 \, ,
\end{align}
which fixes some of the complex structure moduli, that we compute momentarily. 
Let us first consider the change in the total spectrum, which is given by:
\begin{align}
\label{eq:T3T2Change}
\begin{array}{clcl}
   n_{\text{SCP}} &=  + (2 \Kbi - \mathcal{S}_7) \mathcal{Z} \, ,  &
\quad \Delta \mathbf{10}_{(-1/2,0)} &= - (2 \Kbi - \mathcal{S}_7) \mathcal{Z} \, , \\
 \Delta \mathbf{16}_{(-1/4,-1/2)} &= - (2 \Kbi - \mathcal{S}_7) \mathcal{Z}\, , &
\quad  \Delta \mathbf{1}_{(-1,-1)} &= - (2 \Kbi - \mathcal{S}_7) \mathcal{Z}  \, , \\
 \Delta \mathbf{1}_{(0,1)} &= - (2 \Kbi - \mathcal{S}_7) \mathcal{Z} \, ,   &
\quad  \Delta h^{(2,1)} &= - (2 \Kbi - \mathcal{S}_7) \mathcal{Z} \, . 
  \end{array}
\end{align}
Indeed, this sums up to exactly the amount of $29 \times (2 \Kbi - \mathcal{S}_7) \mathcal{Z}$ missing hypermultiplets from ($F_5$, top 2) that are exchanged for $(2 \Kbi - \mathcal{S}_7) \mathcal{Z}$ SCPs. In this matching we have used that we can compute the change in the neutral hypermultiplets by comparing the difference in the Euler numbers similar to the Higgs transition case,
\begin{align}
\Delta \chi =  2 (\Delta h^{(1,1)} - \Delta h^{(2,1)} ) \, ,
\end{align}
which in this case reads
\begin{align}
\chi_{(F_5,\text{ top 3})} -\chi_{(F_5,\text{ top 2})} =  4 (2 \Kbi - \mathcal{S}_7) \Z \, .
\end{align}
 From \eqref{eq:SCMHodge} we deduce that $ \Delta h^{(1,1)}= (2 \Kbi - \mathcal{S}_7) \mathcal{Z}$, since the rank of the gauge group is unchanged. Hence we conclude that the change in the number of complex structure moduli is $\Delta h^{(2,1)} = -(2 \Kbi - \mathcal{S}_7)\Z $, which is exactly the change in the multiplicities of charged  hypermultiplets in \eqref{eq:T3T2Change}. Hence, we conclude that indeed a multiple of $29 \, n_{\text{SCP}}$ hypermultiplets vanish from the perturbative spectrum in order to form $n_{\text{SCP}}$ SCPs that are accounted for by non-toric K\"ahler deformations of the non-flat fiber components.
 
%%%%%%%%%%%%%%%%%%
\subsection{Anomalies in models with superconformal matter}
\label{sec:anomSCP}

After performing the blow-up procedure described in Section~\ref{sec:SCP} the CY threefold is smooth and the theory has a well-defined supergravity limit. In particular, we can check the factorization of the anomaly polynomial, similarly to Section \ref{sec:bianomalycanc}. There are two possible scenarios. Either the original theory with SCPs is directly related to a smooth geometry via a transition, see Section \ref{sec:SCPtrans} above, or it belongs to a separate class of models. In the first case, we can relate the anomaly polynomial of the blown-up geometry to the model without SCPs. In the process we can identify the charged and neutral matter multiplets that vanish in the hypermultiplet sector to account for the additional tensor multiplet after the resolution of the base. This is exemplified for the model ($F_5$, top 3) with relation to ($F_5$, top 2) below. In the second scenario, where there is no such transition, we can nevertheless evaluate the anomaly coefficients after the blow-up by using the fact that the intersection giving rise to SCPs is resolved by the blow-up. This is exemplified using the model ($F_5$, top 1) in \eqref{eq:F5top1}. Moreover, all anomaly coefficients and relations are given in Appendix \ref{App:classification}.

We start with discussing the anomaly cancellation for ($F_3$, top 3) with a transition to ($F_3$, top 2), see Section \ref{sec:SCPtrans}. In order to study the anomalies in the supergravity theory one has to resolve the base. The blow-up, see Section \ref{sec:SCP}, is performed in such a way that the base-independent intersections determining the matter multiplicities remain of the same form with hatted divisors. However, the intersection corresponding to SCPs vanishes. The number of intersection points that lead to SCPs in ($F_5$, top 3) are given by
\begin{align}
(2 K_{\hat{B}}^{-1} - \hat{\mathcal{S}}_7) \hat{\Z} = 0 \,.
\end{align}
All other matter multiplicities differ from that of ($F_5$, top 2) by multiples of this combination. A non-trivial change arises for
\begin{equation}
\begin{split}
\mathbf{10}_{-1/2,0}:& \quad (3 K_{\hat{B}}^{-1} - \hat{\mathcal{S}}_9 - 2 \hat{\Z}) \hat{\Z} - (2 K_{\hat{B}}^{-1} - \hat{\mathcal{S}}_7) \hat{\Z} \,, \\ 
\mathbf{1}_{-1,-1}:& \quad \big[ 6 (K_{\hat{B}}^{-1})^2 + \hat{\mathcal{S}}_7^2 + K_{\hat{B}}^{-1} (-5 \hat{\mathcal{S}}_7 + 4 \hat{\mathcal{S}}_9 - 2 \hat{\Z}) + \hat{\mathcal{S}}_7 (\hat{\mathcal{S}}_9 + \hat{\Z}) - 2 \hat{\mathcal{S}}_9 (\hat{\mathcal{S}}_9 + \hat{\Z}) \big] \\
& \quad - (2 K_{\hat{B}}^{-1} - \hat{\mathcal{S}}_7) \hat{\Z} \\
\mathbf{1}_{0,1}:& \quad \big[ 6 (K_{\hat{B}}^{-1})^2 - 2 \hat{\mathcal{S}}_7^2 - 2 \hat{\mathcal{S}}_9^2 + 2 \hat{\Z}^2 + K_{\hat{B}}^{-1} (4 \hat{\mathcal{S}}_7 + 4 \hat{\mathcal{S}}_9 - 11 \hat{\Z}) - 3 \hat{\mathcal{S}}_9 \hat{\Z} + 2 \hat{\mathcal{S}}_7 \hat{\Z} \big] \\
& \quad - (2 K_{\hat{B}}^{-1} - \hat{\mathcal{S}}_7) \hat{\Z} \\
\mathbf{1}_{0,0}:& \quad \big[ 19 + 11 (K_{\hat{B}}^{-1})^2 + 2 \hat{\mathcal{S}}_7^2 + 2 \hat{\mathcal{S}}_9^2 + 4 \hat{\mathcal{S}}_9 \hat{\Z} + 8 \hat{\Z}^2 - \hat{\mathcal{S}}_7 (\hat{\mathcal{S}}_9 + 2 \hat{\Z}) \\
& \quad - 2 K_{\hat{B}}^{-1} (2 \hat{\mathcal{S}}_7 + 2 \hat{\mathcal{S}}_9 + 7 \hat{\Z}) \big] - (2 K_{\hat{B}}^{-1} - \hat{\mathcal{S}}_7) \hat{\Z} \,,
\end{split}
\end{equation}
which has to be compared with Table \eqref{eq:F5top2}. Hence, we see that the anomaly coefficients are the same as for ($F_5$, top 2) in terms of the new base cohomology basis
\begin{equation}
\begin{split}
& K_{\hat{B}}^{-1} = \sum_{\alpha} \hat{a}^{\alpha} \hat{H}_{\alpha} \,, \enspace - \hat{\Z} = \sum_{\alpha} \hat{b}^{\alpha} \hat{H}_{\alpha} \,, \enspace  - (2 K_{\hat{B}}^{-1} - \tfrac{5}{4} \hat{\Z}) = \sum_{\alpha} \hat{b}_{11}^{\alpha} \hat{H}_{\alpha} \,, \\
&- (K_{\hat{B}}^{-1} - \hat{\mathcal{S}}_7 + \hat{\mathcal{S}}_9 + \tfrac{1}{2} \hat{\Z}) = \sum_{\alpha} \hat{b}_{12}^{\alpha} \hat{H}_{\alpha} \,, \enspace - (2 K_{\hat{B}}^{-1} + 2 \hat{\mathcal{S}}_9 - \hat{\Z}) = \sum_{\alpha} \hat{b}_{22}^{\alpha} \hat{H}_{\alpha} \,.
\end{split}
\end{equation}
Interestingly, the two related models differ by the particle spectrum
\begin{align}
-(2 \Kbi - \Ss) \Z \times (\mathbf{16}_{-1/4,-1/2} \oplus \mathbf{10}_{-1/2,0} \oplus \mathbf{1}_{-1,-1} \oplus \mathbf{1}_{0,1} \oplus \mathbf{1}_{0,0}) \,,
\end{align}
which amounts to a multiple of $29$ degrees of freedom which are now contained in the additional tensor multiplets.

For ($F_5$, top 1) in Table \eqref{eq:F5top1} there is no such transition to a model without SCPs. However, the anomaly factorizes modulo multiples of the SCP intersection, which vanishes after the blow-up,
\begin{align}
(\hat{\mathcal{S}}_7 - \hat{\Z}) \hat{\Z} = 0 \,.
\label{vanishSCP}
\end{align}
To demonstrate that, we evaluate the reducible SO(10) anomaly, which reads
\begin{equation}
\begin{split}
\mathcal{I}_8 &\supset \frac{1}{24} (3 - 3 n[\mathbf{45}] - \tfrac{3}{4} n[\mathbf{16}]) \big( \text{tr} \tilde{F}^2 \big)^2 = - \frac{1}{32} (2 \hat{\Z}^2 - \hat{\mathcal{S}}_7 \hat{\Z}) \big( \text{tr} \tilde{F}^2 \big)^2 \\
&= - \frac{1}{32} \big(\hat{\Z}^2 - (\hat{\mathcal{S}}_7 - \hat{\Z}) \hat{\Z} \big) \big( \text{tr} \tilde{F}^2 \big)^2 = - \frac{1}{32} \hat{\Z}^2 \big( \text{tr} \tilde{F}^2 \big)^2 \,.
\end{split}
\end{equation}
This coincides with the result expected from the geometrical interpretation discussed in Section~\ref{sec:bianomalycanc}. The other terms of the anomaly polynomial can be treated in a similar way, making use of \eqref{vanishSCP}. The anomaly coefficients read
\begin{equation}
\begin{split}
& K_{\hat{B}}^{-1} = \sum_{\alpha} \hat{a}^{\alpha} \hat{H}_{\alpha} \,, \enspace - \hat{\Z} = \sum_{\alpha} \hat{b}^{\alpha} \hat{H}_{\alpha} \,, \enspace  - (2 K_{\hat{B}}^{-1} - \tfrac{5}{4} \hat{\Z}) = \sum_{\alpha} \hat{b}_{11}^{\alpha} \hat{H}_{\alpha} \,, \\
&- (K_{\hat{B}}^{-1} - \hat{\mathcal{S}}_7 + \hat{\mathcal{S}}_9) = \sum_{\alpha} \hat{b}_{12}^{\alpha} \hat{H}_{\alpha} \,, \enspace - (2 K_{\hat{B}}^{-1} + 2 \hat{\mathcal{S}}_9 - 2 \hat{\Z}) = \sum_{\alpha} \hat{b}_{22}^{\alpha} \hat{H}_{\alpha} \,.
\end{split}
\end{equation}
Again, the anomaly coefficients match those given in \cite{Klevers:2014bqa} up to contributions of the resolved GUT divisor $\hat{\Z}$.

%%%%%%%%%%%%%%%%%%%%%%%%%%%%%%%%%%%%%%%%%%%%%%%%%%%%%%%%%%%%%%%%%%%%%%%%%
\section{Towards the Standard Model with high-scale SUSY}
\label{sec:phenomodel}

The analysis of toric F-theory vacua with gauge group $\SO{10}$
has been motivated by the search for string theory embeddings of
six-dimensional grand unified theories. Compactification of
such 6d GUT models to four dimensions on orbifolds with Wilson
lines\footnote{Note, however, that the consistency of such orbifold compactifications with F-theory compactifications to four dimensions is an important open question.}
can yield realistic extensions of the Standard Model
\cite{Asaka:2001eh,Hall:2001xr,Asaka:2003iy}. More recently, it was realized that the inclusion of Abelian magnetic flux
can relate the multiplicity of quark-lepton generations to the scale
of supersymmetry breaking, leading to extensions of the Standard Model
with high-scale SUSY \cite{Buchmuller:2015jna}. In this section we
briefly recall the main features of this 6d model and analyze its
realization in toric F-theory compactifications to six dimensions. 

%%%%%%%%%%%%%%%%%%
\subsection{A 6d supergravity SO(10) GUT model}
\label{sec:6dmodel}

We are particularly interested in Lagrangian 6d models with gauge
group $\sou$ which allow for an Abelian magnetic flux that commutes
with SO(10). Compactifications on $T^2/\mathbb{Z}_2$ with two  Wilson
lines can break $\SO{10}$ to $\SU{3} \times \SU{2} \times
\text{U(1)}^2$, the standard model gauge group supplemented by U(1)$_{B-L}$.
The gauge boson of the additional U(1) factor becomes massive by the
St\"uckelberg mechanism.
Besides the vector multiplets accounting for the gauge fields one
includes hypermultiplets in the $\mathbf{16}$ and $\mathbf{10}$
representation of $\SO{10}$. The standard model Higgs fields arise as
components of neutral $\mathbf{10}$-plets, making use of the standard
doublet-triplet splitting mechanism.
For $\mathbf{16}$-plets charged under the additional U(1) factor the
index theorem implies the appearance of  full matter generations of
fermionic zero modes. The bosonic superpartners on the other hand
receive masses of the order of the compactification scale
\cite{Bachas:1995ik}. Furthermore, uncharged $\mathbf{16}$-plets are
needed to spontaneously break U(1)$_{B-L}$.

The three quark-lepton generations of the Standard Model are
obtained from bulk $\mathbf{16}$-plets with charges $q_i$. For $N$ flux quanta they have to satisfy
the condition
\begin{align}
\sum_i  q_i N  = 3 \,,
\end{align} 
which can be realized in three ways:
\begin{equation}
\begin{array}{ c | c }
\text{charges of } \mathbf{16}_{q_i} & \text{flux quanta}\ N \\ \hline
q_1 = q_2 = q_3 & 1 \\ 
q_1 = 2 q_2 & 1 \\ 
q_1 & 3 
\end{array}
\end{equation}
All additional $\mathbf{16}$-plets have to be uncharged with respect
to the U(1) gauge symmetry containing the flux. The 6d model discussed in
\cite{Buchmuller:2015jna} contains one charged and three uncharged
$\mathbf{16}$-plets. Cancellation of the irreducible gauge anomaly
then requires six $\mathbf{10}$-plets which are all chosen to be uncharged.
One can find a set of charged and uncharged SO(10) singlets such that
all irreducible gauge and gravitational anomalies vanish and
the reducible anomalies can be canceled by a Green-Schwarz term.
The complete matter spectrum realizing this is given by 
\begin{equation}
\begin{array}{ c | c }
\text{representation} & \text{multiplicity} \\ \hline
\mathbf{16}_1 & 1 \\
\mathbf{16}_0 & 3 \\ 
\mathbf{10}_0 & 6 \\ \hline
\mathbf{1}_1 & 80 \\
\mathbf{1}_0 & 86
\end{array}
\label{modelmatterspectrum}
\end{equation}
Specifically one has
\begin{align}
H - V + 29 T - 273 = 0 \,
\end{align}
for a single tensor multiplet $T=1$. The irreducible non-Abelian
anomaly vanishes and the anomaly polynomial factorizes with $\SO{1,1}$
metric and anomaly coefficients given by
\begin{align}
\Omega = \begin{pmatrix} 0 & 1 \\ 1 & 0 \end{pmatrix} \,, \quad a = \begin{pmatrix} 2 \\ 2 \end{pmatrix} \,, \quad b = \begin{pmatrix} -1 \\ 0 \end{pmatrix} \,, \quad b_{11} = \begin{pmatrix} -4 \\ -4 \end{pmatrix} \,.
\end{align}

In the following we analyze the models arising in toric F-theory compactifications that possibly lead to a matter spectrum of the form described above and can be viewed as a starting point for supersymmetric grand unified theories in the flux background.

%%%%%%%%%%%%%%%%%%
\subsection{F-theory realizations}
\label{sec:Fpheno}

We now analyze the list of toric F-theory models with gauge group $\SO{10}$ in terms of their phenomenological viability for flux compactification of 6d GUTs. The various restrictions of the matter spectrum and the required absence of additional non-Abelian gauge groups lead to strong constraints on the allowed base spaces and fibers. The remaining models are then searched for phenomenologically viable realizations.

%%%%%%%%% 
\subsubsection*{F-theory bases}

The phenomenological constraints described above lead to certain restrictions on the geometry of the base manifold. Demanding a Lagrangian description requires  theories with a single tensor multiplet reducing the allowed bases to the
Hirzebruch surfaces $\mathbb{F}_n$ with $n \leq 12$
\cite{Morrison:2012np}. Additionally, in order to avoid non-higgsable
clusters we restrict to $n \leq 2$ and we are left with only three different possible choices for the base of the elliptically fibered Calabi-Yau manifold. However, each of the three bases allows various possibilities to embed the base divisor classes $\Ss$, $\Sn$, and $\Z$ in terms of the irreducible divisors $H_1$ and $H_2$ of~$\mathbb{F}_n$.

The intersection matrix of $\mathbb{F}_n$ is given by 
\begin{equation}
\Omega = \begin{pmatrix} n & 1 \\ 1 & 0 \end{pmatrix} \,,
\label{Hirzeintersection}
\end{equation}
and the anticanonical class reads
\begin{align}
K_{\mathbb{F}_n}^{-1} = 2 H_1 + (2 - n) H_2 \,.
\label{Hirzeantican}
\end{align}
Moreover, in order to avoid additional light fields in the 4d effective action we need the absence of hypermultiplets in the adjoint representation that are connected to the genus of the GUT divisor $\Z$ in the base,
\begin{align}
n[\mathbf{45}] = 1 - \tfrac{1}{2} (K_{\mathbb{F}_n}^{-1} - \Z) \Z \overset{!}{=} 0 \,.
\end{align}
Parametrizing $\Z$ on $\mathbb{F}_n$ in the general form,
\begin{align}
\Z = a_{\Z} H_1 + b_{\Z} H_2 \,,
\label{GUTdivpara}
\end{align}
with positive integer coefficients $a_{\Z}, b_{\Z} \in
\mathbb{Z}^+_0$, we find using \eqref{Hirzeintersection},
\eqref{Hirzeantican} and \eqref{GUTdivpara},
\begin{align}
\tfrac{1}{2} \, n \, a_{\Z} (a_{\Z} - 1) + a_{\Z} + b_{\Z} - a_{\Z} b_{\Z} \overset{!}{=} 1 \,.
\end{align}
This restricts the possible embeddings of the GUT divisor $\Z$ to genus-zero curves in the base $B = \mathbb{F}_n$.

%%%%%%%%%
\subsubsection*{F-theory fibers}

The restriction to F-theory models without additional non-Abelian gauge groups limits the ambient spaces of the fiber to $F_2$, $F_3$, $F_5$ and $F_7$. In all other cases we find either no
Abelian gauge factors or additional non-Abelian ones. Moreover, not all tops lead to $\textbf{10}$-plets that are uncharged with respect to at least one of the Abelian factors, a feature needed to reproduce the Higgs sector of the Standard Model with the required doublet-triplet splitting. In addition, neutral  $\textbf{16}$-plets 
are needed to break $B-L$. Remarkably, this leaves only a single model: ($F_3$, top 4).

Another constraint derived from the fiber is induced by the
effectiveness\footnote{The effectiveness of the sections corresponds to semi-positive coefficients for the divisor classes $[d_i]$ in the basis $\{H_1, H_2\}$ given above.} of the sections $d_i$, see e.g.\ \cite{Cvetic:2013uta}. This leads to a positive volume of the physical matter curves as well as positive intersection numbers. However, it represents a severe restriction on the base divisor classes $\Ss$, $\Sn$, and $\Z$ in $\mathbb{F}_n$. Analogous to \eqref{GUTdivpara} we parametrize
\begin{align}
\Ss = a_7 H_1 + b_7 H_2 \,, \quad \Sn = a_9 H_1 + b_9 H_2 \,.
\end{align}
Plugging this into the expression for the sections $d_i$ one can derive various inequalities for the parameters
$a_{\Z,7,9}$ and $b_{\Z,7,9}$, which leads to a rather restricted set of viable models. These inequalities are:
\begin{equation}
\begin{array}{c | r r}
\text{section} & \multicolumn{2}{c}{\text{inequalities}} \\ \hline
d_1 & 6 - a_7 - a_9 - a_{\Z} \geq 0 \,,& \quad 6 - 3n - b_7 - b_9 - b_{\Z} \geq 0 \\
d_2 & 4 - a_9 - a_{\Z} \geq 0 \,,& \quad 4 - 2n - b_9 - b_{\Z} \geq 0 \\
d_3 & 2 + a_7 - a_9 - 2 a_{\Z} \geq 0 \,,& \quad 2 - n + b_7 - b_9 - 2 b_{\Z} \geq 0 \\
d_4 & 2 a_7 - a_9 - 3 a_{\Z} \geq 0 \,,& \quad 2 b_7 - b_9 - 3 b_{\Z} \geq 0 \\
d_5 & 4 - a_7 - a_{\Z} \geq 0 \,,& \quad 4 - 2n - b_7 - b_{\Z} \geq 0 \\
d_6 & 2 - a_{\Z} \geq 0 \,,& \quad 2 - n - b_{\Z} \geq 0 \\
d_7 & a_7 - a_{\Z} \geq 0 \,,& \quad b_7 - b_{\Z} \geq 0 \\
d_8 & 2 + a_9 - a_7 \geq 0 \,,& \quad 2 - n + b_9 - b_7 \geq 0 \\
d_9 & a_9 \geq 0 \,,& \quad b_9 \geq 0 
\end{array}
\label{InequalitiesScan}
\end{equation}

%%%%%%%%%
\subsubsection*{Specific models}

Let us now examine whether one can obtain the anomaly free model
summarized in \eqref{modelmatterspectrum} as a 6d F-theory vacuum.
The $\SO{1,1}$ metric specifies the base to be $\mathbb{F}_0 =
\mathbb{P}^1 \times \mathbb{P}^1$. 
Moreover, from the anomaly coefficient for the reducible non-Abelian anomaly we can deduce, see Section~\ref{sec:bianomalycanc}
\begin{align}
\Z = H_1
\end{align}
The  matter spectrum containing charged and uncharged
$\mathbf{16}$-plets and neutral $\mathbf{10}$-plets 
and the gauge group $\sou$ singles out the fiber ($F_3$, top 4). The multiplicities of $\mathbf{16}_{-1}$ and $\mathbf{10}_0$ given in \eqref{modelmatterspectrum} further restrict the base divisor classes of $\Ss$ and $\Sn$ to 
\begin{align}
\Ss = a_7 H_1 + 6 H_2 \,, \quad \Sn = a_9 H_1 + H_2 \,,
\label{modelsections}
\end{align}
with $a_{7,9} \in \mathbb{Z}^+_0$. Plugging this into the base-independent multiplicities of the remaining matter representations $\mathbf{16}_0$ and $\mathbf{10}_1$ in \eqref{bimultis} we find
\begin{equation}
\begin{split}
\mathbf{16}_0: \quad & (2 \Kbi - \Sn - \Z) \Z = 3 \,, \\
\mathbf{10}_1: \quad & (3 \Kbi - \Ss - \Z) \Z = 0 \,,
\end{split}
\end{equation}
as desired. The singlet multiplicities are evaluated accordingly, using \eqref{bimultis},
\begin{equation}
\begin{split}
\mathbf{1}_3: \quad & 2 - a_7 - 2 a_9 \,, \\
\mathbf{1}_2: \quad & -6 + 4 a_7 + 16 a_9 \,, \\
\mathbf{1}_1: \quad & 158 - 31 a_7 + 2 a_9 \,.
\end{split}
\end{equation}
One immediately verifies that with positive integer coefficients
$a_7$ and $a_9$ one cannot reproduce the multiplicities of charged
singlets in Table \eqref{modelmatterspectrum}. We conclude that the
model described in Section~\ref{sec:6dmodel} belongs to the `toric swampland', i.e.\ it cannot be reproduced from 6d F-Theory vacua that are obtained from hypersurfaces in toric varieties. 

It is interesting, however, that for a different spectrum of charged
SO(10) singlets one can find a model with the SO(10) matter content of
\eqref{modelmatterspectrum}. Setting $a_7 = 2$ and $a_9 = 0$ yields
the charged singlet multiplicities 
\begin{equation}
\begin{array}{ c | c }
\text{singlets} & \text{multiplicities} \\ \hline
\mathbf{1}_3 & 0 \\ 
\mathbf{1}_2 & 2 \\ 
\mathbf{1}_1 & 96
\end{array}
\end{equation}
This model has the anomaly coefficients
\begin{align}
a = \begin{pmatrix} 2 \\ 2 \end{pmatrix} \,, \quad b = \begin{pmatrix} -1 \\ 0 \end{pmatrix} \,, \quad b_{11} = \begin{pmatrix} -7 \\ -2 \end{pmatrix} \,,
\end{align}
and also solves all anomaly constraints, with $68$ neutral singlets.
However, not all sections $d_i$ are effective,
\begin{equation}
\begin{array}{l l l l}
&[d_1] \sim 3 H_1 - H_2 \,, & [d_2] \sim 3 H_1 + 3 H_2 \,, & [d_3] \sim 2 H_1 + 7 H_2 \,,\\
&[d_4] \sim H_1 + 11 H_2 \,, & [d_5] \sim H_1 - 2 H_2 \,, & [d_6] \sim H_1 + 2 H_2 \,, \\
&[d_7] \sim H_1 + 6 H_2 \,, & [d_8] \sim - 3 H_2 \,, & [d_9] \sim H_2 \,.
\end{array}
\end{equation}
The divisor classes $[d_1]$, $[d_5]$ and $[d_8]$ have one negative
coefficient, which implies that the corresponding geometry cannot be
smoothly realized.

So far we have required that all ${\bf 10}$-plets carry zero U(1)
charge. Phenomenologically, this is not necessary. Two neutral ${\bf 10}$-plets 
are sufficient to realize doublet-triplet splitting. Further charged
${\bf 10}$-plets will lead to zero modes of ${\bf 10}$-plets, but this
is vector-like matter which can obtain mass at orbifold fixed points
or via couplings to singlet fields. One can easily repeat the above
analysis without specifying the number of uncharged ${\bf 10}$-plets.
Instead of Eq.~\eqref{modelsections} one then starts from 
\begin{align}
\Ss = a_7 H_1 + b_7 H_2 \,, \quad \Sn = a_9 H_1 + H_2 \,.
\label{newmodelsections}
\end{align}
One again has one charged and three uncharged ${\bf 16}$-plets whereas
the number of uncharged and charged ${\bf 10}$-plets is now $b_7$ and
$6-b_7$, respectively. 
Using \eqref{bimultis} one obtains for the singlet multiplicities
\begin{equation}
\begin{split}
\mathbf{1}_3: \quad & 2 - a_7 + 4 a_9  - b_7 a_9\,, \\
\mathbf{1}_2: \quad & 42 - 8a_7 -8b_7  + 4 a_9 + 2a_7b_7 + 2b_7a_9 \,, \\
\mathbf{1}_1: \quad & 50 + 17 a_7 + 18 b_7 - 4 a_9 - 8a_7b_7 + b_7a_9 \,.
\end{split}
\end{equation}
It is also straightforward to evaluate the divisor classes $[d_i]$ in
terms of $a_7$, $b_7$ and $a_9$. We find that for $a_7 = 2$, $b_7 = 3$
and $a_9 = 0$ all multiplicities are non-negative and all divisor
classes are effective:
\begin{equation}
\begin{array}{l l l l}
&[d_1] \sim 3 H_1 + 2 H_2 \,, & [d_2] \sim 3 H_1 + 3 H_2 \,, & [d_3] \sim 2 H_1 + 4 H_2 \,,\\
&[d_4] \sim H_1 + 5 H_2 \,, & [d_5] \sim H_1 + H_2 \,, & [d_6] \sim H_1 + 2 H_2 \,, \\
&[d_7] \sim H_1 + 3 H_2 \,, & [d_8] \sim 0 \,, & [d_9] \sim H_2 \,.
\end{array}
\end{equation}
The complete matter spectrum is given by
\begin{equation}
\begin{array}{ c | c }
\text{representation} & \text{multiplicity} \\ \hline
\mathbf{16}_{-1} & 1 \\
\mathbf{16}_0 & 3 \\ 
\mathbf{10}_1 & 3 \\ 
\mathbf{10}_0 & 3 \\ \hline
\mathbf{1}_3 & 0 \\
\mathbf{1}_2 & 14 \\
\mathbf{1}_1 & 90 \\
\mathbf{1}_0 & 62 \\ \hline
\end{array}
\label{modelmatterspectrumalt}
\end{equation}

Since the phenomenological constraints leave only a single model we can scan through all possibilities for effective sections with the general parametrization for $\Ss$, $\Sn$ and $\Z$ satisfying all inequalities \eqref{InequalitiesScan}. Further demanding the presence of at least two uncharged $\mathbf{10}$-plets and two uncharged $\mathbf{16}$-plets for the breaking of the electroweak symmetry and U(1)$_{B-L}$, we find 33 possible realizations. 25 of these contain a single charged $\mathbf{16}$-plet, the remaining 8 models have three charged $\mathbf{16}$-plets. The explicit base divisors and complete matter spectra are summarized in Appendix~\ref{ScanF3Top4}.

%%%%%%%%%%%%%%%%%%
\subsection{Phenomenological aspects}

For completeness we now briefly recall some phenomenologically
attractive features of the 6d supergravity models discussed above
and discuss some connections to 4d F-theory models.

The magnetic flux in the additional U(1) gauge group leads to a
positive energy density and breaks supersymmetry. However, the
positive D-term potential with runaway behavior induced by the flux can be combined with a
nonperturbative superpotential at the orbifold fixed points leading to
metastable de Sitter vacua \cite{Buchmuller:2016dai,
  Buchmuller:2016bgt}. In these models electroweak symmetry breaking still remains to
be investigated, and one may worry that extreme fine-tuning will be
necessary to obtain a Fermi scale far below the GUT scale. On the
other hand, in an Abelian toy model it was recently shown that for
flux compactifications a cancellation of quadratic divergences can be achieved at one-loop
level if the whole Kaluza-Klein tower of Landau levels is taken into
account \cite{Buchmuller:2016gib, Ghilencea:2017jmh}. One can expect
that similar cancellations will also occur for quantum corrections in
the Higgs sector. Furthermore, flux compactifications can lead to a
characteristic flavor structure of quark and lepton mass
matrices. Due to the generation of Yukawa couplings at orbifold fixed
points one starts from mass matrices of rank one, which can be
increased by mixings with additional vector-like matter \cite{Buchmuller:2017vho}.

The flavor structure of fermion mass matrices has been
extensively studied in 4d F-theory models. In the singular limit
Yukawa couplings arise at codimension-three singularities 
and therefore their flavor structure is a matrix of rank one 
\cite{Font:2008id,Heckman:2008qa}. This is an interesting starting
point to obtain hierarchical quark and lepton mass matrices. Necessary
modifications can be realized by an appropriate choice of magnetic
fluxes \cite{Font:2008id} and by taking 
subleading corrections due to finite-size effects of the wavefunctions 
into account \cite{Heckman:2008qa}. Interesting mass matrices are also
obtained by analyzing the dependence of the zero mode wavefunctions on
complex structure moduli \cite{Hayashi:2009bt}. Heavy messenger states
can generate mass matrices of Frogatt-Nielsen type
\cite{Leontaris:2010zd}. Taking these ingredients into account, quark
and lepton mass matrices can be successfully accounted for by
particular local SU(5) F-theory GUTs \cite{Carta:2015eoh} as well as
various global SU(5) F-theory GUTs \cite{Krippendorf:2015kta}. 
Furthermore, also low mass vector-like matter can be incorporated,
with interesting signatures at the LHC \cite{Leontaris:2017vzz}.

%%%%%%%%%%%%%%%%%%%%%%%%%%%%%%%%%%%%%%%%%%%%%%%%%%%%%%%%%%%%%%%%%%%%%%%%%
\section{Summary and outlook}
\label{sec:summary}

In this paper we have classified 6d F-theory vacua with gauge group
SO(10) and additional Mordell-Weil U(1) and discrete gauge factors,
based on toric geometry. To each of the 16 polygons serving as a toric ambient space of a
torus, we have added all possible SO(10) tops. This gave us 36
polytopes from which we constructed $K3$ manifolds with
gauge group SO(10). By  tuning the parameters of an analog construction that gives rise to CY threefolds, we
identified the allowed matter representations and their loci. For these threefolds we computed, solely from geometry, the multiplicities for all SO(10) charged and uncharged multiplets and confirmed, base-independently, cancellation of all gauge and gravitational anomalies.

The various steps of our analysis were explained in detail for a
particular example, the fiber $F_3$. Fibering a torus with
Mordell-Weil group of rank one over a $\mathbb{P}^1$, one obtains
a $K3$ manifold that can be tuned to have an SO(10) singularity.
We resolved this singularity by adding an SO(10) top, ($F_3$, top 1).
This yields five additional $\mathbb{P}^1$s
for the five SO(10) roots with an intersection pattern
corresponding to the extended Dynkin diagram of SO(10).
Particular emphasis has been put on the matter splits. At the
matter loci some SO(10) $\CP$s split into chains of several
new $\CP$s. From these splits and their common $\CP$s one obtains
the extended Dynkin diagrams of the enhanced symmetries SO(12)
or E$_6$. Further tuning leads to codimension-three singularities
where Yukawa couplings occur. At these loci the SO(10) $\CP$s split
further, leading to the extended Dynkin diagram of E$_7$ as well as
other non-ADE intersection patterns.

After specifying a complex two-dimensional base, matter multiplicities 
can be evaluated explicitly. This
is rather straightforward for SO(10) matter but much more involved for
charged and neutral SO(10) singlets. 
Unhiggsing the considered model to another theory with two U(1)
factors, corresponding to the fiber ($F_5$, top 2), the resultant
equations could be solved and the singlet multiplicities were evaluated
base-independently. Moreover, we obtained general expressions for the
Euler numbers in terms of the base divisor classes. This allowed us
to determine the multiplicities of uncharged singlets and to verify
the cancellation of all anomalies base-independently. This procedure has been applied for all 36 models.

Our analysis led to several intriguing insights concerning 6d F-theory vacua.
First, superconformal points appear rather frequently as non-flat fiber points in codimension two in the base. This is not unexpected for
large gauge groups such as SO(10). These points can be removed
by adding exceptional divisors in the base, leading to new tensor
multiplets, corresponding to the tensor branch of the superconformal point. This interpretation is supported by an analysis of full anomaly cancellation which we confirmed in a base-independent way.

Moreover, we discussed the contribution of superconformal points to non-toric K\"ahler deformations in the smooth fibration, which
have to be taken into account when determining the neutral singlets.
Thus, we provide a full description of global 6d theories with superconformal points
coupled to SO(10) and Abelian discrete and continuous gauge factors for the first time. 

We also demonstrated that to a large extent the various SO(10) models
are related by several transitions. First, we relate various SO(10) tops via the
Higgs mechanism realized as conifold transitions in the generic fiber. 
We then considered tensionless string transitions
which are the result of adding an additional vertex to the SO(10) top
 , such that one of the original vertices becomes contained in
the interior of a face. In these transitions we match the change of the full matter spectrum
base-independently and confirm the appearance of non-toric K\"ahler deformations
that are supported by SCPs.

Our work has been motivated by 6d supergravity models with gauge group
$\sou$ which, after orbifold compactifications with flux, can yield
viable extensions of the Standard Model with high-scale
supersymmetry. It appears that the model originally considered belongs
to the `toric swampland'. However, we could identify variants which
are phenomenologically promising. These models contain several charged
and uncharged ${\bf 16}$-plets and ${\bf 10}$-plets together with a
large number of charged and uncharged SO(10) singlets. To explore
these models further requires progress on several conceptually
important questions: the compactification of F-theory on orbifolds, possibly along the lines of \cite{Ludeling:2014oba},
the incorporation of Wilson line symmetry breaking and the consistency
of F-theory vacua with a high scale of flux-induced supersymmetry breaking.

%%%%%%%%%%%%%%%%%%%%%%%%%%%%%%%%%%%%%%%%%%%%%%%%%%%%%%%%%%%%%%%%%%%%%%%%%
\section*{Acknowledgments}
We thank Andreas Braun,  Thomas Grimm, Ling Lin, Wati Taylor, Dami\'an Kaloni Mayorga Pe\~na and Timo Weigand for valuable discussions.
This work was supported by the German Science Foundation (DFG) within the Collaborative Research Center (SFB) 676 ``Particles,
Strings and the Early Universe''. M.D. also acknowledges support from the Studienstiftung des deutschen Volkes. The work of FR is supported by the EPSRC network grant EP/N007158/1.

\clearpage

% #################################
% #            Appendix          #
% #################################
%%%%%%%%%%%%%%%%%%%%%%%%%%%%%%%%%%%%%%%%%%%%%%%%%%%%%%%%%%%%%%%%%%%%%%%%%%%%%%%%%%%%%%%%%%%%%%%%%%%%%%%%%%%%%%%%%%%%%%%%%%%%%%%%
\begin{appendix}

%%%%%%%%%%%%%%%%%%%%%%%%%%%%%%%%%%%%%%%%%%%%%%%%%%%%%%%%%%%%%%%%%%%%%%%%%
\section{More details on the construction of 6d vacua}
\label{a:d6dvac}

For a polytope $\Delta$ with vertices $v_i$, the vertices $m_i$ of the
dual polytope $\Delta^*$ are the solutions of the inequalities  
\begin{align}\label{dualvertices}
\langle m_i,v_j\rangle \geq -1\;\;  \forall \, i,j \,.
\end{align}
The polynomial defining the CY twofold reads\footnote{In Section~\ref{sec:clasvac} this equation is obtained from the polynomial of a non-compact CY two-fold using Batyrev's construction \cite{Batyrev:1994hm} around equations \eqref{pDiamond}-\eqref{pDelta2}.}
\begin{align*}
p_{\Delta} =\sum_{m_i \in \Delta^*} a_i \prod_{v_j \in \Delta}
x_j^{\langle m_i, v_j \rangle + 1 } = \sum_{m_i \in \Delta^*} a_i
\left(\prod_{v_s \in F_3} x_s^{ \langle m_i,v_s\rangle+1}\right)  
\left(\prod_{v_t \in \Delta , v_t \not\in F_3} x_t^{ \langle
    m_i,v_t\rangle+1}\right) \,.
\end{align*}
Its partially  factorized form is preserved once a bottom is
added to the top,
\begin{align}\label{pDelta}
p_{\Delta} = \sum_{m_i \in \Delta^*} a_i \left( \prod_{v_j \in F_3} x_j^{\langle m_i,v_j\rangle +1} \right) \left(
\prod_{v_j \in \Diamond, v_j \not\in F_3} x_j^{\langle m_i,v_j\rangle +1}\right)\left( \prod_{v_j \in \text{base}} x_j^{\langle m_i,v_j\rangle +1} \right)\,.
\end{align}
It is a polynomial in the homogeneous coordinates $x_i$ of the
ambient space with non-negative exponents. The ambient space contains
a Calabi-Yau manifold which is cut out by 
\begin{align}
Y_\Delta = \{p_\Delta = 0\}\,.
\end{align}

In Section~\ref{T6dvac} we start from the polygon $F_3=dP_1$ with
coordinates and vertices given in Table~\ref{tab:1foldexample}. 
The dual polygon has 9 vertices,
\begin{equation}
\begin{array}{l l l l l}
m_1 = (1,0)\,,&\; m_2= (0,1)\,,&\; m_3 = (1,1)\,,&\; m_4 = (1,-1)\,,&\; m_5 = (0,0)\,, \\
m_6 = (-1,0)\,,& \; m_7 = (-1,-1)\,,&\; m_8 = (-2,-1)\,,&\; m_9 = (0,-1)\,.&
\end{array}
\end{equation}
Using Eq.~\eqref{pDelta}, without top and bottom, these vertices determine the polynomial
\eqref{p1} from which one obtains a torus with a non-trivial Mordell-Weil group, giving rise to a U(1) gauge group.

An important piece of data of a polytope, or rather its fan is its associated Stanley-Reisner ideal, which is a set of monomials. Each monomial contains homogeneous coordinates whose corresponding rays do not form a cone. The construction of the ambient space
is such that these monomials cannot vanish, which is important in the
calculation of intersection numbers. For the polygon $F_3$ one has
\begin{align}
\text{SRI}: \{uv, w e_1 \} \,.
\end{align}
If a linear combination of divisors $D_{x_i}$ in an ambient space is a principal
divisor, i.e.\
\begin{align}
D= \sum_i c_i D_{x_i}\;\; \langle c,v_j \rangle = 0\,,\; \forall
v_j \in \Delta\,,
\end{align}
the divisor is called linear dependent, $D \sim 0$. Such linear
combinations of divisors do not affect intersection numbers and can
therefore be conveniently added. 

Extending the ambient space from $dP_1$ to a fibration of $dP_1$ over $\CP$ one 
obtains the CY two-fold $\hat{Y}_2$, i.e.\ a $K3$. The homogeneous coordinates and
vertices of a corresponding three-dimensional polytope are given in
Table~\ref{tab:2foldsingular}. The dual polytope now contains the following 28 vertices,
\begin{equation}
\begin{array}{l l l l}
m_1~ = (1,0,0)\,,&\; m_2~= (0,1,0)\,,&\; m_3~ = (1,1,0)\,,&\; m_4~ = (1,-1,0)\,, \\ 
m_5~ = (0,0,0)\,,& \; m_6~ = (-1,0,0)\,,& \; m_7~ = (-1,-1,0)\,,&\; m_8~ = (-2,-1,0)\,, \\ 
m_9~ = (0,-1,0)\,,&\; m_{10} = (0,1,1)\,,&\; m_{14} = (0,0,1)\,,&\; m_{15} = (-1,0,1)\,,\\
m_{16} = (-1,-1,1)\,,&\; m_{17} = (-2,-1,1)\,,&\; m_{18} = (0,-1,1)\,,&\; m_{21} = (-1,0,2)\,,\\
m_{23} = (-1,-1,2)\,,&\; m_{24} = (-2,-1,3)\,,&\; m_{25} = (1,0,-1)\,,&\; m_{26} = (0,1,-1)\,,\\
m_{27} = (1,1,-1)\,,&\; m_{28} = (1,-1,-1)\,,&\; m_{29} = (0,0,-1)\,,&\; m_{30} = (-1,0,-1)\,,\\
m_{31} = (-1,-1,-1)\,,&\; m_{32} = (-2,-1,-1)\,,&\; m_{33} = (0,-1,-1)\,,&\; m_{34} = (-2,-1,2) \,.
\end{array}
\end{equation}
They determine the polynomial $p_{\hat{Y}_2}$ from which one obtains a
$K3$ manifold.
Since all these vertices have to be orthogonal to the vertices of the
polytope associated with the coordinates $u$, $v$, $w$ and $e_1$, the
various terms of the polynomial factorize into the monomials of $u$, $v$,
$w$ and $e_1$, already present in Eq.~\eqref{p1}, with sections $s_i$, 
$i = 1\ldots 9$, which depend on the base coordinates $z_0$ and $z_1$ as follows:
\begin{equation}
\begin{array} {l l}
s_1= a_4 z_0 + a_{28} z_1\,,&\quad s_2 = a_{18} z_0^2 + a_9 z_0 z_1 + a_{33} z_1^2\,,\\  
s_3= a_{23} z_0^3 + a_{16} z_0^2 z_1 + a_7 z_0 z_1^2 + a_{31} z_1^3\,,& \\
s_4= a_{24} z_0^4 + a_{34} z_0^3 z_1 + a_{17} z_0^2 z_1^2 + a_8 z_0 z_1^3
+ a_{32} z_1^4\,,& \\
s_5= a_1 z_0 + a_{25} z_1\,,&\quad s_6 = a_{14} z_0^2 + a_5 z_0 z_1 + a_{29} z_1^2\,,\\
s_7= a_{21} z_0^3 + a_{15} z_0^2 z_1 + a_6 z_0 z_1^2 + a_{30} z_1^3\,,&\quad s_8 = a_3 z_0 + a_{27} z_1\,,\\
s_9= a_{10} z_0^2 + a_2 z_0 z_1 + a_{26} z_1^2 \,.&
\end{array}
\end{equation}
For generic coefficients $\hat{Y}_2$ is a smooth $K3$ manifold. However,
for a specific tuning of coefficients,
\begin{align}\label{tuning}
a_{28} = a_9 = a_{33} = a_7 = a_{31} = a_{32} = a_8 = a_{17} = a_{29}
= a_{30} = 0\,, 
\end{align}
one obtains a certain factorization of the sections, $s_i =
z_0^{n_i} d_i$, with powers $n_i$ given in Eq.~\eqref{p2a} and coefficients 
\begin{equation}
\begin{array}{ll}\label{diY2}
d_1 = a_4\,,& \quad d_2 = a_{18} \,,\\
d_3 = a_{23} z_0 + a_{16} z_1\,,& \quad d_4 = a_{24} z_0 + a_{34} z_1\,, \\
d_5 = a_1 z_0 + a_{25} z_1\,,&\quad d_6 = a_{14} z_0 + a_5 z_1\,,\\
d_7 = a_{21} z_0^2 + a_{15} z_0 z_1 + a_6 z_1^2\,,&\quad d_8 = a_3 z_0 + a_{27} z_1\,,\\
d_9 = a_{10}z_0^2 + a_2 z_0 z_1 + a_{26} z_1^2\,.&
\end{array}
\end{equation} 
After tuning 10 coefficients to zero, the reduced dual polytope has 18
vertices. A particular triangulation leads to the Stanley-Reissner ideal
\begin{align}
\text{SRI}: \{uv, w e_1, z_0 z_1 \} \,.
\end{align}
As shown in Section~\ref{sing6dvac}, the $K3$ manifold obtained from
the reduced polytope has an SO(10) singularity. The choice of the
particular tuning \eqref{tuning} can be understood from the construction
of a smooth $K3$ with an SO(10) top. 

The reduced polytope with 18 vertices is the dual of a polytope with
an SO(10) top. The vertices are listed in
Table~\ref{tab:2foldsmooth}. In Figure~\ref{fig:SO10Top} the vertices
of the same top together with an enlarged basis $\mathbb{F}_0 = \mathbb{P}_1\times
\mathbb{P}_1$ are given.  The dual polytope now contains 62 vertices,
\small
\begin{equation}
\begin{array}{l l l l}
m_1~ = (0,1,0,0)\,,&\; m_2~= (0,0,0,0)\,,&\; m_3~ = (-1,0,0,0)\,,&\; m_4~ = (0,1,1,0)\,, \\ 
m_5~ = (0,0,1,0)\,,& \; m_6~ = (-1,0,1,0)\,,&\; m_7~ = (-1,-1,1,0)\,,&\; m_8~ = (0,-1,1,0)\,, \\ 
m_9~ = (-1,0,2,0)\,,&\; m_{10} = (-1,-1,2,0)\,,&\; m_{11} = (-2,-1,3,0)\,,&\; m_{12} = (0,1,-1,0)\,,\\
m_{43} = (-2,-1,2,0)\,,&\; m_{13} = (1,0,0,1)\,,&\; m_{14} = (0,1,0,1)\,,&\; m_{15} = (1,1,0,1)\,,\\
m_{16} = (1,-1,0,1)\,,&\; m_{17} = (0,0,0,1)\,,&\; m_{18} = (-1,0,0,1)\,,&\; m_{19} = (0,1,1,1)\,,\\
m_{62} = (0,0,1,1)\,,&\; m_{20} = (-1,0,1,1)\,,&\; m_{21} = (-1,-1,1,1)\,,&\; m_{22} = (0,-1,1,1)\,,\\
m_{23} = (-1,0,2,1)\,,&\; m_{24} = (-1,-1,2,1)\,,&\; m_{25} = (-2,-1,2,1)\,,&\; m_{26} = (-2,-1,3,1)\,,\\
m_{27} = (1,0,-1,1)\,,&\; m_{28} = (0,1,-1,1)\,,&\; m_{29} = (1,1,-1,1)\,,&\; m_{30} = (0,1,0,-1)\,,\\
m_{31} = (0,0,0,-1)\,,&\; m_{32} = (-1,0,0,-1)\,,&\; m_{33} = (0,1,1,-1)\,,&\; m_{34} = (0,0,1,-1)\,,\\
m_{35} = (-1,0,1,-1)\,,&\; m_{36} = (-1,-1,1,-1)\,,&\; m_{37} = (0,-1,1,-1)\,,&\; m_{38} = (-1,0,2,-1)\,,\\
m_{39} = (-1,-1,2,-1)\,,&\; m_{40} = (-2,-1,2,-1)\,,&\; m_{41} =(-2,-1,3,-1)\,,&\; m_{42} = (0,1,-1,-1)\,,\\
m_{44} = (-2,-1,2,-2)\,,&\; m_{45} = (-2,-1,2,-3)\,,&\; m_{46} = (-2,-1,2,-4)\,,&\; m_{47} = (-2,-1,2,-5)\,,\\
m_{48} = (-2,-1,3,-2)\,,&\; m_{49} = (-2,-1,3,-3)\,,&\; m_{50} = (-2,-1,3,-4)\,,&\; m_{51} = (-2,-1,2,-5)\,,\\
m_{52} = (-1,0,1,-2)\,,&\; m_{53} = (-1,0,1,-3)\,,&\; m_{54} = (-1,-1,1,-2)\,,&\; m_{55} = (-1,-1,1,-3)\,,\\
m_{56} = (-1,0,2,-2)\,,&\; m_{57} = (-1,0,2,-3)\,,&\; m_{58} = (-1,-1,2,-2)\,,&\; m_{59} = (-1,-1,2,-3)\,,\\
m_{60} = (-1,0,0,-2)\,,&\; m_{61} = (-1,0,0,-3) \,.&&
\end{array}
\end{equation}
\normalsize
The polynomial $p_{Y_3}$ is again of the
form \eqref{p1}. The sections $s_i$ factorize into powers of $z_0$,
$f_2$,~$\ldots$,~$f_4$ given in Eq.~\eqref{p2b}, and coefficients $d_i$ which
are functions of $z=z_0f_2g_1^2g_2^2f_3f_4$, $z_1$, $z_2$ and $z_3$,
\begin{equation}
\begin{split}\label{diY3}
d_1 =&a_{16} \,,\\
d_2 =&a_{22} z_2^2 + a_8 z_2z_3 + a_{37} z_3^2\,,\\
d_3 =&z(a_{24} z_2^4 + a_{10} z_2^3z_3 + a_{39} z_2^2z_3^2 + a_{58} z_2z_3^3 + a_{59}z_3^4)\\
&+z_1(a_{21} z_2^4 + a_7 z_2^3z_3 + a_{36} z_2^2z_3^2 + a_{54} z_2z_3^3 + a_{59} z_3^4)\,,\\
d_4 =&z(a_{26} z_2^6 + a_{11} z_2^5z_3 + a_{41} z_2^4z_3^2 +
a_{48} z_2^3z_3^3 + a_{49} z_2^2z_3^4 + a_{50} z_2z_3^5 + a_{51} z_3^6)\\
&+z_1(a_{25} z_2^6 + a_{43} z_2^5z_3 + a_{40} z_2^4z_3^2 + a_{44}
z_2^3z_3^3 + a_{45} z_2^2z_3^4 + a_{46} z_2z_3^5 + a_{47} z_3^6)\,,\\
d_5 =&a_{13} z + a_{27} z_1\,,\\
d_6 =&z(a_{62} z_2^2 + a_5 z_2z_3 + a_{34} z_3^2) + z_1(a_{17}z_2^2 +
a_2 z_2z_3 + a_{31} z_3^2)\,,\\
d_7 =&z^2(a_{23} z_2^4 + a_9 z_2^3z_3 + a_{38} z_2^2z_3^2 +
a_{56} z_2z_3^3 + a_{57} z_3^4)\\
&+z z_1(a_{20} z_2^4 + a_6 z_2^3z_3 + a_{35} z_2^2z_3^2 + a_{52}
z_2z_3^3 + a_{53} z_3^4)\\
&+z_1^2 (a_{18} z_2^4 + a_3 z_2^3z_3 + a_{32} z_2^2z_3^2 + a_{60}
z_2z_3^3 + a_{61} z_3^4)\,,\\
d_8 =&a_{15}z + a_{29} z_1\,,\\
d_9 =&z^2(a_{19} z_2^2 + a_4 z_2z_3 + a_{30} z_3^2) + \omega
z_1(a_{14} z_2^2 + a_1 z_2z_3 + a_{30} z_3^2)\\
&+z_1^2(a_{28} z_2^2 + a_{12} z_2z_3 + a_{42} z_3^2)  \,.
\end{split}
\end{equation}
For fixed $z_2$ and $z_3$ the coefficients are identical to the 
expressions given in \eqref{diY2}, with $z_0$ replaced by $z$.

The calculation of intersection numbers is based on a triangulation
which corresponds to a Stanley-Reissner ideal (calculated with SAGE),
\begin{align}
\text{SRI}: \{& uv, w e_1, u g_1, u g_2, u f_3, v f_2, v g_1, v g_2, v f_4,
w z_0, w f_2, w g_1, e_1 f_2,  e_1 g_1, e_1 g_2, \nonumber\\  
&e_1 f_3, e_1 f_4, z_0 g_2, z_0 f_4, f_2 g_2, f_2 f_3, f_3 f_4, z_0 z_1, f_2 z_1, g_1 z_1, g_2 z_1, f_3 z_1, f_4 z_1 \} \, .
\end{align}
Note that the polytope given  in Figure~\ref{fig:SO10Top} admits 81
triangulations.

%%%%%%%%%%%%%%%%%%%%%%%%%%%%%%%%%%%%%%%%%%%%%%%%%%%%%%%%%%%%%%%%%%%%%%%%%
\section{Elliptic curves, divisor classes and Weierstrass form}
\label{sec:WeierstrassForm}

In most cases of our analysis, the elliptic curves are obtained
from the cubic polynomial \eqref{cubic}. 
However, for the fibers $F_2$ and $F_4$ of the genus-one curves are given
by bi-quadratic and quartic polynomials, respectively,
 \begin{align}
 p_{F_2} &= (b_1 y^2+ b_2 s y + b_3 s^2) x^2 + (b_5 y^2+ b_6 s y + b_7
 s^2) x t + (b_8 y^2+ b_9 s y + b_{10} s^2) t^2 \, ,
\label{biquadratic}\\
 p_{F_4} &= c_1 e_1^2 X^4 + c_2 e_1^2 X^3 Y + c_3 e_1^2 X^2 Y^2 + c_4
 e_1^2 X Y^3 + c_5 e_1^2 Y^4 + c_6 e_1 X^2 Z \, . \label{quartic}
 \end{align}
For a given top, the dependence of the sections on the GUT divisor
factorizes in a characteristic way and, following the procedure
described in Section~\ref{sec:polytop}, one obtains for the divisor classes of the
sections $[b_i]$ and $[c_i]$ the dependence on the base divisor classes as follows
\begin{align}
 \text{
\begin{tabular}{c|l}
\text{section} & \multicolumn{1}{c}{\text{divisor class}}\\
\hline
	$[b_1]$&$3[K_B^{-1}]-\cS_7-\cS_9 \rule{0pt}{13pt} - n_i \mathcal{Z}$ \\
	$[b_2]$&$2[K_B^{-1}]-\cS_9 \rule{0pt}{12pt}- n_i \mathcal{Z} $ \\
	$[b_3]$&$[K_B^{-1}]+\cS_7-\cS_9 \rule{0pt}{12pt}- n_i \mathcal{Z}$ \\
	$[b_5]$&$2[K_B^{-1}]-\cS_7 \rule{0pt}{12pt} - n_i \mathcal{Z}$  \\
	$[b_6]$&$[K_B^{-1}] \rule{0pt}{12pt}- n_i \mathcal{Z} $ \\
	$[b_7]$&$\cS_7 \rule{0pt}{12pt} - n_i \mathcal{Z}$  \\
	$[b_8]$&$[K_B^{-1}]+\cS_9-\cS_7 \rule{0pt}{12pt}- n_i \mathcal{Z}$  \\
	$[b_9]$&$\cS_9  \rule{0pt}{12pt} - n_i \mathcal{Z} $ \\
	$[b_{10}]$&$\cS_9+\cS_7-[K_B^{-1}]  \rule{0pt}{12pt} - n_i \mathcal{Z} $
\end{tabular}
}
 \text{
\begin{tabular}{c|l}
\text{section} & \multicolumn{1}{c}{\text{divisor class}}\\
\hline
	$[c_1]$ & $3[K_B^{-1}]-\cS_7-\cS_9   \rule{0pt}{13pt} - n_i \mathcal{Z}$ \\
	$[c_2]$ & $2[K_B^{-1}]-\cS_9  \rule{0pt}{12pt} - n_i \mathcal{Z}$ \\
	$[c_3]$ & $[K_B^{-1}]+\cS_7-\cS_9   \rule{0pt}{12pt} - n_i \mathcal{Z} $ \\ 
	$[c_4]$ & $2\cS_7-\cS_9  \rule{0pt}{12pt} - n_i \mathcal{Z}$ \\ 
	$[c_5]$ & $-[K_B^{-1}]+3\cS_7-\cS_9  \rule{0pt}{12pt}- n_i \mathcal{Z}  $\\
	$[c_6]$ & $2[K_B^{-1}]-\cS_7  \rule{0pt}{12pt}- n_i \mathcal{Z} $ \\ 
	$[c_7]$ & $[K_B^{-1}]  \rule{0pt}{12pt}- n_i \mathcal{Z} $ \\ 
	$[c_8]$ & $\cS_7  \rule{0pt}{12pt} - n_i \mathcal{Z} $ \\ 
	$[c_9]$ & $[K_B^{-1}]-\cS_7+\cS_9  \rule{0pt}{12pt} - n_i \mathcal{Z} $
\end{tabular}
}
 \end{align}
The orders $n_i$ of the factorization are a characteristic feature of
the chosen top. They are listed for each model in Appendix~\ref{App:classification}.

In order to obtain the matter spectrum, i.e.\ representations and loci,
it is best to map the genus-one curves into the short Weierstrass
form. This is achieved by using the expressions for the
functions $f$ and $g$ listed below \cite{An2001304}.
Fiber $F_1$, cubic polynomial \eqref{cubic}:
\begin{equation}
\begin{split}
\label{eq:fcubic}
f&=\frac{1}{48} [-(s_6^2 - 4 (s_5 s_7 + s_3 s_8 + s_2 s_9))^2 + 
    24 (-s_6 (s_{10} s_2 s_3 - 9 s_1 s_{10} s_4 + s_4 s_5 s_8 \\
&\phantom{=}+ s_2 s_7 s_8 + s_3 s_5 s_9 +
          s_1 s_7 s_9) + 
      2 (s_{10} s_3^2 s_5 + s_1 s_7^2 s_8 + s_2 s_3 s_8 s_9 + s_1 s_3 s_9^2 \\
&\phantom{=}+  s_7 (s_{10} s_2^2 - 3 s_1 s_{10} s_3 + s_3 s_5 s_8 + s_2 s_5 s_9) + 
         s_4 (-3 s_{10} s_2 s_5 + s_2 s_8^2 + (s_5^2 - 3 s_1 s_8) s_9)))] \,,
\end{split}
\end{equation}
\begin{equation}
\label{eq:gcubic}
\begin{split}
g&=\frac{1}{864} [(s_6^2 - 4 (s_5 s_7 + s_3 s_8 + s_2 s_9))^3 - 
   36 (s_6^2 - 4 (s_5 s_7 + s_3 s_8 + s_2 s_9)) \\
&\phantom{=}\times (-s_6 (s_{10} s_2 s_3 - 9 s_1 s_{10} s_4 + 
         s_4 s_5 s_8 + s_2 s_7 s_8 + s_3 s_5 s_9 + s_1 s_7 s_9) \\
         &\phantom{=}+  2 (s_{10} s_3^2 s_5 + s_1 s_7^2 s_8 + s_2 s_3 s_8 s_9 + s_1 s_3 s_9^2 + 
         s_7 (s_{10} s_2^2 - 3 s_1 s_{10} s_3 + s_3 s_5 s_8 + s_2 s_5 s_9) \\
&\phantom{=}+  s_4 (-3 s_{10} s_2 s_5 + s_2 s_8^2 + (s_5^2 - 3 s_1 s_8) s_9))) + 
   216 ((s_{10} s_2 s_3 - 9 s_1 s_{10} s_4 + s_4 s_5 s_8 \\
   &\phantom{=}+ s_2 s_7 s_8 + s_3 s_5 s_9 + 
        s_1 s_7 s_9)^2 + 4 (-s_1 s_{10}^2 s_3^3 - s_1^2 s_{10} s_7^3 - 
         s_4^2 (27 s_1^2 s_{10}^2 + s_{10} s_5^3 \\
&\phantom{=}+ s_1 (-9 s_{10} s_5 s_8 + s_8^3)) + 
         s_{10} s_3^2 (-s_2 s_5 + s_1 s_6) s_9 - s_1 s_3^2 s_8 s_9^2 \\
&\phantom{=}-  s_7^2 (s_{10} (s_2^2 s_5 - 2 s_1 s_3 s_5 - s_1 s_2 s_6) + 
            s_1 s_8 (s_3 s_8 + s_2 s_9)) \\
&\phantom{=}-  s_3 s_7 (s_{10} (-s_2 s_5 s_6 + s_1 s_6^2 + s_2^2 s_8 + 
               s_3 (s_5^2 - 2 s_1 s_8) + s_1 s_2 s_9) \\
&\phantom{=}+  s_9 (s_2 s_5 s_8 - s_1 s_6 s_8 + s_1 s_5 s_9)) + 
         s_4 (-s_{10}^2 (s_2^3 - 9 s_1 s_2 s_3) \\
&\phantom{=}+ s_{10} (s_6 (-s_2 s_5 s_6 + s_1 s_6^2 + s_2^2 s_8) + 
               s_3 (s_5^2 s_6 - s_2 s_5 s_8 - 3 s_1 s_6 s_8)) \\
&\phantom{=}+ (s_{10} (2 s_2^2 s_5 + 3 s_1 s_3 s_5 - 
                  3 s_1 s_2 s_6) + 
               s_8 (-s_3 s_5^2 + s_2 s_5 s_6 - s_1 s_6^2 - s_2^2 s_8 + 
                  2 s_1 s_3 s_8)) s_9 \\
&\phantom{=}+ (-s_2 s_5^2 + s_1 s_5 s_6 + 
               2 s_1 s_2 s_8) s_9^2 - s_1^2 s_9^3 + 
            s_7 (s_{10} (2 s_2 s_5^2 - 3 s_1 s_5 s_6 + 3 s_1 s_2 s_8 + 
                  9 s_1^2 s_9) \\
&\phantom{=}- s_8 (s_2 s_5 s_8 - s_1 s_6 s_8 + s_1 s_5 s_9)))))] \,.
\end{split}
\end{equation}
Fiber $F_2$, biquadratic polynomial \eqref{biquadratic}:
\begin{equation} 
\begin{split}\label{eq:f-F2}
f&= \frac{1}{48} [ -(-4 b_1 b_{10} + b_6^2 - 4 (b_5 b_7 + b_3 b_8 + b_2 b_9))^2 
+24 (-b_6 (b_{10} b_2 b_5 + b_2 b_7 b_8 \\
&\phantom{=}+ b_3 b_5 b_9 + b_1 b_7 b_9) 
+ 2 (b_{10}  (b_1 b_5 b_7 + b_2^2 b_8 + b_3 (b_5^2 - 4 b_1 b_8) + b_1 b_2 b_9)  \\
 &\phantom{=}        + b_7 (b_1 b_7 b_8 + b_2 b_5 b_9) +  b_3 (b_5 b_7 b_8 + b_2 b_8 b_9 + b_1 b_9^2)))]\,,
\end{split}
\end{equation}
\begin{equation} 
\begin{split}\label{eq:g-F2}
g&= \frac{1}{864} [ (-4 b_1 b_{10} + b_6^2 - 4 (b_5 b_7 + b_3 b_8 + b_2 b_9))^3 - 36 (-4 b_1 b_{10} + b_6^2 - 
      4 (b_5 b_7\\
&\phantom{=} + b_3 b_8 + b_2 b_9))(-b_6 (b_{10} b_2 b_5 + b_2 b_7 b_8 + b_3 b_5 b_9 + b_1 b_7 b_9) + 
      2 (b_{10} (b_1 b_5 b_7 + b_2^2 b_8 \\
&\phantom{=}+ b_3 (b_5^2 - 4 b_1 b_8) + b_1 b_2 b_9) + b_7 (b_1 b_7 b_8 + b_2 b_5 b_9) + b_3 (b_5 b_7 b_8 + b_2 b_8 b_9 + b_1 b_9^2)))
\\ 
  &\phantom{=} + 216 ((b_{10} b_2 b_5 + b_2 b_7 b_8 + b_3 b_5 b_9 +
  b_1 b_7 b_9)^2 - 4 (b_2 b_3 b_5 b_7 b_8 b_9
\\
  &\phantom{=} + b_1^2 b_{10} (-4 b_{10} b_3 b_8 + b_7^2 b_8 + b_3 b_9^2) + 
         b_{10} (b_3^2 b_5^2 b_8 + b_2^2 b_5 b_7 b_8 + b_2 b_3 (-b_5 b_6 b_8 + b_2 b_8^2 
\\ 
  &\phantom{=} + b_5^2 b_9)) + b_1 (b_{10}^2 (b_3 b_5^2 + b_2^2 b_8) + b_2 b_7^2 b_8 b_9 + b_3^2 b_8 b_9^2 + b_3 b_7 (b_7 b_8^2 - b_6 b_8 b_9 + b_5 b_9^2) \\ 
  &\phantom{=}  + b_{10} (-4 b_3^2 b_8^2 + b_3 b_6 (b_6 b_8 - b_5 b_9) + b_2 b_7 (-b_6 b_8 + b_5 b_9))))) ]\, .
\end{split}
\end{equation}
Fiber $F_4$, quartic polynomial \eqref{quartic}: 
\begin{equation}
\begin{split}
f&=\tfrac{1}{48} [-24 c_9 (-2 c_5 c_6^2 + c_4 c_6 c_7 - 2 c_3 c_6 c_8 + c_2 c_7 c_8 - 
     2 c_1 c_8^2 - 2 c_2 c_4 c_9 + 8 c_1 c_5 c_9) \\
&\phantom{=}
     - (c_7^2 - 4 (c_6 c_8 + c_3 c_9))^2]\,,
\end{split}
\end{equation}
\begin{equation}
\begin{split}
g&=\tfrac{1}{864} [36 c_9 (-2 c_5 c_6^2 + c_4 c_6 c_7 - 2 c_3 c_6 c_8 + c_2 c_7 c_8 - 
     2 c_1 c_8^2 - 2 c_2 c_4 c_9 + 8 c_1 c_5 c_9) \\
&\phantom{=}\times
     (c_7^2 - 4 (c_6 c_8 + c_3 c_9)) \\
&\phantom{=} + (c_7^2 - 4 (c_6 c_8 + c_3 c_9))^3 + 
  216 c_9^2 [4 c_2 c_5 c_6 c_7 - 4 c_1 c_5 c_7^2 
+ c_2^2 c_8^2 + c_4 (-2 c_2 c_6 c_8 + 4 c_1 c_7 c_8) \\
&\phantom{=} - 4 c_2^2 c_5 c_9 + 
     c_4^2 (c_6^2 - 4 c_1 c_9) - 4 c_3 (c_5 c_6^2 + c_1 c_8^2 - 4 c_1 c_5 c_9)]]\,.
\end{split}
\end{equation}

%%%%%%%%%%%%%%%%%%%%%%%%%%%%%%%%%%%%%%%%%%%%%%%%%%%%%%%%%%%%%%%%%%%%%%%%%
\section{Classification of toric SO(10) F-theory models}
\label{App:classification}

In this appendix we classify all toric SO(10) models arising as a single hypersurface in toric ambient spaces. The description of the models is organized as follows.

First, we specify the vertices of the polygons defining the ambient space of the genus-one fibers. The respective gauge groups after inclusion of the SO(10) top are given for the different polygons. A `gauge group$^*$' denotes a non trivial global embedding of the discrete symmetries with respect to the $\mathbb{Z}_4$ center of Spin(10), see Section~\ref{sec:speccomp}. After that we list the different SO(10) tops with their defining vertices and the individual factorization properties of the sections $s_i$, $b_i$, and $c_i$ with respect to $z_0$, respectively. In the subsequent table we classify all matter fields, their representation with respect to the complete gauge group, and their base-independent multiplicity in terms of the base divisor classes $\Kbi$, $\Ss$, $\Sn$, and $\Z$. Matter fields whose representation is marked by $^*$ correspond to half-hypermultiplets. SCPs and their loci are also included. Moreover, the loci of the non-trivial SO(10) matter representations are given (for the loci of the SO(10) singlets we refer to \cite{Klevers:2014bqa} and the explanations given in Section~\ref{sec:speccomp}). We conclude the analysis by the base-independent expression for the Euler number as well as the anomaly coefficients in terms of their base divisor classes whose derivation is described in Section~\ref{sec:speccomp} and \ref{sec:bianomalycanc}. The anomaly coefficients for models with SCPs are obtained for the resolved geometry after base blow-up, and the modified divisor classes are marked by a hat. For singular models with transitions to theories without SCPs we further list the difference in matter representations accounting for the degrees of freedom contained in the additional tensor multiplets after the base blow-up (see Section~\ref{sec:anomSCP}).

%%%%%%%%%%%%%%%%%%
\subsection[Polygon \texorpdfstring{$F_1$}{F1}]{Polygon \texorpdfstring{$\boldsymbol{F_1}$}{F1}}
\vspace{-0.5cm}
{\small
\begin{align*}
\text{vertices:}& \quad u: \, (1,1,0) \,, \enspace v: \, (0,-1,0) \,, \enspace  w: \, (-1,0,0) \\ 
\text{gauge group$^*$:}& \quad \text{SO(10)} \times \mathbb{Z}_3
\end{align*}
}\vspace{-0.5cm}

%%%%%%%%%
\subsubsection*{Top 1}
\vspace{-0.5cm}
{\small
\begin{align*}
\text{vertices:}& \quad z_0: \, (0,0,1) \,, \enspace f_1: \, (0,1,1) \,, \enspace f_2: \, (1,1,1) \,, \enspace f_3: \, (1,0,1) \\ 
& \quad g_1: \, (0,1,2) \,, \enspace g_2: \, (1,1,2) \\ 
\text{factorization:}& \quad s_1 = d_1 \,, \enspace s_2 = d_2 z_0 \,, \enspace s_3 = d_3 z_0 \,, \enspace s_4 = d_4 z_0^2 \,, \enspace s_5 = d_5 \,, \enspace s_6 = d_6 z_0 \,, \\
& \quad s_7 = d_7 z_0 \,, \enspace s_8 = d_8 z_0 \,, \enspace s_9 = d_9 z_0 \,, \enspace s_{10} = d_{10} z_0^2
\end{align*}
\begin{equation}
\begin{array}{|c|c|c|}
\hline
\text{locus} & \text{representation} & \text{multiplicity} \\ \hline \hline
z_0 = d_5 = 0 & \mathbf{16}_{-3/2} & (2 \Kbi - \Ss) \Z \\ \hline
z_0 = d_9 = 0 & \mathbf{10}_{0} & (\Sn - \Z) \Z \\ \hline
z_0 = d_3 d_5 - d_1 d_7 = 0 & \mathbf{10}_{1} & (3 \Kbi - \Sn - \Z) \Z \\ \hline
z_0 = d_7 = 0 & \mathbf{16}_{-1/2} & (\Ss - \Z) \Z \\ \hline
 & \mathbf{1}_{1} & 18 (\Kbi)^2 - 3 \Ss^2 - 3 \Sn^2 + 8 \Z^2 + \Sn \Z  \\ 
 & & + \Kbi (3 \Ss + 3 \Sn - 30 \Z) + \Ss (3 \Sn + 2 \Z) \\ \hline
 & \mathbf{1}_0 & 17 + 11 (\Kbi)^2 + 3 \Ss^2 - 3 \Ss \Sn + 3 \Sn^2 \\ && - 2 \Ss \Z  - \Sn \Z + 8 \Z^2 - 
 3 \Kbi (\Ss + \Sn + 4 \Z)
\\ \hline
\end{array}
\end{equation}
\begin{align*}
\text{Euler number:}& \quad \chi = -24 (\Kbi)^2 + 6 \Kbi \Ss - 6 \Ss^2 + 6 \Kbi \Sn + 6 \Ss \Sn - 6 \Sn^2 + 
  24 \Kbi \Z \\ & \hspace{1.15cm}+ 4 \Ss \Z + 2 \Sn \Z - 16 \Z^2
 \\
\text{anomaly coefficients:}& \quad a \sim \Kbi \,, \enspace b \sim - \Z
\end{align*}
}

%%%%%%%%%
\subsubsection*{Top 2}
\vspace{-0.5cm}
{\small
\begin{align*}
\text{vertices:}& \quad z_0: \, (0,0,1) \,, \enspace f_1: \, (0,1,1) \,, \enspace f_2: \, (1,1,1) \,, \enspace f_3: \,, (1,0,1) \,, \enspace f_4: \, (-1,1,1) \\ 
& \quad g_1: \, (0,1,2) \,, \enspace g_2: \, (1,1,2) \\ 
\text{factorization:}& \quad s_1 = d_1 \,, \enspace s_2 = d_2 z_0 \,, \enspace s_3 = d_3 z_0^2 \,, \enspace s_4 = d_4 z_0^3 \,, \enspace s_5 = d_5 \,, \enspace s_6 = d_6 z_0 \,, \\
& \quad s_7 = d_7 z_0 \,, \enspace s_8 = d_8 z_0 \,, \enspace s_9 = d_9 z_0 \,, \enspace s_{10} = d_{10} z_0^2
\end{align*}
\begin{equation}
\begin{array}{|c|c|c|}
\hline
\text{locus} & \text{representation} & \text{multiplicity} \\ \hline \hline
z_0 = d_5 = 0 & \mathbf{16}_{-3/2} & (2 \Kbi - \Ss) \Z \\ \hline
z_0 = d_9 = 0 & \mathbf{10}_{0} & (\Sn - \Z) \Z \\ \hline
z_0 = d_1 = 0 & \mathbf{10}_{1} & (3 \Kbi - \Ss - \Sn) \Z \\ \hline
z_0 = d_7 = 0 & \text{SCP} & (\Ss - \Z) \Z \\ \hline
 & \mathbf{1}_{1} & 18 (\Kbi)^2 - 3 \Ss^2 - 3 \Sn^2 + 9 \Z^2 + \Sn \Z \\
 & & + \Kbi (3 \Ss + 3 \Sn - 30 \Z) + \Ss (3 \Sn + \Z) \\ \hline
 & \mathbf{1}_0 & 17 + 11 (\Kbi)^2 + 3 \Ss^2 - 3 \Ss \Sn + 3 \Sn^2 \\ &&- 4 \Ss \Z - \Sn \Z + 10 \Z^2 - 
 3 \Kbi (\Ss + \Sn + 4 \Z)
 \\ \hline
\end{array}
\end{equation}
\begin{align*}
\text{Euler number:}& \quad \chi = - 24 (\Kbi)^2 + 6 \Kbi \Ss - 6 \Ss^2 + 6 \Kbi \Sn + 6 \Ss \Sn - 6 \Sn^2 + 
  24 \Kbi \Z\\ & \hspace{1.15cm} + 10 \Ss \Z + 2 \Sn \Z - 22 \Z^2
 \\
\text{anomaly coefficients$^*$:}& \quad \hat{a} \sim K_{\hat{B}}^{-1} \,, \enspace \hat{b} \sim - \hat{\Z} \\
\text{relation to ($F_1$, top 1):}& \quad \mathbf{16}_{-1/2} \, \oplus \mathbf{10}_{1} \, \oplus \mathbf{1}_{1} \,, \oplus \mathbf{1}_0 \, \oplus \mathbf{1}_{0}
\end{align*}
} \vspace{-0.5cm}

%%%%%%%%%
\subsubsection*{Top 3}
\vspace{-0.5cm}
{\small
\begin{align*}
\text{vertices:}& \quad z_0: \, (0,0,1) \,, \enspace f_1: \, (0,1,1) \,, \enspace f_2: \, (1,1,1) \,, \enspace f_3: \,, (1,0,1) \,, \enspace f_4: \, (2,1,1) \\ 
& \quad g_1: \, (1,1,2) \,, \enspace g_2: \, (2,1,2) \\ 
\text{factorization:}& \quad s_1 = d_1 \,, \enspace s_2 = d_2 \,, \enspace s_3 = d_3 z_0 \,, \enspace s_4 = d_4 z_0^2 \,, \enspace s_5 = d_5 \,, \enspace s_6 = d_6 z_0 \,, \\
& \quad s_7 = d_7 z_0 \,, \enspace s_8 = d_8 z_0 \,, \enspace s_9 = d_9 z_0^2 \,, \enspace s_{10} = d_{10} z_0^3
\end{align*}
\begin{equation}
\begin{array}{|c|c|c|}
\hline
\text{locus} & \text{representation} & \text{multiplicity} \\ \hline \hline
z_0 = d_5 = 0 & \mathbf{16}_{2} & (2 \Kbi - \Ss) \Z \\ \hline
z_0 = d_8 = 0 & \mathbf{10}_{1} & (\Kbi - \Ss + \Sn - \Z) \Z \\ \hline
z_0 = d_2 = 0 & \mathbf{10}_{0} & (2 \Kbi - \Sn) \Z \\ \hline
z_0 = d_7 = 0 & \text{SCP} & (\Ss - \Z) \Z \\ \hline
 & \mathbf{1}_{1} & 18 (\Kbi)^2 - 3 \Ss^2 - 3 \Sn^2 + 9 \Z^2 + 4 \Sn \Z \\
 & & + \Kbi (3 \Ss + 3 \Sn - 31 \Z) + \Ss (3 \Sn -2 \Z) \\ \hline
 & \mathbf{1}_0 & 17 + 11 (\Kbi)^2 + 3 \Ss^2 + 3 \Sn^2 - 4 \Sn \Z + 10 \Z^2\\ 
 && - \Ss (3 \Sn + \Z) - 
 \Kbi (3 \Ss + 3 \Sn + 11 \Z) \\ \hline
\end{array}
\end{equation}
\begin{align*}
\text{Euler number:}& \quad \chi =  -24 (\Kbi)^2 + 6 \Kbi \Ss - 6 \Ss^2 + 6 \Kbi \Sn + 6 \Ss \Sn - 6 \Sn^2 + 
  22 \Kbi \Z \\ & \hspace{1.15cm}+ 4 \Ss \Z + 8 \Sn \Z - 22 \Z^2 \\
\text{anomaly coefficients$^*$:}& \quad \hat{a} \sim K_{\hat{B}}^{-1} \,, \enspace \hat{b} \sim - \hat{\Z}
\end{align*}
}\vspace{-0.5cm}

%%%%%%%%%%%%%%%%%%
\subsection[Polygon \texorpdfstring{$F_2$}{F2}]{Polygon \texorpdfstring{$\boldsymbol{F_2}$}{F2}}
\vspace{-0.5cm}
{\small
\begin{align*}
\text{vertices:}& \quad t: \, (1,0,0) \,, \enspace x: \, (-1,0,0) \,, \enspace s: \, (0,-1,0) \,, \enspace y: \, (0,1,0) \\
\text{gauge group$^*$:}& \quad \text{SO(10)} \times \text{U(1)} \times \mathbb{Z}_2
\end{align*}
}\vspace{-0.5cm}

%%%%%%%%%
\subsubsection*{Top 1}
\vspace{-0.5cm}
{\small
\begin{align*}
\text{vertices:}& \quad z_0: \, (0,0,1) \,, \enspace f_1: \, (0,1,1) \,, \enspace f_2: \, (1,0,1) \,, \enspace f_3: \, (1,1,1) \,,\\
&\quad g_1: \, (0,1,2) \,, \enspace g_2: \, (1,1,2) \\
\text{factorization:}& \quad b_1 = d_1 z_0 \,, \enspace b_2 = d_2 z_0 \,, \enspace b_3 = d_3 z_0^2 \,, \enspace b_5 = d_5 \,, \enspace b_6 = d_6 z_0 \,, \\
& \quad b_7 = d_7 z_0 \,, \enspace b_8 = d_8 \,, \enspace b_9 = d_9 z_0 \,, \enspace b_{10} = d_{10} z_0 
\end{align*}
\begin{equation}
\begin{array}{|c|c|c|}
\hline
\text{locus} & \text{representation} & \text{multiplicity} \\ \hline \hline
z_0 = d_5 = 0 & \mathbf{16}_{1/4,1} & (2 \Kbi - \Ss) \Z \\ \hline
z_0 = d_2 = 0 & \mathbf{10}_{1/2,1} & (2 \Kbi - \Sn - \Z) \Z \\ \hline
z_0 = d_{10} d_5 - d_8 d_7 = 0 & \mathbf{10}_{1/2,0} & (\Kbi + \Sn - \Z) \Z \\ \hline
z_0 = d_7 = 0 & \mathbf{16}_{1/4,0} & (\Ss - \Z) \Z \\ \hline
 & \mathbf{1}_{1,0} & 6 (\Kbi)^2 - 2 \Ss^2 + 2 \Sn^2 + 3 \Z^2 + \Sn \Z \\
 & & + \Kbi (4 \Ss - 4 \Sn - 12 \Z) + 2 \Ss \Z \\ \hline
 & \mathbf{1}_{1,1} & 6 (\Kbi)^2 + 2 \Ss^2 - 2 \Sn^2 + 3 \Z^2 - \Sn \Z \\
 & & + \Kbi (- 4 \Ss + 4 \Sn - 5 \Z) - 2 \Ss \Z \\ \hline
 & \mathbf{1}_{0,1} & 6 (\Kbi)^2 - 2 \Ss^2 - 2 \Sn^2 + 3 \Z^2 - \Sn \Z \\
 & & + \Kbi (4 \Ss + 4 \Sn - 13 \Z) + 2 \Ss \Z \\ \hline
 & \mathbf{1}_{0,0} &  18 + 11 (\Kbi)^2 + 2 \Ss^2 + 2 \Sn^2 + 7 \Z^2 + \Sn \Z \\ 
&& - 4 \Kbi (\Ss + \Sn + 3 \Z) - 2 \Ss \Z \\ \hline
\end{array}
\end{equation}
\begin{align*}
\text{Euler number:}& \quad \chi = -24 (\Kbi)^2 + 8 \Kbi \Ss - 4 \Ss^2 + 8 \Kbi \Sn - 4 \Sn^2 + 
  24 \Kbi \Z\\ & \hspace{1.15cm} + 4 \Ss \Z - 2 \Sn \Z - 14 \Z^2
 \\
\text{anomaly coefficients:}& \quad a \sim \Kbi \,, \enspace b \sim - \Z \,, \enspace b_{11} \sim - (2 \Kbi - \tfrac{5}{4} \Z)
\end{align*}
}\vspace{-0.5cm}

%%%%%%%%%
\subsubsection*{Top 2}
\vspace{-0.5cm}
{\small
\begin{align*}
\text{vertices:}& \quad z_0: \, (0,0,1) \,, \enspace f_1: \, (0,1,1) \,, \enspace f_2: \, (1,0,1) \,, \enspace f_3: \, (1,1,1) \,, \enspace f_4: \, (-1,1,1) \,, \\
& \quad g_1: \, (0,1,2) \,, \enspace g_2: \, (1,1,2) \\
\text{factorization:}& \quad b_1 = d_1 z_0 \,, \enspace b_2 = d_2 z_0 \,, \enspace b_3 = d_3 z_0^2 \,, \enspace b_5 = d_5 \,, \enspace b_6 = d_6 z_0 \,, \\
& \quad b_7 = d_7 z_0 \,, \enspace b_8 = d_8 \,, \enspace b_9 = d_9 z_0 \,, \enspace b_{10} = d_{10} z_0^2
\end{align*}
\begin{equation}
\begin{array}{|c|c|c|}
\hline
\text{locus} & \text{representation} & \text{multiplicity} \\ \hline \hline
z_0 = d_5 = 0 & \mathbf{16}_{1/4,1} & (2 \Kbi - \Ss) \Z \\ \hline
z_0 = d_2 = 0 & \mathbf{10}_{1/2,1} & (2 \Kbi - \Sn - \Z) \Z \\ \hline
z_0 = d_8 = 0 & \mathbf{10}_{1/2,0} & (\Kbi - \Ss + \Sn) \Z \\ \hline
z_0 = d_7 = 0 & \text{SCP} & (\Ss - \Z) \Z \\ \hline
 & \mathbf{1}_{1,0} & 6 (\Kbi)^2 - 2 \Ss^2 + 2 \Sn^2 + 3 \Z^2 + \Sn \Z \\
 & & + \Kbi (4 \Ss - 4 \Sn - 12 \Z) + 2 \Ss \Z \\ \hline
 & \mathbf{1}_{1,1} & 6 (\Kbi)^2 + 2 \Ss^2 - 2 \Sn^2 + 4 \Z^2 - \Sn \Z \\
 & & + \Kbi (- 4 \Ss + 4 \Sn - 5 \Z)  - 3 \Ss \Z \\ \hline
 & \mathbf{1}_{0,1} & 6 (\Kbi)^2 - 2 \Ss^2 - 2 \Sn^2 + 4 \Z^2 - \Sn \Z \\ 
 & & + \Kbi (4 \Ss + 4 \Sn - 13 \Z) + \Ss \Z \\ \hline
 & \mathbf{1}_{0,0} & 18 + 11 (\Kbi)^2 + 2 \Ss^2 + 2 \Sn^2 + 8 \Z^2 + \Sn \Z \\ 
&& - 4 \Kbi (\Ss + \Sn + 3 \Z) - 3 \Ss \Z \\ \hline
\end{array}
\end{equation}
\begin{align*}
\text{Euler number:}& \quad \chi = -24 (\Kbi)^2 + 8 \Kbi \Ss - 4 \Ss^2 + 8 \Kbi \Sn - 4 \Sn^2 + 
  24 \Kbi \Z \\ & \hspace{1.15cm}+ 8 \Ss \Z - 2 \Sn \Z - 18 \Z^2
  \\
\text{anomaly coefficients$^*$:}& \quad \hat{a} \sim K_{\hat{B}}^{-1} \,, \enspace \hat{b} \sim - \hat{\Z} \,, \enspace \hat{b}_{11} \sim - (2 K_{\hat{B}}^{-1} - \tfrac{5}{4} \hat{\Z}) \\
\text{relation to ($F_2$, top 1):}& \quad \mathbf{16}_{1/4,0} \, \oplus \mathbf{10}_{1/2,0} \, \oplus \mathbf{1}_{1,1} \, \oplus \mathbf{1}_{0,1} \, \oplus \mathbf{1}_{0,0}
\end{align*}
}\vspace{-0.5cm}

%%%%%%%%%
\subsubsection*{Top 3}
\vspace{-0.5cm}
{\small
\begin{align*}
\text{vertices:}& \quad z_0: \, (0,0,1) \,, \enspace f_1: \, (0,1,1) \,, \enspace f_2: \, (1,0,1) \,, \enspace f_3: \, (1,-1,1) \,, \enspace f_4: \, (2, -1,1) \,, \\
& \quad g_1: \, (1,0,2) \,, \enspace g_2: \, (2,-1,2) \\
\text{factorization:}& \quad b_1 = d_1 z_0^3 \,, \enspace b_2 = d_2 z_0^2 \,, \enspace b_3 = d_3 z_0 \,, \enspace b_5 = d_5 z_0 \,, \enspace b_6 = d_6 z_0 \,, \\
& \quad b_7 = d_7 z_0 \,, \enspace b_8 = d_8 \,, \enspace b_9 = d_9 \,, \enspace b_{10} = d_{10}
\end{align*}
\begin{equation}
\begin{array}{|c|c|c|}
\hline
\text{locus} & \text{representation} & \text{multiplicity} \\ \hline \hline
z_0 = d_8 = 0 & \mathbf{16}_{1/2,1/2} & (\Kbi - \Ss + \Sn) \Z \\ \hline
z_0 = d_5 = 0 & \mathbf{10}_{0,0} & (2 \Kbi - \Ss - \Z) \Z \\ \hline
z_0 = d_9 = 0 & \mathbf{10}_{0,1} & \Sn \Z \\ \hline
z_0 = d_3 = 0 & \text{SCP} & (\Kbi + \Ss - \Sn - \Z) \Z \\ \hline
 & \mathbf{1}_{1,0} & 6 (\Kbi)^2 - 2 \Ss^2 + 2 \Sn^2 + 4 \Z^2 + 4 \Sn \Z \\
 & & + \Kbi (4 \Ss - 4 \Sn - 10 \Z) - 2 \Ss \Z \\ \hline
 & \mathbf{1}_{1,1} & 6 (\Kbi)^2 + 2 \Ss^2 - 2 \Sn^2 + 2 \Z^2 - 2 \Sn \Z \\ 
 & & + \Kbi (- 4 \Ss + 4 \Sn - 6 \Z) \\ \hline
 & \mathbf{1}_{0,1} & 6 (\Kbi)^2 - 2 \Ss^2 - 2 \Sn^2 + 4 \Z^2 - 2 \Sn \Z \\ 
 & & + \Kbi (4 \Ss + 4 \Sn - 10 \Z) - 2 \Ss \Z \\ \hline
 & \mathbf{1}_{0,0} & 18 + 11 (\Kbi)^2 + 2 \Ss^2 + 2 \Sn^2 + 9 \Z^2 + 3 \Sn \Z \\ 
&& - \Kbi (4 \Ss + 4 \Sn + 19 \Z) + \Ss \Z \\ \hline
\end{array}
\end{equation}
\begin{align*}
\text{Euler number:}& \quad \chi =  -24 (\Kbi)^2 + 8 \Kbi \Ss - 4 \Ss^2 + 8 \Kbi \Sn - 4 \Sn^2 + 
  40 \Kbi \Z\\ & \hspace{1.15cm} - 8 \Sn \Z - 20 \Z^2
    \\
\text{anomaly coefficients$^*$:}& \quad \hat{a} \sim K_{\hat{B}}^{-1} \,, \enspace \hat{b} \sim - \hat{\Z} \,, \enspace \hat{b}_{11} \sim - (2 K_{\hat{B}}^{-1} - \hat{\Z})
\end{align*}
}\vspace{-0.5cm}

%%%%%%%%%%%%%%%%%%
\subsection[Polygon \texorpdfstring{$F_3$}{F3}]{Polygon \texorpdfstring{$\boldsymbol{F_3}$}{F3}}
\vspace{-0.5cm}
{\small
\begin{align*}
\text{vertices:}& \quad w: \, (0,1,0) \,, \enspace u: \, (1,-1,0) \,, \enspace e_1: \, (0,-1,0) \,, \enspace v: \, (-1,0,0)  \\ 
\text{gauge group:}& \quad \text{SO(10)} \times \text{U(1)}
\end{align*}
}\vspace{-0.5cm}

%%%%%%%%%
\subsubsection*{Top 1}
\vspace{-0.5cm}
{\small
\begin{align*}
\text{vertices:}& \quad z_0: \, (0,0,1) \,, \enspace f_2: \, (1,0,1) \,, \enspace f_3: \, (0,1,1) \,, \enspace f_4: \, (1,1,1) \,,\\ 
& \quad g_1: \, (1,1,2) \,, \enspace g_2 : \,(1,2,2) \\ 
\text{factorization:}& \quad s_1 = d_1 z_0 \,, \enspace s_2 = d_2 z_0^2 \,, \enspace s_3 = d_3 z_0^2 \,, \enspace s_4 = d_4 z_0^3 \,, \enspace s_5 = d_5 \,, \\
& \quad s_6 = d_6 z_0 \,, \enspace s_7 = d_7 z_0 \,, \enspace s_8 = d_8 \,, \enspace s_9 = d_9
\end{align*}
\begin{equation}
\begin{array}{|c|c|c|}
\hline
\text{locus} & \text{representation} & \text{multiplicity} \\ \hline \hline
z_0 = d_5 = 0 & \mathbf{16}_{3/4} & (2 \Kbi - \Ss) \Z \\ \hline
z_0 = d_9 = 0 & \mathbf{10}_{3/2} & \Sn \Z \\ \hline
z_0 = d_3 d_5 - d_1 d_7 = 0 & \mathbf{10}_{-1/2} & (3\Kbi - \Sn - 2 \Z) \Z \\ \hline
z_0 = d_7 = 0 & \mathbf{16}_{-1/4} & (\Ss - \Z) \Z \\ \hline 
 & \mathbf{1}_3 & (\Kbi - \Ss + \Sn) \Sn \\ \hline
 & \mathbf{1}_2 & 6 (\Kbi)^2 + \Ss^2 + \Kbi (-5 \Ss + 4 \Sn - 2 \Z) \\
 & & + \Ss (2\Sn + \Z) - \Sn (2\Sn + 5 \Z) \\ \hline
 & \mathbf{1}_1 & 12 (\Kbi)^2 - 4 \Ss^2 - \Sn^2 + 6 \Z^2 \\ 
 & & + \Kbi (8 \Ss - \Sn - 25 \Z) + \Ss(\Sn + 4\Z) \\ \hline
 & \mathbf{1}_0 &18 + 11 (\Kbi)^2 + 3 \Ss^2 + 2 \Sn^2 + 10 \Z^2 + 5 \Sn \Z \\
 & & - \Kbi (3 \Ss + 4 \Sn + 15 \Z) - 5 \Ss \Z - 2 \Ss \Sn\\ \hline
\end{array}
\end{equation}
\begin{align*}
\text{Euler number:}& \quad \chi = -24 (\Kbi)^2 + 8 \Kbi \Sn - 4 \Sn^2 + 6 \Kbi \Ss + 4 \Ss \Sn - 6 \Ss^2 + 
   30 \Kbi \Z \\ & \hspace{1.15cm} - 10 \Sn \Z + 10 \Ss \Z - 20 \Z^2
  \\ 
\text{anomaly coefficients:}& \quad a \sim \Kbi \,, \enspace b \sim - \Z \,, \enspace b_{11} \sim - (6 \Kbi - 2 \Ss + 4 \Sn - \tfrac{5}{4} \Z)
\end{align*}
}\vspace{-0.5cm}

%%%%%%%%%
\subsubsection*{Top 2}
\vspace{-0.5cm}
{\small
\begin{align*}
\text{vertices:}& \quad z_0: \, (0,0,1) \,, \enspace f_1: \, (1,0,1) \,, \enspace f_2: \, (1,1,1) \,, \enspace f_3: \, (0,1,1) \,, \enspace f_4: \, (1,2,1) \,, \\ 
& \quad g_1: \, (1,1,2) \,, \enspace g_2 : \,(1,2,2) \\ 
\text{factorization:}& \quad s_1 = d_1 z_0 \,, \enspace s_2 = d_2 z_0^2 \,, \enspace s_3 = d_3 z_0^3 \,, \enspace s_4 = d_4 z_0^4 \,, \enspace s_5 = d_5 \,, \\
& \quad s_6 = d_6 z_0 \,, \enspace s_7 = d_7 z_0 \,, \enspace s_8 = d_8 \,, \enspace s_9 = d_9
\end{align*}
\begin{equation}
\begin{array}{|c|c|c|}
\hline
\text{locus} & \text{representation} & \text{multiplicity} \\ \hline \hline
z_0 = d_5 = 0 & \mathbf{16}_{3/4} & (2 \Kbi - \Ss) \Z \\ \hline
z_0 = d_9 = 0 & \mathbf{10}_{3/2} & \Sn \Z \\ \hline
z_0 = d_1 = 0 & \mathbf{10}_{-1/2} & (3\Kbi - \Ss - \Sn - \Z) \Z \\ \hline
z_0 = d_7 = 0 & $SCP$ & (\Ss - \Z) \Z \\ \hline 
 & \mathbf{1}_3 & (\Kbi - \Ss + \Sn) \Sn \\ \hline
 & \mathbf{1}_2 & 6 (\Kbi)^2 + \Ss^2 + \Kbi (-5 \Ss + 4 \Sn - 2 \Z) \\
 & & + \Ss (2\Sn + \Z) - \Sn (2\Sn + 5 \Z) \\ \hline
 & \mathbf{1}_1 & 12 (\Kbi)^2 - 4 \Ss^2 - \Sn^2 + 7 \Z^2 \\ 
 & & + \Kbi (8 \Ss - \Sn - 25 \Z) + \Ss(\Sn + 3\Z) \\ \hline
 & \mathbf{1}_0 & 18 + 11 (\Kbi)^2 + 3 \Ss^2 + 2 \Sn^2 + 12 \Z^2 + 5 \Sn \Z \\
 & & - \Kbi (3 \Ss + 4 \Sn + 15 \Z) - 7 \Ss \Z - 2 \Ss \Sn \\ \hline
\end{array}
\end{equation}
\begin{align*}
\text{Euler number:}& \quad \chi =  -24 (\Kbi)^2 + 8 \Kbi \Sn - 4 \Sn^2 + 6 \Kbi \Ss + 4 \Ss \Sn - 6 \Ss^2 + 
   30 \Kbi \Z  \\ & \hspace{1.15cm} - 10 \Sn \Z + 16 \Ss \Z - 26 \Z^2 
  \\ 
\text{anomaly coefficients$^*$:}& \quad \hat{a} \sim K_{\hat{B}}^{-1} \,, \enspace \hat{b} \sim - \hat{\Z} \,, \enspace \hat{b}_{11} \sim - (6 K_{\hat{B}}^{-1} - 2 \hat{\mathcal{S}}_7 + 4 \hat{\mathcal{S}}_9 - \tfrac{5}{4} \hat{\Z}) \\ 
\text{relation to ($F_3$, top 1):}& \quad \mathbf{16}_{-1/4} \, \oplus \mathbf{10}_{-1/2} \, \oplus \mathbf{1}_{1} \, \oplus \mathbf{1}_{0} \, \oplus \mathbf{1}_{0}
\end{align*}
}\vspace{-0.5cm}

%%%%%%%%%
\subsubsection*{Top 3}
\vspace{-0.5cm}
{\small
\begin{align*}
\text{vertices:}& \quad z_0: \, (0,0,1) \,, \enspace f_1: \, (1,0,1) \,, \enspace f_2: \, (1,1,1) \,, \enspace f_3: \, (0,1,1) \,, \enspace f_4: \, (0,2,1) \,, \\ 
& \quad g_1: \, (1,1,2) \,, \enspace g_2 : \,(1,2,2) \\ 
\text{factorization:}& \quad s_1 = d_1 z_0^2 \,, \enspace s_2 = d_2 z_0^2 \,, \enspace s_3 = d_3 z_0^2 \,, \enspace s_4 = d_4 z_0^3 \,, \enspace s_5 = d_5 \,, \\
& \quad s_6 = d_6 z_0 \,, \enspace s_7 = d_7 z_0 \,, \enspace s_8 = d_8 \,, \enspace s_9 = d_9
\end{align*}
\begin{equation}
\begin{array}{|c|c|c|}
\hline
\text{locus} & \text{representation} & \text{multiplicity} \\ \hline \hline
z_0 = d_9 = 0 & \mathbf{10}_{3/2} & \Sn \Z \\ \hline
z_0 = d_3 = 0 & \mathbf{10}_{-1/2} & (\Kbi + \Ss - \Sn - 2 \Z) \Z \\ \hline
z_0 = d_7 = 0 & \mathbf{16}_{-1/4} & (\Ss - \Z) \Z \\ \hline 
z_0 = d_5 = 0 & \text{SCP} & (2 \Kbi - \Ss) \Z \\ \hline
 & \mathbf{1}_3 & (\Kbi - \Ss + \Sn) \Sn \\ \hline
 & \mathbf{1}_2 & 6 (\Kbi)^2 + \Ss^2 + \Kbi (-5 \Ss + 4 \Sn - 4 \Z) \\
 & & + 2 \Ss (\Sn + \Z) - \Sn (2\Sn + 5 \Z) \\ \hline
 & \mathbf{1}_1 & 12 (\Kbi)^2 - 4 \Ss^2 - \Sn^2 + 6 \Z^2 \\ 
 & & + \Kbi (8 \Ss - \Sn - 27 \Z) + \Ss(\Sn + 5 \Z) \\ \hline
 & \mathbf{1}_0 & 18 + 11 (\Kbi)^2 + 3 \Ss^2 + 2 \Sn^2 + 10 \Z^2 + 5 \Sn \Z  \\
 && - \Kbi (3 \Ss + 4 \Sn + 17 \Z) - 4 \Ss \Z - 2 \Ss \Sn \\ \hline
\end{array}
\end{equation}
\begin{align*}
\text{Euler number:}& \quad \chi =  -24 (\Kbi)^2 + 8 \Kbi \Sn - 4 \Sn^2 + 6 \Kbi \Ss + 4 \Ss \Sn - 6 \Ss^2 + 
   38 \Kbi \Z \\ & \hspace{1.15cm} - 10 \Sn \Z + 6 \Ss \Z - 20 \Z^2 
 \\ 
\text{anomaly coefficients$^*$:}& \quad \hat{a} \sim K_{\hat{B}}^{-1} \,, \enspace \hat{b} \sim - \hat{\Z} \,, \enspace \hat{b}_{11} \sim - (6 K_{\hat{B}}^{-1} - 2 \hat{\mathcal{S}}_7 + 4 \hat{\mathcal{S}}_9 - \tfrac{5}{4} \hat{\Z}) \\
\text{relation ($F_3$, top 1):}& \quad \mathbf{16}_{3/4} \, \oplus \mathbf{10}_{-1/2} \, \oplus \mathbf{1}_{2} \, \oplus \mathbf{1}_{1} \, \oplus \mathbf{1}_{0}
\end{align*}
}\vspace{-0.5cm}

%%%%%%%%%
\subsubsection*{Top 4}
\vspace{-0.5cm}
{\small
\begin{align*}
\text{vertices:}& \quad z_0: \, (0,0,1) \,, \enspace f_1: \, (1,0,1) \,, \enspace f_2: \, (1,1,1) \,, \enspace f_3: \, (0,1,1) \,, \\ 
& \quad g_1: \, (0,1,2) \,, \enspace g_2 : \,(1,1,2) \\ 
\text{factorization:}& \quad s_1 = d_1 z_0 \,, \enspace s_2 = d_2 z_0 \,, \enspace s_3 = d_3 z_0^2 \,, \enspace s_4 = d_4 z_0^3 \,, \enspace s_5 = d_5 z_0 \,, \\
& \quad s_6 = d_6 z_0 \,, \enspace s_7 = d_7 z_0 \,, \enspace s_8 = d_8 \,, \enspace s_9 = d_9
\end{align*}
\begin{equation}
\begin{array}{|c|c|c|}
\hline
\text{locus} & \text{representation} & \text{multiplicity} \\ \hline \hline
z_0 = d_2 = 0 & \mathbf{16}_{0} & (2 \Kbi - \Sn - \Z) \Z \\ \hline
z_0 = d_7 = 0 & \mathbf{10}_{0} &  (\Ss - \Z) \Z \\ \hline
z_0 = d_2 d_8 - d_1 d_9 = 0 & \mathbf{10}_{1} & (3\Kbi - \Ss - \Z) \Z \\ \hline
z_0 = d_9 = 0 & \mathbf{16}_{-1} & \Sn \Z \\ \hline 
 & \mathbf{1}_3 &  (\Kbi - \Ss + \Sn) \Sn \\ \hline
 & \mathbf{1}_2 & 6 (\Kbi)^2 + \Ss^2 - 2 \Sn^2 + \Z^2 - 4 \Sn \Z \\
 & & + \Kbi (-5 \Ss + 4 \Sn - 5 \Z) + 2 \Ss (\Sn + \Z) \\ \hline
 & \mathbf{1}_1 &  12 (\Kbi)^2 - 4 \Ss^2 - \Sn^2 + 6 \Z^2 \\ 
 & & + \Kbi (8 \Ss - \Sn - 22 \Z) + \Ss (\Sn + 2 \Z) \\ \hline
 & \mathbf{1}_0 & 18 + 11 (\Kbi)^2 + 3 \Ss^2 + 2 \Sn^2 + 9 \Z^2 + 4 \Sn \Z \\ 
 && - \Kbi (3 \Ss + 4 \Sn + 15 \Z) - 4 \Ss \Z - 2 \Ss \Sn \\ \hline
\end{array}
\end{equation}
\begin{align*}
\text{Euler number:}& \quad \chi =  -24 (\Kbi)^2 + 8 \Kbi \Sn - 4 \Sn^2 + 6 \Kbi \Ss + 4 \Ss \Sn - 6 \Ss^2 + 
   30 \Kbi \Z \\ & \hspace{1.15cm} - 8 \Sn \Z + 8 \Ss \Z - 18 \Z^2
 \\ 
\text{anomaly coefficients:}& \quad a \sim \Kbi \,, \quad b \sim - \Z \,, \quad b_{11} \sim - (6 \Kbi - 2 \Ss + 4 \Sn - 2 \Z)
\end{align*}
}\vspace{-0.5cm}

%%%%%%%%%
\subsubsection*{Top 5}
\vspace{-0.5cm}
{\small
\begin{align*}
\text{vertices:}& \quad z_0: \, (0,0,1) \,, \enspace f_1: \, (1,0,1) \,, \enspace f_2: \, (1,1,1) \,, \enspace f_3: \, (0,1,1) \,, \enspace f_4: \, (-1,1,1) \,, \\ 
& \quad g_1: \, (0,1,2) \,, \enspace g_2 : \,(1,1,2) \\ 
\text{factorization:}& \quad s_1 = d_1 z_0^2 \,, \enspace s_2 = d_2 z_0 \,, \enspace s_3 = d_3 z_0^2 \,, \enspace s_4 = d_4 z_0^3 \,, \enspace s_5 = d_5 z_0 \,, \\
& \quad s_6 = d_6 z_0 \,, \enspace s_7 = d_7 z_0 \,, \enspace s_8 = d_8 \,, \enspace s_9 = d_9
\end{align*}
\begin{equation}
\begin{array}{|c|c|c|}
\hline
\text{locus} & \text{representation} & \text{multiplicity} \\ \hline \hline
z_0 = d_7 = 0 & \mathbf{10}_{0} &  (\Ss - \Z) \Z \\ \hline
z_0 = d_8 = 0 & \mathbf{10}_{1} & (\Kbi - \Ss + \Sn) \Z \\ \hline
z_0 = d_9 = 0 & \mathbf{16}_{-1} & \Sn \Z \\ \hline 
z_0 = d_2 = 0 & \text{SCP} & (2 \Kbi - \Sn - \Z) \Z \\ \hline
 & \mathbf{1}_3 &  (\Kbi - \Ss + \Sn) \Sn \\ \hline
 & \mathbf{1}_2 & 6 (\Kbi)^2 + \Ss^2 - 2 \Sn^2 + \Z^2 - 4 \Sn \Z \\
 & & + \Kbi (-5 \Ss + 4 \Sn - 5 \Z) + 2 \Ss (\Sn + \Z) \\ \hline
 & \mathbf{1}_1 &  12 (\Kbi)^2 - 4 \Ss^2 - \Sn^2 + 8 \Z^2 + \Ss \Sn \\ 
 & & + \Kbi (8 \Ss - \Sn - 26 \Z) + 2 (\Ss + \Sn) \Z \\ \hline
 & \mathbf{1}_0 & 18 + 11 (\Kbi)^2 + 3 \Ss^2 + 2 \Sn^2 + 10 \Z^2 + 5 \Sn \Z \\ 
 && - \Kbi (3 \Ss + 4 \Sn + 17 \Z) - 4 \Ss \Z - 2 \Ss \Sn \\ \hline
\end{array}
\end{equation}
\begin{align*}
\text{Euler number:}& \quad \chi = -24 (\Kbi)^2 + 8 \Kbi \Sn - 4 \Sn^2 + 6 \Kbi \Ss + 4 \Ss \Sn - 6 \Ss^2 + 
   38 \Kbi \Z \\ & \hspace{1.15cm} - 12 \Sn \Z + 8 \Ss \Z - 22 \Z^2
 \\ 
\text{anomaly coefficients$^*$:}& \quad \hat{a} \sim K_{\hat{B}}^{-1} \,, \quad \hat{b} \sim - \hat{\Z} \,, \quad \hat{b}_{11} \sim - (6 K_{\hat{B}}^{-1} - 2 \hat{\mathcal{S}}_7 + 4 \hat{\mathcal{S}}_9 - 2 \hat{\Z}) \\ 
\text{relation to ($F_4$, top 4):}& \quad \mathbf{16}_{0} \, \oplus \mathbf{10}_{1} \, \oplus \mathbf{1}_{1} \, \oplus \mathbf{1}_{1} \, \oplus \mathbf{1}_{0}
\end{align*}
}\vspace{-0.5cm}

%%%%%%%%%
\subsubsection*{Top 6}
\vspace{-0.5cm}
{\small
\begin{align*}
\text{vertices:}& \quad z_0: \, (0,0,1) \,, \enspace f_1: \, (1,0,1) \,, \enspace f_2: \, (2,0,1) \,, \enspace f_3: \, (1,1,1) \,, \enspace f_4: \, (0,1,1) \,, \\ 
& \quad g_1: \, (1,1,2) \,, \enspace g_2 : \,(2,1,2) \\ 
\text{factorization:}& \quad s_1 = d_1 z_0 \,, \enspace s_2 = d_2 z_0 \,, \enspace s_3 = d_3 z_0^2 \,, \enspace s_4 = d_4 z_0^4 \,, \enspace s_5 = d_5 \,, \\
& \quad s_6 = d_6 z_0 \,, \enspace s_7 = d_7 z_0^2 \,, \enspace s_8 = d_8 \,, \enspace s_9 = d_9
\end{align*}
\begin{equation}
\begin{array}{|c|c|c|}
\hline
\text{locus} & \text{representation} & \text{multiplicity} \\ \hline \hline
z_0 = d_2 = 0 & \mathbf{16}_{1/2} &  (2 \Kbi - \Sn - \Z) \Z \\ \hline
z_0 = d_3 = 0 & \mathbf{10}_{0} & (\Kbi + \Ss - \Sn - 2 \Z) \Z \\ \hline
z_0 = d_5 = 0 & \mathbf{10}_{1} & (2 \Kbi - \Ss) \Z \\ \hline 
z_0 = d_9 = 0 & \text{SCP} & \Sn \Z \\ \hline
 & \mathbf{1}_3 & (\Kbi - \Ss + \Sn) \Sn \\ \hline
 & \mathbf{1}_2 & 6 (\Kbi)^2 + \Ss^2 - 2 \Sn^2 - 6 \Sn \Z \\
 & & + \Kbi (-5 \Ss + 4 \Sn - 2 \Z) + \Ss (2 \Sn + \Z) \\ \hline
 & \mathbf{1}_1 & 12 (\Kbi)^2 - 4 \Ss^2 - \Sn^2 + 4 \Z^2 - 2 \Sn \Z \\ 
 & & + \Kbi (8 \Ss - \Sn - 26 \Z) + \Ss (\Sn + 6 \Z) \\ \hline
 & \mathbf{1}_0 & 18 + 11 (\Kbi)^2 + 3 \Ss^2 + 2 \Sn^2 + 12 \Z^2 + 5 \Sn \Z \\ 
 & & - \Kbi (3 \Ss + 4 \Sn + 14 \Z) - 7 \Ss \Z - 2 \Ss \Sn \\ \hline
\end{array}
\end{equation}
\begin{align*}
\text{Euler number:}& \quad \chi = -24 (\Kbi)^2 + 8 \Kbi \Sn - 4 \Sn^2 + 6 \Kbi \Ss + 4 \Ss \Sn - 6 \Ss^2 + 
   28 \Kbi \Z \\ & \hspace{1.15cm} - 8 \Sn \Z + 14 \Ss \Z - 24 \Z^2
 \\ 
\text{anomaly coefficients$^*$:}& \quad \hat{a} \sim K_{\hat{B}}^{-1} \,, \quad \hat{b} \sim - \hat{\Z} \,, \quad \hat{b}_{11} \sim - (6 K_{\hat{B}}^{-1} - 2 \hat{\mathcal{S}}_7 + 4 \hat{\mathcal{S}}_9 - \hat{\Z})
\end{align*}
}\vspace{-0.5cm}

%%%%%%%%%
\subsubsection*{Top 7}
\vspace{-0.5cm}
{\small
\begin{align*}
\text{vertices:}& \quad z_0: \, (0,0,1) \,, \enspace f_1: \, (1,0,1) \,, \enspace f_2: \, (2,0,1) \,, \enspace f_3: \, (1,1,1) \,, \enspace f_4: \, (2,1,1) \,, \\ 
& \quad g_1: \, (2,1,2) \,, \enspace g_2 : \,(3,1,2) \\ 
\text{factorization:}& \quad s_1 = d_1 \,, \enspace s_2 = d_2 z_0 \,, \enspace s_3 = d_3 z_0^3 \,, \enspace s_4 = d_4 z_0^5 \,, \enspace s_5 = d_5 \,, \\
& \quad s_6 = d_6 z_0 \,, \enspace s_7 = d_7 z_0^2 \,, \enspace s_8 = d_8 \,, \enspace s_9 = d_9
\end{align*}
\begin{equation}
\begin{array}{|c|c|c|}
\hline
\text{locus} & \text{representation} & \text{multiplicity} \\ \hline \hline
z_0 = d_2 = 0 & \mathbf{16}_{0} &  (2 \Kbi - \Sn - \Z) \Z \\ \hline
z_0 = d_7 = 0 & \mathbf{10}_{0} & (\Ss - 2 \Z) \Z \\ \hline
z_0 = d_1 = 0 & \mathbf{10}_{1} & (3 \Kbi - \Ss - \Sn) \Z \\ \hline 
z_0 = d_9 = 0 & \text{SCP} & \Sn \Z \\ \hline
 & \mathbf{1}_3 & (\Kbi - \Ss + \Sn) \Sn \\ \hline
 & \mathbf{1}_2 & 6 (\Kbi)^2 + \Ss^2 - 2 \Sn^2 - 6 \Sn \Z \\
 & & + \Kbi (-5 \Ss + 4 \Sn) + 2 \Ss \Sn \\ \hline
 & \mathbf{1}_1 & 12 (\Kbi)^2 - 4 \Ss^2 - \Sn^2 - 2 \Sn \Z \\ 
 & & + \Kbi (8 \Ss - \Sn - 30 \Z) + \Ss \Sn + 10 \Ss \Z \\ \hline
 & \mathbf{1}_0 & 18 + 11 (\Kbi)^2 + 3 \Ss^2 + 2 \Sn^2 + 16 \Z^2 + 5 \Sn \Z \\ 
 & & - \Kbi (3 \Ss + 4 \Sn + 12 \Z) - 10 \Ss \Z - 2 \Ss \Sn \\ \hline
\end{array}
\end{equation}
\begin{align*}
\text{Euler number:}& \quad \chi = -24 (\Kbi)^2 + 8 \Kbi \Sn - 4 \Sn^2 + 6 \Kbi \Ss + 4 \Ss \Sn - 6 \Ss^2 + 
   24 \Kbi \Z \\ & \hspace{1.15cm} - 8 \Sn \Z + 20 \Ss \Z - 32 \Z^2
    \\ 
\text{anomaly coefficients$^*$:}& \quad \hat{a} \sim K_{\hat{B}}^{-1} \,, \enspace \hat{b} \sim - \hat{Z} \,, \enspace \hat{b}_{11} \sim - (6 K_{\hat{B}}^{-1} - 2 \hat{\mathcal{S}}_7 + 4 \hat{\mathcal{S}}_9)
\end{align*}
}\vspace{-0.5cm}

%%%%%%%%%%%%%%%%%%
\subsection[Polygon \texorpdfstring{$F_4$}{F4}]{Polygon \texorpdfstring{$\boldsymbol{F_4}$}{F4}}
\vspace{-0.5cm}
{\small
\begin{align*}
\text{vertices:}& \quad Y: \, (-1,-1,0) \,, \enspace X: \, (-1,1,0) \,, \enspace Z: \, (1,0,0) \,, \enspace e_1: \, (-1,0,0) \\ 
\text{gauge group$^*$:}& \quad \text{SO(10)} \times \text{SU(2)} \times \mathbb{Z}_2
\end{align*}
}

%%%%%%%%%
\subsubsection*{Top 1}
\vspace{-0.5cm}
{\small
\begin{align*}
\text{vertices:}& \quad z_0: \, (0,0,1) \,, \enspace f_1: \, (0,-2,1) \,, \enspace f_2: \, (0,-1,1) \,, \enspace f_3: \, (1,-1,1) \,, \enspace f_4: \, (1,0,1) \,, \\
& \quad g_1: \, (1,-2,2) \,, \enspace g_2: \, (1,-1,2) \\
\text{factorization:}& \quad c_1 = d_1 z_0^4 \,, \enspace c_2 = d_2 z_0^2 \,, \enspace c_3 = d_3 z_0 \,, \enspace c_4 = d_4 z_0 \,, \enspace c_5 = d_5 z_0 \,, \enspace c_6 = d_6 z_0^2 \,, \\
& \quad c_7 = d_7 z_0 \,, \enspace c_8 = d_8 \,, \enspace c_9 = d_9 
\end{align*}
\begin{equation}
\begin{array}{|c|c|c|}
\hline
\text{locus} & \text{representation} & \text{multiplicity} \\ \hline \hline
z_0 = d_8 = 0 & (\mathbf{10},\mathbf{1})_{1} & \Ss \Z \\ \hline
z_0 = d_2 = 0 & (\mathbf{10},\mathbf{1})_{0} & (2 \Kbi - \Sn - 2 \Z) \Z \\ \hline
z_0 = d_3 = 0 & (\mathbf{16},\mathbf{1})_{1/2} & (\Kbi + \Ss - \Sn - \Z) \Z \\ \hline
z_0 = d_9 = 0 & \text{SCP} & (\Kbi - \Ss + \Sn) \Z \\ \hline
 & (\mathbf{1},\mathbf{3})_0 & 1 - \tfrac{1}{2} (\Kbi - \Ss + \Sn) (\Ss - \Sn) \\ \hline
 & (\mathbf{1},\mathbf{2})_{1/2} & (\Kbi - \Ss + \Sn) (6 \Kbi + 2 \Ss - 2 \Sn - 6 \Z) \\ \hline
 & (\mathbf{1},\mathbf{1})_{1} & 6 (\Kbi)^2 - 3 \Ss^2 + \Sn^2 + 4 \Z^2 +4 \Sn \Z \\
 & & + \Kbi (13 \Ss - 5 \Sn - 10 \Z) - 2 \Ss (\Sn + 5 \Z) \\ \hline
 & (\mathbf{1},\mathbf{1})_0 & 18 + 11 (\Kbi)^2 + 6 \Ss^2 + 2 \Sn^2 + 5 \Sn \Z + 12 \Z^2\\
 && + \Ss (-4 \Sn + \Z) - \Kbi (4 \Ss + 4 \Sn + 23 \Z) \\ \hline
\end{array}
\end{equation}
\begin{align*}
\text{Euler number:}& \quad \chi =  -24 (\Kbi)^2 + 8 \Kbi \Ss - 12 \Ss^2 + 8 \Kbi \Sn + 8 \Ss \Sn - 4 \Sn^2 + 
   48 \Kbi \Z\\ & \hspace{1.15cm} - 4 \Ss \Z - 8 \Sn \Z - 24 \Z^2
    \\ 
\text{anomaly coefficients$^*$:}& \quad \hat{a} \sim K_{\hat{B}}^{-1} \,, \enspace \hat{b} \sim - \hat{\Z} \,, \enspace \hat{b}_{\text{SU(2)}} \sim - (K_{\hat{B}}^{-1} - \hat{\mathcal{S}}_7 + \hat{\mathcal{S}}_9)
\end{align*}
}\vspace{-0.5cm}

%%%%%%%%%
\subsubsection*{Top 2}
\vspace{-0.5cm}
{\small
\begin{align*}
\text{vertices:}& \quad z_0: \, (0,0,1) \,, \enspace f_1: \, (0,1,1) \,, \enspace f_2: \, (1,0,1) \,, \enspace f_3: \, (1,1,1) \,,\\
& \quad g_1: \, (1,1,2) \,, \enspace g_2: \, (2,1,2) \\
\text{factorization:}& \quad c_1 = d_1 z_0 \,, \enspace c_2 = d_2 z_0 \,, \enspace c_3 = d_3 z_0^2 \,, \enspace c_4 = d_4 z_0^2 \,, \enspace c_5 = d_5 z_0^3 \,, \enspace c_6 = d_6 \,, \\
& \quad c_7 = d_7 z_0 \,, \enspace c_8 = d_8 z_0 \,, \enspace c_9 = d_9 
\end{align*}
\begin{equation}
\begin{array}{|c|c|c|}
\hline
\text{locus} & \text{representation} & \text{multiplicity} \\ \hline \hline
z_0 = d_6 = 0 & (\mathbf{16},\mathbf{1})_{-1/4} & (2 \Kbi - \Ss) \Z \\ \hline
z_0 = d_9 = 0 & (\mathbf{10},\mathbf{2})^*_{1} & (\Kbi - \Ss + \Sn) \Z \\ \hline
z_0 = d_4 d_6 - d_2 d_8 = 0 & (\mathbf{10}, \mathbf{1})_{1/2} & (2 \Kbi + \Ss - \Sn - 2 \Z) \Z \\ \hline
z_0 = d_8 = 0 & (\mathbf{16},\mathbf{1})_{3/4} & (\Ss - \Z) \Z \\ \hline
 & (\mathbf{1},\mathbf{3})_0 & 1 - \tfrac{1}{2} (\Kbi - \Ss + \Sn) (\Ss - \Sn) \\ \hline
 & (\mathbf{1},\mathbf{2})_{1/2} & (\Kbi - \Ss + \Sn) (6 \Kbi + 2 \Ss - 2 \Sn - 5 \Z) \\ \hline
 & (\mathbf{1},\mathbf{1})_{1} & 6 (\Kbi)^2 - 3 \Ss^2 + \Sn^2 + 6 \Z^2 + 5 \Sn \Z\\ 
 & & + \Kbi (13 \Ss - 5 \Sn -20 \Z) - \Ss \Z - 2 \Ss \Sn \\ \hline
 & (\mathbf{1},\mathbf{1})_0 & 18 + 11 (\Kbi)^2 + 6 \Ss^2 + 2 \Sn^2 + 10 \Z^2 + 5 \Sn \Z\\
 & & - 4 \Kbi (\Ss + \Sn + 3 \Z) - 9 \Ss \Z - 4 \Ss \Sn \\ \hline
\end{array}
\end{equation}
\begin{align*}
\text{Euler number:}& \quad \chi =-24 (\Kbi)^2 + 8 \Kbi \Ss - 12 \Ss^2 + 8 \Kbi \Sn + 8 \Ss \Sn - 4 \Sn^2 + 
  24 \Kbi \Z\\ & \hspace{1.15cm} + 18 \Ss \Z - 10 \Sn \Z - 20 \Z^2 \quad \\
\text{anomaly coefficients:}& \quad a \sim \Kbi \,, \enspace b \sim - \Z \,, \enspace b_{\text{SU(2)}} \sim - (\Kbi - \Ss + \Sn)
\end{align*}
}\vspace{-0.5cm}

%%%%%%%%%
\subsubsection*{Top 3}
\vspace{-0.5cm}
{\small
\begin{align*}
\text{vertices:}& \quad z_0: \, (0,0,1) \,, \enspace f_1: \, (0,1,1) \,, \enspace f_2: \, (1,0,1) \,, \enspace f_3: \, (1,1,1) \,, \enspace f_4: \, (2,0,1) \,, \\
& \quad g_1: \, (1,1,2) \,, \enspace g_2: \, (2,1,2) \\
\text{factorization:}& \quad c_1 = d_1 z_0^2 \,, \enspace c_2 = d_2 z_0^2 \,, \enspace c_3 = d_3 z_0^2 \,, \enspace c_4 = d_4 z_0^2 \,, \enspace c_5 = d_5 z_0^3 \,, \enspace c_6 = d_6 \,, \\
& \quad c_7 = d_7 z_0 \,, \enspace c_8 = d_8 z_0 \,, \enspace c_9 = d_9 
\end{align*}
\begin{equation}
\begin{array}{|c|c|c|}
\hline
\text{locus} & \text{representation} & \text{multiplicity} \\ \hline \hline
z_0 = d_8 = 0 & (\mathbf{16},\mathbf{1})_{3/4} & (\Ss - \Z) \Z \\ \hline
z_0 = d_4 = 0 & (\mathbf{10},\mathbf{1})_{1/2} & (2 \Ss - \Sn - 2 \Z) \Z \\ \hline
z_0 = d_9 = 0 & (\mathbf{10},\mathbf{2})_1^* & (\Kbi - \Ss + \Sn) \Z \\ \hline
z_0 = d_6 = 0 & \text{SCP} & (2 \Kbi - \Ss) \Z \\ \hline
 & (\mathbf{1},\mathbf{3})_0 & 1 - \tfrac{1}{2} (\Kbi - \Ss + \Sn) (\Ss - \Sn) \\ \hline
 & (\mathbf{1},\mathbf{2})_{1/2} & (\Kbi - \Ss + \Sn) (6 \Kbi + 2 \Ss - 2 \Sn - 5 \Z) \\ \hline
 & (\mathbf{1},\mathbf{1})_{1} & 6 (\Kbi)^2 - 3 \Ss^2 + \Sn^2 + 6 \Z^2 + 5 \Sn \Z \\
 & & + \Kbi (13 \Ss - 5 \Sn - 22 \Z) - 2 \Ss \Sn \\ \hline
 & (\mathbf{1},\mathbf{1})_0 & 18 + 11 (\Kbi)^2 + 6 \Ss^2 + 2 \Sn^2 + 10 \Z^2 + 5 \Sn \Z\\
 & & - 4 \Kbi (\Ss + \Sn + 4 \Z) - 7 \Ss \Z - 4 \Ss \Sn \\ \hline
\end{array}
\end{equation}
\begin{align*}
\text{Euler number:}& \quad \chi = -24 (\Kbi)^2 + 8 \Kbi \Ss - 12 \Ss^2 + 8 \Kbi \Sn + 8 \Ss \Sn - 4 \Sn^2 + 
  36 \Kbi \Z \\ & \hspace{1.15cm}+ 12 \Ss \Z - 10 \Sn \Z - 20 \Z^2  \\
\text{anomaly coefficients$^*$:}& \quad \hat{a} \sim K_{\hat{B}}^{-1} \,, \enspace \hat{b} \sim - \hat{\Z} \,, \enspace \hat{b}_{\text{SU(2)}} \sim - (K_{\hat{B}}^{-1} - \hat{\mathcal{S}}_7 + \hat{\mathcal{S}}_9) \\
\text{relation to ($F_4$, top 2):}& \quad (\mathbf{16},\mathbf{1})_{-1/4} \, \oplus (\mathbf{10},\mathbf{1})_{1/2} \, \oplus (\mathbf{1},\mathbf{1})_1 \, \oplus (\mathbf{1},\mathbf{1})_0 \, \oplus (\mathbf{1},\mathbf{1})_0
\end{align*}
}\vspace{-0.5cm}

%%%%%%%%%%%%%%%%%%
\subsection[Polygon \texorpdfstring{$F_5$}{F5}]{Polygon \texorpdfstring{$\boldsymbol{F_5}$}{F5}}
\vspace{-0.5cm}
{\small
\begin{align*}
\text{vertices:}& \quad w: \, (1,0,0) \,, \enspace v: \, (0,-1,0) \,, \enspace u: \, (-1,1,0) \,, \\
&\quad e_1: \, (-1,0,0) \,, \enspace e_2: \, (0,1,0) \\ 
\text{gauge group:}& \quad \text{SO(10)} \times \text{U(1)}^2
\end{align*}
}\vspace{-0.5cm}

%%%%%%%%%
\subsubsection*{Top 1}
\vspace{-0.5cm}
{\small
\begin{align*}
\text{vertices:}& \quad z_0: \, (0,0,1) \,, \enspace f_1: \, (0,1,1) \,, \enspace f_2: \, (-1,1,1) \,, \enspace f_3: \, (1,0,1) \,, \enspace f_4: \, (1,1,1) \,, \\ 
& \quad g_1: \, (0,1,2) \,, \enspace g_2 : \,(1,1,2) \\ 
\text{factorization:}& \quad s_1 = d_1 z_0 \,, \enspace s_2 = d_2 z_0 \,, \enspace s_3 = d_3 z_0^2 \,, \enspace s_5 = d_5 \,, \enspace s_6 = d_6 z_0 \,, \enspace s_7 = d_7 z_0 \,, \\
& \quad s_8 = d_8 \,, \enspace s_9 = d_9 z_0
\end{align*}
\begin{equation}
\begin{array}{|c|c|c|}
\hline
\text{locus} & \text{representation} & \text{multiplicity} \\ \hline \hline
z_0 = d_5 = 0 & \mathbf{16}_{1/4,0} & (2 \Kbi - \Ss) \Z \\ \hline
z_0 = d_2 = 0 & \mathbf{10}_{1/2,0} & (2 \Kbi - \Sn - \Z) \Z \\ \hline
z_0 = d_8 = 0 & \mathbf{10}_{1/2,1} & (\Kbi - \Ss + \Sn) \Z \\ \hline
z_0 = d_7 = 0 & \text{SCP} & (\Ss - \Z) \Z \\ \hline
 & \mathbf{1}_{1,-1} & (\Kbi + \Ss - \Sn - 2 \Z) (\Ss - \Z) \\ \hline
 & \mathbf{1}_{1,2} & (\Kbi - \Ss + \Sn) (\Sn - \Z) \\ \hline
 & \mathbf{1}_{0,2} & (\Sn - \Z) (\Ss - \Z) \\ \hline
 & \mathbf{1}_{-1,-1} & 6 (\Kbi)^2 + \Ss^2 - 2 \Sn^2 + 2 \Z^2 - 2 \Sn \Z \\
 & & + \Kbi (-5 \Ss + 4 \Sn - 4 \Z) + \Ss \Sn \\ \hline
 & \mathbf{1}_{1,0} & 6 (\Kbi)^2 - 2 \Ss^2 + \Sn^2 + 3 \Z^2 + 2 \Sn \Z \\
 & & + \Kbi (4 \Ss - 5 \Sn - 11 \Z) + \Ss (\Sn + \Z) \\ \hline
 & \mathbf{1}_{0,1} & 6 (\Kbi)^2 - 2 \Ss^2 - 2 \Sn^2 + 4 \Z^2 - \Sn \Z \\
 & & + \Kbi (4 \Ss + 4 \Sn - 13 \Z) + \Ss \Z \\ \hline
 & \mathbf{1}_{0,0} & 19 + 11 (\Kbi)^2 + 2 \Ss^2 + 2 \Sn^2 + 2 \Sn \Z + 7 \Z^2  \\
 &&- \Ss (\Sn + 2 \Z) -  4 \Kbi (\Ss + \Sn + 3 \Z) \\ \hline
\end{array}
\label{eq:F5top1}
\end{equation}
\begin{align*}
\text{Euler number:}& \quad \chi = - 24 (\Kbi)^2 - 4 \Ss^2 - 4 \Sn^2 - 16 \Z^2 - 4 \Sn \Z \\
& \hspace{1.15cm} + 8 \Kbi (\Ss + \Sn + 4 \Z) + 2 \Ss \Sn + 6 \Ss \Z \\
\text{anomaly coefficients$^*$:}& \quad \hat{a} \sim K_{\hat{B}}^{-1} \,, \enspace \hat{b} \sim - \hat{\Z} \,, \enspace \hat{b}_{11} \sim - (2 K_{\hat{B}}^{-1} - \tfrac{5}{4} \hat{\Z}) \,, \\
& \quad \hat{b}_{12} \sim - (K_{\hat{B}}^{-1} - \hat{\mathcal{S}}_7 + \hat{\mathcal{S}}_9) \,, \enspace \hat{b}_{22} \sim - (2 K_{\hat{B}}^{-1} + 2 \hat{\mathcal{S}}_9 - 2 \hat{\Z})
\end{align*}
}\vspace{-0.5cm}

%%%%%%%%%
\subsubsection*{Top 2}
\vspace{-0.5cm}
{\small
\begin{align*}
\text{vertices:}& \quad z_0: \, (0,0,1) \,, \enspace f_1: \, (0,1,1) \,, \enspace f_2: \, (1,0,1) \,, \enspace f_3: \, (1,1,1) \,, \\ 
& \quad g_1: \, (1,1,2) \,, \enspace g_2 : \,(2,1,2) \\ 
\text{factorization:}& \quad s_1 = d_1 z_0 \,, \enspace s_2 = d_2 z_0^2 \,, \enspace s_3 = d_3 z_0^2 \,, \enspace s_5 = d_5 \,, \enspace s_6 = d_6 z_0 \,, \enspace s_7 = d_7 z_0 \,, \\
& \quad s_8 = d_8 \,, \enspace s_9 = d_9
\end{align*}
\begin{equation}
\begin{array}{|c|c|c|}
\hline
\text{locus} & \text{representation} & \text{multiplicity} \\ \hline \hline
z_0 = d_5 = 0 & \mathbf{16}_{-1/4,-1/2} &  (2 \Kbi - \Ss) \Z \\ \hline
z_0 = d_9 = 0 & \mathbf{10}_{1/2, 1} & \Sn \Z \\ \hline
z_0 = d_3 d_5 - d_1 d_7 = 0 & \mathbf{10}_{-1/2,0} & (3 \Kbi - \Sn - 2 \Z) \Z \\ \hline 
z_0 = d_7 = 0 & \mathbf{16}_{-1/4, 1/2} & (\Ss - \Z) \Z \\ \hline
 & \mathbf{1}_{1,-1} & (\Kbi + \Ss - \Sn - 2 \Z) (\Ss - \Z) \\ \hline
 & \mathbf{1}_{1,2} & (\Kbi - \Ss + \Sn) \Sn \\ \hline
 & \mathbf{1}_{0,2} & \Sn (\Ss - \Z) \\ \hline
 & \mathbf{1}_{-1,-1} & 6 (\Kbi)^2 + \Ss^2 + \Kbi (-5 \Ss + 4 \Sn - 2 \Z) \\
 & & + \Ss (\Sn + \Z) - 2 \Sn (\Sn + 2 \Z) \\ \hline 
 & \mathbf{1}_{1,0} & 6 (\Kbi)^2 - 2 \Ss^2 + \Sn^2 + 4 \Z^2 + 3 \Sn \Z \\
 & & + \Kbi (4 \Ss - 5 \Sn - 14 \Z) + \Ss (\Sn + 2 \Z) \\ \hline 
 & \mathbf{1}_{0,1} & 6 (\Kbi)^2 - 2 \Ss^2 - 2 \Sn^2 + 2 \Z^2 + 2 \Ss \Z \\
 & & + \Kbi (4 \Ss + 4 \Sn - 11 \Z) - 3 \Sn \Z \\ \hline
 & \mathbf{1}_{0,0} & 19 + 11 (\Kbi)^2 + 2 \Ss^2 + 2 \Sn^2 + 4 \Sn \Z + 8 \Z^2 \\ 
 &&- \Ss (\Sn + 2 \Z) -  2 \Kbi (2 \Ss + 2 \Sn + 7 \Z) \\ \hline
\end{array}
\label{eq:F5top2}
\end{equation}
\begin{align*}
\text{Euler number:}& \quad \chi =  - 24 (\Kbi)^2 - 4 \Ss^2 - 4 \Sn^2 - 16 \Z^2 - 8 \Sn \Z \\
& \hspace{1.15cm} + 4 \Kbi (2 \Ss + 2 \Sn + 7 \Z) + 2 \Ss \Sn + 4 \Ss \Z \\
\text{anomaly coefficients:}& \quad a \sim \Kbi \,, \enspace b \sim - \Z \,, \enspace b_{11} \sim - (2 \Kbi - \tfrac{5}{4} \Z) \,, \\
& \quad b_{12} \sim - (\Kbi - \Ss + \Sn + \tfrac{1}{2} \Z) \,, \enspace b_{22} \sim - (2 \Kbi + 2 \Sn - \Z)
\end{align*}
}\vspace{-0.5cm}

%%%%%%%%%
\subsubsection*{Top 3}
\vspace{-0.5cm}
{\small
\begin{align*}
\text{vertices:}& \quad z_0: \, (0,0,1) \,, \enspace f_1: \, (0,1,1) \,, \enspace f_2: \, (1,0,1) \,, \enspace f_3: \, (1,1,1) \,, \enspace f_4: \, (2,0,1) \,, \\ 
& \quad g_1: \, (1,1,2) \,, \enspace g_2 : \,(2,1,2) \\ 
\text{factorization:}& \quad s_1 = d_1 z_0^2 \,, \enspace s_2 = d_2 z_0^2 \,, \enspace s_3 = d_3 z_0^2 \,, \enspace s_5 = d_5 \,, \enspace s_6 = d_6 z_0 \,, \enspace s_7 = d_7 z_0 \,, \\
& \quad s_8 = d_8 \,, \enspace s_9 = d_9
\end{align*}
\begin{equation}
\begin{array}{|c|c|c|}
\hline
\text{locus} & \text{representation} & \text{multiplicity} \\ \hline \hline
z_0 = d_9 = 0 &\mathbf{10}_{1/2,1} & \Sn \Z \\ \hline
z_0 = d_3 = 0 & \mathbf{10}_{-1/2,0} & (\Kbi + \Ss - \Sn - 2 \Z) \Z \\ \hline
z_0 = d_7 = 0 & \mathbf{16}_{-1/4,1/2} & (\Ss - \Z) \Z \\ \hline
z_0 = d_5 = 0 & \text{SCP} & (2 \Kbi - \Ss) \Z \\ \hline
 & \mathbf{1}_{1,-1} & (\Kbi + \Ss - \Sn - 2 \Z) (\Ss - \Z) \\ \hline
 & \mathbf{1}_{1,2} & (\Kbi - \Ss + \Sn) \Sn \\ \hline
 & \mathbf{1}_{0,2} & \Sn (\Ss - \Z) \\ \hline
 & \mathbf{1}_{-1,-1} & 6 (\Kbi)^2 + \Ss^2 + \Kbi (-5 \Ss + 4 \Sn - 4 \Z) \\ 
 & & + \Ss (\Sn + 2 \Z) - 2 \Sn (\Sn + 2 \Z) \\ \hline
 & \mathbf{1}_{1,0} & 6 (\Kbi)^2 - 2 \Ss^2 + \Sn^2 + 4 \Z^2 + 3 \Sn \Z \\
 & & + \Kbi (4 \Ss - 5 \Sn - 14 \Z) + \Ss (\Sn + 2 \Z) \\ \hline
 & \mathbf{1}_{0,1} & 6 (\Kbi)^2 - 2 \Ss^2 - 2 \Sn^2 + 2 \Z^2 - 3 \Sn \Z \\
 & & + \Kbi (4 \Ss + 4 \Sn - 13 \Z) + 3 \Ss \Z \\ \hline
 & \mathbf{1}_{0,0} & 19 + 11 (\Kbi)^2 + 2 \Ss^2 + 2 \Sn^2 + 8 \Z^2 + 4 \Sn \Z\\
 && - \Ss (\Sn + \Z) -  4 \Kbi (\Ss + \Sn + 4 \Z)\\ \hline
\end{array}
\label{eq:F5top3}
\end{equation}
\begin{align*}
\text{Euler number:}& \quad \chi =  - 24 (\Kbi)^2 - 4 \Ss^2 - 4 \Sn^2 - 16 \Z^2 - 8 \Sn \Z \\
& \hspace{1.15cm} + 4 \Kbi (2 \Ss + 2 \Sn + 9 \Z) + 2 \Ss \Sn \\
\text{anomaly coefficients$^*$:}& \quad \hat{a} \sim K_{\hat{B}}^{-1} \,, \enspace \hat{b} \sim - \hat{\Z} \,, \enspace \hat{b}_{11} \sim -(2 K_{\hat{B}}^{-1} - \tfrac{5}{4} \hat{\Z}) \,, \\
& \quad \hat{b}_{12} \sim - (K_{\hat{B}}^{-1} - \hat{\mathcal{S}}_7 + \hat{\mathcal{S}}_9 + \tfrac{1}{2} \hat{\Z}) \,, \enspace \hat{b}_{22} \sim -(2 K_{\hat{B}}^{-1} + 2 \hat{\mathcal{S}}_9 - \hat{\Z}) \\ 
\text{relation to ($F_5$, top 2):}& \quad \mathbf{16}_{-1/4,-1/2} \, \oplus \mathbf{10}_{1/2,0} \, \oplus \mathbf{1}_{-1,-1} \, \oplus \mathbf{1}_{0,1} \, \oplus \mathbf{1}_{0,0}
\end{align*}
}\vspace{-0.5cm}

%%%%%%%%%
\subsubsection*{Top 4}
\vspace{-0.5cm}
{\small
\begin{align*}
\text{vertices:}& \quad z_0: \, (0,0,1) \,, \enspace f_1: \, (0,1,1) \,, \enspace f_2: \, (1,1,1) \,, \enspace f_3: \, (1,0,1) \,, \enspace f_4: \, (2,1,1) \,, \\ 
& \quad g_1: \, (1,1,2) \,, \enspace g_2 : \,(2,1,2) \\ 
\text{factorization:}& \quad s_1 = d_1 z_0 \,, \enspace s_2 = d_2 z_0^2 \,, \enspace s_3 = d_3 z_0^3 \,, \enspace s_5 = d_5 \,, \enspace s_6 = d_6 z_0 \,, \enspace s_7 = d_7 z_0 \,, \\
& \quad s_8 = d_8 \,, \enspace s_9 = d_9
\end{align*}
\begin{equation}
\begin{array}{|c|c|c|}
\hline
\text{locus} & \text{representation} & \text{multiplicity} \\ \hline \hline
z_0 = d_5 = 0 & \mathbf{16}_{-1/4,-1/2} & (2 \Kbi - \Ss) \Z \\ \hline
z_0 = d_9 = 0 & \mathbf{10}_{-1/2,-1} & \Sn \Z \\ \hline
z_0 = d_1 = 0 & \mathbf{10}_{1/2,0} & (3 \Kbi - \Ss - \Sn - \Z) \Z \\ \hline
z_0 = d_7 = 0 & \text{SCP} & (\Ss - \Z) \Z \\ \hline
 & \mathbf{1}_{1,-1} & (\Kbi + \Ss - \Sn - 3 \Z) (\Ss - \Z) \\ \hline
 & \mathbf{1}_{1,2} & (\Kbi - \Ss + \Sn) \Sn \\ \hline
 & \mathbf{1}_{0,2} & \Sn (\Ss - \Z) \\ \hline
 & \mathbf{1}_{-1,-1} & 6 (\Kbi)^2 + \Ss^2 + \Kbi (-5 \Ss + 4 \Sn - 2 \Z) \\
 & & + \Ss (\Sn + \Z) - 2 \Sn (\Sn + 2 \Z) \\ \hline
 & \mathbf{1}_{1,0} & 6 (\Kbi)^2 - 2 \Ss^2 + \Sn^2 + 4 \Z^2 + 3 \Sn \Z \\ 
 & & + \Kbi (4 \Ss - 5 \Sn - 14 \Z) + \Ss (\Sn + 2 \Z) \\ \hline 
 & \mathbf{1}_{0,1} & 6 (\Kbi)^2 - 2 \Ss^2 - 2 \Sn^2 + 3 \Z^2 - 3 \Sn \Z \\ 
 & & + \Kbi (4 \Ss + 4 \Sn - 11 \Z) + \Ss \Z \\ \hline
 & \mathbf{1}_{0,0} & 19 + 11 (\Kbi)^2 + 2 \Ss^2 + 2 \Sn^2 + 4 \Sn \Z + 9 \Z^2 \\
 && - \Ss (\Sn + 3 \Z) -  2 \Kbi (2 \Ss + 2 \Sn + 7 \Z) \\ \hline
\end{array}
\end{equation}
\begin{align*}
\text{Euler number:}& \quad \chi = -24 (\Kbi)^2 + 8 \Kbi \Ss - 4 \Ss^2 + 8 \Kbi \Sn + 2 \Ss \Sn - 4 \Sn^2 + 
   28 \Kbi \Z \\ & \hspace{1.15cm}  + 8 \Ss \Z - 8 \Sn \Z - 20 \Z^2
 \\ 
\text{anomaly coefficients$^*$:}& \quad \hat{a} \sim K_{\hat{B}}^{-1} \,, \enspace \hat{b} \sim - \hat{\Z} \,, \enspace \hat{b}_{11} \sim - (2 K_{\hat{B}}^{-1} - \tfrac{5}{4} \hat{\Z}) \,, \\
& \quad \hat{b}_{12} \sim - (K_{\hat{B}}^{-1} - \hat{\mathcal{S}}_7 + \hat{\mathcal{S}}_9 + \tfrac{1}{2} \hat{\Z}) \,, \enspace \hat{b}_{22} \sim -(2 K_{\hat{B}}^{-1} + 2 \hat{\mathcal{S}}_9 - \hat{\Z}) \\ 
\text{relation to ($F_5$, top 2):}& \quad \mathbf{16}_{-1/4,1/2} \, \oplus \mathbf{10}_{1/2,0} \, \oplus \mathbf{1}_{1,-1} \, \oplus \mathbf{1}_{0,1} \, \oplus \mathbf{1}_{0,0}
\end{align*}
}\vspace{-0.5cm}

%%%%%%%%%
\subsubsection*{Top 5}
\vspace{-0.5cm}
{\small
\begin{align*}
\text{vertices:}& \quad z_0: \, (0,0,1) \,, \enspace f_1: \, (0,1,1) \,, \enspace f_2: \, (1,-1,1) \,, \enspace f_3: \, (1,0,1) \,, \enspace f_4: \, (2,-1,1) \,, \\ 
& \quad g_1: \, (1,0,2) \,, \enspace g_2 : \,(2,-1,2) \\ 
\text{factorization:}& \quad s_1 = d_1 z_0^3 \,, \enspace s_2 = d_2 z_0^2 \,, \enspace s_3 = d_3 z_0 \,, \enspace s_5 = d_5 z_0 \,, \enspace s_6 = d_6 z_0 \,, \enspace s_7 = d_7 z_0 \,, \\
& \quad s_8 = d_8 \,, \enspace s_9 = d_9
\end{align*}
\begin{equation}
\begin{array}{|c|c|c|}
\hline
\text{locus} & \text{representation} & \text{multiplicity} \\ \hline \hline
z_0 = d_5 = 0 & \mathbf{10}_{0,0} & (2 \Kbi - \Ss - \Z) \Z \\ \hline
z_0 = d_9 = 0 & \mathbf{10}_{0,1} & \Sn \Z \\ \hline
z_0 = d_8 = 0 & \mathbf{16}_{-1/2,-1/2} & (\Kbi - \Ss + \Sn) \Z \\ \hline
z_0 = d_3 = 0 & \text{SCP} & (\Kbi + \Ss - \Sn - \Z) \Z \\ \hline
 & \mathbf{1}_{1,-1} & (\Kbi + \Ss - \Sn - \Z) (\Ss - \Z) \\ \hline
 & \mathbf{1}_{1,2} & (\Kbi - \Ss + \Sn) \Sn \\ \hline
 & \mathbf{1}_{0,2} & (\Ss - \Z) \Sn \\ \hline
 & \mathbf{1}_{-1,-1} & 6 (\Kbi)^2 + \Ss^2 - 2 \Sn^2 + \Z^2 - 3 \Sn \Z \\
 & & + \Kbi (-5 \Ss + 4 \Sn - 5 \Z) + \Ss (\Sn + 2 \Z) \\ \hline
 & \mathbf{1}_{1,0} & 6 (\Kbi)^2 - 2 \Ss^2 + \Kbi (4 \Ss - 5 \Sn - 10 \Z) \\ 
 & & + \Ss (\Sn - 2 \Z) + (\Sn + 2 \Z)^2 \\ \hline 
 & \mathbf{1}_{0,1} & 6 (\Kbi)^2 - 2 \Ss^2 - 2 \Sn^2 + 4 \Z^2 - 2 \Sn \Z \\ 
 & & + \Kbi (4 \Ss + 4 \Sn - 10 \Z) - 2 \Ss \Z \\ \hline
 & \mathbf{1}_{0,0} & 19 + 11 (\Kbi)^2 + 2 \Ss^2 + 2 \Sn^2 + 4 \Sn \Z + 9 \Z^2 \\
 && + \Ss (-\Sn + \Z) - \Kbi (4 \Ss + 4 \Sn + 19 \Z) \\ \hline
\end{array}
\end{equation}
\begin{align*}
\text{Euler number:}& \quad \chi = -24 (\Kbi)^2 + 8 \Kbi \Ss - 4 \Ss^2 + 8 \Kbi \Sn + 2 \Ss \Sn - 4 \Sn^2\\ & \hspace{1.15cm}   + 
  40 \Kbi \Z - 10 \Sn \Z - 20 \Z^2 \\
\text{anomaly coefficients$^*$:}& \quad \hat{a} \sim K_{\hat{B}}^{-1} \,, \enspace \hat{b} \sim - \hat{\Z} \,, \enspace \hat{b}_{11} \sim - (2 K_{\hat{B}}^{-1} - \hat{\Z}) \,, \\
& \quad \hat{b}_{12} \sim - (K_{\hat{B}}^{-1} - \hat{\mathcal{S}}_7 + \hat{\mathcal{S}}_9) \,, \enspace \hat{b}_{22} \sim - (2 K_{\hat{B}}^{-1} + 2 \hat{\mathcal{S}}_9 - \hat{\Z})
\end{align*}
}\vspace{-0.5cm}

%%%%%%%%%%%%%%%%%%
\subsection[Polygon \texorpdfstring{$F_6$}{F6}]{Polygon \texorpdfstring{$\boldsymbol{F_6}$}{F6}}
\vspace{-0.5cm}
{\small
\begin{align*}
\text{vertices:}& \quad w: \, (0,1,0) \,, \enspace v: \, (1,-1,0) \,, \enspace u: \, (-1,0,0) \,, \\
& \quad e_1: \, (0,-1,0) \,, \enspace e_2: \, (-1,-1,0)\\ 
\text{gauge group:}& \quad \text{SO(10)} \times \text{SU(2)} \times \text{U(1)}
\end{align*}
}\vspace{-0.5cm}

%%%%%%%%%
\subsubsection*{Top 1}
\vspace{-0.5cm}
{\small
\begin{align*}
\text{vertices:}& \quad z_0: \, (0,0,1) \,, \enspace f_1: \, (0,1,1) \,, \enspace f_2: \, (1,0,1) \,, \enspace f_3: \, (1,1,1) \,, \enspace f_4: \, (2,0,1) \,, \\
& \quad g_1: \, (1,1,2) \,, \enspace g_2: \, (2,1,2) \\
\text{factorization:}& \quad s_1 = d_1 z_0^4 \,, \enspace s_2 = d_2 z_0^2 \,, \enspace s_3 = d_3 z_0 \,, \enspace s_4 = d_4 z_0 \,, \enspace s_5 = d_5 z_0^2 \,, \\
& \quad s_6 = d_6 z_0 \,, \enspace s_7 = d_7 \,, \enspace s_8 = d_8
\end{align*}
\begin{equation}
\begin{array}{|c|c|c|}
\hline
\text{locus} & \text{representation} & \text{multiplicity} \\ \hline \hline
z_0 = d_3 = 0 & (\mathbf{16},\mathbf{1})_{-1/2} & (\Kbi + \Ss - \Sn - \Z) \Z \\ \hline 
z_0 = d_2 = 0 & (\mathbf{10},\mathbf{1})_0 & (2 \Kbi - \Sn - 2 \Z) \Z \\ \hline
z_0 = d_7 = 0 & (\mathbf{10},\mathbf{1})_1 & \Ss \Z \\ \hline
z_0 = d_8 = 0 & \text{SCP} & (\Kbi - \Ss + \Sn) \Z \\ \hline
 & (\mathbf{1},\mathbf{3})_0 & 1 - \tfrac{1}{2} (\Ss - \Sn) (\Kbi - \Ss + \Sn) \\ \hline
 & (\mathbf{1},\mathbf{2})_{3/2} & (\Kbi - \Ss + \Sn) \Ss \\ \hline
 & (\mathbf{1},\mathbf{2})_{1/2} & (\Kbi - \Ss + \Sn) (6 \Kbi + \Ss - 2 \Sn - 6\Z) \\ \hline
 & (\mathbf{1},\mathbf{1})_2 & (2 \Ss - \Sn - \Z) \Ss \\ \hline
 & (\mathbf{1},\mathbf{1})_1 & 6 (\Kbi)^2 - 3 \Ss^2 + \Kbi (13 \Ss - 5 \Sn - 10 \Z) \\
 & & - (2 \Sn + 10 \Z) \Ss + (\Sn + 2 \Z)^2 \\ \hline
 & (\mathbf{1},\mathbf{1})_0 & 19 + 11 (\Kbi)^2 + 4 \Ss^2 - 3 \Ss \Sn + 2 \Sn^2 + 2 \Ss \Z \\
 &&+ 5 \Sn \Z +  12 \Z^2 - \Kbi (4 \Ss + 4 \Sn + 23 \Z) \\ \hline
\end{array}
\end{equation}
\begin{align*}
\text{Euler number:}& \quad \chi = -24 (\Kbi)^2 + 8 \Kbi \Ss - 8 \Ss^2 + 8 \Kbi \Sn + 6 \Ss \Sn - 4 \Sn^2 + 
   48 \Kbi \Z \\ & \hspace{1.15cm}    - 6 \Ss \Z - 8 \Sn \Z - 24 \Z^2 
 \\
\text{anomaly coefficients$^*$:}& \quad \hat{a} \sim K_{\hat{B}}^{-1} \,, \enspace \hat{b} \sim - \hat{\Z} \,, \enspace \hat{b}_{\text{SU(2)}} \sim  - (K_{\hat{B}}^{-1} - \hat{\mathcal{S}}_7 + \hat{\mathcal{S}}_9) \\ 
& \quad \hat{b}_{11} \sim - (\tfrac{3}{2} K_{\hat{B}}^{-1} + \tfrac{5}{2} \hat{\mathcal{S}}_7 - \tfrac{1}{2} \hat{\mathcal{S}}_9 - \hat{\Z})
\end{align*}
}\vspace{-0.5cm}

%%%%%%%%%
\subsubsection*{Top 2}
\vspace{-0.5cm}
{\small
\begin{align*}
\text{vertices:}& \quad z_0: \, (0,0,1) \,, \enspace f_1: \, (1,0,1) \,, \enspace f_2: \, (1,1,1) \,, \enspace f_3: \, (2,0,1) \,, \enspace f_4: \, (2,1,1) \,, \\
& \quad g_1: \, (2,1,2) \,, \enspace g_2: \, (3,1,2) \\
\text{factorization:}& \quad s_1 = d_1 z_0^5 \,, \enspace s_2 = d_2 z_0^3 \,, \enspace s_3 = d_3 z_0 \,, \enspace s_4 = d_4 \,, \enspace s_5 = d_5 z_0^2 \,, \\
& \quad s_6 = d_6 z_0 \,, \enspace s_7 = d_7 \,, \enspace s_8 = d_8
\end{align*}
\begin{equation}
\begin{array}{|c|c|c|}
\hline
\text{locus} & \text{representation} & \text{multiplicity} \\ \hline \hline
z_0 = d_3 = 0 & (\mathbf{16},\mathbf{1})_{0} & (\Kbi + \Ss - \Sn - \Z) \Z \\ \hline 
z_0 = d_4 = 0 & (\mathbf{10},\mathbf{1})_{1} & (2 \Ss - \Sn) \Z \\ \hline
z_0 = d_5 = 0 & (\mathbf{10},\mathbf{1})_{0} & (2 \Kbi - \Ss - 2 \Z) \Z \\ \hline
z_0 = d_8 = 0 & \text{SCP} & (\Kbi - \Ss + \Sn) \Z \\ \hline
 & (\mathbf{1},\mathbf{3})_0 & 1 - \tfrac{1}{2} (\Ss - \Sn) (\Kbi - \Ss + \Sn) \\ \hline
 & (\mathbf{1},\mathbf{2})_{3/2} & (\Kbi - \Ss + \Sn) \Ss \\ \hline
 & (\mathbf{1},\mathbf{2})_{1/2} & (\Kbi - \Ss + \Sn) (6 \Kbi + \Ss - 2 \Sn - 6 \Z) \\ \hline
 & (\mathbf{1},\mathbf{1})_{2} & (2 \Ss - \Sn) \Ss \\ \hline
 & (\mathbf{1},\mathbf{1})_{1} & 6 (\Kbi)^2 - 3 \Ss^2 + \Sn^2 + 4 \Sn \Z \\
 & & + \Kbi (13 \Ss - 5 \Sn - 6 \Z) -  (2 \Sn + 14 \Z) \Ss \\ \hline
 & (\mathbf{1},\mathbf{1})_0 & 19 + 11 (\Kbi)^2 + 4 \Ss^2 - 3 \Ss \Sn + 2 \Sn^2 + 5 \Ss \Z \\&&+ 5 \Sn \Z +  16 \Z^2 - \Kbi (4 \Ss + 4 \Sn + 27 \Z) \\ \hline
\end{array}
\end{equation}
\begin{align*}
\text{Euler number:}& \quad \chi =-24 (\Kbi)^2 + 8 \Kbi \Ss - 8 \Ss^2 + 8 \Kbi \Sn + 6 \Ss \Sn - 4 \Sn^2 \\ & \hspace{1.15cm} +   
  56 \Kbi \Z - 12 \Ss \Z - 8 \Sn \Z - 32 \Z^2 \\ 
\text{anomaly coefficients$^*$:}& \quad \hat{a} \sim K_{\hat{B}}^{-1} \,, \enspace \hat{b} \sim - \Z \,, \enspace \hat{b}_{\text{SU(2)}} \sim  - (K_{\hat{B}}^{-1} - \hat{\mathcal{S}}_7 + \hat{\mathcal{S}}_9) \,, \\
& \quad \hat{b}_{11} \sim - (\tfrac{3}{2} K_{\hat{B}}^{-1} + \tfrac{5}{2} \hat{\mathcal{S}}_7 - \tfrac{1}{2} \hat{\mathcal{S}}_9) 
\end{align*}
}\vspace{-0.5cm}

%%%%%%%%%
\subsubsection*{Top 3}
\vspace{-0.5cm}
{\small
\begin{align*}
\text{vertices:}& \quad z_0: \, (0,0,1) \,, \enspace f_1: \, (0,1,1) \,, \enspace f_2: \, (1,0,1) \,, \enspace f_3: \, (1,1,1) \,, \\
& \quad g_1: \, (1,1,2) \,, \enspace g_2: \, (1,2,2) \\
\text{factorization:}& \quad s_1 = d_1 z_0^3 \,, \enspace s_2 = d_2 z_0^2 \,, \enspace s_3 = d_3 z_0^2 \,, \enspace s_4 = d_4 z_0 \,, \enspace s_5 = d_5 z_0 \,, \\
& \quad s_6 = d_6 z_0 \,, \enspace s_7 = d_7 \,, \enspace s_8 = d_8
\end{align*}
\begin{equation}
\begin{array}{|c|c|c|}
\hline
\text{locus} & \text{representation} & \text{multiplicity} \\ \hline \hline
z_0 = d_7 = 0 & (\mathbf{16},\mathbf{1})_{-3/4} & \Ss \Z \\ \hline 
z_0 = d_8 = 0 & (\mathbf{10},\mathbf{2})^*_0 & (\Kbi - \Ss + \Sn) \Z \\ \hline
z_0 = d_4 d_5 - d_2 d_7 = 0 & (\mathbf{10},\mathbf{1})_{1/2} & (2 \Kbi + \Ss - \Sn - 2 \Z) \Z \\ \hline
z_0 = d_5 = 0 & (\mathbf{16},\mathbf{1})_{1/4} & (2 \Kbi - \Ss - \Z) \Z \\ \hline
 & (\mathbf{1},\mathbf{3})_0 & 1 - \tfrac{1}{2} (\Ss - \Sn) (\Kbi - \Ss + \Sn) \\ \hline
 & (\mathbf{1},\mathbf{2})_{-3/2} & (\Kbi - \Ss + \Sn) \Ss \\ \hline
 & (\mathbf{1},\mathbf{2})_{1/2} & (\Kbi - \Ss + \Sn) (6 \Kbi + \Ss - 2 \Sn - 5 \Z) \\ \hline
 & (\mathbf{1},\mathbf{1})_{2} & (2 \Ss - \Sn - \Z) \Ss \\ \hline
 & (\mathbf{1},\mathbf{1})_{1} & 6 (\Kbi)^2 - 3 \Ss^2 + \Sn^2 + 6 \Z^2 + 5 \Sn \Z \\
 & & + \Kbi (13 \Ss - 5 \Sn - 12 \Z) - (2 \Sn + 9 \Z) \Ss \\ \hline
 & (\mathbf{1},\mathbf{1})_0 & 19 + 11 (\Kbi)^2 + 4 \Ss^2 - 3 \Ss \Sn + 2 \Sn^2 \\
 && + 5 \Sn \Z + 10 \Z^2 -  4 \Kbi (\Ss + \Sn + 5 \Z) \\ \hline
\end{array}
\end{equation}
\begin{align*}
\text{Euler number:}& \quad \chi =-24 (\Kbi)^2 + 8 \Kbi \Ss - 8 \Ss^2 + 8 \Kbi \Sn + 6 \Ss \Sn - 4 \Sn^2 \\ & \hspace{1.15cm}    + 
  40 \Kbi \Z - 10 \Sn \Z - 20 \Z^2 \\
\text{anomaly coefficients:}& \quad a \sim \Kbi \,, \enspace b \sim - \Z \,, \enspace b_{\text{SU(2)}} \sim - (\Kbi - \Ss + \Sn) \,, \\
& \quad b_{11} \sim - (\tfrac{3}{2} \Kbi + \tfrac{5}{2} \Ss - \tfrac{1}{2} \Sn - \tfrac{5}{4} \Z)
\end{align*}
}\vspace{-0.5cm}

%%%%%%%%%
\subsubsection*{Top 4}
\vspace{-0.5cm}
{\small
\begin{align*}
\text{vertices:}& \quad z_0: \, (0,0,1) \,, \enspace f_1: \, (0,1,1) \,, \enspace f_2: \, (1,0,1) \,, \enspace f_3: \, (1,1,1) \,, \enspace f_4: \, (1,2,1) \,, \\
& \quad g_1: \, (1,1,2) \,, \enspace g_2: \, (1,2,2) \\
\text{factorization:}& \quad s_1 = d_1 z_0^4 \,, \enspace s_2 = d_2 z_0^3 \,, \enspace s_3 = d_3 z_0^2 \,, \enspace s_4 = d_4 z_0 \,, \enspace s_5 = d_5 z_0 \,, \\
& \quad s_6 = d_6 z_0 \,, \enspace s_7 = d_7 \,, \enspace s_8 = d_8
\end{align*}
\begin{equation}
\begin{array}{|c|c|c|}
\hline
\text{locus} & \text{representation} & \text{multiplicity} \\ \hline \hline
z_0 = d_7 = 0 & (\mathbf{16},\mathbf{1})_{-3/4} & \Ss \Z \\ \hline 
z_0 = d_8 = 0 & (\mathbf{10},\mathbf{2})^*_0 & (\Kbi - \Ss + \Sn) \Z \\ \hline
z_0 = d_4 = 0 & (\mathbf{10},\mathbf{1})_{1/2} & (2 \Ss - \Sn - \Z) \Z \\ \hline
z_0 = d_5 = 0 & \text{SCP} & (2 \Kbi - \Ss - \Z) \Z \\ \hline
 & (\mathbf{1},\mathbf{3})_0 & 1 - \tfrac{1}{2} (\Ss - \Sn) (\Kbi - \Ss + \Sn) \\ \hline
 & (\mathbf{1},\mathbf{2})_{-3/2} & (\Kbi - \Ss + \Sn) \Ss \\ \hline
 & (\mathbf{1},\mathbf{2})_{1/2} & (\Kbi - \Ss + \Sn) (6 \Kbi + \Ss - 2 \Sn - 5 \Z) \\ \hline
 & (\mathbf{1},\mathbf{1})_{2} & (2 \Ss - \Sn - \Z) \Ss \\ \hline
 & (\mathbf{1},\mathbf{1})_{1} & 6 (\Kbi)^2 - 3 \Ss^2 + \Sn^2 + 7 \Z^2 + 5 \Sn \Z \\
 & & + \Kbi (13 \Ss - 5 \Sn - 14 \Z) - (2 \Sn + 8 \Z) \Ss \\ \hline
 & (\mathbf{1},\mathbf{1})_0 & 19 + 11 (\Kbi)^2 + 4 \Ss^2 - 3 \Ss \Sn + 2 \Sn^2 + 2 \Ss \Z \\&&+ 5 \Sn \Z +  12 \Z^2 - 4 \Kbi (\Ss + \Sn + 6 \Z) \\ \hline
\end{array}
\end{equation}
\begin{align*}
\text{Euler number:}& \quad \chi = -24 (\Kbi)^2 + 8 \Kbi \Ss - 8 \Ss^2 + 8 \Kbi \Sn + 6 \Ss \Sn - 4 \Sn^2 + 
  52 \Kbi \Z \\ & \hspace{1.15cm}    - 6 \Ss \Z - 10 \Sn \Z - 26 \Z^2\\
\text{anomaly coefficients$^*$:}& \quad \hat{a} \sim K_{\hat{B}}^{-1} \,, \enspace \hat{b} \sim - \hat{\Z} \,, \enspace \hat{b}_{\text{SU(2)}} \sim - (K_{\hat{B}}^{-1} - \hat{\mathcal{S}}_7 + \hat{\mathcal{S}}_9) \,, \\
& \quad \hat{b}_{11} \sim - (\tfrac{3}{2} K_{\hat{B}}^{-1} + \tfrac{5}{2} \hat{\mathcal{S}}_7 -\tfrac{1}{2} \hat{\mathcal{S}}_9 - \tfrac{5}{4} \hat{\Z}) \\
\text{relation to ($F_6$, top 3):}& \quad (\mathbf{16},\mathbf{1})_{1/4} \, \oplus (\mathbf{10},\mathbf{1})_{1/2} \, \oplus (\mathbf{1},\mathbf{1})_{1} \, \oplus (\mathbf{1},\mathbf{1})_{0} \, \oplus (\mathbf{1},\mathbf{1})_{0}
\end{align*}
}\vspace{-0.5cm}

%%%%%%%%%
\subsubsection*{Top 5}
\vspace{-0.5cm}
{\small
\begin{align*}
\text{vertices:}& \quad z_0: \, (0,0,1) \,, \enspace f_1: \, (0,1,1) \,, \enspace f_2: \, (0,2,1) \,, \enspace f_3: \, (1,0,1) \,, \enspace f_4: \, (1,1,1) \\
& \quad g_1: \, (1,1,2) \,, \enspace g_2: \, (1,2,2) \\
\text{factorization:}& \quad s_1 = d_1 z_0^3 \,, \enspace s_2 = d_2 z_0^2 \,, \enspace s_3 = d_3 z_0^2 \,, \enspace s_4 = d_4 z_0^2 \,, \enspace s_5 = d_5 z_0 \,, \\
& \quad s_6 = d_6 z_0 \,, \enspace s_7 = d_7 \,, \enspace s_8 = d_8
\end{align*}
\begin{equation}
\begin{array}{|c|c|c|}
\hline
\text{locus} & \text{representation} & \text{multiplicity} \\ \hline \hline
z_0 = d_5 = 0 & (\mathbf{16},\mathbf{1})_{1/4} & (2 \Kbi - \Ss - \Z) \Z \\ \hline
z_0 = d_2 = 0 & (\mathbf{10},\mathbf{1})_{1/2} & (2 \Kbi - \Sn - 2 \Z) \Z \\ \hline
z_0 = d_8 = 0 & (\mathbf{10},\mathbf{2})_0^* & (\Kbi - \Ss + \Sn) \Z \\ \hline
z_0 = d_7 = 0 & \text{SCP} & \Ss \Z \\ \hline
 & (\mathbf{1},\mathbf{3})_0 & 1 - \tfrac{1}{2} (\Ss - \Sn) (\Kbi - \Ss + \Sn) \\ \hline
 & (\mathbf{1},\mathbf{2})_{-3/2} & (\Kbi - \Ss + \Sn) \Ss \\ \hline
 & (\mathbf{1},\mathbf{2})_{1/2} & (\Kbi - \Ss + \Sn) (6 \Kbi + \Ss - 2 \Sn - 5 \Z) \\ \hline
 & (\mathbf{1},\mathbf{1})_{2} & (2 \Ss - \Sn - 2 \Z) \Ss \\ \hline
 & (\mathbf{1},\mathbf{1})_{1} & 6 (\Kbi)^2 - 3 \Ss^2 + \Sn^2 + 6 \Z^2 + 5 \Sn \Z \\
 & & + \Kbi (13 \Ss - 5 \Sn -12 \Z) - (2 \Sn + 10 \Z) \Ss \\ \hline
 & (\mathbf{1},\mathbf{1})_0 & 
19 + 11 (\Kbi)^2 + 4 \Ss^2 + 2 \Sn^2 + 10 \Z^2 + 5 \Sn \Z \\ 
&& - 4 \Kbi (\Ss + \Sn + 5 \Z) - \Ss \Z - 3 \Ss \Sn
 \\ \hline
\end{array}
\end{equation}
\begin{align*}
\text{Euler number:}& \quad \chi =-24 (\Kbi)^2 + 8 \Kbi \Ss - 8 \Ss^2 + 8 \Kbi \Sn + 6 \Ss \Sn - 4 \Sn^2 + 
  40 \Kbi \Z  \\ & \hspace{1.15cm} + 4 \Ss \Z - 10 \Sn \Z - 20 \Z^2 \\
\text{anomaly coefficients$^*$:}& \quad \hat{a} \sim K_{\hat{B}}^{-1} \,, \enspace \hat{b} \sim - \hat{\Z} \,, \enspace \hat{b}_{\text{SU(2)}} \sim - (K_{\hat{B}}^{-1} - \hat{\mathcal{S}}_7 + \hat{\mathcal{S}}_9) \,, \\
& \quad \hat{b}_{11} \sim - (\tfrac{3}{2} K_{\hat{B}}^{-1} + \tfrac{5}{2} \hat{\mathcal{S}}_7 -\tfrac{1}{2} \hat{\mathcal{S}}_9 - \tfrac{5}{4} \hat{\Z}) \\
\text{relation to ($F_6$, top 3):}& \quad (\mathbf{16},\mathbf{1})_{-3/4} \, \oplus (\mathbf{10},\mathbf{1})_{1/2} \, \oplus (\mathbf{1},\mathbf{1})_{2} \, \oplus (\mathbf{1},\mathbf{1})_{1} \, \oplus (\mathbf{1},\mathbf{1})_{0}
\end{align*}
}\vspace{-0.5cm}

%%%%%%%%%%%%%%%%%%
\subsection[Polygon \texorpdfstring{$F_7$}{F7}]{Polygon \texorpdfstring{$\boldsymbol{F_7}$}{F7}}
\vspace{-0.5cm}
{\small
\begin{align*}
\text{vertices:}& \quad u: \, (1,1,0) \,, \enspace w: \, (0,-1,0) \,, \enspace v: (-1,0,0) \\
& \quad e_1: \, (0,1,0) \,, \enspace e_2: \, (-1,-1,0) \,, \enspace e_3: \, (1,0,0) \\
\text{gauge group:}& \quad \text{SO(10)} \times \text{U(1)}^3
\end{align*}
}\vspace{-0.5cm}

%%%%%%%%%
\subsubsection*{Top 1}
\vspace{-0.5cm}
{\small
\begin{align*}
\text{vertices:}& \quad z_0: \, (0,0,1) \,, \enspace f_1: \, (1,0,1) \,, \enspace f_2: \, (1,1,1) \,, \enspace f_3: \, (2,1,1) \,, \enspace f_4: \, (0,1,1) \,, \\
& \quad g_1: \, (1,1,2) \,, \enspace g_2: \, (2,1,2) \\
\text{factorization:}& \quad s_2 = d_2 \,, \enspace s_3 = d_3 z_0 \,, \enspace s_5 = d_5 \,, \enspace s_6 = d_6 z_0 \,, \enspace s_7 = d_7 z_0 \,,\\
& \quad s_8 = d_8 z_0 \,, \enspace s_9 = d_9 z_0^2
\end{align*}
\begin{equation}
\begin{array}{|c|c|c|}
\hline
\text{locus} & \text{representation} & \text{multiplicity} \\ \hline \hline
z_0 = d_5 = 0 & \mathbf{16}_{-1/4, 1/4, 0} & (2 \Kbi - \Ss) \Z \\ \hline
z_0 = d_8 = 0 & \mathbf{10}_{1/2,1/2,1} & (\Kbi - \Ss + \Sn - \Z) \Z \\ \hline
z_0 = d_2 = 0 & \mathbf{10}_{-1/2,-1/2,0} & (2 \Kbi - \Sn) \Z \\ \hline
z_0 = d_7 = 0 & \text{SCP} & (\Ss - \Z) \Z \\ \hline
 & \mathbf{1}_{1,1,0} & (2 \Kbi - \Sn) (\Kbi + \Ss - \Sn - \Z) \\ \hline
 & \mathbf{1}_{0,-1,0} & (2 \Kbi - \Ss) (2 \Kbi - \Sn) \\ \hline
 & \mathbf{1}_{2,1,1} & (\Kbi + \Ss - \Sn - \Z) (\Ss - \Z) \\ \hline
 & \mathbf{1}_{0,1,1} & (\Kbi - \Ss + \Sn - \Z) (2 \Kbi - \Ss) \\ \hline
 & \mathbf{1}_{2,1,2} & (\Sn - 2 \Z) (\Ss - \Z) \\ \hline
 & \mathbf{1}_{1,1,2} & (\Kbi - \Ss + \Sn - \Z) (\Sn - 2 \Z) \\ \hline
 & \mathbf{1}_{1,0,0} & 4 (\Kbi)^2 - 2 \Ss^2 + 2 \Z^2 - \Sn \Z \\ 
 & & + \Kbi (2 \Ss - 2 \Sn - 4 \Z) + 2 \Ss \Sn \\ \hline
 & \mathbf{1}_{0,0,1} & 4 (\Kbi)^2 - 2 \Sn^2 + 2 \Z^2 + 2 \Sn \Z \\ 
 & & + \Kbi (-2 \Ss + 2 \Sn - 6 \Z) + 2 \Ss (\Sn - \Z) \\ \hline
 & \mathbf{1}_{1,0,1} & 2 (\Kbi)^2 + \Kbi (2 \Ss + 2 \Sn - 7 \Z) \\
 & & + \Ss (-2 \Sn + \Z) + (\Sn + \Z) \Z \\ \hline
 & \mathbf{1}_{1,1,1} & 4 (\Kbi)^2 - 2 \Ss^2 -2 \Sn^2 + 2 \Z^2 + 2 \Sn \Z \\
 & & + \Kbi (2 \Ss + 2 \Sn - 10 \Z) + 2 \Ss \Sn\\ \hline
 & \mathbf{1}_{0,0,0} & 20 + 7 (\Kbi)^2 + 2 \Ss^2 + 2 \Sn^2 - 2 \Sn \Z + 7 \Z^2 \\
  && - \Ss (2 \Sn + \Z) -  2 \Kbi (\Ss + \Sn + 4 \Z)
 \\ \hline
\end{array}
\end{equation}
\begin{align*}
\text{Euler number:}& \quad \chi =  -16 (\Kbi)^2 + 4 \Kbi \Ss - 4 \Ss^2 + 4 \Kbi \Sn + 4 \Ss \Sn - 4 \Sn^2  \\ & \hspace{1.15cm} + 
  16 \Kbi \Z + 4 \Ss \Z + 4 \Sn \Z - 16 \Z^2
 \\
\text{anomaly coefficients$^*$:}& \quad \hat{a} \sim K_{\hat{B}}^{-1} \,, \enspace \hat{b} \sim - \hat{\Z} \,, \enspace \hat{b}_{11} \sim - (2 K_{\hat{B}}^{-1} + 2 \hat{\mathcal{S}}_7 - \tfrac{13}{4} \hat{\Z}) \,, \\
& \quad \hat{b}_{12} \sim - (K_{\hat{B}}^{-1} + \hat{\mathcal{S}}_7 - \tfrac{7}{4} \hat{\Z}) \,, \enspace \hat{b}_{13} \sim - (K_{\hat{B}}^{-1} + \hat{\mathcal{S}}_7 + \hat{\mathcal{S}}_9 - 3 \hat{\Z}) \,, \\
& \quad \hat{b}_{22} \sim - (2 K_{\hat{B}}^{-1} - \tfrac{5}{4} \hat{\Z}) \,, \enspace \hat{b}_{23} \sim - (K_{\hat{B}}^{-1} + \hat{\mathcal{S}}_9 - 2 \hat{\Z}) \,, \\
& \quad \hat{b}_{33} \sim - (2 K_{\hat{B}}^{-1} + 2 \hat{\mathcal{S}}_9 - 4 \hat{\Z})
\end{align*}
}\vspace{-0.5cm}

%%%%%%%%%%%%%%%%%%
\subsection[Polygon \texorpdfstring{$F_8$}{F8}]{Polygon \texorpdfstring{$\boldsymbol{F_8}$}{F8}}
\vspace{-0.5cm}
{\small
\begin{align*}
\text{vertices:}& \quad u: \, (0,-1,0) \,, \enspace v: \, (-1,1,0) \,, \enspace w: \, (1,0,0) \,, \enspace e_1: \, (-1,0,0) \\
& \quad e_2: \, (-1,-1,0) \,, \enspace e_3: \, (1,-1,0) \\
\text{gauge group:}& \quad \text{SO(10)} \times \text{SU(2)}^2 \times \text{U(1)}
\end{align*}
}\vspace{-0.5cm}

%%%%%%%%%
\subsubsection*{Top 1}
\vspace{-0.5cm}
{\small
\begin{align*}
\text{vertices:}& \quad z_0: \, (0,0,1) \,, \enspace f_1: \, (1,0,1) \,, \enspace f_2: \, (0,1,1) \,, \enspace f_3: \, (1,1,1) \,, \enspace f_4: \, (2,0,1) \\
& \quad g_1: \, (1,1,2) \,, \enspace g_2: \, (2,1,2) \\
\text{factorization:}& \quad s_1 = d_1 z_0^3 \,, \enspace s_2 = d_2 z_0^2 \,, \enspace s_3 = d_3 z_0^2 \,, \enspace s_5 = d_5 z_0 \,, \enspace s_6 = d_6 z_0 \,, \\
& \quad s_7 = d_7 \,, \enspace s_8 = d_8 
\end{align*}
\begin{equation}
\begin{array}{|c|c|c|}
\hline
\text{locus} & \text{representation} & \text{multiplicity} \\ \hline \hline
z_0 = d_5 = 0 & (\mathbf{16},\mathbf{1},\mathbf{1})_{1/4} & (2 \Kbi - \Ss - \Z) \Z \\ \hline
z_0 = d_2 = 0 & (\mathbf{10},\mathbf{1},\mathbf{1})_{1/2} & (2 \Kbi - \Sn - 2 \Z) \Z \\ \hline
z_0 = d_8 = 0 & (\mathbf{10},\mathbf{1},\mathbf{2})_0^* & (\Kbi - \Ss + \Sn) \Z \\ \hline
z_0 = d_7 = 0 & \text{SCP} & \Ss \Z \\ \hline
 & (\mathbf{1},\mathbf{3},\mathbf{1})_0 & 1 - \tfrac{1}{2} (\Kbi - \Ss) \Ss \\ \hline
 & (\mathbf{1},\mathbf{1},\mathbf{3})_0 & 1 + \tfrac{1}{2} (\Kbi - \Ss + \Sn) (\Sn - \Ss) \\ \hline
 & (\mathbf{1},\mathbf{2},\mathbf{2})_{1/2} & (\Kbi - \Ss + \Sn) \Ss \\ \hline
 & (\mathbf{1},\mathbf{2},\mathbf{1})_1 & (\Kbi + \Ss - \Sn - 2 \Z) \Ss \\ \hline
 & (\mathbf{1},\mathbf{1},\mathbf{2})_{1/2} & (\Kbi - \Ss + \Sn) (6 \Kbi - 2 \Sn - 5 \Z) \\ \hline
 & (\mathbf{1},\mathbf{2},\mathbf{1})_0 & (5 \Kbi - \Ss - \Sn - 4 \Z) \Ss \\ \hline
 & (\mathbf{1},\mathbf{1},\mathbf{1})_1 & 6 (\Kbi)^2 - \Ss^2 + \Sn^2 + 6 \Z^2 + 5 \Sn \Z \\
 & & + \Kbi (3 \Ss - 5 \Sn - 12 \Z) - 2 \Ss \Z \\ \hline 
 & (\mathbf{1},\mathbf{1},\mathbf{1})_0 & 20 + 11 (\Kbi)^2 + 3 \Ss^2 + 2 \Sn^2 + 5 \Sn \Z + 10 \Z^2 \\ &&+  \Ss (-2 \Sn + \Z) - \Kbi (5 \Ss + 4 (\Sn + 5 \Z)) \\ \hline
\end{array}
\end{equation}
\begin{align*}
\text{Euler number:}& \quad \chi =  -24 (\Kbi)^2 + 10 \Kbi \Ss - 6 \Ss^2 + 8 \Kbi \Sn + 4 \Ss \Sn - 4 \Sn^2  \\ & \hspace{1.15cm} + 
  40 \Kbi \Z - 10 \Sn \Z - 20 \Z^2
 \\
\text{anomaly coefficients$^*$:}& \quad \hat{a} \sim K_{\hat{B}}^{-1} \,, \enspace \hat{b} \sim - \hat{\Z} \,, \enspace \hat{b}_{\text{SU(2)}_1} \sim - \hat{\mathcal{S}}_7 \,, \\
& \quad \hat{b}_{\text{SU(2)}_2} \sim -(K_{\hat{B}}^{-1} - \hat{\mathcal{S}}_7 + \hat{\mathcal{S}}_9) \,, \enspace \hat{b}_{11} \sim - (\tfrac{3}{2} K_{\hat{B}}^{-1} + \tfrac{1}{2} \hat{\mathcal{S}}_7 - \tfrac{1}{2} \hat{\mathcal{S}}_9 - \tfrac{5}{4} \hat{\Z} )
\end{align*}
}\vspace{-0.5cm}

%%%%%%%%%
\subsubsection*{Top 2}
\vspace{-0.5cm}
{\small
\begin{align*}
\text{vertices:}& \quad z_0: \, (0,0,1) \,, \enspace f_1: \, (1,0,1) \,, \enspace f_2: \, (0,1,1) \,, \enspace f_3: \, (1,1,1) \,, \enspace f_4: \, (0,2,1) \\
& \quad g_1: \, (1,1,2) \,, \enspace g_2: \, (1,2,2) \\
\text{factorization:}& \quad s_1 = d_1 z_0^4 \,, \enspace s_2 = d_2 z_0^2 \,, \enspace s_3 = d_3 z_0 \,, \enspace s_5 = d_5 z_0^2 \,, \enspace s_6 = d_6 z_0 \,, \\
& \quad s_7 = d_7 \,, \enspace s_8 = d_8 
\end{align*}
\begin{equation}
\begin{array}{|c|c|c|}
\hline
\text{locus} & \text{representation} & \text{multiplicity} \\ \hline \hline
z_0 = d_3 = 0 & (\mathbf{16},\mathbf{1},\mathbf{1})_{-1/2} & (\Kbi + \Ss - \Sn - \Z) \Z \\ \hline
z_0 = d_2 = 0 & (\mathbf{10},\mathbf{1},\mathbf{1})_{0} & (2 \Kbi - \Sn - 2 \Z) \Z \\ \hline
z_0 = d_7 = 0 & (\mathbf{10},\mathbf{2},\mathbf{1})_0^* & \Ss \Z \\ \hline
z_0 = d_8 = 0 & \text{SCP} & (\Kbi - \Ss + \Sn) \Z \\ \hline
 & (\mathbf{1},\mathbf{3},\mathbf{1})_0 & 1 - \tfrac{1}{2} (\Kbi - \Ss) \Ss \\ \hline
 & (\mathbf{1},\mathbf{1},\mathbf{3})_0 & 1 + \tfrac{1}{2} (\Kbi - \Ss + \Sn) (\Sn - \Ss) \\ \hline
 & (\mathbf{1},\mathbf{2},\mathbf{2})_{1/2} & (\Kbi - \Ss + \Sn) \Ss \\ \hline
 & (\mathbf{1},\mathbf{2},\mathbf{1})_1 & (\Kbi + \Ss - \Sn - \Z) \Ss \\ \hline
 & (\mathbf{1},\mathbf{1},\mathbf{2})_{1/2} & (\Kbi - \Ss + \Sn) (6 \Kbi - 2 \Sn - 6 \Z) \\ \hline
 & (\mathbf{1},\mathbf{2},\mathbf{1})_0 & (5 \Kbi - \Ss - \Sn - 4 \Z) \Ss \\ \hline
 & (\mathbf{1},\mathbf{1},\mathbf{1})_1 & 6 (\Kbi)^2 - \Ss^2 + \Sn^2 + 4 \Z^2 + 4 \Sn \Z \\
 & & + \Kbi (3 \Ss - 5 \Sn - 10 \Z) - 2 \Ss \Z \\ \hline 
 & (\mathbf{1},\mathbf{1},\mathbf{1})_0 & 20 + 11 (\Kbi)^2 + 3 \Ss^2 - 2 \Ss \Sn + 2 \Sn^2 + 3 \Ss \Z  \\
 &&+ 5 \Sn \Z +  12 \Z^2 - \Kbi (5 \Ss + 4 \Sn + 23 \Z) \\ \hline
\end{array}
\end{equation}
\begin{align*}
\text{Euler number:}& \quad \chi = -24 (\Kbi)^2 + 10 \Kbi \Ss - 6 \Ss^2 + 8 \Kbi \Sn + 4 \Ss \Sn - 4 \Sn^2 + 
  48 \Kbi \Z  \\ & \hspace{1.15cm}  - 8 \Ss \Z - 8 \Sn \Z - 24 \Z^2 \\
\text{anomaly coefficients$^*$:}& \quad \hat{a} \sim K_{\hat{B}}^{-1} \,, \enspace \hat{b} \sim - \hat{\Z} \,, \enspace \hat{b}_{\text{SU(2)}_1} \sim - \hat{\mathcal{S}}_7 \,, \\
& \quad \hat{b}_{\text{SU(2)}_2} \sim -(K_{\hat{B}}^{-1} - \hat{\mathcal{S}}_7 + \hat{\mathcal{S}}_9) \,, \enspace \hat{b}_{11} \sim - (\tfrac{3}{2} K_{\hat{B}}^{-1} + \tfrac{1}{2} \hat{\mathcal{S}}_7 - \tfrac{1}{2} \hat{\mathcal{S}}_9 - \hat{\Z} )
\end{align*}
}\vspace{-0.5cm}

%%%%%%%%%%%%%%%%%%
\subsection[Polygon \texorpdfstring{$F_9$}{F9}]{Polygon \texorpdfstring{$\boldsymbol{F_9}$}{F9}}
\vspace{-0.5cm}
{\small
\begin{align*}
\text{vertices:}& \quad u: \, (-1,1,0) \,, \enspace w: \, (1,0,0) \,, \enspace v: \, (0,-1,0) \\
& \quad e_1: \, (0,1,0) \,, \enspace e_2: \, (-1,0,0) \,, \enspace e_3: \, (-1,-1,0) \\
\text{gauge group:}& \quad \text{SO(10)} \times \text{SU(2)} \times \text{U(1)}^2
\end{align*}
}\vspace{-0.5cm}

%%%%%%%%%
\subsubsection*{Top 1}
\vspace{-0.5cm}
{\small
\begin{align*}
\text{vertices:}& \quad z_0: \, (0,0,1) \,, \enspace f_1: \, (0,1,1) \,, \enspace f_2: \, (1,0,1) \,, \enspace f_3: \, (1,1,1) \\
& \quad g_1: \, (1,1,2) \,, \enspace g_2: \, (2,1,2) \\
\text{factorization:}& \quad s_1 = d_1 z_0 \,, \enspace s_2 = d_2 z_0^2 \,, \enspace s_3 = d_3 z_0^2 \,, \enspace s_5 = d_5 \,, \enspace s_6 = d_6 z_0 \\
& \quad s_7 = d_7 z_0 \,, \enspace s_9 = d_9
\end{align*}
\begin{equation}
\begin{array}{|c|c|c|}
\hline
\text{locus} & \text{representation} & \text{multiplicity} \\ \hline \hline
z_0 = d_5 = 0 & (\mathbf{16},\mathbf{1})_{1/4,1/2} & (2 \Kbi - \Ss) \Z \\ \hline
z_0 = d_7 = 0 & (\mathbf{16},\mathbf{1})_{-3/4,-1/2} & (\Ss - \Z) \Z \\ \hline
z_0 = d_9 = 0 & (\mathbf{10},\mathbf{2})_{0,0}^* & \Sn \Z \\ \hline
z_0 = d_3 d_5 - d_1 d_7 = 0 & (\mathbf{10},\mathbf{1})_{1/2,0} & (3 \Kbi - \Sn - 2 \Z) \Z \\ \hline
 & (\mathbf{1},\mathbf{3})_{0,0} & 1 - \tfrac{1}{2} (\Kbi - \Sn) \Sn \\ \hline
 & (\mathbf{1},\mathbf{2})_{3/2,1} & (\Ss - \Z) \Sn \\ \hline
 & (\mathbf{1},\mathbf{2})_{-1/2,-1} & (2 \Kbi - \Ss) \Sn \\ \hline
 & (\mathbf{1},\mathbf{2})_{-1/2,0} & (6 \Kbi - 2 \Sn - 4\Z) \Sn \\ \hline
 & (\mathbf{1},\mathbf{1})_{2,1} & (\Kbi + \Ss - \Sn - 2 \Z) (\Ss - \Z) \\ \hline
 & (\mathbf{1},\mathbf{1})_{0,1} & (3 \Kbi - \Ss - \Sn - \Z) (2 \Kbi - \Ss) \\ \hline
 & (\mathbf{1},\mathbf{1})_{1,0} & 6 (\Kbi)^2 - 2 \Ss^2 + \Kbi (4 \Ss - 5 \Sn - 14 \Z) \\ 
 & & + 2 \Ss \Z + (\Sn + 2 \Z)^2 \\ \hline
 & (\mathbf{1},\mathbf{1})_{1,1} & 6 (\Kbi)^2 - 2 \Ss^2 + \Kbi (4 \Ss - 2 \Sn - 11 \Z) \\
 & & + 2 \Ss \Z + (\Sn + 2 \Z) \Z \\ \hline
 & (\mathbf{1},\mathbf{1})_{0,0} & 20 + 11 (\Kbi)^2 - 2 \Kbi (2 \Ss + 3 \Sn + 7 \Z)\\
 && + 2 (\Ss^2 + \Sn^2 - \Ss \Z + 2 \Sn \Z + 4 \Z^2) \\ \hline
\end{array}
\end{equation}
\begin{align*}
\text{Euler number:}& \quad \chi = -24 (\Kbi)^2 + 8 \Kbi \Ss - 4 \Ss^2 + 12 \Kbi \Sn - 4 \Sn^2 + 28 \Kbi \Z \\ & \hspace{1.15cm}  + 
 4 \Ss \Z - 8 \Sn \Z - 16 \Z^2 \\
\text{anomaly coefficients:}& \quad a \sim \Kbi \,, \enspace b \sim -\Z \,, \enspace b_{\text{SU(2)}} \sim -\Sn \,, \enspace b_{22} \sim -(2 \Kbi - \Z) \,, \\
& \quad b_{12} \sim -(\Kbi + \Ss - \tfrac{3}{2} \Z) \,, \enspace b_{11} \sim - (2 \Kbi + 2 \Ss - \tfrac{1}{2} \Sn - \tfrac{13}{4} \Z)
\end{align*}
}\vspace{-0.5cm}

%%%%%%%%%
\subsubsection*{Top 2}
\vspace{-0.5cm}
{\small
\begin{align*}
\text{vertices:}& \quad z_0: \, (0,0,1) \,, \enspace f_1: \, (0,1,1) \,, \enspace f_2: \, (1,0,1) \,, \enspace f_3: \, (1,1,1) \,, \enspace f_4: \, (2,0,1) \,, \\
& \quad g_1: \, (1,1,2) \,, \enspace g_2: \, (2,1,2) \\
\text{factorization:}& \quad s_1 = d_1 z_0^2 \,, \enspace s_2 = d_2 z_0^2 \,, \enspace s_3 = d_3 z_0^2 \,, \enspace s_5 = d_5 \,, \enspace s_6 = d_6 z_0 \\
& \quad s_7 = d_7 z_0 \,, \enspace s_9 = d_9
\end{align*}
\begin{equation}
\begin{array}{|c|c|c|}
\hline
\text{locus} & \text{representation} & \text{multiplicity} \\ \hline \hline
z_0 = d_7 = 0 & (\mathbf{16},\mathbf{1})_{-3/4,-1/2} & (\Ss - \Z) \Z \\ \hline
z_0 = d_9 = 0 & (\mathbf{10},\mathbf{2})_{0,0}^* & \Sn \Z \\ \hline
z_0 = d_3 = 0 & (\mathbf{10},\mathbf{1})_{1/2,0} & (\Kbi + \Ss - \Sn - 2 \Z) \Z  \\ \hline
z_0 = d_5 = 0 & \text{SCP} & (2 \Kbi - \Ss) \Z \\ \hline
 & (\mathbf{1},\mathbf{3})_{0,0} & 1 - \tfrac{1}{2} (\Kbi - \Sn) \Sn \\ \hline
 & (\mathbf{1},\mathbf{2})_{3/2,1} & (\Ss - \Z) \Sn \\ \hline
 & (\mathbf{1},\mathbf{2})_{-1/2,-1} & (2 \Kbi - \Ss) \Sn \\ \hline
 & (\mathbf{1},\mathbf{2})_{-1/2,0} & (6 \Kbi - 2 \Sn - 4 \Z) \Sn \\ \hline
 & (\mathbf{1},\mathbf{1})_{2,1} & (\Kbi + \Ss - \Sn - 2 \Z) (\Ss - \Z) \\ \hline
 & (\mathbf{1},\mathbf{1})_{0,1} & (3 \Kbi - \Ss - \Sn - 2 \Z) (2 \Kbi - \Ss) \\ \hline
 & (\mathbf{1},\mathbf{1})_{1,0} & 6 (\Kbi)^2 - 2 \Ss^2 + \Kbi (4 \Ss - 5 \Sn - 14 \Z) \\
 & & + 2 \Ss \Z + (\Sn + 2 \Z)^2 \\ \hline
 & (\mathbf{1},\mathbf{1})_{1,1} & 6 (\Kbi)^2 - 2 \Ss^2 + \Kbi (4 \Ss - 2 \Sn - 13 \Z) \\
 & & + 3 \Ss \Z + (\Sn + 2 \Z) \Z \\ \hline
 & (\mathbf{1},\mathbf{1})_{0,0} & 20 + 11 (\Kbi)^2 + 2 \Ss^2 + 2 \Sn^2 - \Ss \Z \\
 && + 4 \Sn \Z + 8 \Z^2 -  2 \Kbi (2 \Ss + 3 \Sn + 8 \Z)\\ \hline
\end{array}
\end{equation}
\begin{align*}
\text{Euler number:}& \quad \chi = -24 (\Kbi)^2 + 8 \Kbi \Ss - 4 \Ss^2 + 12 \Kbi \Sn - 4 \Sn^2 \\ & \hspace{1.15cm} + 
  36 \Kbi \Z - 8 \Sn \Z - 16 \Z^2 \\
\text{anomaly coefficients$^*$:}& \quad \hat{a} \sim K_{\hat{B}}^{-1} \,, \enspace \hat{b} \sim - \hat{\Z} \,, \enspace \hat{b}_{\text{SU(2)}} \sim - \hat{\mathcal{S}}_9 \,, \enspace \hat{b}_{22} \sim - (2 K_{\hat{B}}^{-1} - \hat{\Z}) \,, \\
& \quad \hat{b}_{12} \sim - (K_{\hat{B}}^{-1} + \hat{\mathcal{S}}_7 - \tfrac{3}{2} \hat{\Z}) \,, \enspace \hat{b}_{11} \sim - (2 K_{\hat{B}}^{-1} + 2 \hat{\mathcal{S}}_7 - \tfrac{1}{2} \hat{\mathcal{S}}_9 - \tfrac{13}{4} \hat{\Z}) \\
\text{relation to ($F_9$, top 1):}& \quad (\mathbf{16},\mathbf{1})_{1/4,1/2} \, \oplus (\mathbf{10},\mathbf{1})_{1/2,0} \, \oplus (\mathbf{1},\mathbf{1})_{0,1} \, \oplus (\mathbf{1},\mathbf{1})_{1,1} \, \oplus (\mathbf{1},\mathbf{1})_{0,0}
\end{align*}
}\vspace{-0.5cm}

%%%%%%%%%
\subsubsection*{Top 3}
\vspace{-0.5cm}
{\small
\begin{align*}
\text{vertices:}& \quad z_0: \, (0,0,1) \,, \enspace f_1: \, (0,1,1) \,, \enspace f_2: \, (0,2,1) \,, \enspace f_3: \, (1,0,1) \,, \enspace f_4: \, (1,1,1) \\
& \quad g_1: \, (1,1,2) \,, \enspace g_2: \, (1,2,2) \\ 
\text{factorization:}& \quad s_1 = d_1 z_0 \,, \enspace s_2 = d_2 z_0 \,, \enspace s_3 = d_3 z_0^2 \,, \enspace s_5 = d_5 \,, \enspace s_6 = d_6 z_0 \,, \\
& \quad s_7 = d_7 z_0^2 \,, \enspace s_9 = d_9
\end{align*}
\begin{equation}
\begin{array}{|c|c|c|}
\hline
\text{locus} & \text{representation} & \text{multiplicity} \\ \hline \hline
z_0 = d_2 = 0 & (\mathbf{16},\mathbf{1})_{0,-1/4} & (2 \Kbi - \Sn - \Z) \Z \\ \hline
z_0 = d_3 = 0 & (\mathbf{10},\mathbf{1})_{1,1/2} & (\Kbi + \Ss - \Sn - 2 \Z) \Z \\ \hline
z_0 = d_5 = 0 & (\mathbf{10},\mathbf{1})_{0,1/2} & (2 \Kbi - \Ss) \Z \\ \hline
z_0 = d_9 = 0 & \text{SCP} & \Sn \Z \\ \hline
 & (\mathbf{1},\mathbf{3})_{0,0} & 1 - \tfrac{1}{2} (\Kbi - \Sn) \Sn \\ \hline
 & (\mathbf{1},\mathbf{2})_{3/2,1} & (\Ss - 2 \Z) \Sn \\ \hline
 & (\mathbf{1},\mathbf{2})_{-1/2,-1} & (2 \Kbi - \Ss) \Sn \\ \hline
 & (\mathbf{1},\mathbf{2})_{-1/2,0} & (6 \Kbi - 2 \Sn - 4 \Z) \Sn \\ \hline
 & (\mathbf{1},\mathbf{1})_{2,1} & (\Kbi + \Ss - \Sn - 2 \Z) (\Ss - 2 \Z) \\ \hline
 & (\mathbf{1},\mathbf{1})_{0,1} & (2 \Kbi - \Ss) (3 \Kbi - \Ss - \Sn - \Z) \\ \hline
 & (\mathbf{1},\mathbf{1})_{1,0} & 6 (\Kbi)^2 - 2 \Ss^2 + \Sn^2 + \Kbi (4 \Ss - 5 \Sn - 13 \Z) \\
 & & + 3 \Ss \Z + 3 \Sn \Z + 2 \Z^2 \\ \hline
 & (\mathbf{1},\mathbf{1})_{1,1} & 6 (\Kbi)^2 - 2 \Ss^2 + \Kbi (4 \Ss - 2 \Sn - 13 \Z) \\ 
 & & + 3 \Ss \Z + (\Sn + 2 \Z) \Z \\ \hline
 & (\mathbf{1},\mathbf{1})_{0,0} & 20 + 11 (\Kbi)^2 + 2 \Ss^2 + 2 \Sn^2 - 3 \Ss \Z \\
 && + 3 \Sn \Z + 8 \Z^2 -  2 \Kbi (2 \Ss + 3 \Sn + 6 \Z) \\ \hline
\end{array}
\end{equation}
\begin{align*}
\text{Euler number:}& \quad \chi =  -24 (\Kbi)^2 + 8 \Kbi \Ss - 4 \Ss^2 + 12 \Kbi \Sn - 4 \Sn^2 + 
   24 \Kbi \Z \\ & \hspace{1.15cm}  + 6 \Ss \Z - 4 \Sn \Z - 16 \Z^2 \\
\text{anomaly coefficients$^*$:}& \quad \hat{a} \sim K_{\hat{B}}^{-1} \,, \enspace \hat{b} \sim - \hat{\Z} \,, \enspace \hat{b}_{\text{SU(2)}} \sim - \hat{\mathcal{S}}_9 \,, \enspace \hat{b}_{22} \sim - (2 K_{\hat{B}}^{-1} - \tfrac{5}{4} \hat{\Z}) \,, \\
& \quad \hat{b}_{12} \sim - (K_{\hat{B}}^{-1} + \hat{\mathcal{S}}_7 - 2 \hat{\Z}) \,, \enspace \hat{b}_{11} \sim - (2 K_{\hat{B}}^{-1} + 2 \hat{\mathcal{S}}_7 - \tfrac{1}{2} \hat{\mathcal{S}}_9 -  4 \hat{\Z})
\end{align*}
}\vspace{-0.5cm}

%%%%%%%%%%%%%%%%%%
\subsection[Polygon \texorpdfstring{$F_{10}$}{F10}]{Polygon \texorpdfstring{$\boldsymbol{F_{10}}$}{F10}}
\vspace{-0.5cm}
{\small
\begin{align*}
\text{vertices:}& \quad u: \, (-1,-1,0) \,, \enspace v: \, (0,1,0) \,, \enspace w: \, (1,0,0) \,, \\
& \quad e_1: \, (-1,0,0) \,, \enspace e_2: \, (-2,-1,0) \,, \enspace e_3: \, (-3,-2,0) \\
\text{gauge group:}& \quad \text{SO(10)} \times \text{SU(3)} \times \text{SU(2)}
\end{align*}
}\vspace{-0.5cm}

%%%%%%%%%
\subsubsection*{Top 1}
\vspace{-0.5cm}
{\small
\begin{align*}
\text{vertices:}& \quad z_0: \, (0,0,1) \,, \enspace f_1: \, (1,1,1) \,, \enspace f_2: \, (2,2,1) \,, \enspace f_3: \, (2,1,1) \,, \enspace f_4: \, (3,2,1) \,, \\
& \quad g_1: \, (3,2,2) \,, \enspace g_2: \, (4,3,2) \\
\text{factorization:} & \quad s_1 = d_1 z_0^5 \, \enspace s_2 = d_2 z_0^3 \,, \enspace s_3 = d_3 z_0 \,, \enspace s_4 = d_4 \,, \enspace s_5 = d_5 z_0^2 \,, \\
& \quad s_6 = d_6 z_0 \,, \enspace s_8 = d_8 
\end{align*}
\begin{equation}
\begin{array}{|c|c|c|}
\hline
\text{locus} & \text{representation} & \text{multiplicity} \\ \hline \hline
z_0 = d_5 = 0 & (\mathbf{10},\mathbf{1},\mathbf{1}) & (2 \Kbi - \Ss - 2 \Z) \Z \\ \hline
z_0 = d_3 = 0 & (\mathbf{16},\mathbf{1},\mathbf{1}) & (\Kbi + \Ss - \Sn - \Z) \Z \\ \hline
z_0 = d_4 = 0 & (\mathbf{10},\mathbf{1},\mathbf{2})^* & (2 \Ss - \Sn) \Z \\ \hline
z_0 = d_8 = 0 & \text{SCP} & (\Kbi - \Ss + \Sn) \Z \\ \hline
 & (\mathbf{1},\mathbf{8},\mathbf{1}) & 1 - \tfrac{1}{2} (\Ss - \Sn) (\Kbi - \Ss + \Sn) \\ \hline
 & (\mathbf{1},\mathbf{1},\mathbf{3}) & 1 - \tfrac{1}{2} (\Kbi - 2 \Ss + \Sn) (2 \Ss - \Sn) \\ \hline
 & (\mathbf{1},\mathbf{3},\mathbf{2}) & (2 \Ss - \Sn) (\Kbi - \Ss + \Sn) \\ \hline
 & (\mathbf{1},\mathbf{3},\mathbf{1}) & (\Kbi - \Ss + \Sn) (6 \Kbi - \Ss - \Sn - 6 \Z) \\ \hline
 & (\mathbf{1},\mathbf{1},\mathbf{2}) & (2 \Ss - \Sn) (5 \Kbi - \Ss - \Sn - 5 \Z) \\ \hline
 & (\mathbf{1},\mathbf{1},\mathbf{1}) & 20 + 11 (\Kbi)^2 + 6 \Ss^2 - 6 \Ss \Sn + 3 \Sn^2 + 5 \Ss \Z \\ &&+ 5 \Sn \Z +  16 \Z^2 - 3 \Kbi (2 \Ss + \Sn + 9 \Z) \\ \hline
\end{array}
\end{equation}
\begin{align*}
\text{Euler number:}& \quad \chi = - 24 (\Kbi)^2 - 12 \Ss^2 - 6 \Sn^2 - 32 \Z^2 - 12 \Ss \Z - 8 \Sn \Z \\
& \hspace{1.15cm} + \Kbi (12 \Ss + 6 \Sn + 56 \Z) + 12 \Ss \Sn \\
\text{anomaly coefficients$^*$:}& \quad \hat{a} \sim K_{\hat{B}}^{-1} \,, \enspace \hat{b} \sim - \hat{Z} \,, \enspace \hat{b}_{\text{SU(3)}} \sim - (K_{\hat{B}}^{-1} - \hat{\mathcal{S}}_7 + \hat{\mathcal{S}}_9) \,, \\
& \quad \hat{b}_{\text{SU(2)}} \sim - (2 \hat{\mathcal{S}}_7 - \hat{\mathcal{S}}_9)
\end{align*}
}\vspace{-0.5cm}

%%%%%%%%%%%%%%%%%%
\subsection[Polygon \texorpdfstring{$F_{11}$}{F11}]{Polygon \texorpdfstring{$\boldsymbol{F_{11}}$}{F11}}

\vspace{-0.5cm}
{\small
\begin{align*}
\text{vertices:}& \quad u: \, (-1,-1,0) \,, \enspace v: \, (1,0,0) \,, \enspace w: \, (0,1,0) \,, \\
& \quad e_1: \, (-1,0,0) \,, \enspace e_2: \, (0,-1,0) \,, \enspace e_3: \, (1,-1,0) \,, \enspace e_4: \, (-2,-1,0)  \\
\text{gauge group:}& \quad \text{SO(10)} \times \text{SU(3)} \times \text{SU(2)} \times \text{U(1)}
\end{align*}
}\vspace{-0.5cm}

%%%%%%%%%
\subsubsection*{Top 1}
\vspace{-0.5cm}
{\small
\begin{align*}
\text{vertices:}& \quad z_0: \, (0,0,1) \,, \enspace f_1: \, (1,0,1) \,, \enspace f_2: \, (1,1,1) \,, \enspace f_3: \, (2,0,1) \,, \enspace f_4: \, (2,1,1) \\
& \quad g_1: (2,1,2) \,, \enspace g_2: \, (3,1,2) \\  
\text{factorization:}& \quad s_1 = d_1 z_0^3 \,, \enspace s_2 = d_2 z_0 \,, \enspace s_3 = d_3 \,, \enspace s_5 = d_5 z_0^2 \,, \enspace s_6 = d_6 z_0 \,, \enspace s_9 = d_9
\end{align*}
\begin{equation}
\begin{array}{|c|c|c|}
\hline
\text{locus} & \text{representation} & \text{multiplicity} \\ \hline \hline
z_0 = d_5 = 0 & (\mathbf{10},\mathbf{1},\mathbf{1})_{-1/2} & (2 \Kbi - \Ss - 2 \Z) \Z \\ \hline
z_0 = d_2 = 0 & (\mathbf{16},\mathbf{1},\mathbf{1})_{1/4} & (2 \Kbi - \Sn - \Z) \Z \\ \hline
z_0 = d_3 = 0 & (\mathbf{10},\mathbf{1},\mathbf{2})_0^* & (\Kbi + \Ss - \Sn) \Z \\ \hline
z_0 = d_9 = 0 & \text{SCP} & \Sn \Z \\ \hline
 & (\mathbf{1},\mathbf{8},\mathbf{1})_0 & 1 - \tfrac{1}{2} (\Kbi - \Sn) \Sn \\ \hline 
 & (\mathbf{1},\mathbf{1},\mathbf{3})_0 & 1 - \tfrac{1}{2} (\Sn - \Ss) (\Kbi + \Ss - \Sn) \\ \hline
 & (\mathbf{1},\mathbf{3},\mathbf{2})_{-1/6} & (\Kbi + \Ss - \Sn) \Sn \\ \hline
 & (\mathbf{1},\mathbf{3},\mathbf{1})_{-2/3} & (2 \Kbi - \Ss - 2 \Z) \Sn \\ \hline
 & (\mathbf{1},\mathbf{3},\mathbf{1})_{1/3} & (5 \Kbi - \Ss - \Sn - 4 \Z) \Sn \\ \hline
 & (\mathbf{1},\mathbf{1},\mathbf{2})_{1/2} & (\Kbi + \Ss - \Sn) (6 \Kbi - 2 \Ss - \Sn - 5 \Z) \\ \hline
 & (\mathbf{1},\mathbf{1},\mathbf{1})_{-1} & (3 \Kbi - \Ss - \Sn - 3 \Z) (2 \Kbi - \Ss - 2 \Z) \\ \hline
 & (\mathbf{1},\mathbf{1},\mathbf{1})_{0} & 21 + 11 (\Kbi)^2 + 2 \Ss^2 - \Ss \Sn + 3 \Sn^2 + 5 \Ss \Z \\ &&+ 3 \Sn \Z +  10 \Z^2 - \Kbi (4 \Ss + 7 \Sn + 20 \Z) \\ \hline
\end{array}
\end{equation}
\begin{align*}
\text{Euler number:}& \quad \chi = - 24 (\Kbi)^2 - 4 \Ss^2 - 6 \Sn^2 - 20 \Z^2 - 4 \Sn \Z - 10 \Ss \Z \\
& \hspace{1.15cm} + \Kbi (8 \Ss + 14 \Sn + 40 \Z) + 2 \Ss \Sn \\
\text{anomaly coefficients$^*$:}& \quad \hat{a} \sim K_{\hat{B}}^{-1} \,, \enspace \hat{b} \sim - \hat{\Z} \,, \enspace \hat{b}_{\text{SU(2)}} \sim - (K_{\hat{B}}^{-1} + \hat{\mathcal{S}}_7 - \hat{\mathcal{S}}_9) \,, \\
& \quad \hat{b}_{\text{SU(3)}} \sim - \hat{\mathcal{S}}_9 \,, \enspace \hat{b}_{11} \sim - (\tfrac{3}{2} K_{\hat{B}}^{-1} - \tfrac{1}{2} \hat{\mathcal{S}}_7 - \tfrac{1}{6} \hat{\mathcal{S}}_9 - \tfrac{5}{4} \hat{\Z})
\end{align*}
}\vspace{-0.5cm}

%%%%%%%%%%%%%%%%%%
\subsection[Polygon \texorpdfstring{$F_{12}$}{F12}]{Polygon \texorpdfstring{$\boldsymbol{F_{12}}$}{F12}}

\vspace{-0.5cm}
{\small
\begin{align*}
\text{vertices:}& \quad u: \, (1,-1,0) \,, \enspace v: \, (0,1,0) \,, \enspace w: \, (-1,0,0) \,, \enspace e_1: \, (0,-1,0) \\
& \quad e_2: \, (1,0,0) \,, \enspace e_3: \, (1,1,0) \,, \enspace e_4: \, (-1,-1,0) \\
\text{gauge group:}& \quad \text{SO(10)} \times \text{SU(2)}^2 \times \text{U(1)}^2
\end{align*}
}\vspace{-0.5cm}

%%%%%%%%%
\subsubsection*{Top 1}
\vspace{-0.5cm}
{\small
\begin{align*}
\text{vertices:}& \quad z_0: \, (0,0,1) \,, \enspace f_1: \, (1,0,1) \,, \enspace f_2: \, (1,1,1) \,, \enspace f_3: \, (2,0,1) \,, \enspace f_4: \, (2,1,1) \\
& \quad g_1: (2,1,2) \,, \enspace g_2: (3,1,2) \\
\text{factorization:}& \quad s_1 = d_1 \,, \enspace s_2 = d_2 \,, \enspace s_5 = d_5 z_0 \,, \enspace s_6 = d_6 z_0 \,, \enspace s_7 = d_7 \,, \enspace s_9 = d_9 z_0^2
\end{align*}
\begin{equation}
\begin{array}{|c|c|c|}
\hline
\text{locus} & \text{representation} & \text{multiplicity} \\ \hline \hline
z_0 = d_5 = 0 & (\mathbf{16},\mathbf{1},\mathbf{1})_{-1/2,-1/4} & (2 \Kbi - \Ss - \Z) \Z \\ \hline
z_0 = d_1 = 0 & (\mathbf{10},\mathbf{1},\mathbf{1})_{0,1/2} & (3 \Kbi - \Ss - \Sn) \Z \\ \hline
z_0 = d_9 = 0 & (\mathbf{10},\mathbf{1},\mathbf{2})_0^* & (\Sn - 2 \Z) \Z \\ \hline
z_0 = d_7 = 0 & \text{SCP} & \Ss \Z \\ \hline
 & (\mathbf{1},\mathbf{3},\mathbf{1})_0 & 1 - \tfrac{1}{2} (\Kbi - \Ss) \Ss \\ \hline
 & (\mathbf{1},\mathbf{1},\mathbf{3})_0 & 1 - \tfrac{1}{2} (\Kbi - \Sn + 2 \Z) (\Sn - 2 \Z) \\ \hline
 & (\mathbf{1},\mathbf{2},\mathbf{2})_{1/2,1/2} & (\Sn - 2 \Z) \Ss \\ \hline
 & (\mathbf{1},\mathbf{1},\mathbf{2})_{-1,-1/2} & (2 \Kbi - \Ss - \Z) (\Sn - 2 \Z) \\ \hline
 & (\mathbf{1},\mathbf{2},\mathbf{1})_{-1/2,-1} & (2 \Kbi - \Sn) \Ss \\ \hline
 & (\mathbf{1},\mathbf{1},\mathbf{2})_{0,-1/2} & (6 \Kbi - \Ss - 2 \Sn) (\Sn - 2 \Z) \\ \hline
 & (\mathbf{1},\mathbf{2},\mathbf{1})_{-1/2,0} & (6 \Kbi - 2 \Ss - \Sn - 2 \Z) \Ss \\ \hline
 & (\mathbf{1},\mathbf{1},\mathbf{1})_{1,0} & (3 \Kbi - \Ss - \Sn) (2 \Kbi - \Ss - \Z) \\ \hline
 & (\mathbf{1},\mathbf{1},\mathbf{1})_{0,1} & (2 \Kbi - \Sn) (3 \Kbi - \Ss -\Sn) \\ \hline
 & (\mathbf{1},\mathbf{1},\mathbf{1})_{1,1} & 6 (\Kbi)^2 + \Kbi (-2 \Ss - 2 \Sn - 3 \Z) \\
 & & + \Sn \Z + \Ss (\Sn - \Z) \\ \hline
 & (\mathbf{1},\mathbf{1},\mathbf{1})_{0,0} & 21 + 11 (\Kbi)^2 + 2 \Ss^2 + 2 \Sn^2 - 4 \Sn \Z \\ && + 8 \Z^2 + \Ss (\Sn + \Z) -  6 \Kbi (\Ss + \Sn + \Z) \\ \hline
\end{array}
\end{equation}
\begin{align*}
\text{Euler number:}& \quad \chi =-24 (\Kbi)^2 + 12 \Kbi \Ss - 4 \Ss^2 + 12 \Kbi \Sn - 2 \Ss \Sn - 4 \Sn^2 \\ & \hspace{1.15cm}  + 
 12 \Kbi \Z + 8 \Sn \Z - 16 \Z^2 \\  
\text{anomaly coefficients$^*$:}& \quad \hat{a} \sim K_{\hat{B}}^{-1} \,, \enspace \hat{b} \sim - \hat{\Z} \,, \enspace \hat{b}_{\text{SU(2)}_1} \sim - \hat{\mathcal{S}}_7 \,, \enspace \hat{b}_{\text{SU(2)}_2} \sim -(\hat{\mathcal{S}}_9 - 2 \hat{\Z}) \,, \\
& \quad \hat{b}_{11} \sim - (2 K_{\hat{B}}^{-1} - \tfrac{1}{2} \hat{\mathcal{S}}_7 - \hat{\Z}) \,, \enspace \hat{b}_{12} \sim - (K_{\hat{B}}^{-1} - \tfrac{1}{2} \hat{\Z}) \,, \\
& \quad \hat{b}_{22} \sim - (2 K_{\hat{B}}^{-1} - \tfrac{1}{2} \hat{\mathcal{S}}_9 - \tfrac{1}{4} \hat{\Z})
\end{align*}
}\vspace{-0.5cm}

%%%%%%%%%%%%%%%%%%
\subsection[Polygon \texorpdfstring{$F_{14}$}{F14}]{Polygon \texorpdfstring{$\boldsymbol{F_{14}}$}{F14}}

\vspace{-0.5cm}
{\small
\begin{align*}
\text{vertices:}& \quad u: \, (1,-1,0) \,, \enspace v: \, (0,1,0) \,, \enspace w: \, (-1,0,0) \,, \enspace e_1: \, (0,-1,0) \,, \\
& \quad e_2: \, (1,0,0) \,, \enspace e_3: \, (1,1,0) \,, \enspace e_4: \, (-1,-1,0) \,, \enspace e_5: \, (-2,-1,0) \\
\text{gauge group:}& \quad \text{SO(10)} \times \text{SU(3)} \times \text{SU(2)}^2 \times \text{U(1)}
\end{align*}
}\vspace{-0.5cm}

%%%%%%%%%
\subsubsection*{Top 1}
\vspace{-0.5cm}
{\small
\begin{align*}
\text{vertices:}& \quad z_0: \, (0,0,1) \,, \enspace f_1: \, (1,0,1) \,, \enspace f_2: \, (1,1,1) \,, \enspace f_3: \, (2,0,1) \,, \enspace f_4: \, (2,1,1) \\ 
& \quad g_1: \, (2,1,2) \,, \enspace g_2: \, (3,1,2) \\
\text{factorization:} & \quad s_1 = d_1 \,, \enspace s_5 = d_5 z_0 \,, \enspace s_6 = d_6 z_0 \,, \enspace s_7 = d_7 \,, \enspace s_9 = d_9 z_0^2
\end{align*}
\begin{equation}
\begin{array}{|c|c|c|}
\hline
\text{locus} & \text{representation} & \text{multiplicity} \\ \hline \hline
z_0 = d_1 = 0 & (\mathbf{10},\mathbf{1},\mathbf{2},\mathbf{1})_0^* & (3 \Kbi - \Ss - \Sn) \Z \\ \hline
z_0 = d_9 = 0 & (\mathbf{10},\mathbf{1},\mathbf{1},\mathbf{2})_0^* & (\Sn - 2 \Z) \Z \\ \hline
z_0 = d_5 = 0 & (\mathbf{16},\mathbf{1},\mathbf{1},\mathbf{1})_{1/4} & (2 \Kbi - \Ss - \Z) \Z \\ \hline
z_0 = d_7 = 0 & \text{SCP} & \Ss \Z \\ \hline
 & (\mathbf{1},\mathbf{8},\mathbf{1},\mathbf{1})_0 & 1 - \tfrac{1}{2} (\Kbi - \Ss) \Ss \\ \hline
 & (\mathbf{1},\mathbf{1},\mathbf{3},\mathbf{1})_0 & 1 + \tfrac{1}{2} (2 \Kbi - \Ss - \Sn) (3 \Kbi - \Ss - \Sn) \\ \hline
 & (\mathbf{1},\mathbf{1},\mathbf{1},\mathbf{3})_0 & 1 - \tfrac{1}{2} (\Kbi - \Sn + 2 \Z) (\Sn - 2 \Z) \\ \hline
 & (\mathbf{1},\mathbf{1},\mathbf{2},\mathbf{1})_{1/2} & (3 \Kbi - \Ss - \Sn) (2 \Kbi - \Ss - \Z) \\ \hline
 & (\mathbf{1},\mathbf{3},\mathbf{1},\mathbf{1})_{-1/3} & (3 \Kbi - \Ss - 2 \Z) \Ss \\ \hline
 & (\mathbf{1},\mathbf{1},\mathbf{1},\mathbf{2})_{1/2} & (2 \Kbi - \Ss - \Z) (\Sn - 2 \Z) \\ \hline
 & (\mathbf{1},\mathbf{3},\mathbf{2},\mathbf{1})_{1/6} & (3 \Kbi - \Ss - \Sn) \Ss \\ \hline
 & (\mathbf{1},\mathbf{1},\mathbf{2},\mathbf{2})_0 & (3 \Kbi - \Ss - \Sn) (\Sn - 2 \Z) \\ \hline
 & (\mathbf{1},\mathbf{3},\mathbf{1},\mathbf{2})_{1/6} & (\Sn - 2 \Z) \Ss \\ \hline
 & (\mathbf{1},\mathbf{1},\mathbf{1},\mathbf{1})_{0}  & 22 + 11 (\Kbi)^2 + 3 \Ss^2 + 2 \Ss \Sn + 2 \Sn^2 + \Ss \Z \\ &&  - 4 \Sn \Z  + 8 \Z^2 - 3 \Kbi (3 \Ss + 2 (\Sn + \Z)) \\ \hline
\end{array}
\end{equation}
\begin{align*}
\text{Euler number:}& \quad \chi = - 24 (\Kbi)^2 - 6 \Ss^2 -4 \Sn^2 - 16 \Z^2 + 8 \Sn \Z - 4 \Ss \Sn \\
& \hspace{1.15cm} + \Kbi (18 \Ss + 12 \Sn + 12 \Z)\\
\text{anomaly coefficients$^*$:}& \quad \hat{a} \sim K_{\hat{B}}^{-1} \,, \enspace \hat{b} \sim - \hat{\Z} \,, \enspace \hat{b}_{\text{SU(3)}} \sim - \hat{\mathcal{S}}_7 \,, \enspace \hat{b}_{\text{SU(2)}} \sim -(\hat{\mathcal{S}}_9 - 2 \hat{\Z}) \,, \\
& \quad \hat{b}_{\text{SU(2)}} \sim - (3 K_{\hat{B}}^{-1} - \hat{\mathcal{S}}_7 - \hat{\mathcal{S}}_9)\,, \enspace \hat{b}_{11} \sim - (\tfrac{1}{2} K_{\hat{B}}^{-1} - \tfrac{1}{6} \hat{\mathcal{S}}_7 - \tfrac{1}{4} \hat{\Z} )
\end{align*}
}

%%%%%%%%%%%%%%%%%%%%%%%%%%%%%%%%%%%%%%%%%%%%%%%%%%%%%%%%%%%%%%%%%%%%%%%%%
 \section{Phenomenologically viable models}
 \label{ScanF3Top4}
 
 In this appendix we summarize the matter spectra of the phenomenologically viable models discussed in Section~\ref{sec:Fpheno}. The criteria are that the 6d theory has a Lagrangian description and gauge group SO(10)$\times$U(1) implying the presence of a single tensor multiplet, i.e.\ the base is $\mathbb{F}_n$ with $n \leq 2$ and the fiber is embedded in $F_3$ with top 4. Moreover, for the Higgs sector we demand at least two uncharged $\mathbf{10}$-plets as well as at least two uncharged $\mathbf{16}$-plets for the breaking of the U(1)$_{B-L}$ symmetry.

Two scenarios are possible according to the analysis in Section~\ref{sec:6dmodel}. First, the particle spectrum contains three minimally charged $\mathbf{16}$-plets that lead to three generations of fermion zero modes in the background of a single flux quantum. We find eight possible realizations, whose base dependence and complete matter spectrum is given in \eqref{3chargedspectrum}. Second, the 6d theory includes a single charged $\mathbf{16}$-plet and the generations are due to three flux quanta. For this we find 25 possibilities given in \eqref{1chargedspectrum}.
\begin{equation}
\begin{array}{|@{} c@{} ||@{} c @{}|}
\hline
\text{coefficients} & \text{mutliplicities} \\ \hline \hline
\begin{array}{c || c | c | c | c | c | c}
n & a_{\Z} & b_{\Z} & a_7 & b_7 & a_9 & b_9 \\ \hline \hline
1 & 1 & 0 & 2 & 1 & 1 & 2 \\ \hline
0 & 1 & 1 & 2 & 3 & 1 & 2 \\ \hline
1 & 1 & 0 & 3 & 1 & 1 & 2 \\ \hline
0 & 1 & 1 & 3 & 3 & 1 & 2 \\ \hline
1 & 1 & 0 & 3 & 1 & 2 & 1 \\ \hline
1 & 1 & 0 & 3 & 2 & 2 & 1 \\ \hline
0 & 1 & 1 & 3 & 2 & 2 & 1 \\ \hline
0 & 1 & 1 & 3 & 3 & 2 & 1
\end{array}
&
\begin{array}{c | c | c | c | c | c | c | c }
\mathbf{16}_{-1} & \mathbf{16}_0 & \mathbf{10}_1 & \mathbf{10}_0 & \mathbf{1}_3 & \mathbf{1}_2
& \mathbf{1}_1 & \mathbf{1}_0 \\ \hline \hline
3 & 2 & 5 & 2 & 5 & 28 & 69 & 38 \\ \hline
3 & 3 & 5 & 3 & 3 & 20 & 59 & 32 \\ \hline
3 & 2 & 4 & 3 & 2 & 28 & 70 & 40 \\ \hline
3 & 3 & 4 & 4 & 1 & 22 & 55 & 36 \\ \hline
3 & 2 & 4 & 3 & 5 & 28 & 67 & 40 \\ \hline
3 & 2 & 3 & 4 & 3 & 30 & 63 & 44 \\ \hline
3 & 3 & 5 & 3 & 3 & 20 & 59 & 32 \\ \hline
3 & 3 & 4 & 4 & 1 & 22 & 55 & 36
\end{array}
\\ \hline
\end{array}
\label{3chargedspectrum}
\end{equation}
\begin{equation}
\begin{array}{|@{} c @{}||@{} c@{} |}
\hline
\text{coefficients} & \text{multiplicities} \\ \hline \hline
\begin{array}{c || c | c | c | c | c | c}
n & a_{\Z} & b_{\Z} & a_7 & b_7 & a_9 & b_9 \\ \hline \hline
0 & 0 & 1 & 2 & 2 & 1 & 0 \\ \hline
0 & 0 & 1 & 2 & 2 & 1 & 1 \\ \hline
0 & 0 & 1 & 2 & 3 & 1 & 1 \\ \hline
0 & 0 & 1 & 2 & 3 & 1 & 2 \\ \hline
0 & 0 & 1 & 3 & 2 & 1 & 0 \\ \hline
0 & 0 & 1 & 3 & 2 & 1 & 1 \\ \hline
0 & 0 & 1 & 3 & 3 & 1 & 1 \\ \hline
0 & 0 & 1 & 3 & 3 & 1 & 2 \\ \hline
1 & 1 & 0 & 2 & 1 & 0 & 1 \\ \hline
0 & 1 & 0 & 2 & 2 & 0 & 1 \\ \hline
1 & 1 & 0 & 2 & 2 & 0 & 1 \\ \hline
0 & 1 & 0 & 2 & 3 & 0 & 1 \\ \hline
0 & 1 & 1 & 2 & 2 & 0 & 1 \\ \hline
0 & 1 & 1 & 2 & 3 & 0 & 1 \\ \hline
1 & 1 & 0 & 2 & 1 & 1 & 0 \\ \hline
0 & 1 & 0 & 2 & 2 & 1 & 1 \\ \hline
0 & 1 & 0 & 2 & 3 & 1 & 1 \\ \hline
0 & 1 & 1 & 2 & 2 & 1 & 0 \\ \hline
1 & 1 & 0 & 3 & 0 & 1 & 0 \\ \hline
1 & 1 & 0 & 3 & 1 & 1 & 0 \\ \hline
0 & 1 & 0 & 3 & 2 & 1 & 1 \\ \hline
0 & 1 & 0 & 3 & 3 & 1 & 1 \\ \hline
0 & 1 & 1 & 3 & 2 & 1 & 0 \\ \hline
0 & 1 & 0 & 3 & 2 & 2 & 1 \\ \hline
0 & 1 & 0 & 3 & 3 & 2 & 1
\end{array}
&
\begin{array}{c | c | c | c | c | c | c | c }
\mathbf{16}_{-1} & \mathbf{16}_0 & \mathbf{10}_1 & \mathbf{10}_0 & \mathbf{1}_3 & \mathbf{1}_2
& \mathbf{1}_1 & \mathbf{1}_0 \\ \hline \hline
1 & 3 & 4 & 2 & 0 & 18 & 88 & 60 \\ \hline
1 & 3 & 4 & 2 & 2 & 26 & 86 & 52 \\ \hline
1 & 3 & 4 & 2 & 1 & 22 & 87 & 56 \\ \hline
1 & 3 & 4 & 2 & 3 & 30 & 85 & 48 \\ \hline
1 & 3 & 3 & 3 & 0 & 14 & 90 & 62 \\ \hline
1 & 3 & 3 & 3 & 1 & 24 & 89 & 52 \\ \hline
1 & 3 & 3 & 3 & 0 & 22 & 82 & 62 \\ \hline
1 & 3 & 3 & 3 & 1 & 32 & 81 & 52 \\ \hline
1 & 4 & 5 & 2 & 0 & 16 & 74 & 50 \\ \hline
1 & 3 & 4 & 2 & 0 & 18 & 88 & 60 \\ \hline
1 & 4 & 4 & 3 & 0 & 12 & 76 & 52 \\ \hline
1 & 3 & 3 & 3 & 0 & 14 & 90 & 62 \\ \hline
1 & 5 & 6 & 2 & 0 & 14 & 60 & 40 \\ \hline
1 & 5 & 5 & 3 & 0 & 10 & 62 & 42 \\ \hline
1 & 4 & 5 & 2 & 1 & 20 & 73 & 46 \\ \hline
1 & 3 & 4 & 2 & 2 & 26 & 86 & 52 \\ \hline
1 & 3 & 3 & 3 & 1 & 24 & 89 & 52 \\ \hline
1 & 5 & 6 & 2 & 0 & 14 & 60 & 40 \\ \hline
1 & 4 & 5 & 2 & 1 & 16 & 77 & 46 \\ \hline
1 & 4 & 4 & 3 & 0 & 16 & 72 & 52 \\ \hline
1 & 3 & 4 & 2 & 1 & 22 & 87 & 56 \\ \hline
1 & 3 & 3 & 3 & 0 & 22 & 82 & 62 \\ \hline
1 & 5 & 5 & 3 & 0 & 10 & 62 & 42 \\ \hline
1 & 3 & 4 & 2 & 3 & 30 & 85 & 48 \\ \hline
1 & 3 & 3 & 3 & 1 & 32 & 81 & 52
\end{array}
\\ \hline
\end{array}
\label{1chargedspectrum}
\end{equation}

\end{appendix}
\clearpage

% #################################
% #           Bibliography        #
% #################################

%\bibliographystyle{JHEP}
%\bibliography{biblio}

\providecommand{\href}[2]{#2}\begingroup\raggedright\endgroup

\end{document}